\pdfoutput=1

\documentclass[useAMS,usenatbib]{mn2e}
\usepackage{times}
\usepackage{graphicx}
\usepackage{amssymb}
\usepackage{longtable}

\title[Bulge mass is king]{Bulge mass is king: The dominant role of the bulge in determining the fraction of passive galaxies in the {\it Sloan Digital Sky Survey}}
\author[Asa F. L. Bluck et al.]{Asa F. L. Bluck$^{1,*}$, J. Trevor Mendel$^{2}$, Sara L. Ellison$^{1}$, Jorge Moreno$^{1,3}$, Luc Simard$^{4}$, 
\newauthor David R. Patton$^{5}$, and Else Starkenburg$^{1,3,6}$  \footnotemark[0]\\
\\$^1$ University of Victoria, Department of Physics and Astronomy, Victoria, British Columbia, V8P 1A1, Canada
\\$^2$ Max-Planck-Institut fur Extraterrestrische Physik, Giessenbachstraße, D-85748 Garching, Germany 
\\$^3$ CITA National Fellow
\\$^4$ National Research Council of Canada, Herzberg Institute of Astrophysics, 5071 West Saanich Road, Victoria, British Columbia, V9E 2E7, Canada
\\$^5$ Department of Physics \& Astronomy, Trent University, 1600 West Bank Drive, Peterborough, Ontario, K9J 7B8, Canada 
\\$^6$ CIfAR Global Scholar
\\$*$ Email: abluck@uvic.ca }

\begin{document}

\maketitle

\begin{abstract}

We investigate the origin of galaxy bimodality by quantifying the relative role of intrinsic and environmental drivers to the cessation (or `quenching') of star formation in over half a million local Sloan Digital Sky Survey (SDSS) galaxies. Our sample contains a wide variety of galaxies at $z$=0.02-0.2, with stellar masses of 8$< \rm{log(M_{*}/M_{\odot}})<$12, spanning the entire morphological range from pure disks to spheroids, and over four orders of magnitude in local galaxy density and halo mass. We utilise published star formation rates and add to this recent GIM2D photometric and stellar mass bulge + disk decompositions from our group. We find that the passive fraction of galaxies increases steeply with stellar mass, halo mass, and bulge mass, with a less steep dependence on local galaxy density and bulge-to-total stellar mass ratio (B/T). At fixed internal properties, we find that central and satellite galaxies have different passive fraction relationships.
For centrals, we conclude that there is less variation in the passive fraction at a fixed bulge mass, than for any other variable, including total stellar mass, halo mass, and B/T. This implies that the quenching mechanism must be most tightly coupled to the bulge. We argue that radio-mode AGN feedback offers the most plausible explanation of the observed trends. 

\end{abstract}

\section{Introduction}

\begin{figure*}
\includegraphics[width=0.49\textwidth,height=0.49\textwidth]{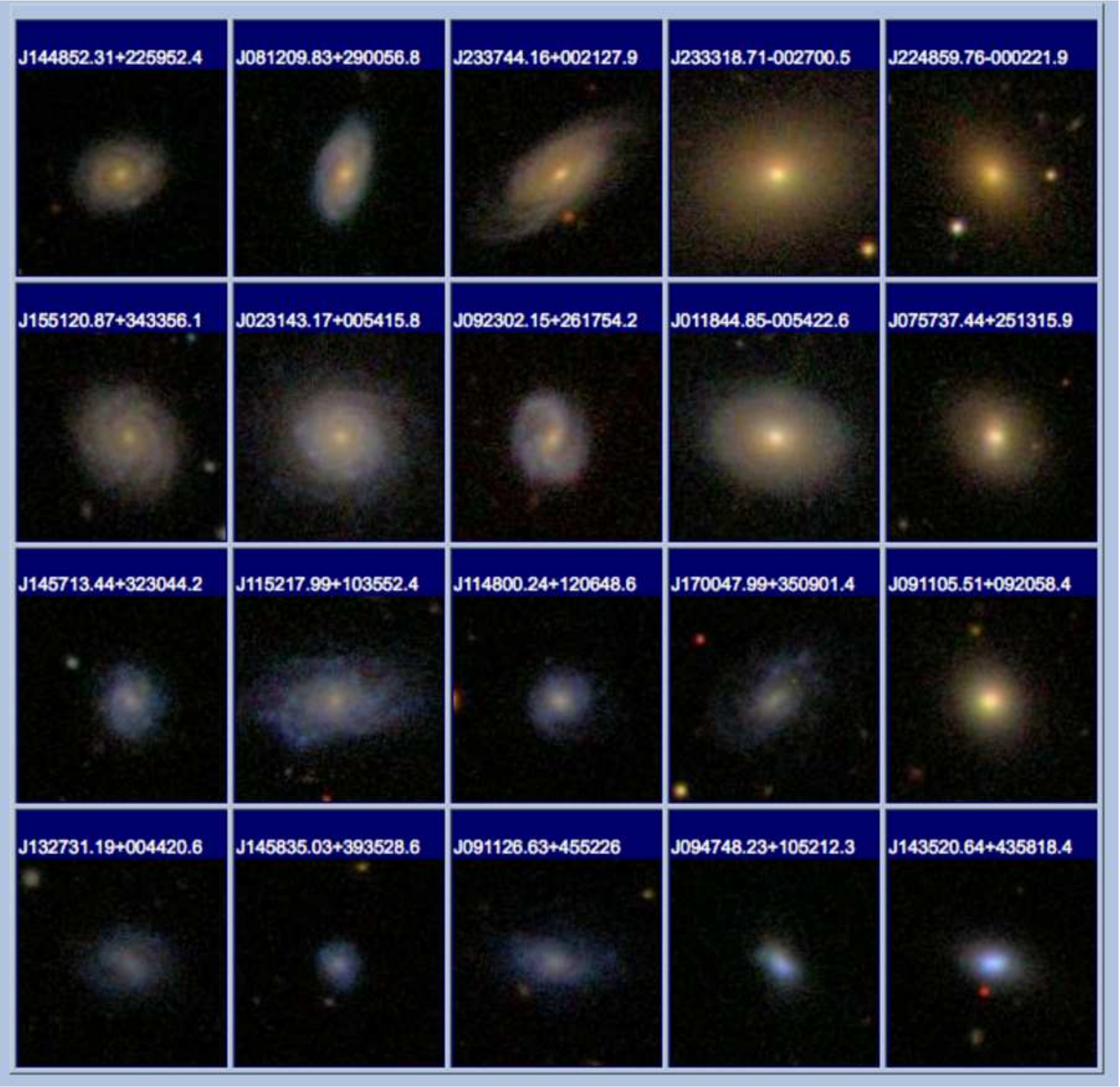}
\includegraphics[width=0.49\textwidth,height=0.49\textwidth]{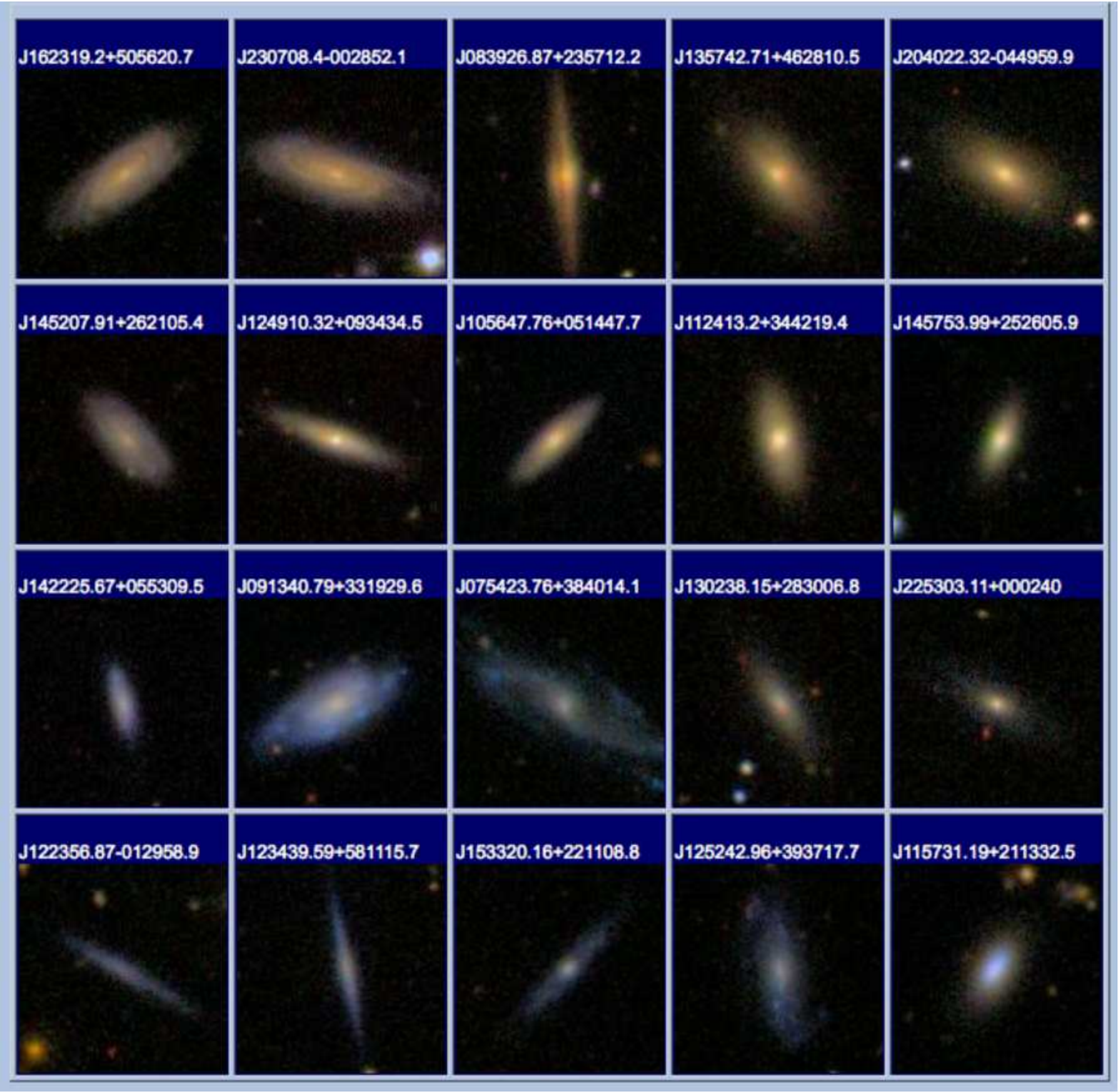}
\caption{A collage of face-on (inclination $<$ 30 degrees, {\it left panel}) and edge-on 
(inclination $>$ 60 degrees, {\it right panel}) galaxy images from our SDSS sample in ugriz wavebands, with z $<$ 0.1. In both panels, from {\it bottom to top} galaxies increase in stellar mass with log($M_{*}/M_{\odot}$) = 8 -- 9, 9 -- 10, 10 -- 11, 11 -- 12 in each row. From {\it left to right} galaxies increase in bulge-to-total stellar mass ratio, with B/T = 0 -- 0.2, 0.2 -- 0.4, 0.4 -- 0.6, 0.6 -- 0.8, 0.8 -- 1 in each column. Postage stamps are 40$''$ $\times$ 40$''$, which corresponds to $\sim$ 15 -- 70 kpc on each side across the redshift range of these images. Not all types of galaxy represented here are equally likely, e.g. high B/T high $M_{*}$ galaxies are much more common than low B/T high $M_{*}$ galaxies, conversely, high B/T low $M_{*}$ galaxies are much rarer than low B/T low $M_{*}$ galaxies (see Fig. 2 for the distributions). Note how galaxies appear redder in colour as we increase stellar mass at a fixed B/T ratio (moving up a column), and as we increase B/T ratio at a fixed stellar mass 
(moving along a row).}
\end{figure*}

One of the most important unresolved questions in modern cosmology is how galaxies form and evolve over cosmic time (see e.g. Conselice 2013 for a review). The existence of two principal types of galaxies may be inferred from the bimodality of several fundamental galaxy properties, most notably for colour - magnitude (or stellar mass), colour - concentration, and colour - morphology (Strateva et al. 2001, Kauffmann et al. 2003, Brinchmann et al. 2004, Baldry et al. 2004, Baldry et al. 2006, Driver et al. 2006, Bamford et al. 2009). The galaxy population is divided into an actively star forming `blue cloud' where galaxies exhibit a tight correlation between stellar mass and star formation rate, SFR, known as the star forming main sequence (e.g. Brinchmann et al. 2004), and a principally passive `red sequence' where there is no positive SFR - stellar mass correlation. 
For the red sequence there is a relationship between colour and stellar mass (or luminosity), such that more massive (or luminous) galaxies become progressively redder (e.g. Strateva et al. 2001). Relatively few galaxies lie midway between the two peaks, and those that do are given the designation of `green valley', or transition, galaxies. Understanding the fundamental reason behind the galaxy bimodality, and particularly the transition between the blue cloud and red sequence, is the goal of much active research in contemporary extragalactic astrophysics.

A common attempt to explain the origin of galaxy bimodality appeals to the dual component nature of galaxies, i.e. that galaxies are in effect structurally bulge + disk systems (to first order). In general disks are found to be bluer in colour than bulges (e.g. Peletier \& Balcells 1996) which has led, for example, Driver et al. (2006) to argue for the necessity of routine bulge + disk decompositions of galaxy structural components when investigating the evolution of galaxy properties through star formation. 
The primary use of colour in the bimodal distributions is best understood as a proxy for global star formation rate per unit stellar mass, or sSFR, which indicates the relative importance of ongoing star formation to established stellar populations in the overall integrated light distribution of a galaxy. Thus, galaxies separate out in terms of whether or not they are experiencing substantial ongoing formation or not.

In most cases, lower stellar mass and lower S\'{e}rsic index galaxies tend to be bluer (and hence have higher sSFRs) than higher stellar mass and higher S\'{e}rsic index systems (Baldry et al. 2004, Driver et al. 2006, Baldry et al. 2006, Bamford et al. 2009). Similar relationships are found for luminosity and stellar light concentration (e.g. Strateva et al. 2001, Driver et al. 2006). These relationships are complicated by the impact of local environment on the process of star formation, where there is a pronounced effect for star formation to be suppressed preferentially in higher density environments (e.g. Butcher \& Oemler 1984, Balogh et al. 2004, Cooper et al. 2006, van den Bosch et al. 2008, Peng et al. 2010, Peng et al. 2012). 
Additionally there is a tendency for galaxies at higher densities to be more bulge dominated and hence have higher S\'{e}rsic index and concentration parameters (Dressler 1984, Moran et al. 2007, Tasca et al. 2009). Disentangling the various effects on current formation efficiency is essential in coming to understand how and why galaxies become the way they are today.

Many proposed mechanisms for inducing the observed cessation in star formation and subsequent build up of the red sequence have been suggested. In particular, Peng et al. (2010, 2012) argue for the separate influence of galaxy stellar mass and local environment on the formation of passive galaxies. This implies that there must be at least two distinct routes by which galaxies can reduce their SFRs and become passive, one coupled to the local environment and one to internal properties. 
Further, Peng et al. (2010, 2012) construct a phenomenological model, which fits the observed data extremely well; however, this does not necessarily reveal the base physical process(es) which cause the transition of galaxies from the blue cloud (of actively star forming main sequence galaxies) to the red sequence (of passive galaxies). 

Theoretically, a number of physical mechanisms for quenching \footnote{The term `quenching' is commonly used in the literature to express the fact or process of star formation coming to an end in a galaxy (e.g. Peng et al. 2010, 2012, Mendel et al. 2013, Woo et al. 2013). In this sense it is interchangeable with `cessation'. However, it is not a universally accepted as a useful or desirable term. This is largely due to the implicit assumption that some mechanism is needed to {\it cause} the cessation in star formation, i.e. that galaxies would otherwise be actively forming stars. In this paper we shall use the term `quenching' from time to time to refer to those proposed mechanisms which do cause the shutting down of star formation, and try to avoid its usage in the more general sense.} have been suggested. These can be, loosely, divided into environmental and intrinsic (mass-correlating) processes. 
Environmentally, effects such as ram pressure stripping, galaxy tidal interaction and harassment, stifling of the gas supply routes to galaxies from hot gas in massive halos, and preprocessing of gas into stars from extensive merging in groups prior to cluster formation can all lead to the shutting down of star formation in galaxies (Balogh et al. 2004, Boselli et al. 2006, Cooper et al. 2006, Cortese et al. 2006, Moran et al. 2007, Font et al. 2008, van den Bosch et al. 2008, Berrier et al. 2009, Tasca et al. 2009, Peng et al. 2010, 2012, Hirschmann et al. 2013, Wetzel et al. 2013). 

Proposed intrinsic drivers of quenching in models include quasar-mode AGN feedback (Hopkins et al. 2006a,b, Hopkins et al. 2008), and radio-mode AGN feedback (Croton et al. 2006, Bower et al. 2008, Guo et al. 2011). Quasar-mode AGN feedback may be thought of as the blowing out of cold gas from the centre of galaxies in cataclysmic events such as during enhanced periods of nuclear activity throughout the major merging of gas rich galaxies (see Nulsen et al. 2005, Dunn et al. 2010, Feruglio et al. 2010, Fabian 2012, and Cicone et al. 2014 for observational evidence). Alternatively, radio-mode AGN feedback involves the gentle heating of gas in the halos of massive galaxies from jets over long timescales (up to several Gyrs) resulting in disruption of supply routes for replenishment of cold gas needed as fuel for star formation (e.g. McNamara et al. 2000, McNamara et al. 2007, Fabian et al. 2012). 

Other intrinsic mechanisms for quenching may arise from the virial shock heating of infalling accreted cold gas onto galaxies in halos above some critical dynamical mass (often cited as around $M_{\rm crit} \sim 10^{12-13} M_{\odot}$) leading to the shutting off of star formation in those galaxies (Dekel \& Birnboim 2006 and Dekel et al. 2009). Furthermore, the potential of the bulge structural component to stabilise giant molecular cloud collapse (which are the sites of star formation) through tidal torques has been argued for in Martig et al. (2009). Finally, major and minor galaxy mergers and interactions can give rise to enhanced star formation rates, and hence elevated gas depletion (e.g. Ellison et al. 2008, Darg et al. 2010, Patton et al. 2013, Hung et al. 2013), as well as increased supernova and stellar feedback which may also accelerate the transition of galaxies from the blue cloud to the red sequence. 
Galaxy mergers and interactions are also found to enhance nuclear activity (e.g. Ellison et al. 2011, Silverman et al. 2011, Ellison et al. 2013) and thus couple to increased AGN feedback as well. 

There is currently some debate as to whether residing in a galaxy pair will increase (as argued for in Robotham et al. 2013) or decrease (as argued for in Wild et al. 2009) the star formation of galaxies. Most probably the effect is varied depending on the timescale and stage of the interaction. Widely separated (non-interacting) pairs will merely trace local environment (e.g. Patton et al. 2013) so one might expect that since being in a pair is more probable at higher local densities (up to a point), residing in a pair might lead to lower star formation rates due to environmental effects. Closer (interacting) pairs are found to provoke enhanced star formation (Ellison et al. 2008,2010, Scudder et al. 2012, Patton et al. 2013), and the final coalescence of galaxies at the end of the merger sequence leads initially to the highest enhancements in star formation (e.g. Ellison et al. 2013). Thereafter, most probably, depletion in gas supplies will result in reduced star formation later in the post-merger 
phase (as in Wild et al. 2009). However, in this scenario, in order for a galaxy to remain passive some additional process(es) will be required to prevent cold gas accretion re-invigorating the galaxy, such as AGN feedback (e.g. Bower et al. 2006, 2008) or halo-mass-quenching (e.g. Dekel \& Birnboim 2006).

Most of the intrinsically-driven routes for galaxies to become passive are expected to couple to the bulge component primarily. For example, AGN and the central black holes have well documented scaling relations with the bulge properties (mass, velocity dispersion, luminosity), e.g. Magorrian et al. (1998), Ferrarese \& Merritt (2000), Gebhardt et al. (2000), Haring \& Rix (2004). At a fixed stellar mass, galaxies with higher bulge-to-total stellar mass ratios (B/T) tend to reside in higher mass halos, and hence will be affected more by shock heating of accreted gas from cold streams into their halos (Dekel et al. 2009, Woo et al. 2013). 
Also, bulge mass and B/T will couple with the capacity of a bulge to stabilize the collapse of giant molecular clouds in surrounding disks, leading to a relation between the bulge and the passive fraction (Martig et al. 2009). Finally, galaxy merging tends to result in growing the bulge component (e.g. Toomre \& Toomre 1972) and thus gas depletion may also correlate with the extent of the bulge structure.

Since most of the theoretically favoured mechanisms to achieve the cessation of star formation are predicted to correlate more strongly with the bulge structure than that of the whole galaxy, it is desirable to look for these predicted relationships explicitly in observations. Indirect observational evidence of the importance of the bulge has been noted for several years now. Kauffmann et al. (2003) and Brinchmann et al. (2004) both find indirect evidence for the importance of the bulge in star formation cessation, noting strong correlations between SFR and stellar mass density and concentration of light, which are effectively proxies for `bulgeness' of galaxy. Drory \& Fisher (2007) demonstrate that the type of bulge (as well as the bulge-to-total light ratio) is important in forming galaxy bimodality in colour, with classical pressure supported bulges being required to make a galaxy red, and pseudo-bulges being insufficient. Bell et al. (2008) find that high S\'{e}rsic index bulges are always present 
in truly passive systems, further implicating the role of the classical bulge in star formation cessation.

Two recent papers, Cheung et al. (2012) and Fang et al. (2013), also look for the effect of indirect indicators of the bulge (central stellar mass surface density within 1 kpc) coupling to the star forming properties of the whole galaxy. They find that at a fixed stellar mass there is a differential impact on the red (passive) fraction due to central stellar mass surface density, which suggests that mass-quenching (as in Peng et al. 2010, 2012) is a process perhaps more directly linked with the bulge. Woo et al. (2013), however, have argued for stellar-mass-quenching being better understood as halo-mass-quenching for central galaxies, with virialised shock heating of infalling gas in massive halos being the root cause of star formation cessation (e.g. Dekel \& Birnboim 2006).

In this paper we investigate the role of the bulge in constraining star formation directly, utilising the largest photometric and stellar mass bulge + disk decomposition catalogue in existence (Simard et al. 2011, Mendel et al. 2014). In Fig. 1 we show an illustration of SDSS galaxies separated into bins of stellar mass and bulge-to-total stellar mass ratio (B/T) from our decomposition, for both face-on and edge-on systems. We examine the role of bulge-to-total stellar mass ratio (B/T), bulge mass, and disk mass in addition to halo mass, total stellar mass, and local density which have previously been studied (e.g. Peng et al. 2010,2012, Woo et al. 2013). We find that bulge mass correlates most tightly with the passive fraction for central galaxies.

This paper is structured as follows. Section 2 outlines the sources of our data, including details on the derivation of star formation rates, stellar masses, halo masses, and bulge + disk decompositions. Section 3 presents our methods and techniques, particularly in reference to our metric for defining passive, $\Delta$SFR. In section 4 we present our results, including the dominant role of the bulge in constraining the passive fraction. Section 5 provides a discussion of our results in terms of the context of the substantial body of literature on this subject, and we go on to present a feasibility argument for the explanation of our observed trends through radio-mode AGN feedback. We conclude in Section 6. An appendix is also included where we discuss the various tests and checks we have performed on the data to assess the reliability of our bulge + disk decompositions. 

Throughout this paper we assume a (concordance) $\Lambda$CDM Cosmology with: H$_{0}$ = 70 km s$^{-1}$ Mpc$^{-1}$, $\Omega_{m}$ = 0.3, $\Omega_{\Lambda}$ = 0.7, and adopt AB magnitude units.

\section{Data}

In this study we utilise photometry and spectroscopy from the Sloan Digital Sky Survey (SDSS) data release seven (DR7), see Abazajian et al. (2009) for full details on this release. For the original data release of the SDSS please see Abazajian et al. (2003) which contains full information on the survey design. The SDSS DR7 gives ugriz band photometry and 3800 - 9200 \AA \hspace{0.1cm} spectroscopic coverage. We restrict our sample to those galaxies with spectroscopic redshifts of z = 0.02 - 0.2 and with stellar masses in the range 8 ${\rm < log(M_{*}/M_{\odot}) <}$ 12, leaving us with a total of 538,046 galaxies in our full parent sample. 
This sample is the largest collection of galaxies at (spectroscopically confirmed) low redshifts in existence, and provides an excellent route to study the differential effects of stellar mass, halo mass, morphology, and environment on the passive galaxy fraction. Since many of the derived quantities used in this study are taken from previous published work, we will give only an outline of the key features of the data here. Readers who are not interested in the full details of the sample may like to skip ahead to the method in \S 3 or the results in \S 4.

\subsection{Star Formation Rates}

The star formation rates used in this paper are taken from Brinchmann et al. (2004), with updates following these same methods to the later data releases\footnote{www.mpa-garching.mpg.de/SDSS/DR7}. In this paper, we use two distinct kinds of star formation rates, both of which use the spectra and photometric imaging available for each galaxy. The first type require galaxies to have emission line fluxes in at least some of the BPT (Baldwin, Phillips \& Terlevich 1981) lines, [NII], H$\alpha$, H$\beta$, and [OIII]. 
For those galaxies with a S/N $>$ 3 in each of these lines ($\sim$ 40 \%), the first step is to classify where they lie on the BPT diagram in terms of the Kauffmann et al. (2003) and Kewley et al. (2001) demarcations, which separate out emission line galaxies into those that are predominantly star forming, composite, or AGN dominated. A reduced set of emission lines is then used for the remainder as is explained in detail in Brinchmann et al. (2004) \S 3. For the BPT selected star forming galaxies, SFRs are determined via fitting of their emission lines with Charlot \& Longhetti (2001) models. 

The remainder of the sample, i.e. those which do not have BPT emission lines above the S/N thresholds and those which do but are classified as AGN or composite, have SFRs determined via an empirical relation between the 4000\AA \hspace{0.05cm} break strength (D$_{n}$4000) and the sSFR of the galaxy (Balogh et al. 1999). This allows for passive galaxies and AGN to be included in our SFR analysis. However, to demonstrate the robustness of our results, we consider the effect of removing AGN from our analysis in \S App C. None of the conclusions of this paper change due to the choice of including or excluding AGN from the sample.

Both of the methods used for assigning SFRs to galaxies discussed above are derived from the spectra of these galaxies, thus they are restricted to the fibre aperture. At the median redshift and stellar mass of the sample (z = 0.1, log($M_{*}/M_{\odot}$) = 10.8), this corresponds to $\sim$ 1/3 of the light being contained within the spectroscopic fibre. Therefore, in order to calculate the global (or total) SFR for galaxies in the SDSS an aperture correction must be made. This is discussed in detail in Brinchmann et al. (2004) \S 5. The fibre corrections we use are constructed by fitting stochastic models to the photometry outside of the fibre, which gives an improved recovery of SFRs (e.g. Salim et al. 2007) \footnote{www.mpa-garching.mpg.de/SDSS/DR7/sfrs.html}. 

We test that the fibre correction does not unduly bias any of our results by recomputing the passive fraction as purely a function of global photometry (colour), in effect making a mapping from passive fraction to red fraction (see \S App C). This changes none of the main features of the subsequent plots, and all of our conclusions remain invariant. Thus, we conclude that the fibre corrections are generally successful for our sample, introducing no systematic effects which might affect our conclusions in any significant way. This is an important point to establish because a main result of this work is the tight correlation between bulge mass (and hence the central region) and the passive fraction, requiring that a successful transition from the fibre to the global is achieved.

\begin{figure*}
\rotatebox{270}{\includegraphics[height=0.49\textwidth]{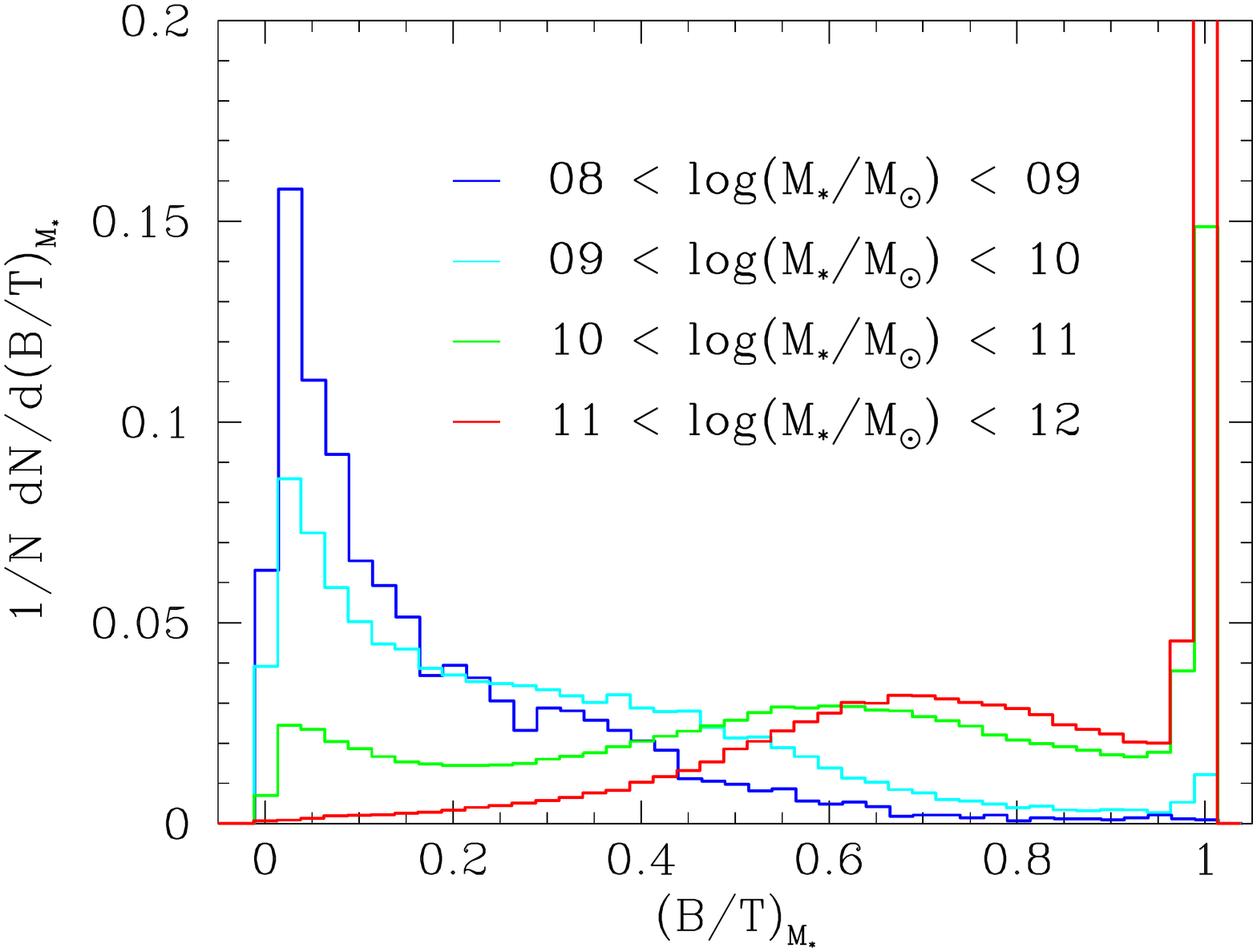}}
\rotatebox{270}{\includegraphics[height=0.49\textwidth]{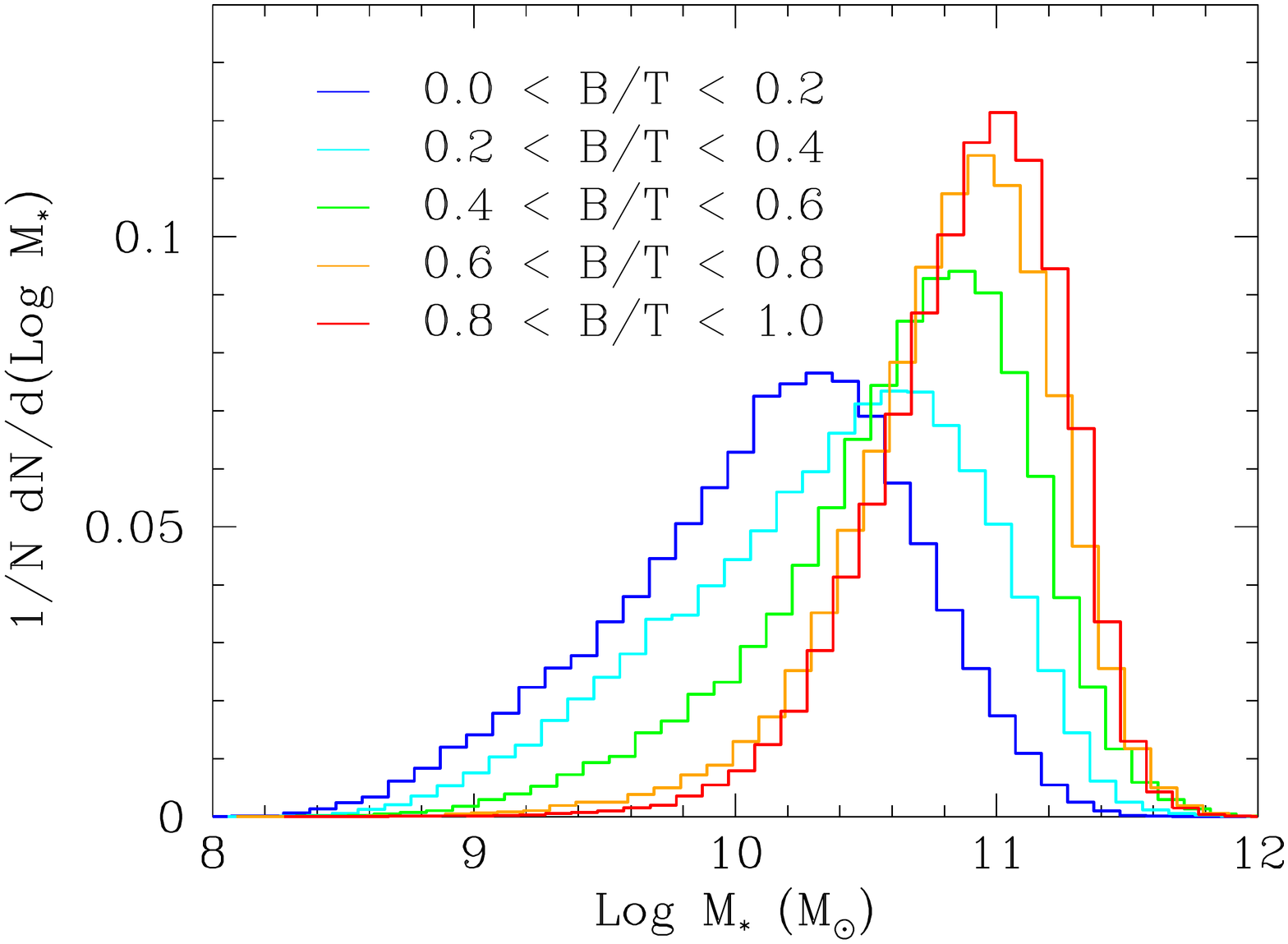}}
\caption{{\it Left}: Normalised distribution of bulge-to-total stellar mass ratio (B/T), split by stellar mass range as indicated. {\it Right}: Normalised distribution of stellar mass split by B/T, as indicated. Higher stellar mass galaxies tend to have higher B/T ratios compared to lower stellar mass galaxies. The considerable variation in these parameters with each other highlights the necessity for carefully binning by stellar mass and morphology when considering the passive fraction evolution, as done in the latter plots of this paper.}
\end{figure*}

\subsection{Stellar Masses of Bulge and Disk Components}

The stellar masses used in this paper for bulge, disk, and total all come from Mendel et al. (2014), and are based on updated SDSS photometry and bulge + disk photometric decompositions in Simard et al. (2011). Full details of the GIM2D photometric decomposition of galaxy light profiles in the ugriz bands into bulge + disk systems is provided in Simard et al. (2011). The basic procedure is to fit galaxy light profiles in the ugriz bands as a dual S\'{e}rsic profile, with $n_{s}$ = 4 for the bulge component and $n_{s}$ = 1 for the disk component, where the total intensity of light at radius r, $I_{r}$, is given by (S\'{e}rsic 1963, Simard et al. 2002):

\begin{equation}
I_{r} = I_{e} \hspace{0.1cm} {\rm exp} \big( -b_{n_{s}} \big[ \big( \frac{r}{r_{e}} \big)^{1/n_{s}} - 1 \big] \big)
\end{equation}

\noindent where $I_{e}$ is the intensity of light at the effective radii, $r_{e}$ is the (projected) effective radius from the best fit, and $b_{n_{s}}$ is a constant of the fit. The best dual S\'{e}rsic fit is computed by minimizing the residuals between the model light distributions and the observations, in each waveband. A bulge-to-total {\it light} ratio is output by the GIM2D bulge + disk fitting code (Simard et al. 2002) for each waveband. Taken together, the colour and magnitude of each component (bulge and disk) is obtained. 

A single (free) S\'{e}rsic profile is also fit to each galaxy. The relative success of the single and the two-component GIM2D fit is assessed with the probability that a given galaxy is best fit by a pure single S\'{e}rsic profile, $P_{pS}$, thus prescribing a method to ascertain when a single fit is apparently sufficient and, conversely, when a two-component fit is statistically favoured. The $P_{pS}$ is constructed using the F-statistic, which provides a quantitative means for calculating the relative success of each fit from comparison of the $\chi^{2}$ residuals (accounting for a dual component fit being expected to do better), see Simard et al. (2011) \S 4.2 for a full description. 
In principle this allows us to select galaxies where two component fits are statistically preferred. We also consider the use of $n_{s}$ = 4 classical bulge fitting in \S App. D by comparison to a free S\'{e}rsic index bulge model. We conclude that there would be no significant differences to any of our results if we fit with the free model instead of the classical bulge.

In Mendel et al. (2014) the decomposed bulge and disk photometry from the GIM2D fitting in Simard et al. (2011) is converted to stellar masses, based on the ugriz spectral energy distributions (SEDs) for each component in isolation, and for the total light of the galaxy. A large grid of synthetic photometry is utilised using the flexible stellar population synthesis (FSPS) code of Conroy, Gunn \& White (2009). 
These incorporate a range of star formation histories, dust content, age, and metalicity. A Chabrier (2003) stellar initial mass function (IMF) is employed. A wide variety of smoothly declining star formation histories are constructed, with $\psi_{*}(t) \propto \tau^{-1}{\rm exp}(-t/\tau)$, where $\tau$ is the e-folding time. The estimated mass for each component is taken as the median of the probability density function, with errors quoted as the 16th and 84th percentiles of this distribution. Typical errors on the components are $\sim$ 0.2 dex from the model parameter fitting. 
The total error, summing over the various possible systematics, such as choice of FSPS code, choice of IMF, uncertainties in stellar evolution, and choice of dust extinction law, may lead to a slightly larger value. For a detailed consideration of all of these issues, and for a full description of the techniques used in determining the component masses see Mendel et al. (2014), in particular \S 3 and \S 4. The photometric and stellar mass bulge + disk decompositions are publicly available (see Mendel et al. 2014). 

An illustration of how galaxies from the SDSS look in bins of derived B/T and $M_{*}$ is shown in Fig. 1, for both face-on and edge-on galaxies. We generally recover a very good relation between visual morphology and structural B/T determination. We show the distribution of B/T ratio in bins of stellar mass, and the stellar mass distribution in bins of B/T in Fig. 2. It is particularly interesting to note that there is considerable interdependence between these variables, such that higher stellar mass galaxies tend to have higher B/T ratios, and lower stellar mass galaxies tend to have lower B/T ratios (as seen in, e.g., Bernardi et al. 2010). This strongly motivates the need to carefully disentangle these two potential drivers to quenching in the following sections.  

We present the various tests and calibrations we have performed on the bulge + disk decompositions in the appendix (particularly \S App A \& B), including via fitting model galaxies. One important result of this work is that a correction for `false' disks must be made in order to correct for a tendency of high S\'{e}rsic index galaxies to erroneously be split into bulge + disk systems. This is completely avoided if a $P_{pS}$ cut is made (see Simard et al. 2011). However, since the intention of this paper is to be as inclusive as possible, we construct an alternative in \S App B to culling the sample. When the correction is applied we achieve excellent recovery of model galaxy structural components for both composite and single profile cases. We also find that if we restrict our sample to z $<$ 0.1 (instead of 0.2 used throughout the paper), and if we use a free S\'{e}rsic as opposed to classical $n_{s}$ = 4 bulge, all of our results and conclusions remain unchanged (see \S App. D), which strongly 
suggests that our results are not biased by any difficulties with fitting GIM2D bulge + disk models to low resolution data.

\subsection{Total Stellar Masses}

Due to the variety of techniques used to model the light profiles of galaxies in our sample, there are a number of possible ways to define the total stellar mass. The main one we will use in this work is simply:

\begin{equation}
M_{*} = M_{\rm bulge} + M_{\rm disk}
\end{equation}

\noindent i.e. the total mass is the sum of the two component masses fit separately. Other definitions include $M_{B+D}$, which is the stellar mass of the bulge and disk photometric components added together and fit as a single entity with the stellar mass codes outlined above. For most of the galaxies in our sample there is very good agreement between these mass estimates, however, there are a few cases where this is not true.  

Following the recommendation in \S 5 of Mendel et al. (2014) we will discard from our final sample any stellar masses where there is disagreement between $M_{*}$ and $M_{B+D}$ at a level above 1 $\sigma$ ($\sim$ 10 \% of galaxies). For these galaxies the main issue seems to be that the bulge component appears very red without having a high photometric B/T in any band. Thus, a qualitatively similar route to achieve this selection is to require that $(B/T)|_{M*} \sim (B/T)|_{i}$, or any other of the redder bands, for a suitably calibrated comparison. A plausible explanation for this discrepancy is that the error on the exact colour gradient becomes very high compared to the measured value of the colour itself, incorporating unphysically red bulges (which lead to very high bulge mass estimates) in some predominantly bulge-less systems. In the following analyses, these unphysical cases are removed. However, even if we include those cases our primary results and conclusions remain unchanged.

\subsection{Measures of Environment}

Throughout this paper we use two complementary measures of environment: First, we use the Yang et al. (2007, 2009) SDSS group catalogue to classify whether a given galaxy is central to, or a satellite of, its host dark matter halo. In practice, the assignment of the label central is based on whether a galaxy is the most massive (in terms of stellar component) in the group in which it resides. The groups are constructed from a friends-of-friends algorithm with a linking length constrained from N-body simulations, as is outlined in detail in Yang et al. (2007). Satellite galaxies are defined as any galaxy assigned to a group in which they are not the most massive member.

Second, we use a measure of local surface (over-)density of galaxies (as in, e.g., Dressler 1980, Baldry et al. 2006, Chuter et al. 2011). We define our measure of local surface over-density, $\delta_{n}$, thus:

\begin{equation}
\delta_{n} = \frac{\Sigma_{n}}{\langle \Sigma_{n}(z \pm \delta_{z}) \rangle} \hspace{0.2cm} ,
\end{equation}

\noindent where 

\begin{equation}
\Sigma_{n} = \frac{n}{\pi r^{2}_{p,n}}
\end{equation}

\noindent and $r_{p,n}$ is the projected physical (h=0.7) separation to the nth nearest neighbour, given in kpc.    
$\langle \Sigma_{n}(z \pm \delta_{z}) \rangle$ is the mean value of the local surface density in the redshift range (z $-$ $\delta z$) to (z + $\delta z$), where $\delta z$ = 0.01. This term normalizes our environmental measure and accounts for differential redshift effects due to variation in the luminosity of identifiable galaxies as a function of redshift in the SDSS, a consequence of the flux limit of the survey. Thus, we can compare the measures of over-density, $\delta_{n}$, across different redshifts, whereas, the direct density measure, $\Sigma_{n}$, is only meaningful at a given redshift slice. 
The value, $n$, is a free parameter, and in this work we consider various values. In a sense, $n$ selects the scale of interest, or the definition of `local' in the local environment measure. We investigate the use of 3rd, 5th, and 10th nearest neighbours and find generally very strong correlations between these environmental indicators. In the end we opt to use the 5th nearest neighbour in all subsequent analyses, however, none of our results are strongly affected by this choice.

In Fig. 3 we show the normalised distribution of the over-density parameter evaluated for the 5th nearest neighbour, $\delta_{5}$. We split this distribution into the the contribution from central and satellite galaxies. We find that satellites are skewed to higher local over-densities, as expected. This figure illustrates the inter-relation of our two measures of environment for the SDSS sample.  

\begin{figure}
\rotatebox{270}{\includegraphics[height=0.49\textwidth]{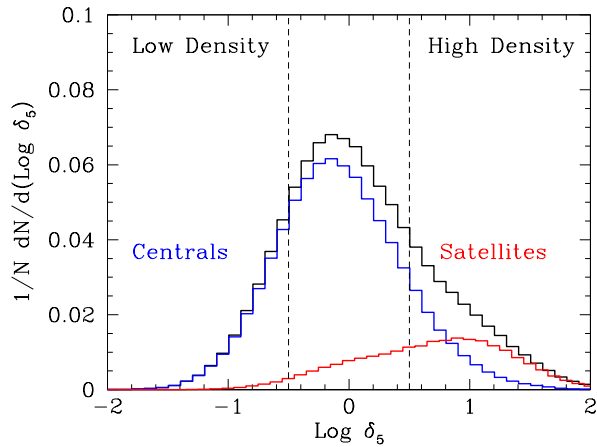}}
\caption{Normalised distribution of the logarithm of the over-density parameter evaluated for the 5th nearest neighbour, $\delta_{5}$ (defined in eqs. 3 \& 4). The distribution is sub-divided into the contribution from centrals and satellites as indicated. The full distribution (black line) has a log-Gaussian form with a power law tail to higher over-densities determined by gravitational clustering effects. Satellites have a skewed distribution, peaking at higher over-densities than for centrals. This indicates the inter-dependence of these two different measures of environment. The dashed lines indicate the regions of low density, intermediate density (middle), and high density, which are used later to separate out environment.}
\end{figure}

\subsection{Halo Masses}

\begin{figure}
\rotatebox{270}{\includegraphics[height=0.49\textwidth]{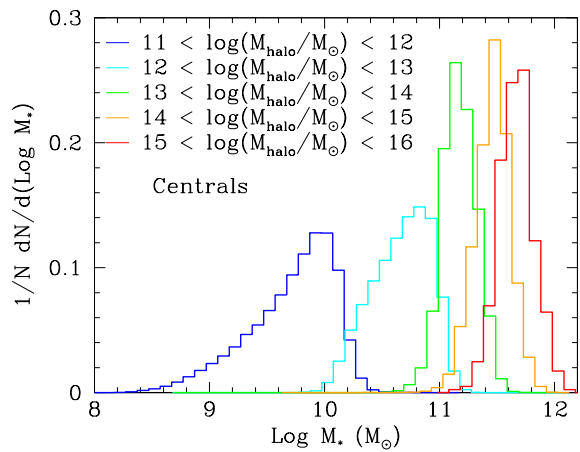}}
\caption{Normalised distribution of stellar masses, sub-divided by halo mass as indicated for central galaxies. Higher stellar mass central galaxies live in higher mass halos, however, there is a greater range in halo mass available at a given stellar mass at higher masses. Conversely, at low masses, there is a greater range in stellar mass at a fixed halo mass. This illustrates the inter-dependence of galaxy stellar and halo mass for centrals.}
\end{figure}

The halo masses we use in this paper come from the group catalogue of Yang et al. (2007, 2012). An abundance matching technique is used whereby groups are ordered in terms of their total group stellar mass and equated to halo masses by comparison to the halo mass function of Warren et al. (2006), with a transfer function from Eisenstein \& Hu (1998). It is also possible to rank order galaxies according to their luminosities, however, this is found to be less successful than with stellar mass. Testing of the group finding algorithm with mock galaxy catalogues revealed that over 90 \% of halos with $M_{\rm halo} > 10^{12} M_{\odot}$ are successfully identified. The stellar masses are also calibrated to observational methods from X-ray measurements of clusters and velocity dispersions of galaxies within groups and clusters. This constitutes the largest observationally constrained sample of group halo masses in existence today. 

In Fig. 4 we show the distribution of stellar masses as a function of group halo mass for centrals. At low masses there is a broad range of stellar masses associated with a given halo mass, yet at higher masses the converse is true. This is a direct consequence of the shape of the $M_{*} - M_{\rm halo}$ relation, which starts off very steep at low masses and become much shallower at high masses, beginning around $M^{*}$ (see e.g. Moster et al. 2010). The implication of this for our study is that, even for central galaxies, it is possible to probe a range of halo masses at a fixed stellar mass, although the range is mass dependent. However, there are configurations of halo and stellar mass which are not possible to probe for centrals. For satellites, a broader range of parent halo mass is permitted at any given stellar mass, up to the limit at which the satellite would become massive enough to be defined as the central of that halo.

\begin{figure*}
\rotatebox{270}{\includegraphics[height=0.33\textwidth]{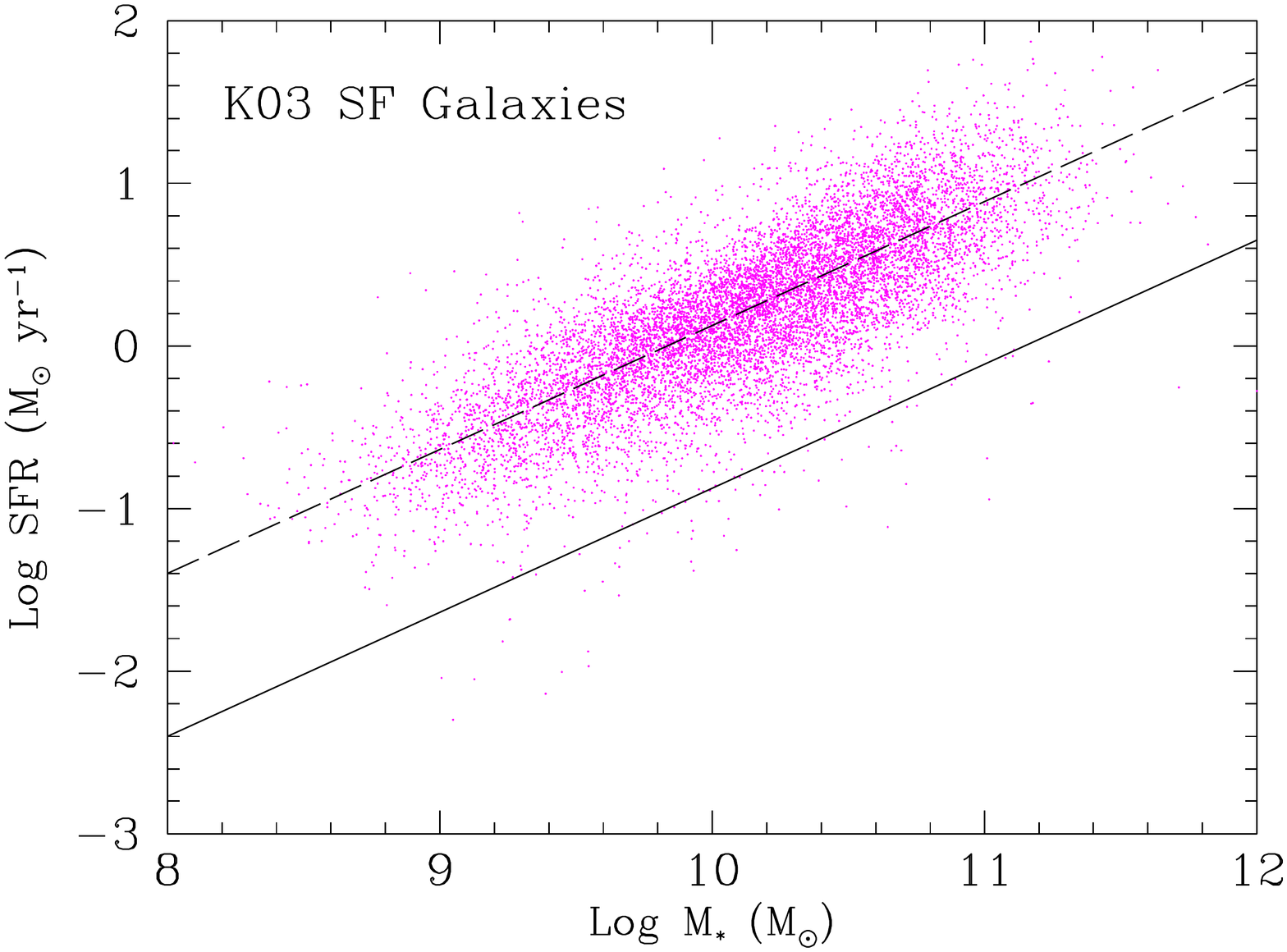}}
\rotatebox{270}{\includegraphics[height=0.33\textwidth]{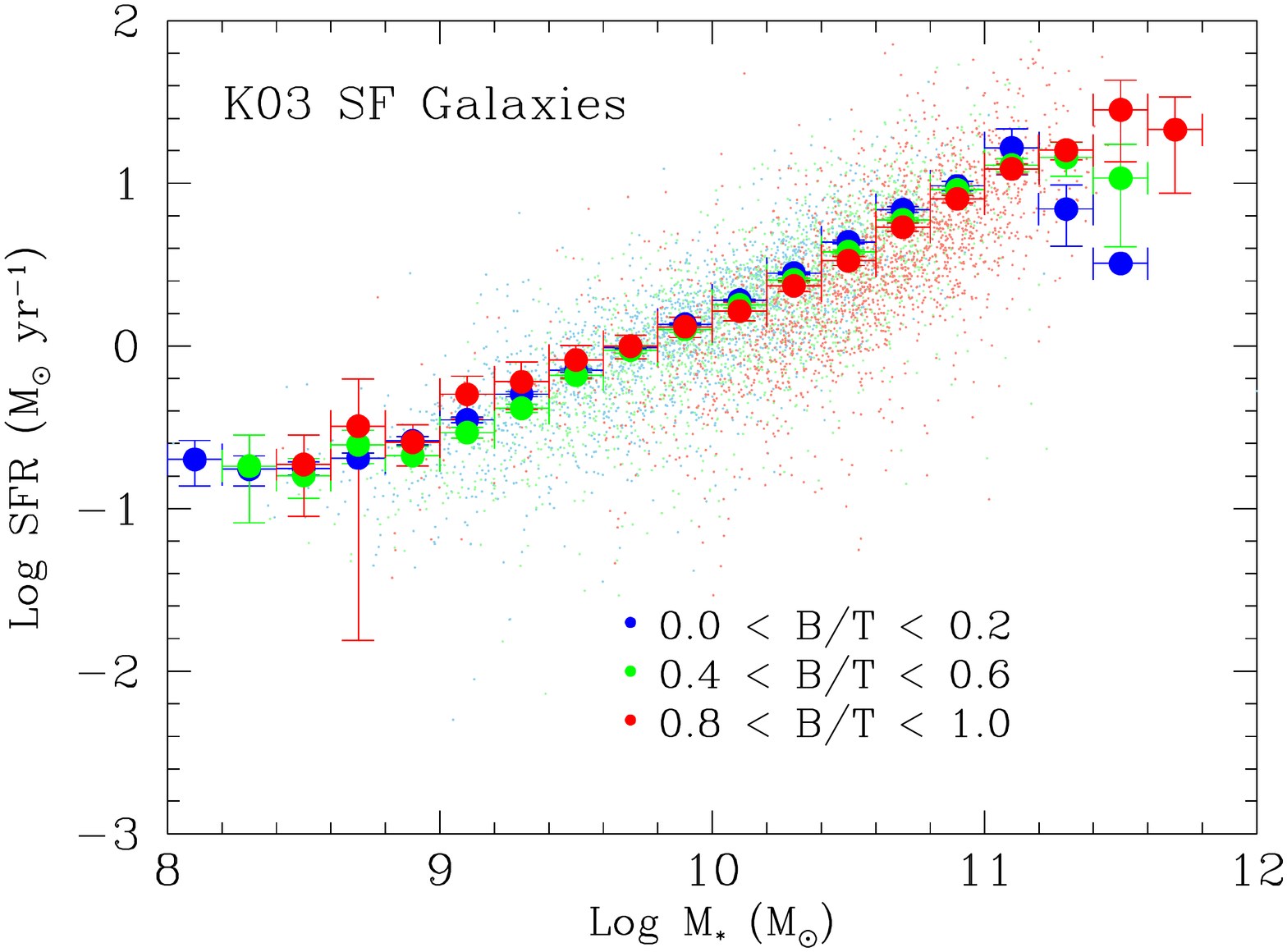}}
\rotatebox{270}{\includegraphics[height=0.33\textwidth]{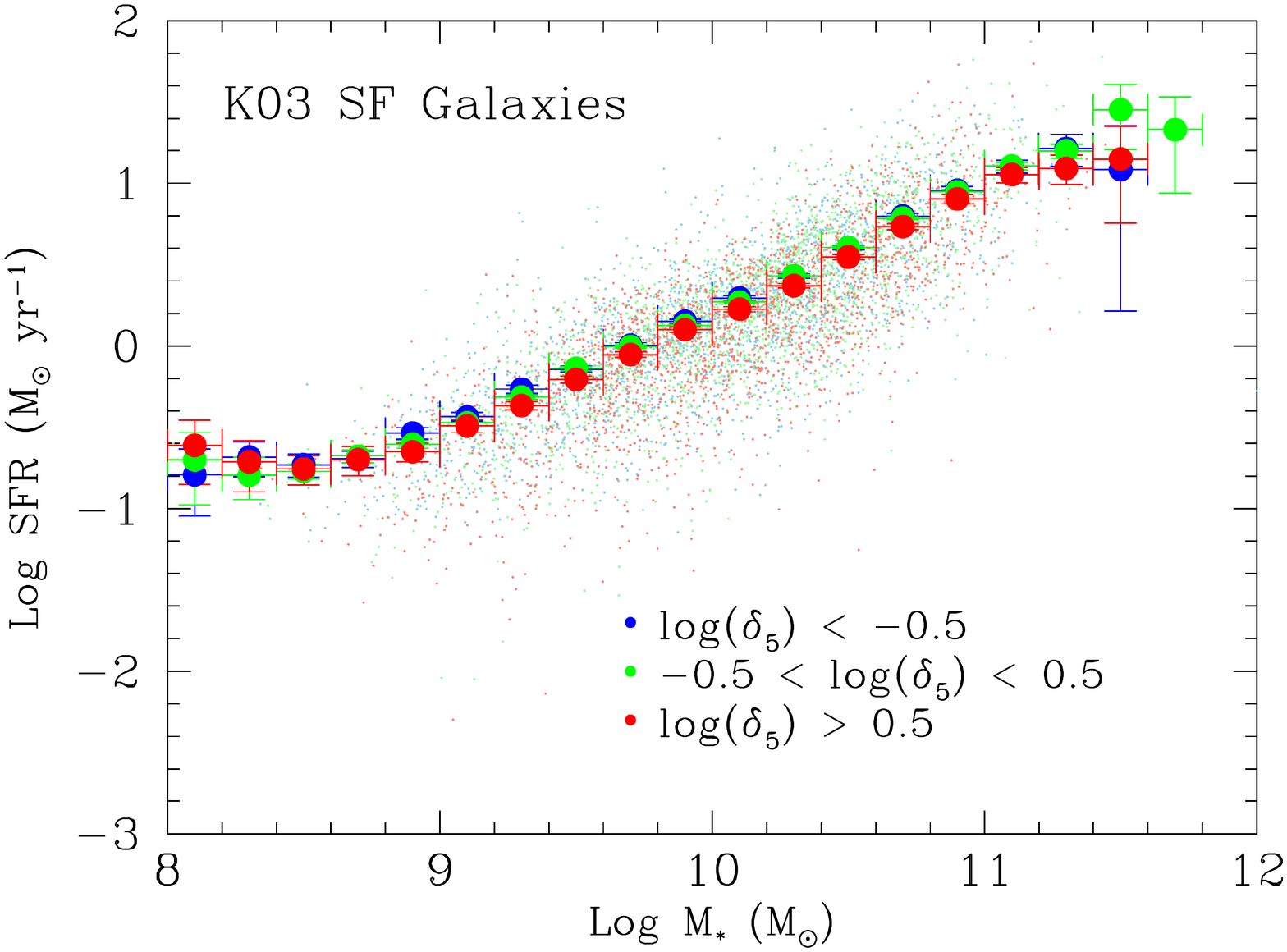}}
\caption{{\it Left panel}: Star formation rate (SFR) vs. total stellar mass ($M_{*}$) for star forming galaxies selected by the Kauffmann et al. (2003) BPT line cut at a S/N $>$ 3 in each of the spectral lines. The dashed line represents a simple power-law fit to the median of the data, with the solid line indicating the region a factor of ten lower than the median. {\it Middle panel}: Same as for top left but with galaxies subdivided by their B/T (by stellar mass) morphology, note the lack of variation in the SFR - $M_{*}$ relation for star forming galaxies across a wide range of morphologies. 
{\it Right panel}: Same as for top left but with galaxies subdivided by local density evaluated to the 5th nearest neighbour, $\delta_{5}$. Note the distinct lack of variation in the SFR - $M_{*}$ relation for star forming galaxies across a wide range of local densities. The large points are mean values, with the error bars shown being (2$\times$) the standard error on the mean. For clarity, only one in every twenty points are shown on each plot displayed here.}
\end{figure*}

\subsection{Sample Selection}

\begin{table}
\caption{Sample Details}
 \label{tab1}
 \begin{tabular}{@{}ccccc}
  \hline
\hline
$z_{\rm spec}$ & log($M_{*}/M_{\odot}$) & log($M_{\rm halo}/M_{\odot}$) & $N_{\rm tot}$ & $N_{\rm cen}$ \\ 
\hline
0.02 -- 0.2 & 8 -- 12 & 11 -- 15.5 & 538046 & 423480 \\
\hline
\end{tabular}
\end{table}

The basic sample details are shown in Tab. 1. We make no selection based on the morphology, colour, or SFR of galaxies, and no initial selection based on their environmental properties. In order to be in line with the suggested usage of the stellar mass decompositions, we follow all of the detailed suggestions in \S 5 of Mendel et al. (2014). As stated earlier, we consider only those cases where $M_{\rm bulge} + M_{\rm disk} \sim M_{B+D}$ to within 1 $\sigma$, assuring that the available stellar mass estimates are consistent (selecting $\sim$ 90 \% of galaxies). We also remove all galaxies for which the photometric and/or the stellar mass decompositions fail to give an output (this is for $<$ 3 \% of the sample).

Using our prescription to calculate the probability of a given disk in a decomposition being false, we correct the sample statistically as is explained in detail in \S App B. Briefly, we redefine the B/T ratio to unity and the bulge mass to the total stellar mass in all cases where the probability of the disk being false $Pr_{\rm FD} >$ 0.5. We check that our results are not adversely impacted by this correction in \S App C. Further, we check that our results are unchanged if we restrict the sample to z $<$ 0.1, and also if we use a free S\'{e}rsic (as opposed to $n_{s}$ = 4) bulge in \S App. D. None of these changes make any significant difference to any of our conclusions.

In the latter part of this paper, we assess the role of intrinsic observables as drivers of the passive fraction. Here we select only central galaxies, which comprise $\sim$ 80 \% of the parent sample (423,480 galaxies).

\section{Method: The Passive Fraction}

Most galaxies can be divided cleanly by whether or not they are forming a significant amount of new stars (e.g. Brinchmann et al. 2004). This bimodality is also seen in terms of colour (e.g. Strateva et al. 2001, Baldry et al. 2004, Driver et al. 2006). For those galaxies which are actively star forming, there is a well known and documented tight relation between the SFR and total stellar mass of galaxies (see Fig. 5 left panel, and Brinchmann et al. 2004), which is frequently referred to as the star forming main sequence. In this figure we include only those galaxies which are designated as star forming by the Kauffman et al. (2003) line cut on the BPT emission line diagram, with a S/N $>$ 3 in each of the BPT emission lines. This removes passive galaxies as well as AGN and composites, and allows us to focus on the properties of the star forming sample before considering how to define `passive'. 

We find that the correlation between stellar mass and SFR is not morphology dependent for star forming galaxies (See Fig. 5 middle panel). This fact is important because it implies that stellar mass alone is sufficient to assign a fairly accurate SFR to a given galaxy at some redshift, if it is established to be on the star forming main sequence. For galaxies in general, morphology is expected to play a much more significant role, and we will look at this is some detail in the subsequent sections. However, morphology only affects whether or not a galaxy lies on the main sequence, and does not modify the sequence itself. 
Further, environmental effects from varying local density are also found to have a negligible effect on the star forming main sequence (see Fig. 5 right panel). Finally, the star forming main sequence is found to evolve with redshift (e.g. Madau 1996, Noeske et al. 2007, Daddi et al. 2007), and as such it is important to consider its functional form as a redshift dependent quantity as well as a function of stellar mass. 

For the full sample (see Fig. 6 top left panel) we see that only some galaxies are on the main sequence, and many more lie below it. We show the distribution of star formation rates divided into stellar mass range in Fig. 6 (bottom left panel). This is complicated because with SFR there are two competing effects: On the one hand, for actively star forming galaxies, there is a strong positive correlation between SFR and stellar mass, however, the fraction of galaxies which are not on the star forming main sequence also increases with stellar mass. Thus, the two peaks in the SFR distribution are overlapping at various stellar masses, and this makes it very difficult to separate out star forming from passive galaxies on the basis of SFR measurements alone. 

The usual way around this is to use sSFR (= SFR/M$_{*}$) to define the division between passive and star forming systems. However, since the SFR - M$_{*}$ relation has an exponent a little less than one, the sSFR itself varies as a function of stellar mass, and in fact slightly declines with increasing stellar mass for star forming galaxies (see Fig. 6 top middle panel). The distribution of sSFR (Fig. 6 bottom middle panel) shows a clearer separation between star forming and passive systems than for SFR, although the peaks of the distribution still vary as a function of stellar mass. Thus, it is not trivial to place a division between passive and star forming galaxies using this metric.

In order to simplify the division between star forming and passive galaxies we construct a new statistic. We define the distance from the star forming main sequence that each galaxy in our full sample resides at ($\Delta$SFR) by calculating the logarithmic difference between the median SFR of a star forming control sample (at similar total stellar mass and redshift) and the SFR of each galaxy. For our star forming (main sequence) controls we include only star forming galaxies as identified by their BPT emission lines. Here we use the Kauffmann et al. (2003) cut at a S/N $>$ 3, but we find no significant change by moving to the Stasinska et al. (2006) definition of star forming dominated galaxies, or by varying the signal-to-noise threshold modestly. 
Specifically, for each galaxy in our sample we calculate:

\begin{figure*}
\rotatebox{270}{\includegraphics[height=0.33\textwidth]{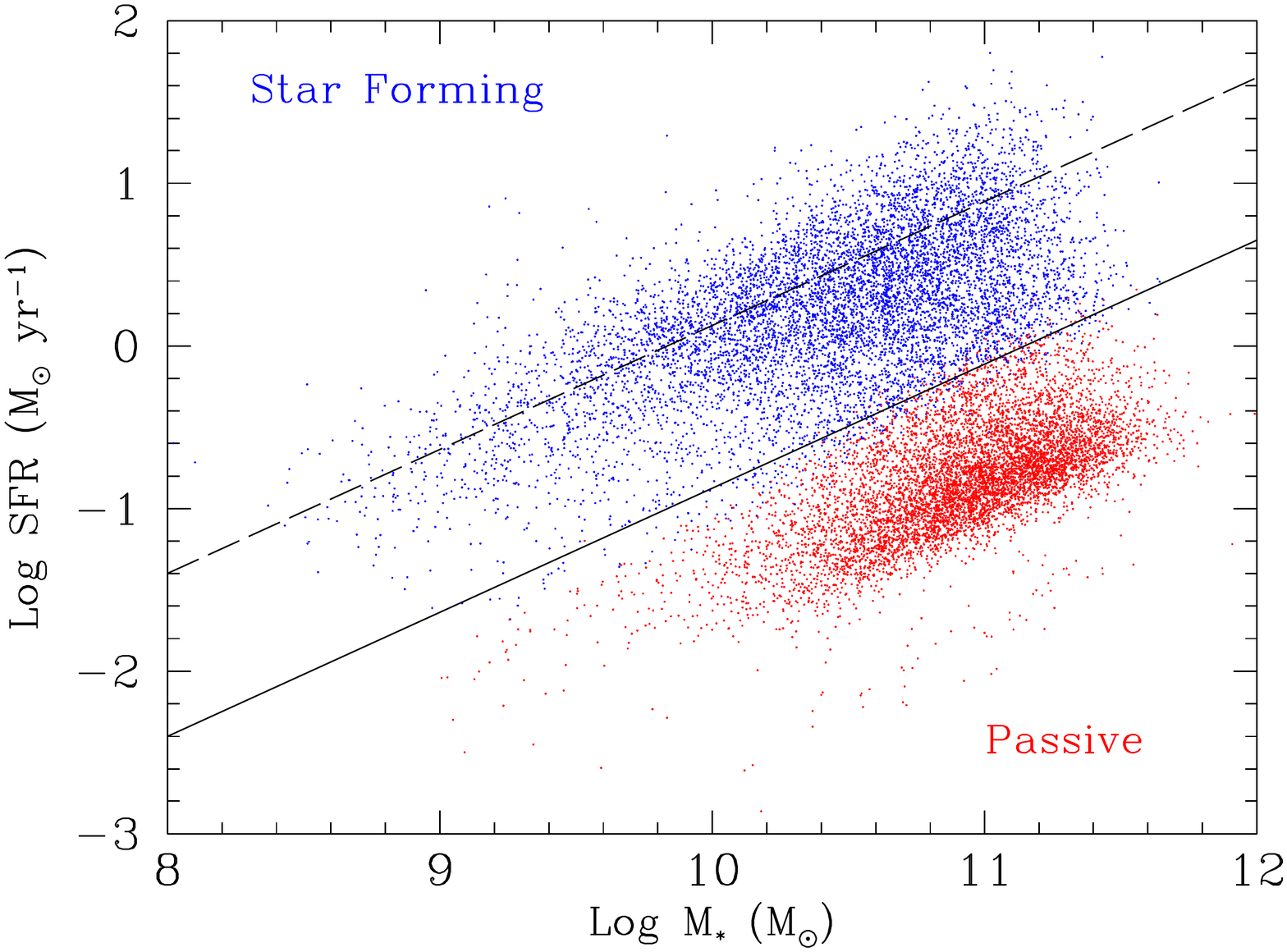}}
\rotatebox{270}{\includegraphics[height=0.33\textwidth]{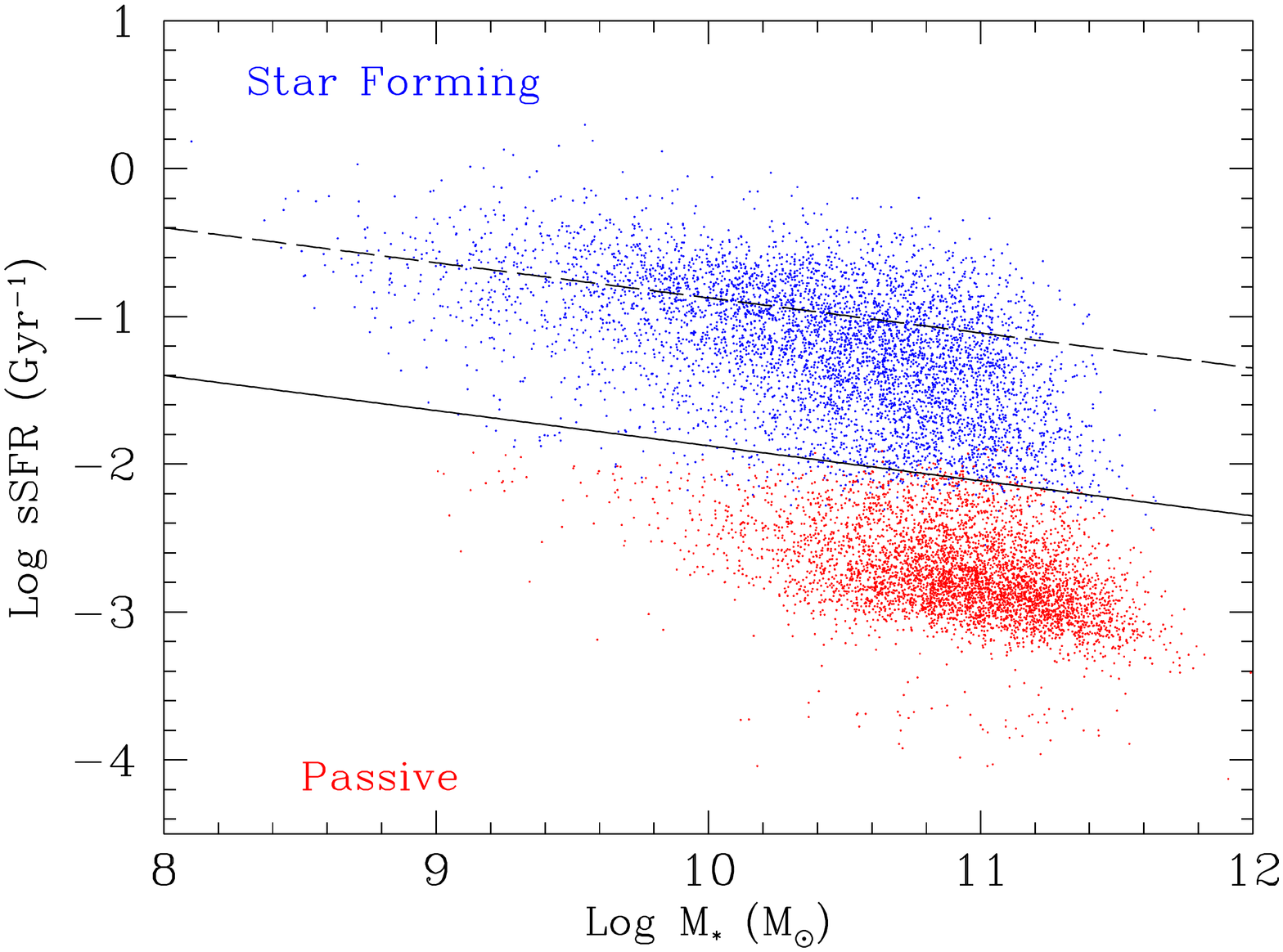}}
\rotatebox{270}{\includegraphics[height=0.33\textwidth]{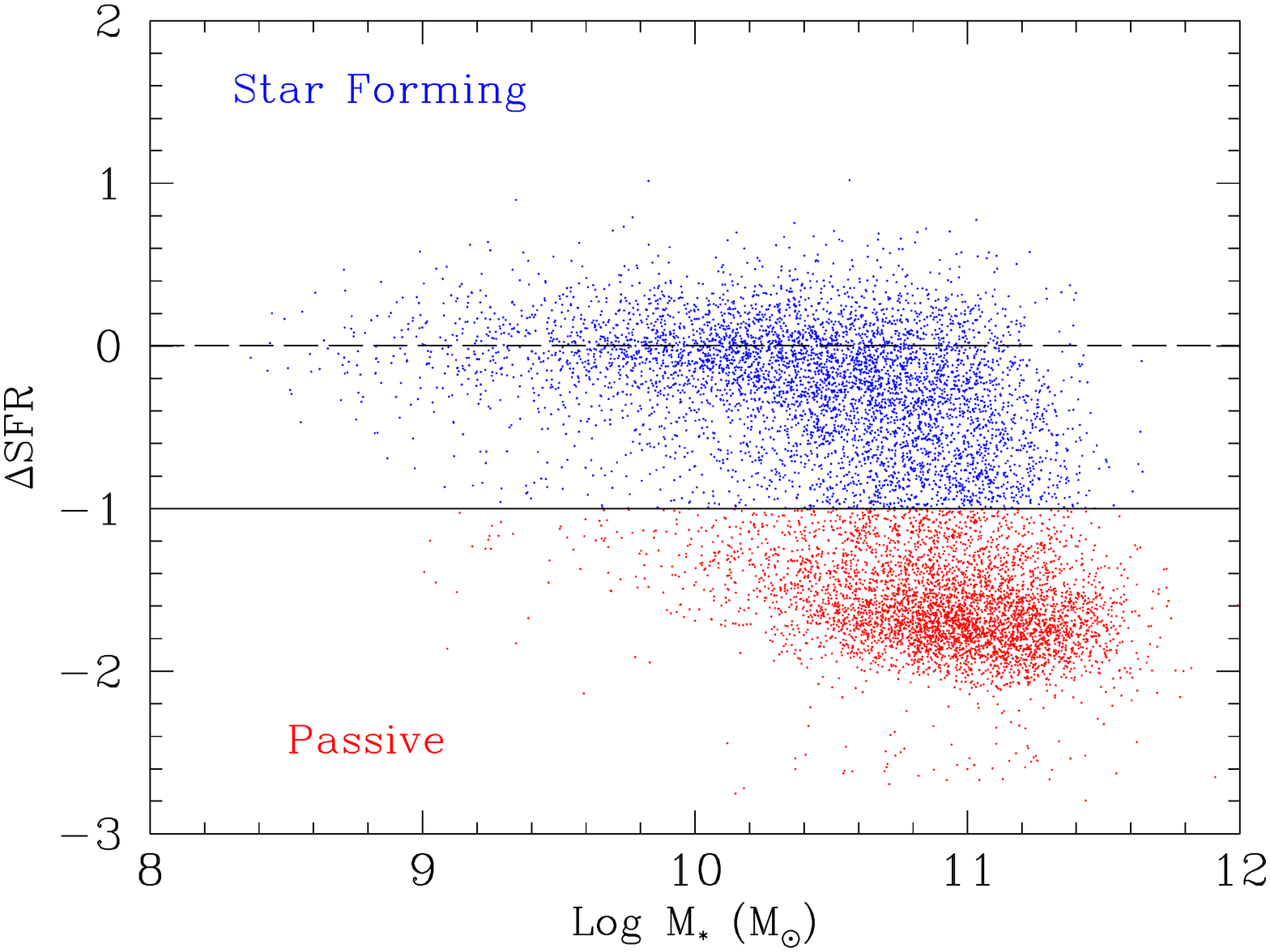}}
\rotatebox{270}{\includegraphics[height=0.33\textwidth]{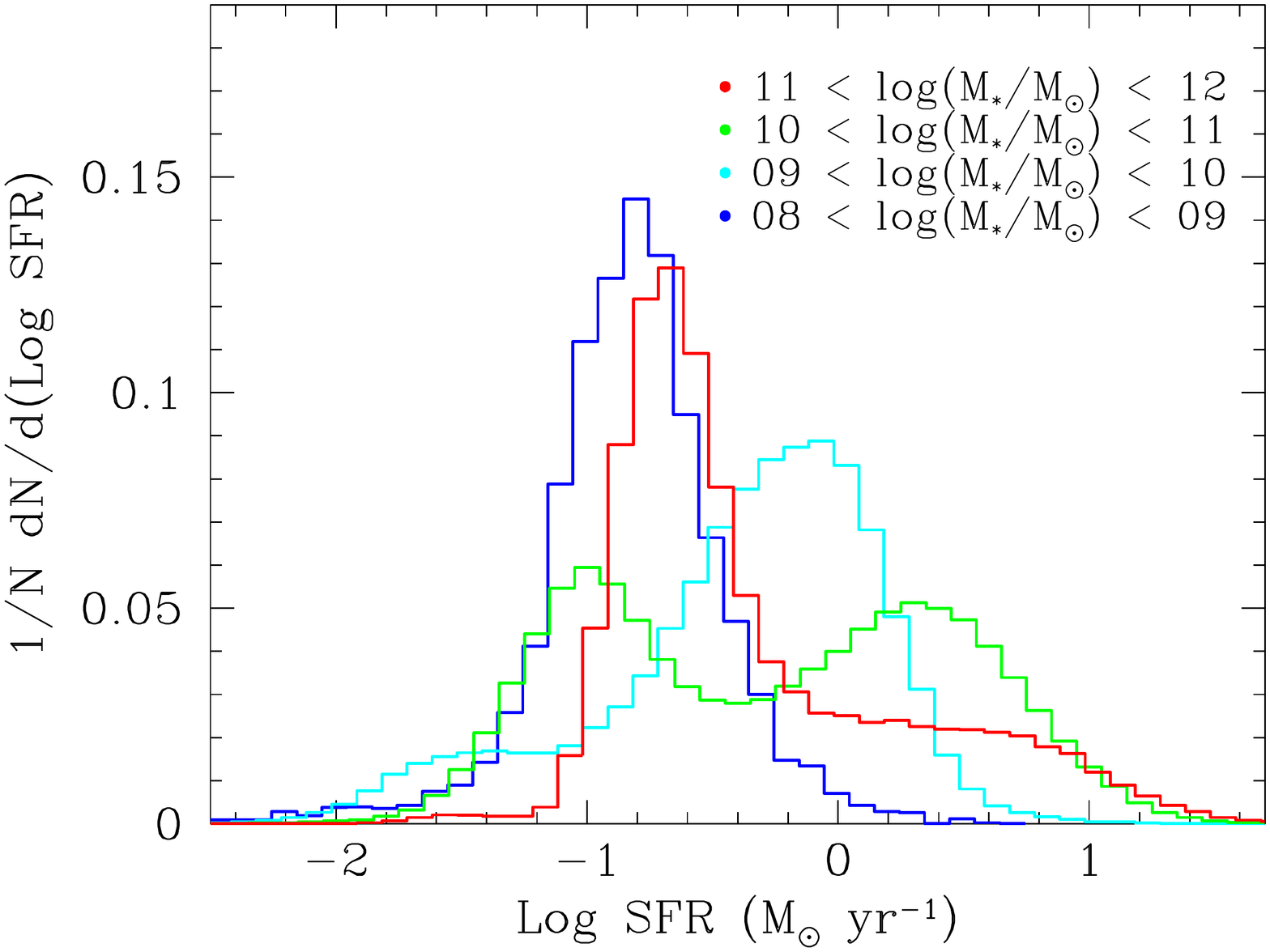}}
\rotatebox{270}{\includegraphics[height=0.33\textwidth]{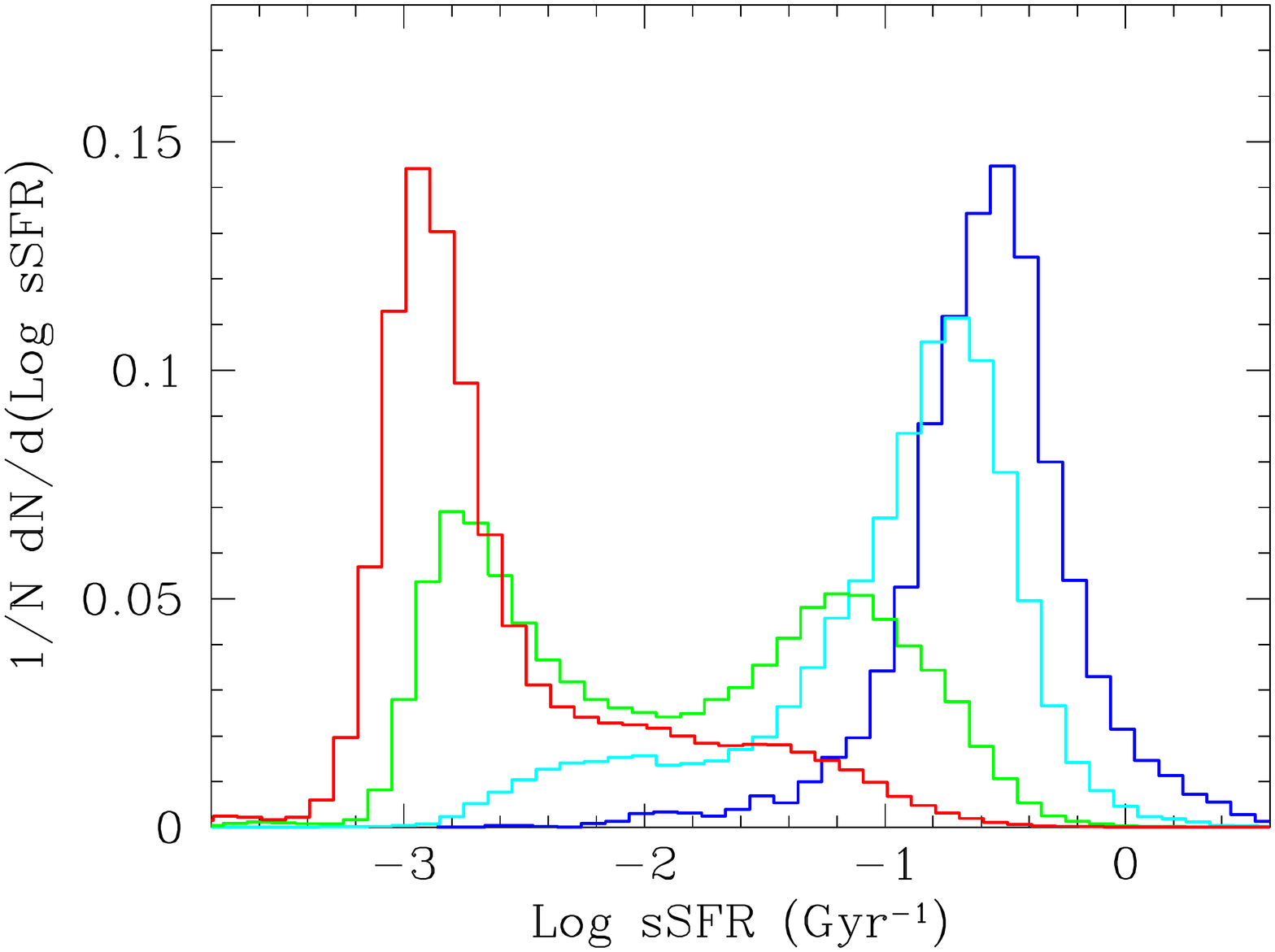}}
\rotatebox{270}{\includegraphics[height=0.33\textwidth]{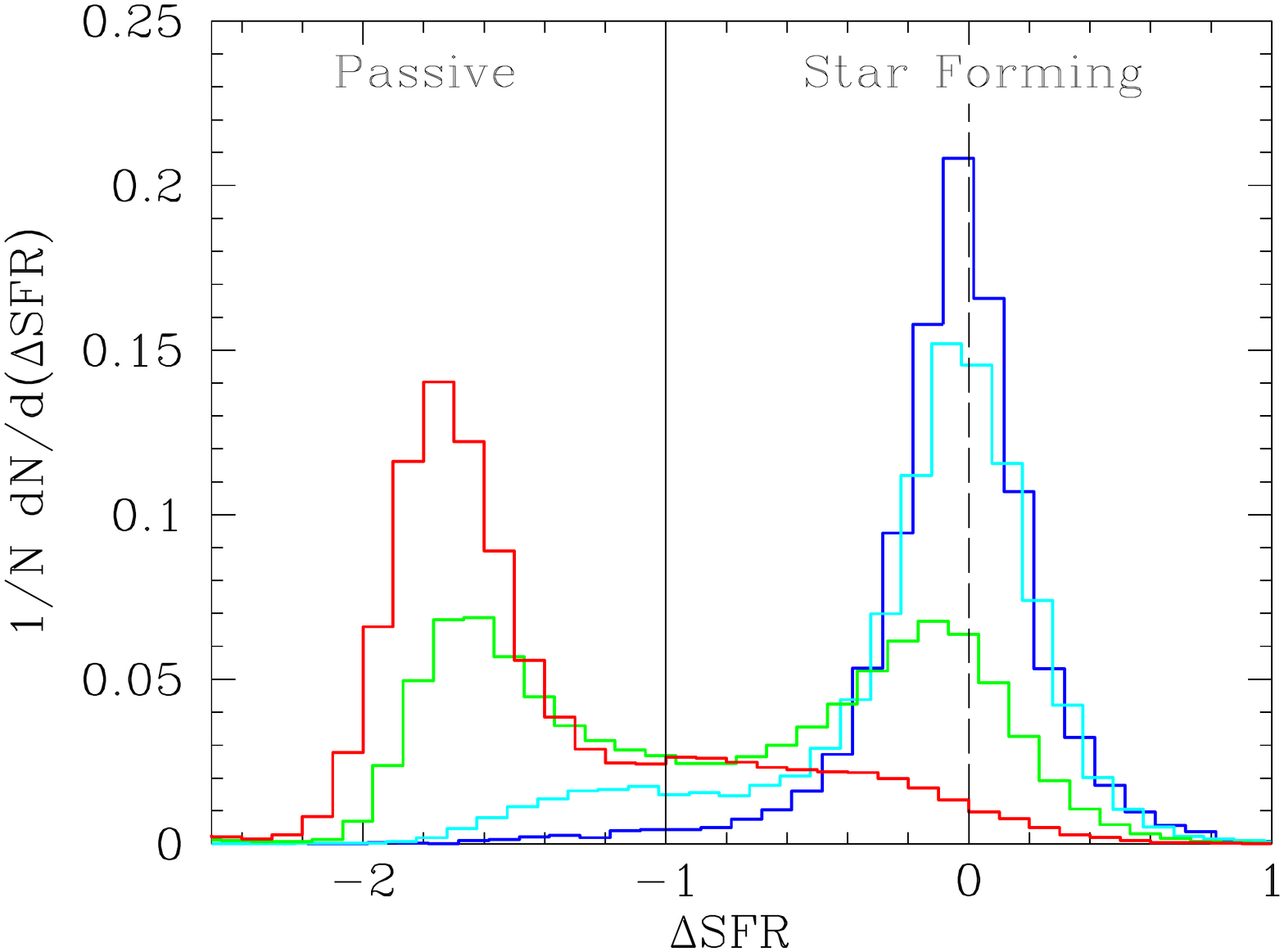}}
\caption{{\it Top panels}: The whole sample colour coded by a $\Delta$SFR cut, where galaxies with SFRs less than a factor of ten below the star forming main sequence are classified as being passive (red points), and those with SFRs greater than that are defined as star forming (blue points). From {\it left to right}: SFR, sSFR, $\Delta$SFR vs. stellar mass. The dotted and solid lines are the same as in Fig. 5. Note that, by construction, star forming galaxies are centered around $\Delta$SFR = 0 (dashed line). Since the cut in $\Delta$SFR requires binning by redshift and stellar mass some points will not be cleanly divided by the solid line in the SFR and sSFR plots.
{\it Bottom panels}: Normalised distributions of three star forming metrics, from {\it left to right}: SFR, sSFR, $\Delta$SFR, all sub-divided by stellar mass range, as indicated on the left (SFR) panel. We operationally define a passive galaxy to have $\Delta$SFR $<$ -1, which cleanly divides the two peaks of the bimodal distribution in $\Delta$SFR (solid line). Higher stellar mass galaxies have a more dominant passive peak than star forming peak, whereas lower mass galaxies have a more dominant star forming peak than passive peak. This plot demonstrates the integrated stellar mass dependence of galaxy bimodality in terms of $\Delta$SFR. The existence of the star forming main sequence relation (SFR - $M_{*}$, see Fig. 5) results in an easier separation of galaxies by $\Delta$SFR than by SFR or sSFR.}
\end{figure*}

\begin{equation}
\Delta{\rm SFR} = {\rm log} \big( \frac{{\rm SFR}(M_{*},z)} {\rm{median}(\rm{SFR_{SF}}(M_{*} \pm \delta M_{*},z \pm \delta z))} \big) 
\end{equation}

\noindent The tolerance on stellar mass and redshift matching ($\delta M_{*}$ and $\delta z$, respectively) are initially set to 0.005 for redshift and 0.1 dex for stellar mass. We require that there be a minimum of five controls per galaxy in order for us to utilise a $\Delta$SFR measurement. In the event that there are less than five controls available from the initial tolerances we increase the search range to two times the initial tolerances, and we continue in this fashion until either five controls are found or the hard limits (of 0.02 and 0.3) are reached. 
In most cases we end up with many more than five controls, with a median value of number of controls $\sim$ 200 galaxies. Further, over 95 \% of all control groups require at most one growth of the tolerances, and less than 1\% of galaxies are not assigned a $\Delta$SFR value with these criteria.

\begin{figure*}
\rotatebox{270}{\includegraphics[height=0.33\textwidth]{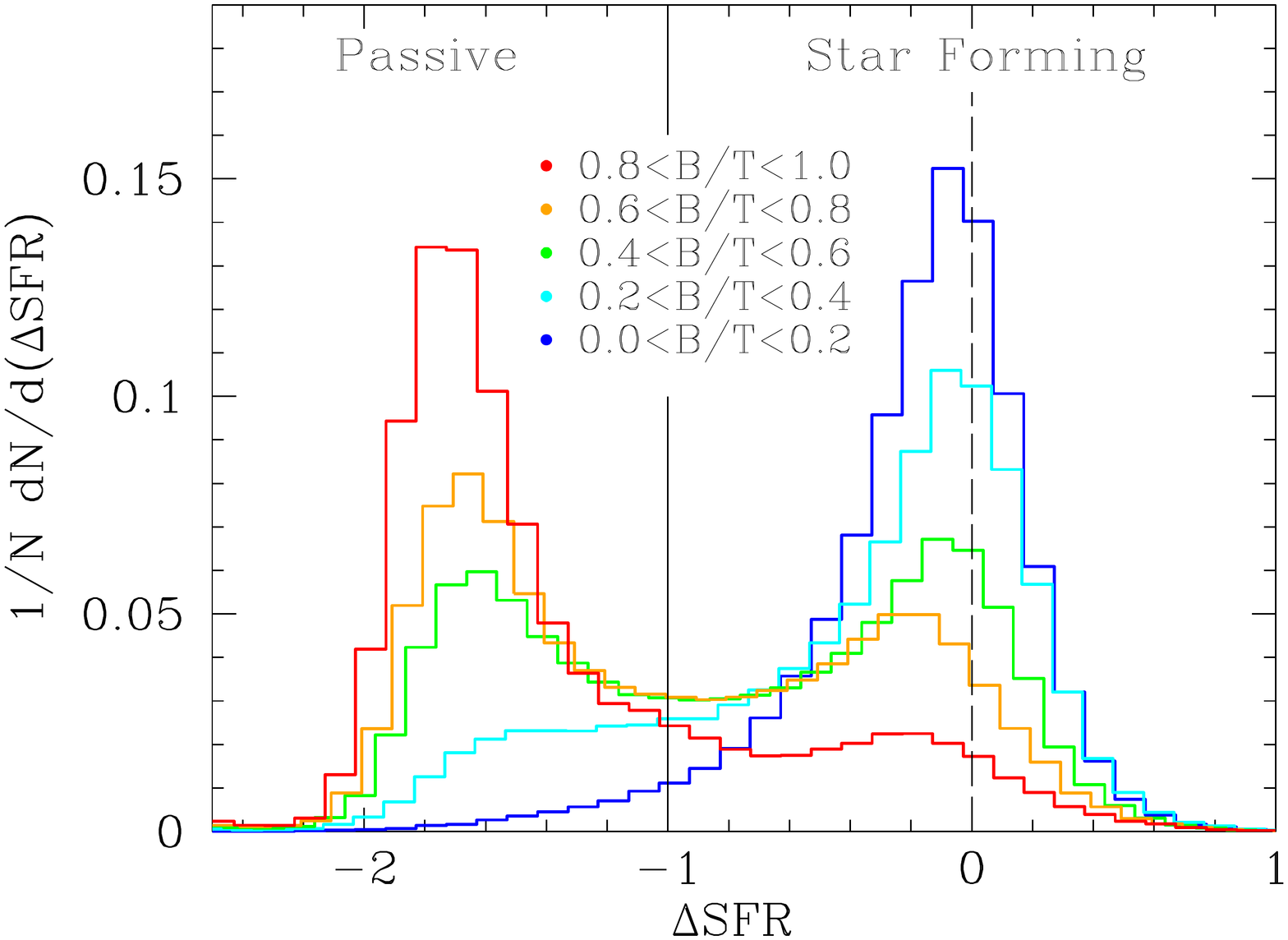}}
\rotatebox{270}{\includegraphics[height=0.33\textwidth]{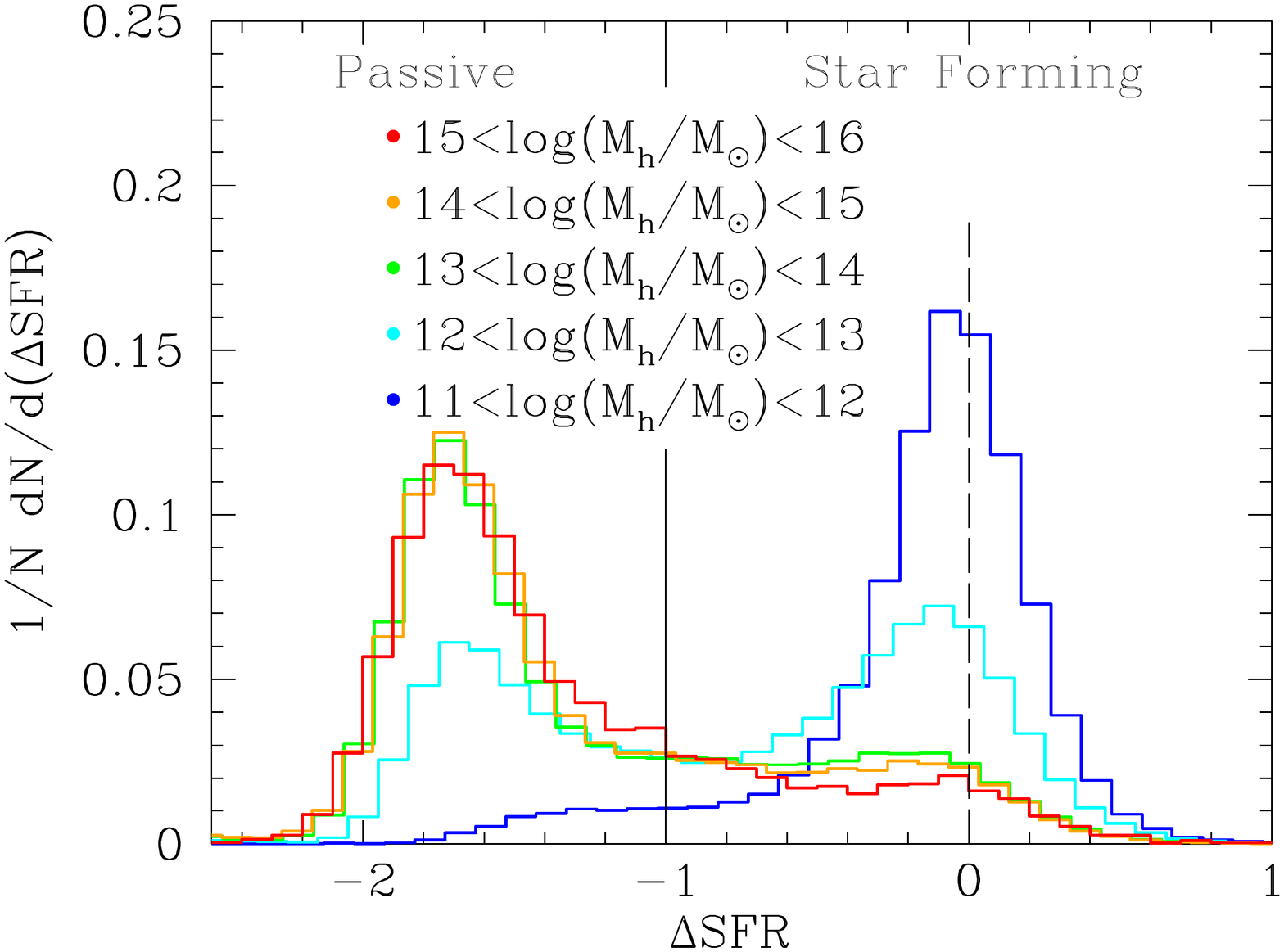}}
\rotatebox{270}{\includegraphics[height=0.33\textwidth]{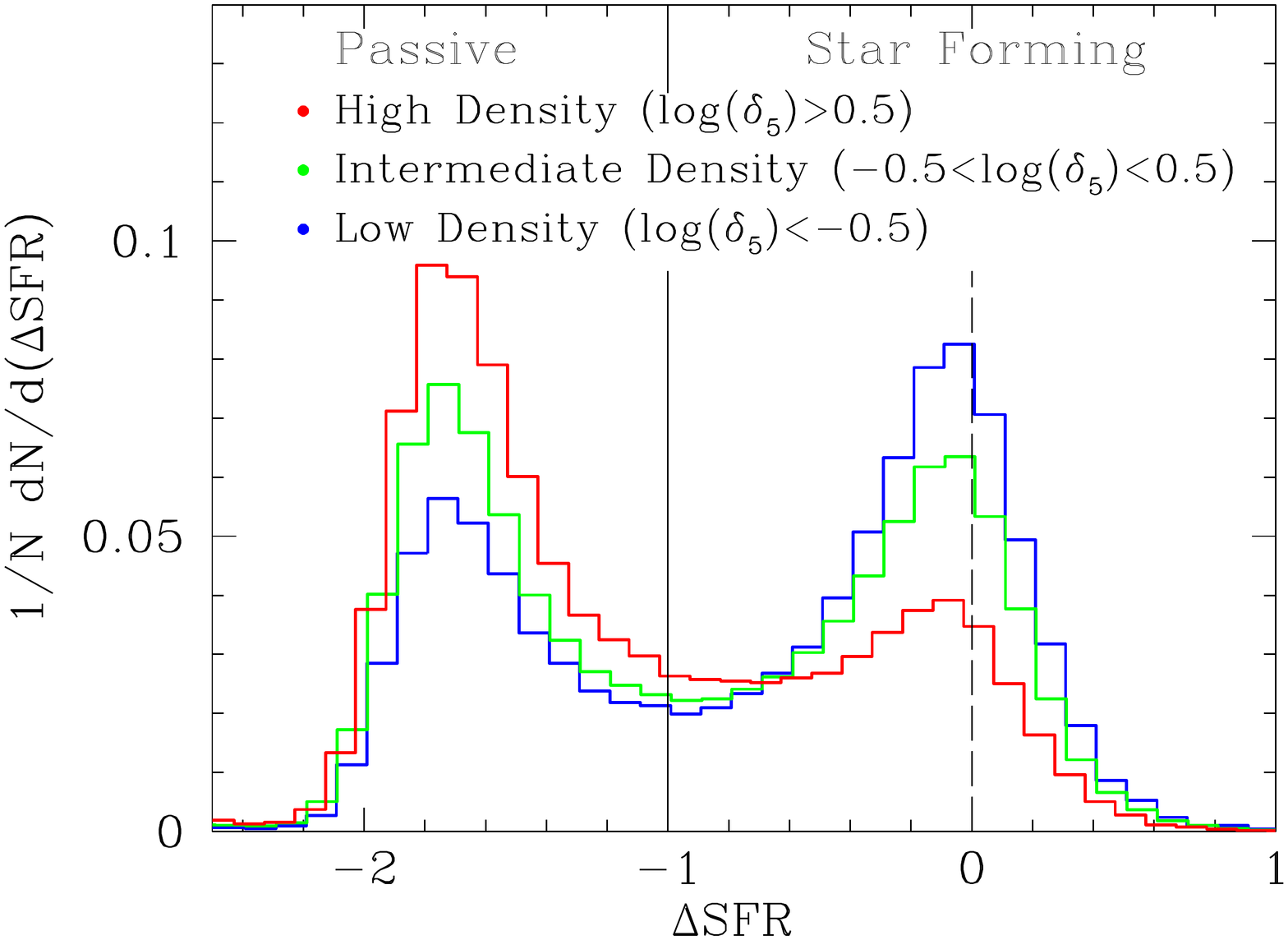}}
\caption{Normalised distributions of $\Delta$SFR for the whole sample, split by {\it left panel}: bulge-to-total stellar mass ratio (B/T); {\it middle panel}: halo mass; and {\it right panel}: over-density evaluated at the 5th nearest neighbour, $\delta_{5}$. As in Fig. 6, we operationally define a passive galaxy to have $\Delta$SFR $<$ -1, which cleanly divides the two peaks of the bimodal distribution in $\Delta$SFR (solid line). Star forming galaxies are centered around $\Delta$SFR $\sim$ 0 (dashed line). Galaxies with higher B/T ratios, higher halo masses, and in higher density environments (shown in redder colours) are more frequently found on the passive peak, whereas galaxies with lower B/T ratios, lower halo masses, and in lower density environments (shown in bluer colours) are more frequently found on the star forming peak. This indicates the integrated morphological, halo mass, and local galaxy density effect on the star formation of galaxies.}
\end{figure*}

We show the dependence of $\Delta$SFR on stellar mass at the top right of Fig. 6. We find that the distribution of $\Delta$SFR is highly bimodal, like with the specific star formation rate (sSFR), and much more obviously so than with SFR (see Fig. 6, bottom right panel). Our $\Delta$SFR measure is defined such that there is no stellar mass (or redshift) dependence on it for the main sequence, i.e. all star forming (main sequence) galaxies should lie around zero by construction, with passive systems lying significantly below zero. This greatly simplifies the separation of star forming and passive systems. We find a natural break between the star forming and passive peaks at around a factor of ten below the star forming main sequence. We use this break to operationally define a passive galaxy as a galaxy with a SFR less than a factor of ten than the median SFR of its (mass and redshift matched star forming) control 
group. That is, we define a galaxy to be passive if:

\begin{equation}
\Delta{\rm SFR} \leq -1 \hspace{0.2cm} .
\end{equation}

\noindent The line denoting this division is displayed on the histogram displaying $\Delta$SFR in Fig. 6 (bottom right panel). We find no significant change in our results by varying the threshold anywhere between -0.5 and -1.5, but the definition breaks down as a useful tool if one of the peaks is sub-divided by the value chosen. 

The SFRs of the actively star forming systems are known robustly through BPT line classifications whereas the SFRs of passive systems, and AGN, are determined via the 4000 \AA \hspace{0.1cm} break - sSFR relation. In Brinchmann et al. (2004) a lower limit of log(sSFR) = -12 is imposed, which results in the mass-scaling of the lower (red) sequence seen in the top panels of Fig. 6. The scatter around this threshold is due to the fibre correction. The implication of this is that the star forming peak is truly a peak (as seen), whereas the passive `peak' is only peaked due to the lower limit imposed, resulting in a collapse of the expected long power law tail to arbitrarily low SFRs. Ultimately, for passive systems, AGN, and composites, it is more reliable as to which peak a galaxy belongs to than its specific SFR value. Hence, for the remainder of the analyses presented in this paper we will consider the passive fraction as our primary indicator of the global star forming properties of SDSS galaxies.

We define the passive fraction as:

\begin{equation}
f_{\rm passive} = \frac{{\sum\limits_{i=1}^{N_{\rm passive}}} 1/V_{\rm max,i}({\rm passive)}}{{\sum\limits_{j=1}^{N_{\rm all}}} 1/V_{\rm max,j}({\rm all})}
\end{equation}

\noindent where $V_{\rm max}$ is the total volume over which a galaxy of given absolute ugriz magnitudes and colours would be visible in the SDSS, which is used here to weight the fractional contributions from various sources, as in Geha et al. (2012). See also Mendel et al. (2014) Fig. 9 for an illustration of the maximum redshift dependence on stellar mass and g-r colour for SDSS galaxies. The $i$-summation is computed for passive only galaxies (defined above in eq. 6), whereas the $j$-summation is calculated for all galaxies in our sample regardless of star formation rate. This expression gives a volume corrected passive fraction with which we can investigate the potential drivers of star formation quenching in the next section.

We estimate the error on the passive fraction for each binning of our data in the next section via the jack-knife technique (e.g. Nichol et al. 2006, Cabre et al. 2007). Briefly, the volume weighted passive fraction is computed for each binning, then recomputed $n$-times excluding each of the members of the dataset in turn. The error is taken as the variance in the passive fraction across the full range of measurements. In most cases for computing combined statistics on a dataset with unknown or complex likelihood distributions, the jack-knife error is shown to perform very well, and will asymptotically approach the actual error as in Monte-Carlo simulation. We show values for the passive fraction in a given bin if there are at least 20 galaxies present, which we find is sufficient to provide an estimate of the passive fraction. However, we note here that most bins contain far more galaxies than this, with a median value $>$ 500 galaxies.

\section{Results}

In this section we examine both the integrated and differential effects of total stellar mass, halo mass, local environment (measured from nearest neighbour over-densities and central - satellite divisions), bulge-to-total stellar mass ratio (B/T), and bulge and disk mass on the passive fraction.

\subsection{The integrated effect of stellar mass, B/T structure, halo mass, and local density on star formation}

Many previous studies have investigated the role of stellar mass, morphology or structure, and environment on the star forming properties of galaxies (e.g. Kauffmann et al. 2003, Brinchmann et al. 2004, Baldry et al. 2004, Baldry et al. 2006, Driver et al. 2006, Drory \& Fisher 2007, Bell et al. 2008, Bamford et al. 2009, Peng et al. 2010, Peng et al. 2012, Mendel et al. 2013, Woo et al. 2013). In this sub-section we review the dominant trends with these variables for our sample before delving into the detailed inter-dependencies. We explore the differential effects of these potential drivers to star formation cessation in \S 4.2 - \S 4.6.

\begin{figure*}
\rotatebox{270}{\includegraphics[height=0.33\textwidth]{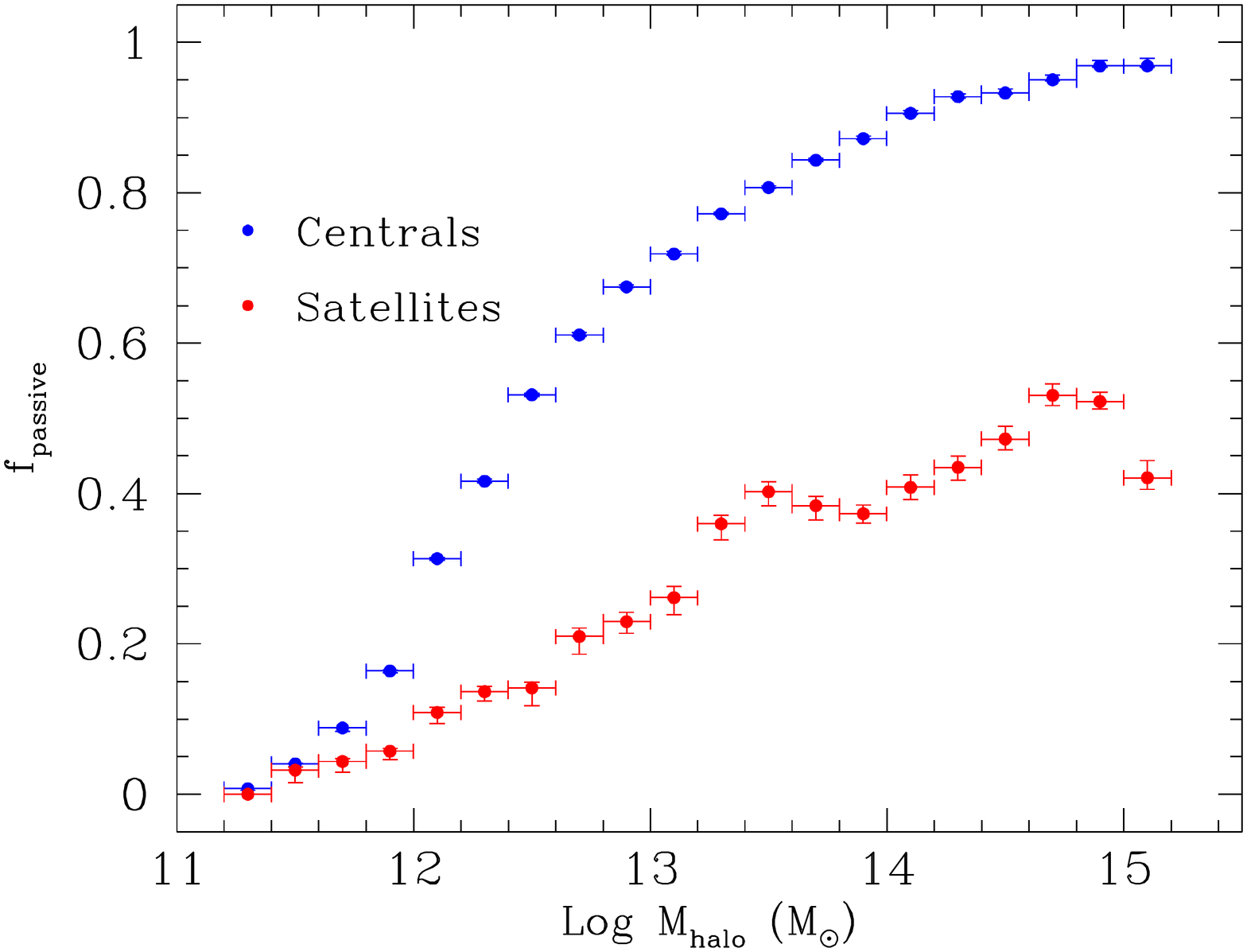}}
\rotatebox{270}{\includegraphics[height=0.33\textwidth]{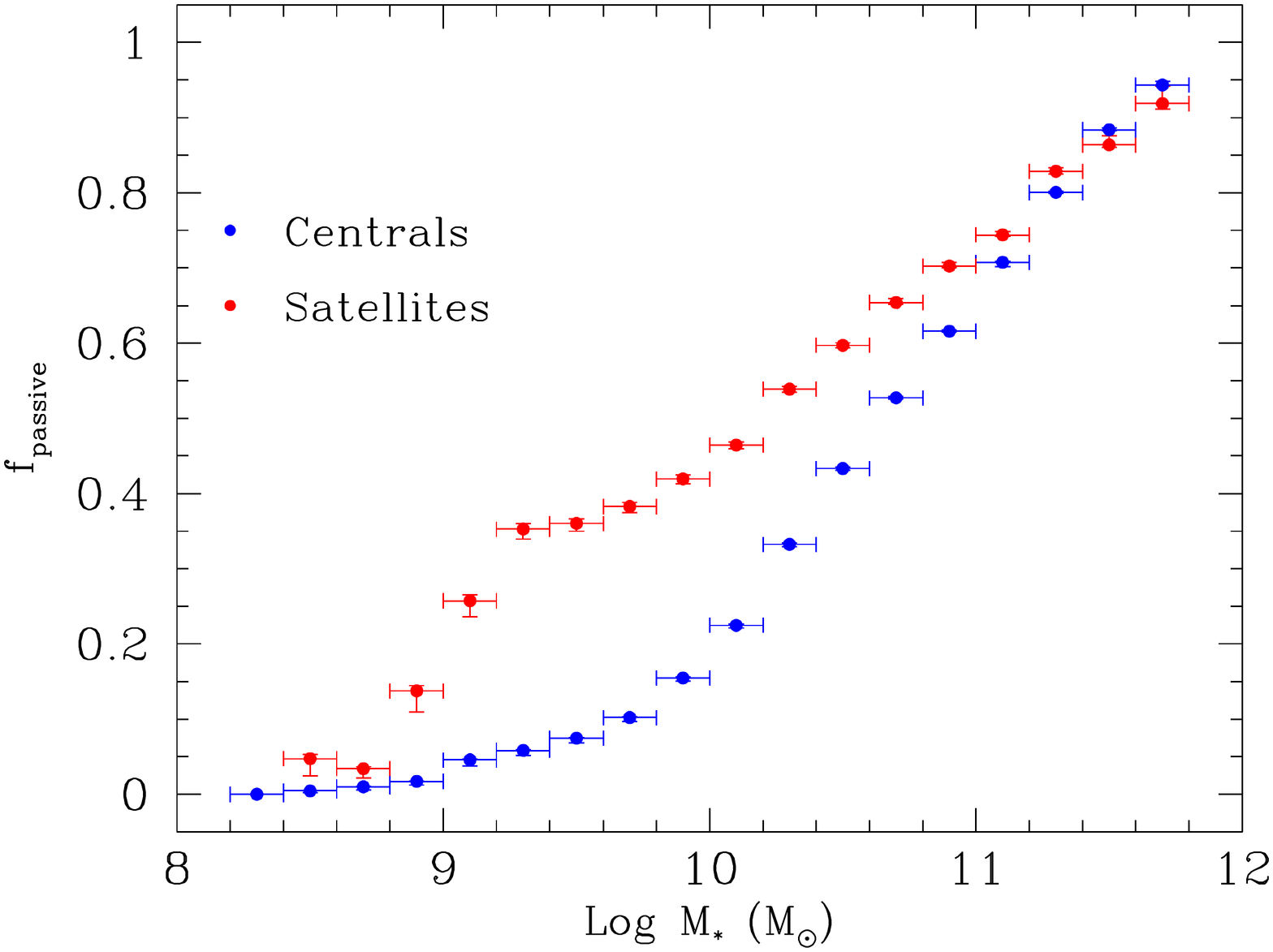}}
\rotatebox{270}{\includegraphics[height=0.33\textwidth]{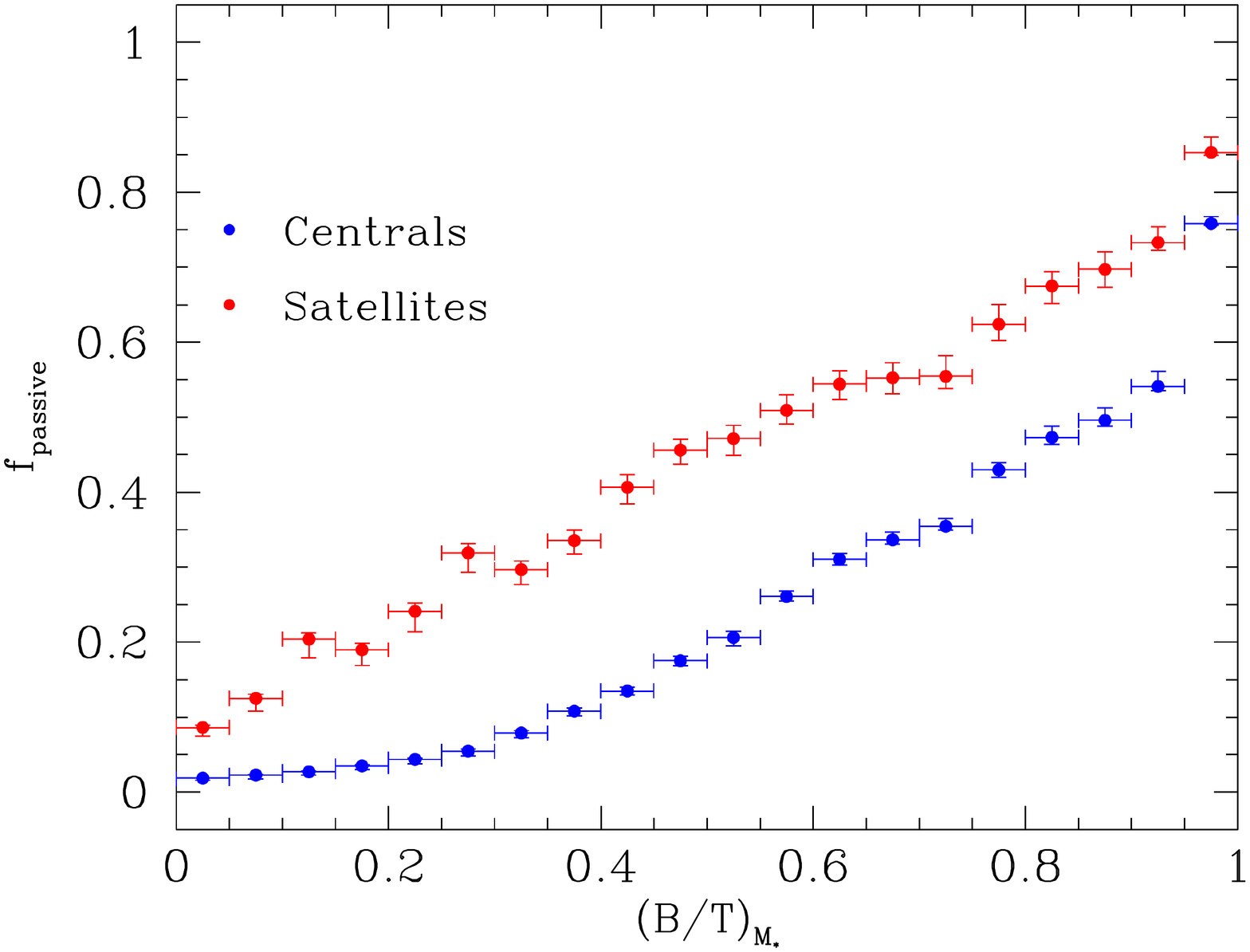}}
\caption{Passive fraction vs. halo mass ({\it left panel}), total stellar mass ({\it middle panel}), and bulge-to-total stellar mass ratio, B/T ({\it right panel}). In each plot the passive fraction is split into the contribution from central (blue points) and satellite (red points) galaxies. The 1$\sigma$ error on the passive fraction computed via the jack-knife technique, and bin size, are displayed as the error bars. At a fixed halo mass central galaxies are more passive than satellites, whereas at a fixed stellar mass and B/T ratio satellites are more passive than central galaxies. For central galaxies, the passive fraction experiences the steepest gradient with higher stellar masses, and lower halo masses, whereas the dependence on B/T is generally less steep for all values.}
\end{figure*}

In Fig. 6 (bottom right panel) we show the normalised distribution of $\Delta$SFR for all galaxies in our full parent sample, as first introduced in \S 3. We divide this distribution by stellar mass ($M_{*}$) range, as indicated on the plot, with redder colours indicating higher stellar masses. We find that lower stellar mass galaxies are preferentially found on the star forming peak, whereas higher stellar mass galaxies are preferentially passive. There is a smooth transition from the star forming peak to the passive peak as we increase stellar mass. In the stellar mass range 10 $< {\rm log}(M_{*}/M_{\odot}) <$ 11 we find that galaxies are roughly evenly divided between the passive and star forming peak. This result is qualitatively very similar to that of Baldry et al. (2006) and  Peng et al. (2010, 2012), who both find that higher stellar mass systems are more frequently passive (or red). 

In Fig. 7 (left panel) we show the normalised $\Delta$SFR distribution split by B/T. Throughout the results section, when we refer to B/T we mean bulge-to-total stellar mass ratio, and any other usage will be made explicit. Here we note that higher B/T (more spheroidal) galaxies are more frequently passive than lower B/T (more disk-like) galaxies. An analogous result has been previously seen for S\'{e}rsic index and concentration in, e.g., Driver et al. (2006), for a smaller sample size and over a smaller range in stellar masses. We notice that the transition from the star forming to passive peak is monotonic with B/T, as with stellar mass. At intermediate B/T values (0.4 - 0.6) the galaxy population is fairly evenly divided between the two peaks. However, given the high level of inter-correlation between $M_{*}$ and B/T, it is not clear which, if either, is primary in constraining the passive fraction (see Fig. 2 for an illustration of this inter-dependence). 
Clearly, it is desirable to look at the differential effects of, say, stellar mass at a fixed B/T and vice versa. This is shown later in \S 4.3 \& \S 4.5.

In Fig. 7 (middle panel) we show the normalised $\Delta$SFR distribution for all galaxies split by halo mass (from the Yang et al. 2009 group catalogue, see \S 2.5 for details). Galaxies in lower mass halos tend to be actively star forming, whereas galaxies in higher mass halos tend to be preferentially passive. However, the effect of increasing halo mass beyond $M_{\rm halo} \sim 10^{13} M_{\odot}$ does not result in an increase in the passive fraction for the whole population. The notion of halo mass is quite different for centrals and satellites, and we explore these differences in the  next section. For now it is sufficient to note that there appears to be a threshold in halo mass above which varying this parameter further leads to a negligible effect. For central galaxies, there is a strong correlation between $M_{\rm halo}$ and $M_{*}$ which motivates looking at the differential effect of varying halo mass at a fixed stellar mass (as in Woo et al. 2013) and vice versa. The relation between the 
range in these variables is shown in Fig. 4 for central galaxies. We examine the differential impact of $M_{\rm halo}$ and $M_{*}$ in detail in \S 4.4 \& \S 4.5.

Finally, we show the normalised distribution in $\Delta$SFR split by the density parameter, $\delta_{5}$ (defined in \S 2.4). See Fig. 6, right panel. Galaxies residing in higher local densities tend to be more passive than galaxies in lower local densities. At intermediate densities, corresponding to the average galaxy density at z $\sim$ 0, there is an approximately even split between the passive and star forming peaks. This result is essentially the same as seen in van den Bosch et al. (2008) and Peng et al. (2010), and is qualitatively similar to Butcher \& Oemler (1984). Since centrals and satellites reside in distinct regions of the $\delta_{5}$ parameter space (see back to Fig. 3), it is desirable to assess the role of environment separately for centrals and satellites (as in Peng et al. 2012 and Woo et al. 2013). 
We investigate this in the next section. Moreover, the inter-relations between $M_{*}$, $M_{\rm halo}$, B/T, and $\delta_{5}$ (see back to Figs. 2 - 4 for an illustration of some of these) emphasize the need to look at the differential effects of these variables at fixed values of the remaining correlators, as well as the total integrated effects discussed in this section.

\subsection{Centrals vs. Satellites}

The location of galaxies within their parent dark matter halos is an important consideration when assessing the origin and evolution of galaxy bimodality (e.g. van den Bosch et al. 2007, 2008, Peng et al. 2010, 2012, Woo et al. 2013). Satellite galaxies are expected to have additional quenching mechanisms applicable to them than central galaxies, due to the extra physics associated with motion relative to the hot gas halo (e.g. Balogh et al. 2004, van den Bosch et al. 2008). In Fig. 8 we show the passive fraction dependence on halo mass, total stellar mass, and B/T structure, subdivided into the contribution from centrals and satellites. Note that halo mass refers to the total mass of the host halo in which each galaxy resides, which for satellite galaxies is not the dark matter halo of the galaxy itself (which we shall refer to as its sub-halo). In each of these cases there is a strong positive correlation between the passive fraction and the galaxy/ halo property. Additionally, for each relation there is a 
clear separation between central and satellite galaxies, which strongly motivates the need for considering these two types of galaxies separately (as in e.g. Peng et al. 2012, Woo et al. 2013).

The $f_{\rm passive}$ - $M_{\rm halo}$ relation for centrals is very steep from $M_{\rm halo}$ $\sim$ $10^{11} - 10^{13}$, but levels off significantly at higher masses. Satellites exhibit a generally less steep dependence on halo mass than centrals, but show no strong sign of a flattening of this dependence at higher masses. Interestingly, central galaxies are found to be more passive than satellites at a fixed halo mass. The most probable explanation for this observation is that centrals tend to have higher stellar masses than satellites in the same halo (by definition, see \S 2.4 \& \S 2.5) which leads to the difference between satellites and centrals at a fixed halo mass being driven by differences in their stellar masses (e.g. Peng et al. 2012). Since the difference in passive fraction between centrals and satellites is in general greater at fixed halo masses than at fixed stellar masses, this suggests that intrinsic effects win over environmental effects in assigning the passive fraction. However, this 
dependence is highly mass dependent, and for low stellar mass galaxies and low mass halos this trend is actually reversed. 

The $f_{\rm passive}$ - $M_{*}$ relation is steep for centrals from $M_{*}$ $\sim$ $10^{10} - 10^{12}$, but at lower stellar masses it is very flat, with most low stellar mass centrals being predominantly star forming. Satellites are found to be more passive at a fixed stellar mass than centrals at low stellar masses, but these differences vanish at higher stellar masses. This implies that the process(es) which drive the cessation of star formation in the most massive systems are likely independent of environment, and, hence, happen equally in the field, groups, and clusters. This further suggests that it may be a process coupled to intrinsic galaxy properties, such as e.g. total stellar mass and/or morphology/ internal kinematics which drives the increase of the passive fraction for high stellar mass galaxies (a similar conclusion to Baldry et al. 2006, Peng et al. 2010, 2012). We consider this possibility in more detail in the next section.

\begin{figure}
\rotatebox{270}{\includegraphics[height=0.49\textwidth]{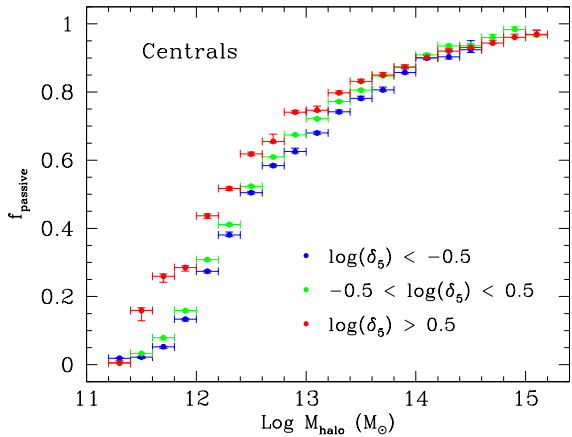}}
\caption{Passive fraction vs. halo mass, binned by the over-density of galaxies evaluated at the 5th nearest neighbour, $\delta_{5}$, as indicated on the plot. The 1$\sigma$ error on the passive fraction computed via the jack-knife technique, and bin size, are displayed as the error bars. Note the relative lack of variation in the passive fraction with $\delta_{5}$ at a fixed halo mass.}
\end{figure}

We also examine the $f_{\rm passive}$ - B/T relationship for centrals and satellites. This relationship is noticeably less steep than for halo mass and stellar mass, perhaps suggesting that B/T is sub-dominant to these other potential indicators of passivity. Satellites are found to be more passive at a fixed B/T ratio than central galaxies, and this difference appears roughly constant throughout the full range of B/T, from 0 - 1. This strongly suggests that whatever process(es) are giving rise to the increase in the passive fraction for satellites do not have to remove the disk, or engender significant morphological evolution of any kind. This is particularly interesting because it implies that the morphology - density and colour - density relationships (e.g. Dressler et al. 1980, Butcher \& Oemler 1984) are not necessarily coupled as is often assumed, i.e. galaxies can become redder (with lower sSFR) in denser environments without becoming more bulge dominated.

\begin{figure*}
\rotatebox{270}{\includegraphics[height=0.33\textwidth]{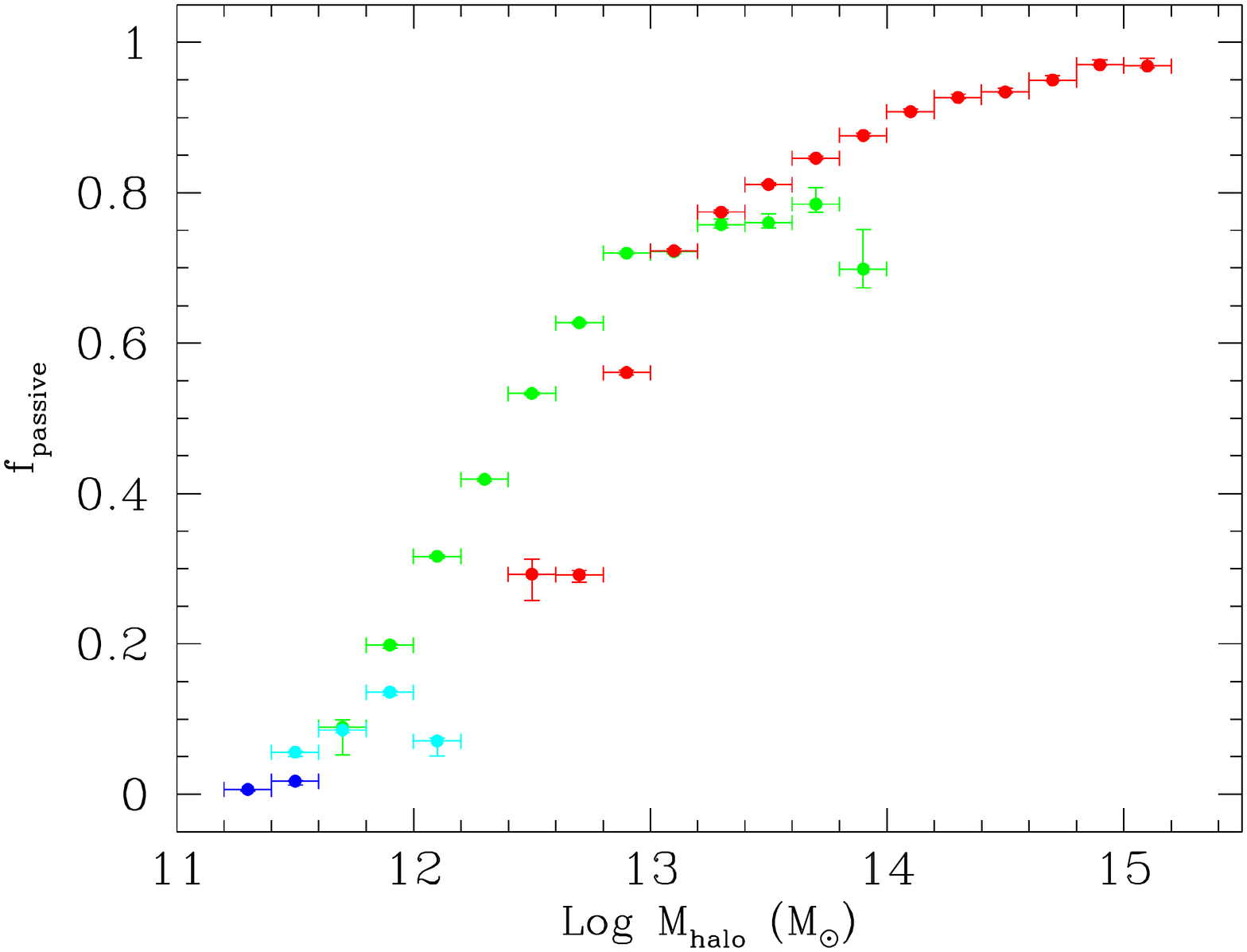}}
\rotatebox{270}{\includegraphics[height=0.33\textwidth]{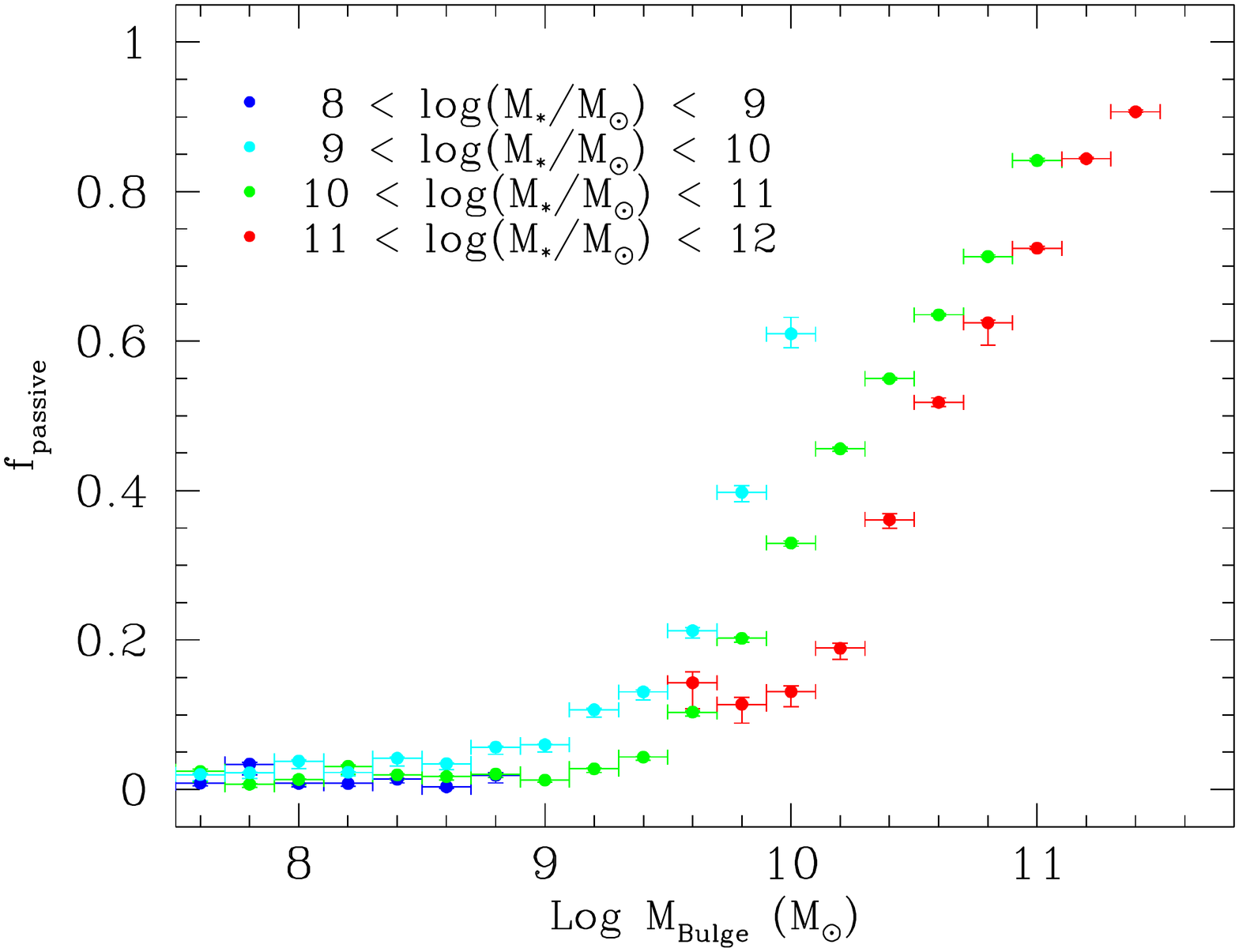}}
\rotatebox{270}{\includegraphics[height=0.33\textwidth]{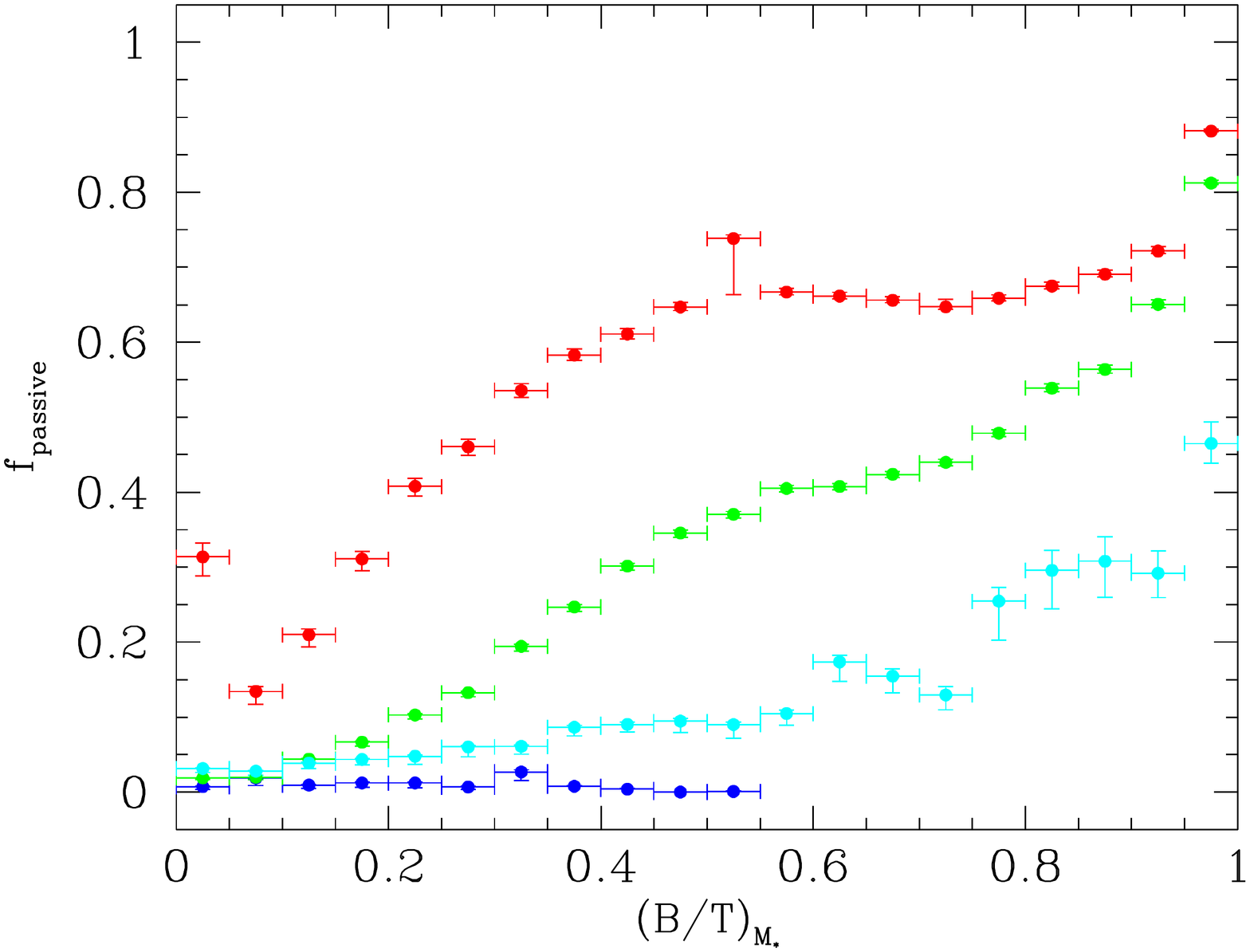}}
\rotatebox{270}{\includegraphics[height=0.33\textwidth]{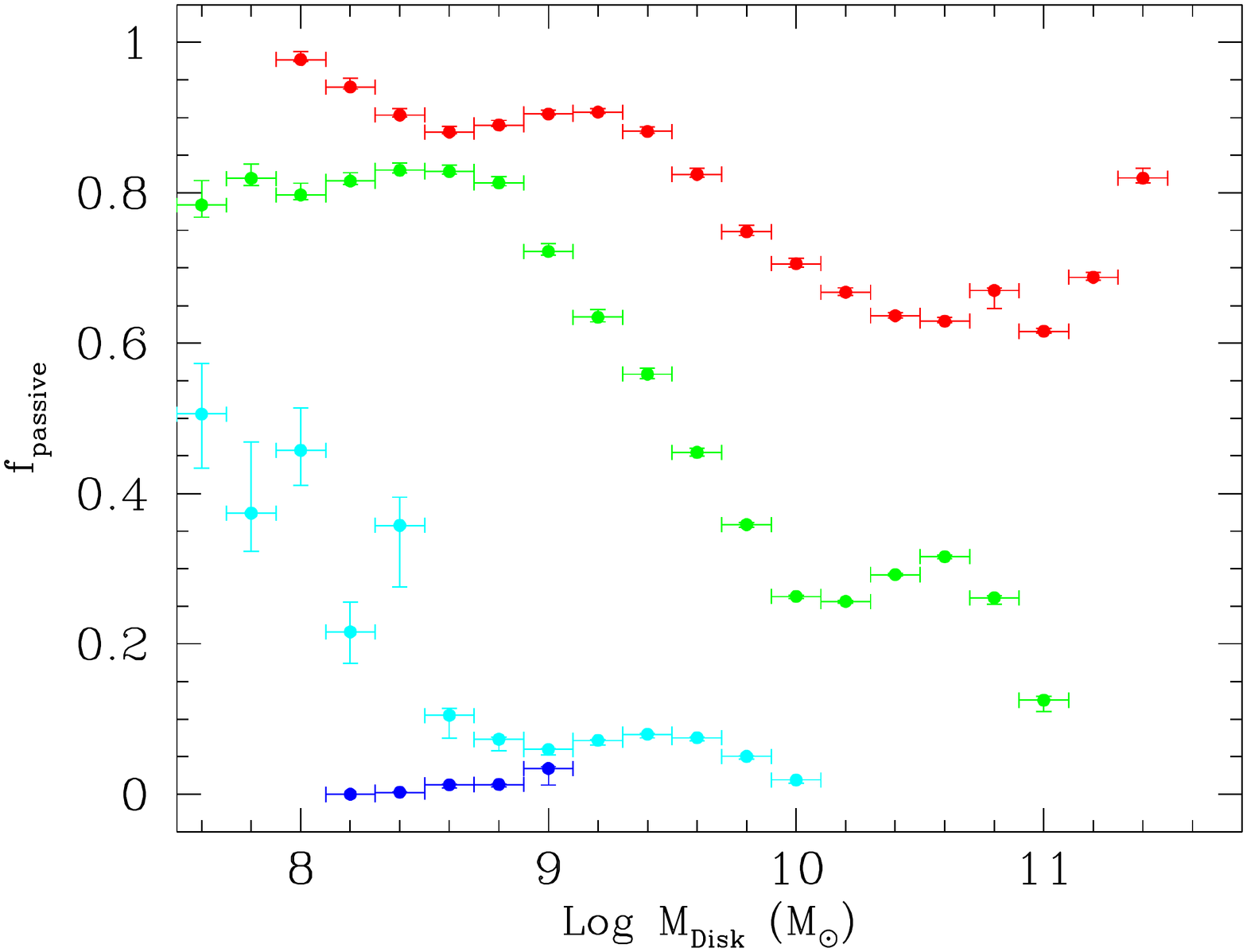}}
\rotatebox{270}{\includegraphics[height=0.33\textwidth]{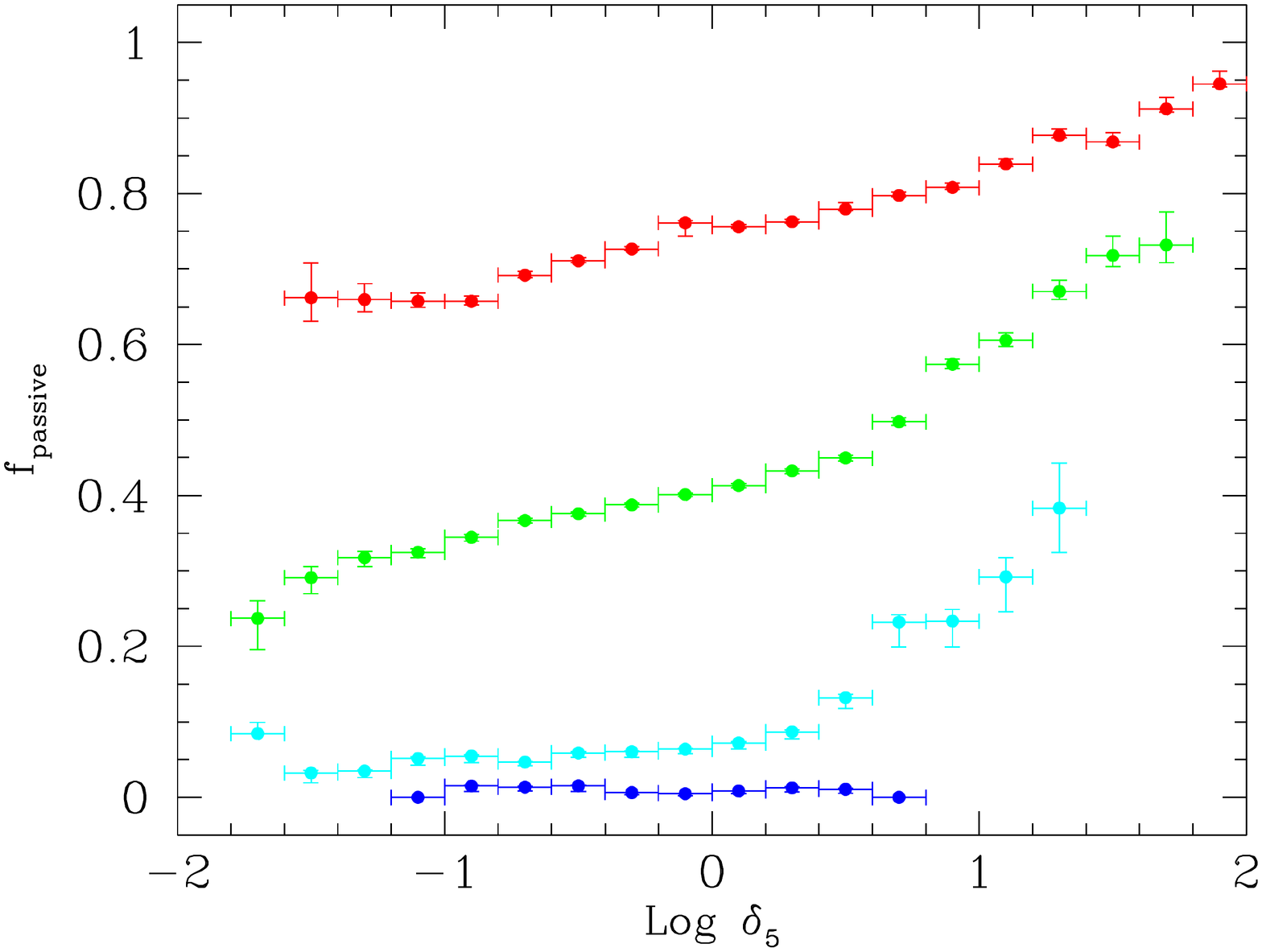}}
\caption{Passive fraction split into stellar mass range as indicated on the top middle panel, for (from {\it top left to bottom right}): halo mass, bulge mass, bulge-to-total stellar mass ratio (B/T), disk mass, and the over-density of galaxies evaluated at the 5th nearest neighbour ($\delta_{5}$). The 1$\sigma$ error on the passive fraction computed via the jack-knife technique, and bin size, are displayed as the error bars. The plots are organised in terms of the amount of variation experienced in the passive fraction across stellar mass ranges at fixed other galaxy properties, from least ({\it top left}) to most ({\it bottom right}) variation. Of particular note in this figure is the level to which the passive fraction at a fixed B/T ratio, disk mass, and $\delta_{5}$ is subdivided by stellar mass range, whereas halo mass and bulge mass show distinctly less variation.}
\end{figure*}

The fact that central and satellite galaxies exhibit different relationships between their passive fractions and various fundamental galaxy properties (i.e. halo mass, total stellar mass, B/T structure) requires that these two classes of galaxies be considered separately. In this paper we will predominantly concentrate on the drivers of quenching for central galaxies in the subsequent sections. However, we are currently preparing a follow up paper specifically related to understanding the quenching of satellite galaxies (Bluck et al. in prep.). 

For central galaxies the dependence on local galaxy over-density is found to be noticeably weaker than that for satellites (see e.g. Peng et al. 2012, Woo et al. 2013). In Fig. 9 we show explicitly the passive fraction - halo mass relation for central only galaxies, and divide this relationship in bins of over-density. We find that there is little variation in this relationship across a wide range in local over-density at a fixed halo mass. This immediately allows us to disregard differential environmental effects as a primary driver of quenching for central galaxies, at least insofar as these apply to differences over and above that of halo mass. Differences in local galaxy over-density produce only a small deviation in the $f_{\rm passive}$ - $M_{*}$ and B/T relations as well, although they do exhibit somewhat larger differences than that of halo mass. Thus, for central galaxies, if one knows the halo mass additional environmental indices are not required to accurately constrain the passive fraction.

In this paper we do not consider the role of galaxy pairs on star formation directly. However, we direct interested readers to other publications from our group (Ellison et al. 2008, Ellison et al. 2011, Scudder et al. 2012, Patton et al. 2013, and Ellison et al. 2013) as well as other references (Darg et al. 2010, Silverman et al. 2011, Hung et al. 2013, Robotham et al. 2013) which together provide a substantial body of evidence which suggests that the interaction of galaxies in pairs, and subsequent merging, can lead to enhanced star formation rates, as well as elevated nuclear activity, bluer colours, and lower metalicities. We also note here that there is some debate as to the final impact of merging and interactions on the global star formation of galaxies, with e.g. Wild et al. (2009) arguing for interactions leading to star formation cessation or quenching. However, even if interaction and merging does ultimately cause galaxies to shut off their star formation and turn red, they cannot fully 
explain the existence of passive galaxies. This is because eventually star formation would re-ignite due to cold gas inflows unless some additional process(es) prevent this from happening (e.g. Bower et al. 2008). In this paper our primary concern is to better constrain and understand the root cause(s) of why passive galaxies exist.

\subsection{Differential effects on the passive fraction for centrals at fixed stellar mass}

The impact on the passive fraction of galaxies from varying stellar mass and environment has been the subject of several papers in the field (e.g. Baldry et al. 2006 in terms of galaxy colour, and more recently Peng et al. 2010 and 2012 in terms of the passive fraction). The fundamental result of this prior work is that both of these properties are found to positively correlate with the passive fraction, and moreover, that the effects of stellar mass and environment are separable, and hence presumably of different underlying physical origin. In this section we investigate the effects of varying stellar mass range at fixed galaxy properties (halo mass, bulge mass, B/T, $\delta_{5}$, disk mass) for central galaxies. The basic idea behind this approach is that if a variable experiences little variation in terms of its passive fraction dependence with the other variables considered, e.g. stellar mass in this case, then it is shown to be more fundamental than if it shows a large amount of variation. 

The volume corrected passive fraction (defined in eq. 7) varies as a strong function of total stellar mass (see Fig. 8, middle panel) for central galaxies, after an initial threshold at $M_{*} \sim 10^{10} M_{\odot}$. In Fig. 10 we show the passive fraction relation with (from top left to bottom right) halo mass, bulge mass, B/T, disk mass, and $\delta_{5}$. In each case we split these relations into four ranges of stellar mass, from $M_{*} = 10^{8} M_{\odot}  - 10^{12} M_{\odot}$. This allows us to assess the effect of varying the stellar mass whilst keeping other galaxy properties fixed, and determine which, if any, are primarily unaffected by differences in stellar mass. The ordering in Fig. 10 is such that the least variation with stellar mass is shown at the top left, and increasing variation progresses down to the bottom right. Initially this ordering is determined visually. Obviously, not all stellar mass ranges are allowable at each binning of the other properties. 
For example, we cannot probe an arbitrary range in bulge mass, disk mass, or halo mass for centrals in terms of their stellar masses. However, there is sufficient variation in these properties as a function of stellar mass that there is almost always at least two stellar mass ranges possible to probe at each binning of the other variables, and frequently more than this.

By visual inspection of Fig. 10, we find that there is less variation across stellar mass ranges at a fixed halo mass, than for any other galaxy property. However, this is at least partly due to the restrictive range of halo masses allowable at a fixed stellar mass for centrals (see back to Fig. 4 for an illustration). This result is qualitatively similar to Woo et al. (2013) who use this fact to argue for halo mass-quenching (e.g. Dekel et al. 2009) being the most likely physical driver for quenching. We also find that there is fairly little variation at a fixed bulge mass across varying stellar mass ranges, which can lead to an alternative explanation for the quenching mechanism through bulge correlating processes, e.g. AGN feedback. Interestingly, there is huge variation across stellar mass ranges at a fixed B/T stellar mass ratio, indicating that B/T cannot be the primary indicator for the passive fraction. 
This rules out morphological quenching (e.g. Martig et al. 2009) as being a primary driver to the cessation in star formation. The variation in $f_{\rm passive}$ at a fixed B/T with stellar mass range is also contrary to the simplistic view that bulges are red and disks are blue, which would imply that the relative contribution of each (i.e. B/T) would be the primary observable correlator with star formation (see e.g. Driver et al. 2006). Finally, we find a huge variation at fixed disk mass and $\delta_{5}$ with stellar mass, which implies that the existence and extent of the disk component, and differential environmental effects, are to first order unimportant in constraining the passive fraction of central galaxies.

The relative variation with stellar mass at fixed other galaxy properties may be easily read off the plots in Fig. 10, however, it is desirable to be a little more quantitative about this effect. We calculate the maximum difference, $\delta f_{\rm p}^{\rm max}$, between the highest and lowest available range in stellar mass across the entire range of values in the other galaxy properties (i.e. halo mass, bulge mass, B/T, disk mass, $\delta_{5}$). Specifically we calculate:

\begin{equation}
\delta f_{\rm p}^{\rm max} = \delta f_{\rm p}(\alpha,\beta) \big|_{\rm max} = \big( f_{\rm p}(\alpha |_{\beta_{\rm max}}) - f_{\rm p}(\alpha |_{\beta_{\rm min}}) \big)\big|_{\rm max}
\end{equation}

\noindent where $\alpha$ indicates the variable on the horizontal axis (halo mass, bulge mass, etc.) in, e.g., Fig. 10. In this case $\beta$ indicates stellar mass, i.e the range by which we split the initial variable, $\alpha$ (various colours in, e.g., Fig. 10). We show this in its general form because we will investigate the use of different metrics as our $\beta$-variable in the subsequent sections. These values are presented in the first column of Tab. 2. $\delta f_{\rm p}^{\rm max}$ gives the maximum variation at a fixed galaxy property, $\alpha$, from varying another galaxy property, $\beta$, in this case stellar mass. High values indicate that the stellar mass has a significant effect at a fixed property, whereas lower values indicate that stellar mass has a less significant differential effect. The errors on $\delta f_{\rm p}^{\rm max}$ are computed from adding in quadrature the errors on the input $f_{\rm p}$ values, which lead to the maximum difference.

We also consider the total impact of varying stellar mass over the entire range of values of the other galaxy properties in, e.g., Fig. 10. Specifically, we calculate the (normalised) area enclosed between the highest and lowest binnings of variable $\beta$, in this case stellar mass, across the full range in each of the other galaxy properties, $\alpha$:

\begin{eqnarray}
A(\delta f_{\rm p}) = \frac{1}{\Delta \alpha}  \int^{\alpha_{\rm max}}_{\alpha_{\rm min}} \delta f_{\rm p}(\alpha,\beta) \hspace{0.1cm} d\alpha \nonumber \\
= \frac{1}{\Delta \alpha} \int^{\alpha_{\rm max}}_{\alpha_{\rm min}} \big( f_{\rm p}(\alpha |_{\beta_{\rm max}}) - f_{\rm p}(\alpha |_{\beta_{\rm min}}) \big)  \hspace{0.1cm} d\alpha
\end{eqnarray}

\noindent where $\alpha$ and $\beta$ are defined as above. We show this in its most general form because we will investigate the use of different metrics as our $\beta$-variable in the subsequent sections (i.e. halo mass in \S 4.4, and B/T in \S 4.5). We divide by the range in $\alpha$ ($\Delta \alpha$ = $\alpha_{\rm max} - \alpha_{\rm min}$) so that maximum variation, and hence area enclosed, between the highest and lowest bins of $\beta$ across the full range in $\alpha$ has a value of unity. No variation in the passive fraction from varying $\beta$ across the full range in $\alpha$ will lead to a normalised area equal to zero. Note that this is similar to taking a mean of $\delta f_{\rm p}$, however, crucially we do not weight by the number of galaxies in each bin. This is extremely important because we are not attempting to construct a likelihood of any given configuration, instead we are aiming to assess the maximum allowable effect on the passive fraction of varying $\beta$ (in this case stellar mass) 
across $\alpha$ (each of the other variables in our study). In this sense it does not matter if, for example, low B/T values are unlikely at high stellar masses (which happens to be true, see Fig. 2), since we are attempting only to deduce the total impact of varying stellar mass at a $\it fixed$ B/T ratio. The errors on $A(\delta f_{\rm p})$ are evaluated from propagating the individual errors on each $\delta f_{\rm p}$ value through the summation.

\begin{figure*}
\rotatebox{270}{\includegraphics[height=0.33\textwidth]{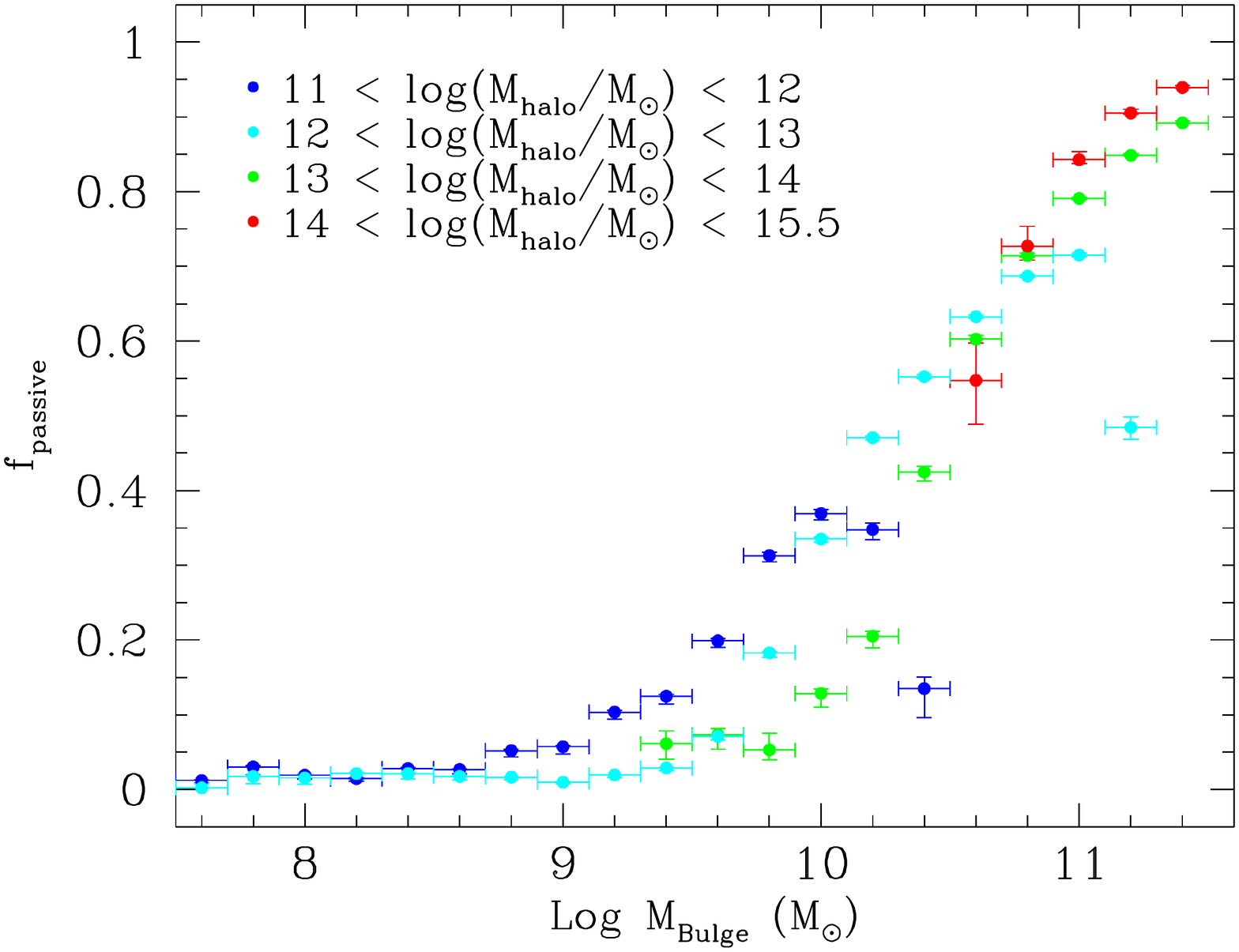}}
\rotatebox{270}{\includegraphics[height=0.33\textwidth]{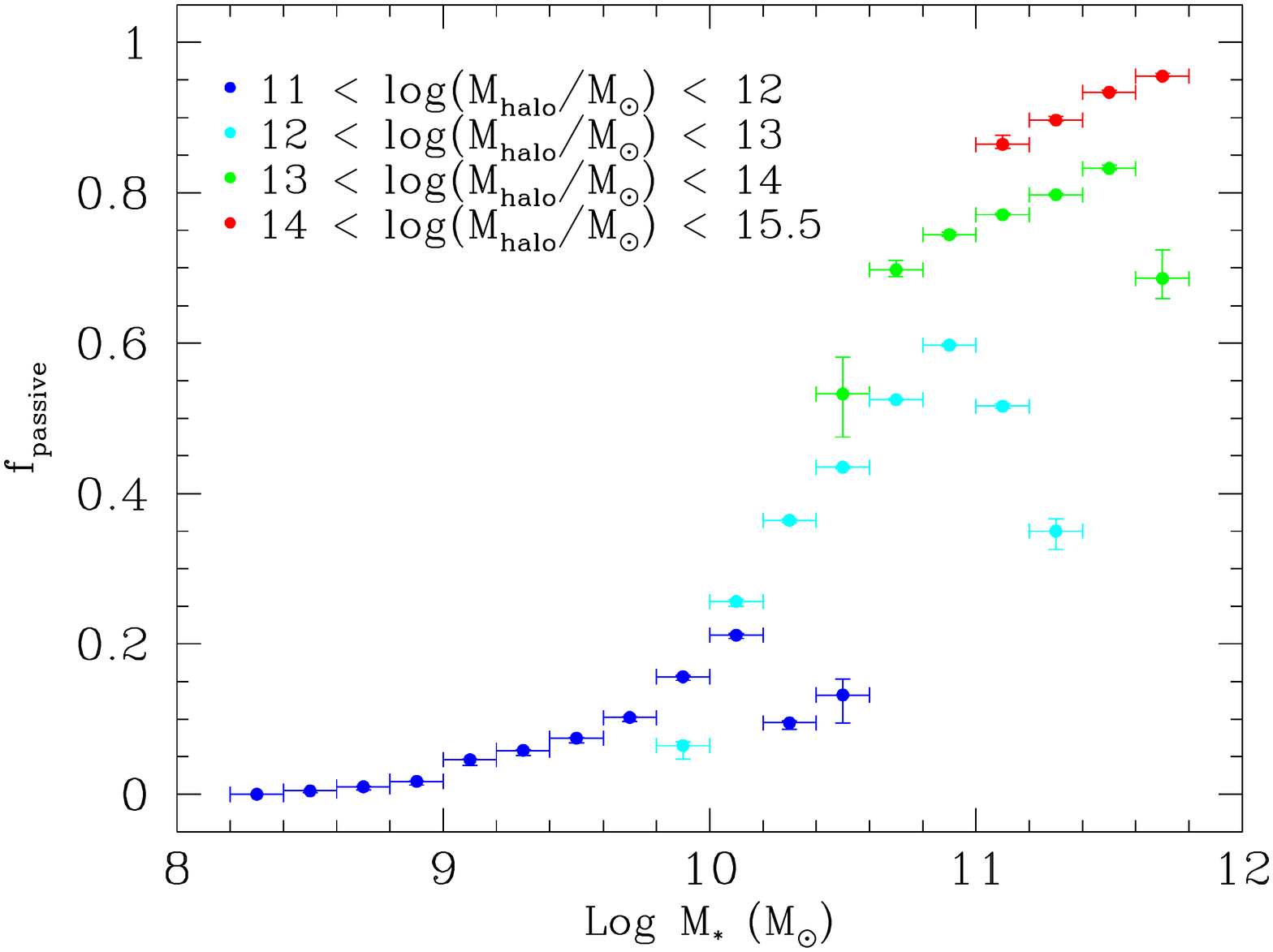}}
\rotatebox{270}{\includegraphics[height=0.33\textwidth]{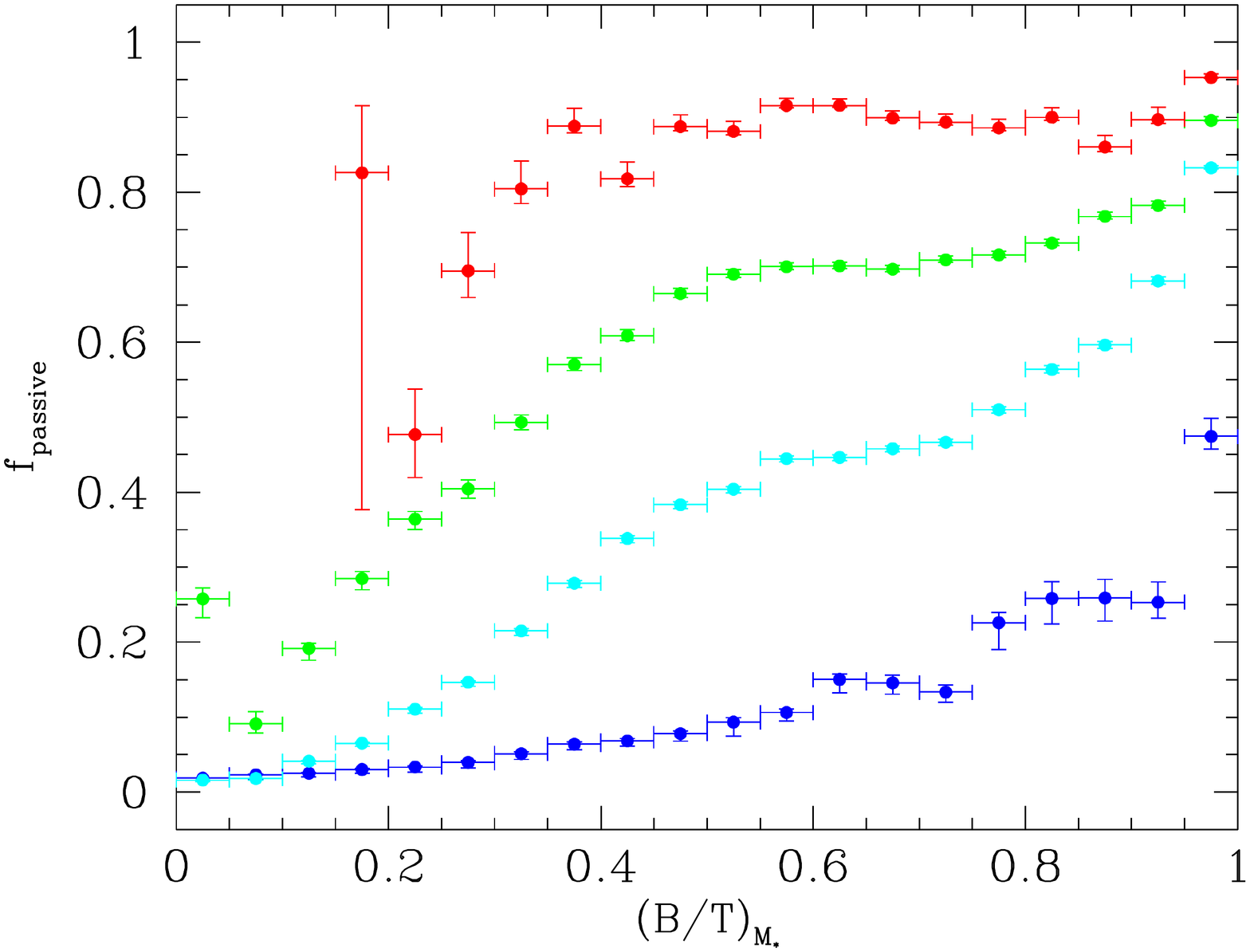}}
\rotatebox{270}{\includegraphics[height=0.33\textwidth]{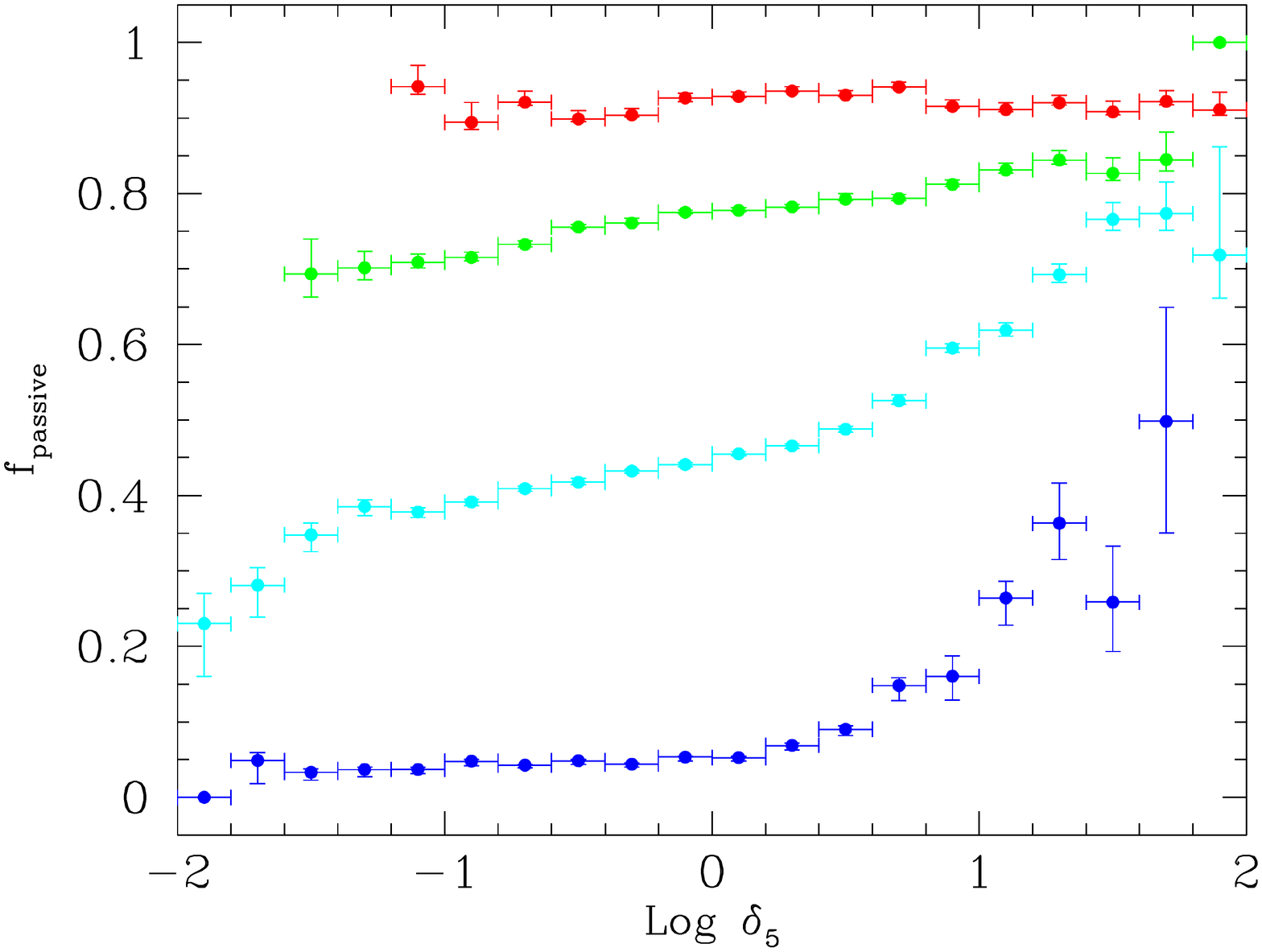}}
\rotatebox{270}{\includegraphics[height=0.33\textwidth]{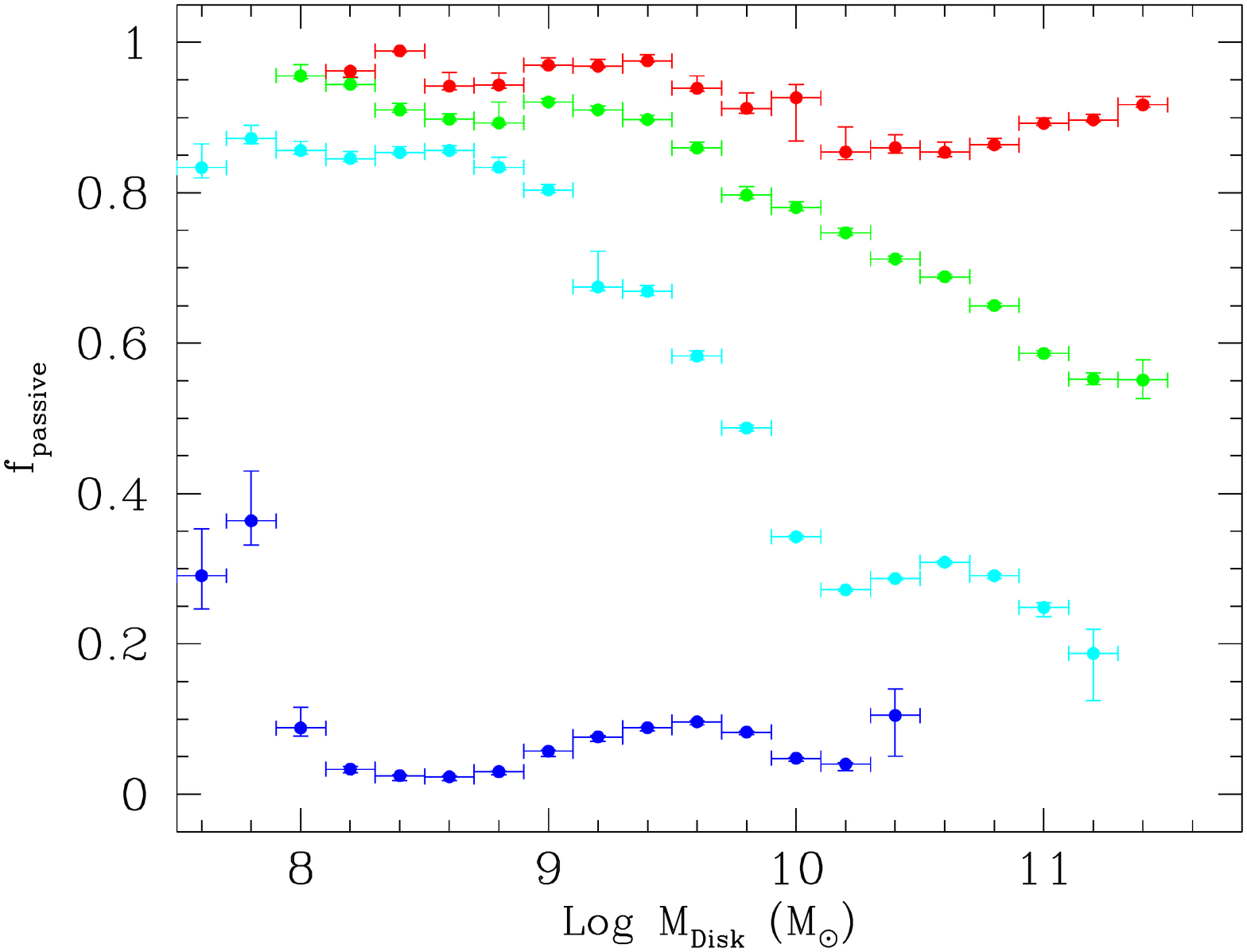}}
\caption{Passive fraction split into halo mass range as indicated on the top left panel, for (from {\it top left to bottom right}): bulge mass, stellar mass, bulge-to-total stellar mass ratio (B/T), the over-density of galaxies evaluated at the 5th nearest neighbour ($\delta_{5}$), and disk mass. The 1$\sigma$ error on the passive fraction computed via the jack-knife technique, and bin size, are displayed as the error bars. The plots are organised in terms of the amount of variation experienced in the passive fraction across halo mass ranges at fixed other galaxy properties, from least ({\it top left}) to most ({\it bottom right}) variation. Of particular note in this figure is the level to which the passive fraction at a fixed B/T ratio, disk mass, and $\delta_{5}$ is subdivided by halo mass range, whereas bulge mass and total stellar mass show distinctly less variation, with bulge mass in particular showing very little differential effect from varying halo mass.}
\end{figure*}

\begin{figure*}
\rotatebox{270}{\includegraphics[height=0.33\textwidth]{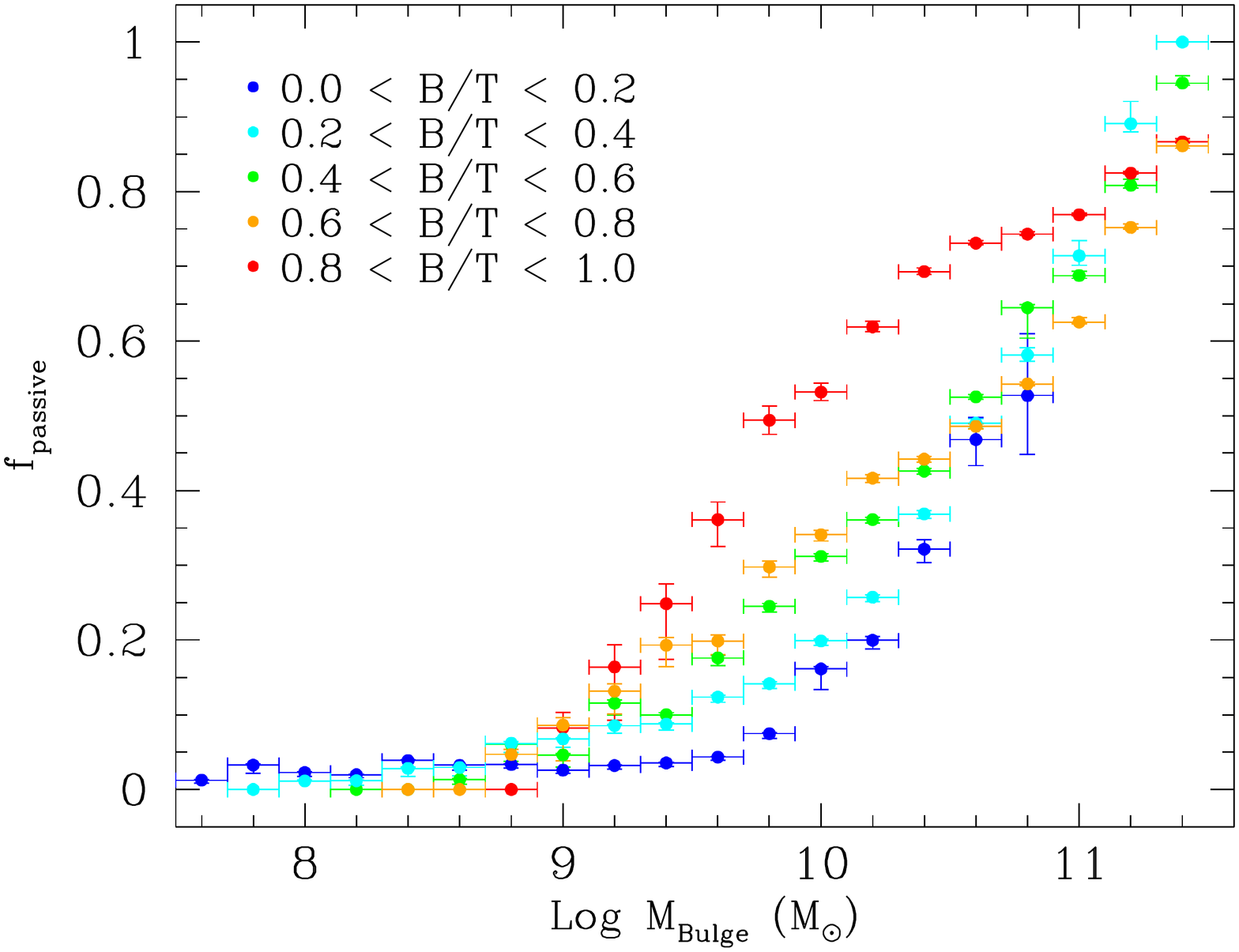}}
\rotatebox{270}{\includegraphics[height=0.33\textwidth]{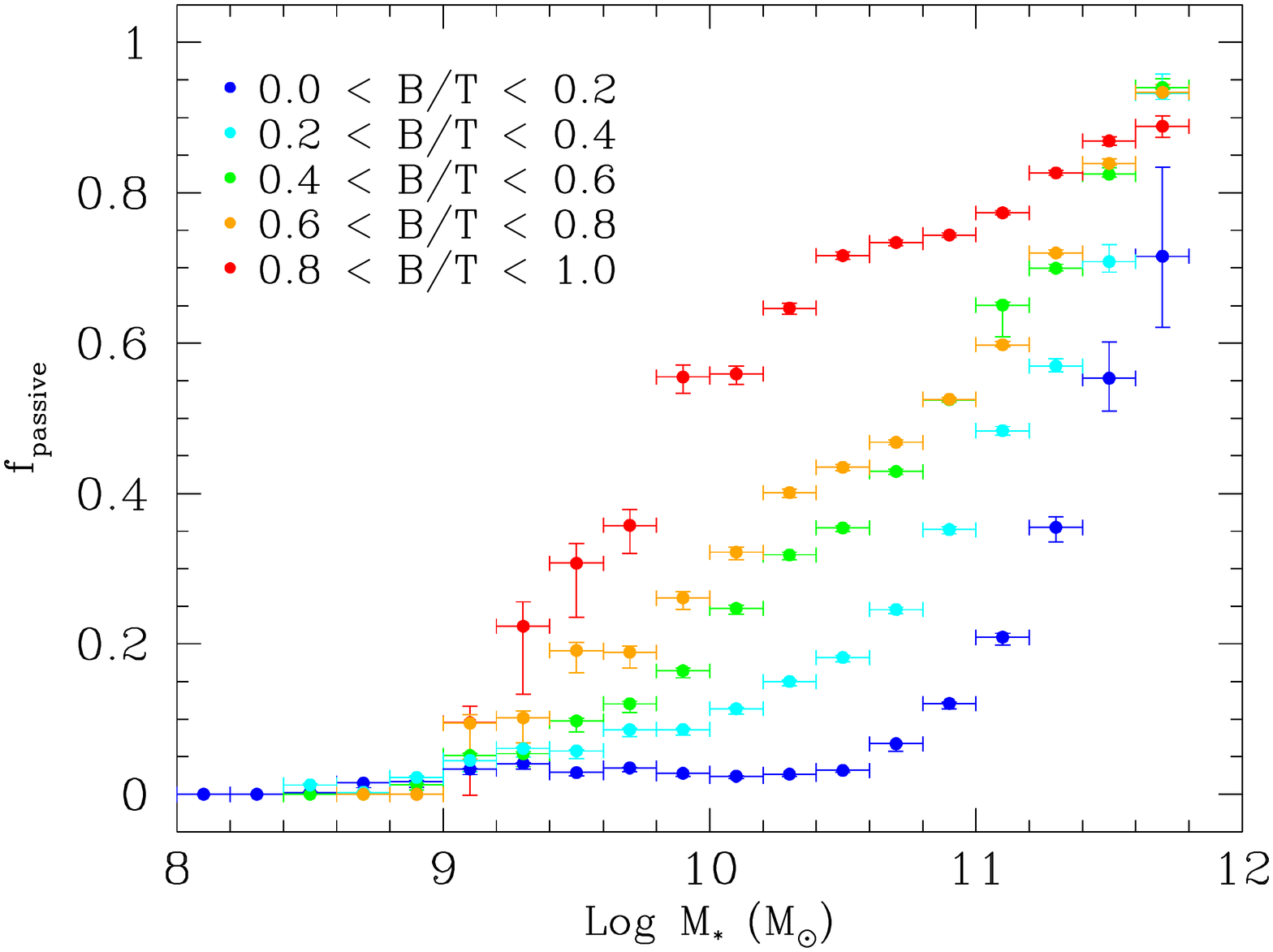}}
\rotatebox{270}{\includegraphics[height=0.33\textwidth]{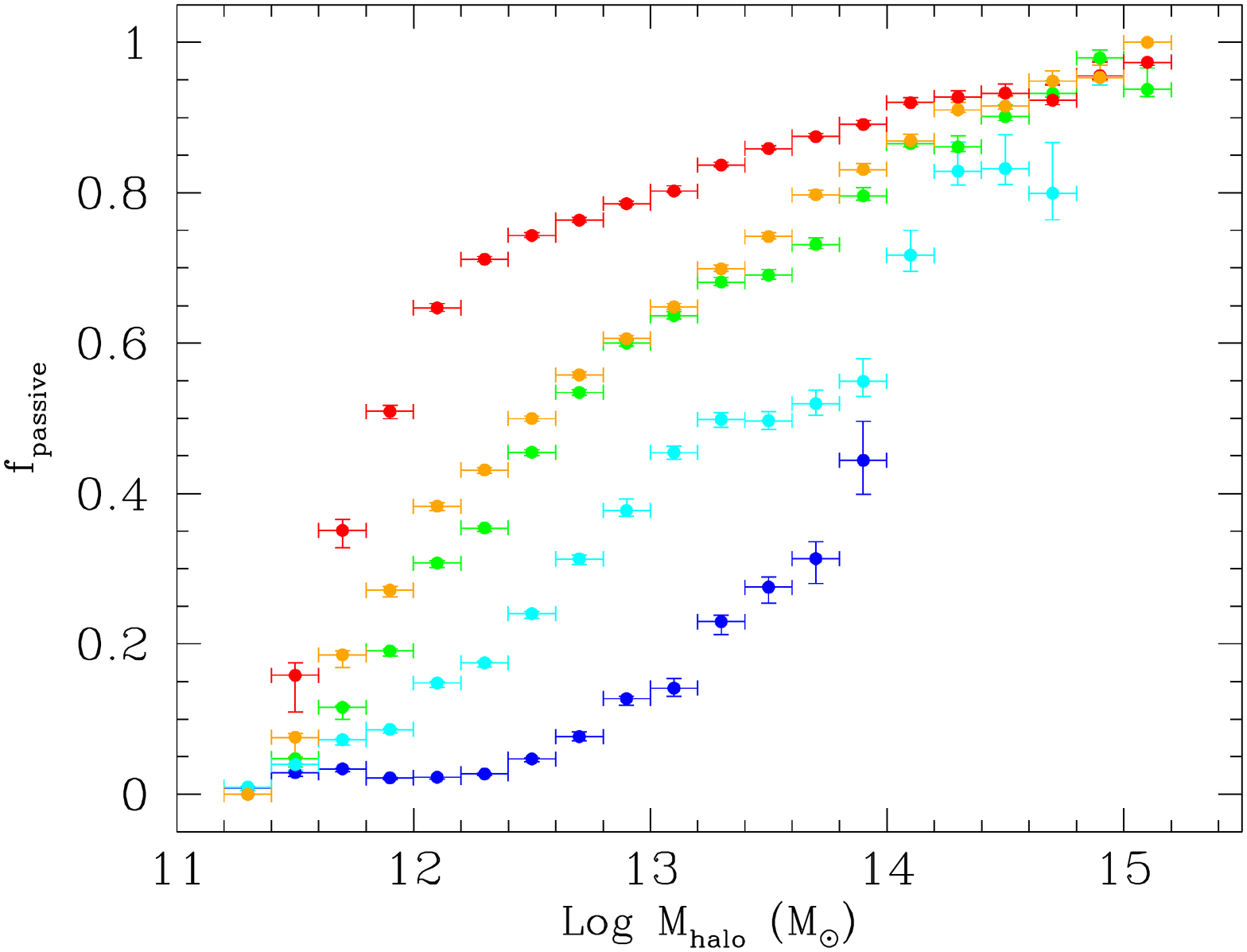}}
\rotatebox{270}{\includegraphics[height=0.33\textwidth]{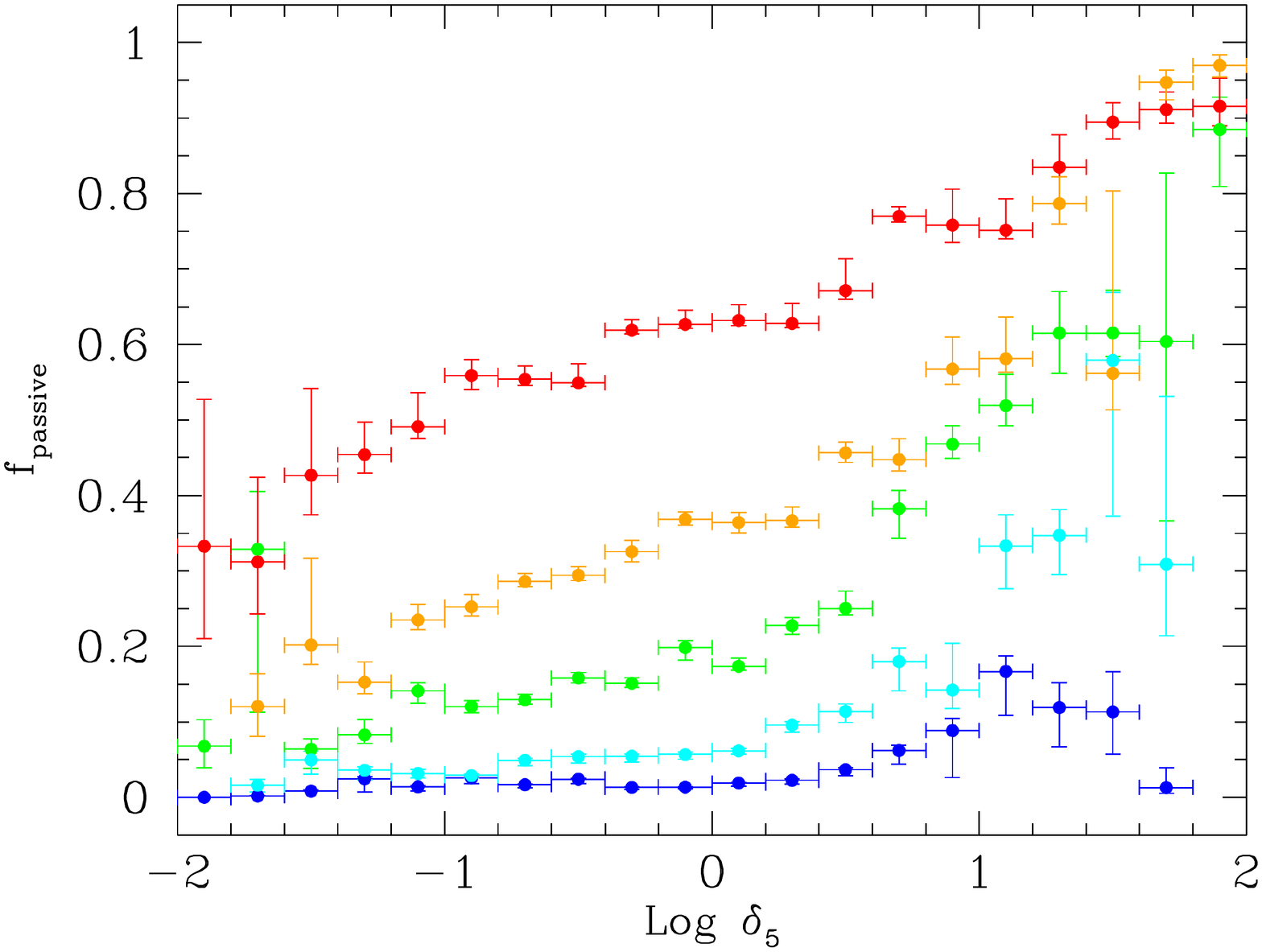}}
\rotatebox{270}{\includegraphics[height=0.33\textwidth]{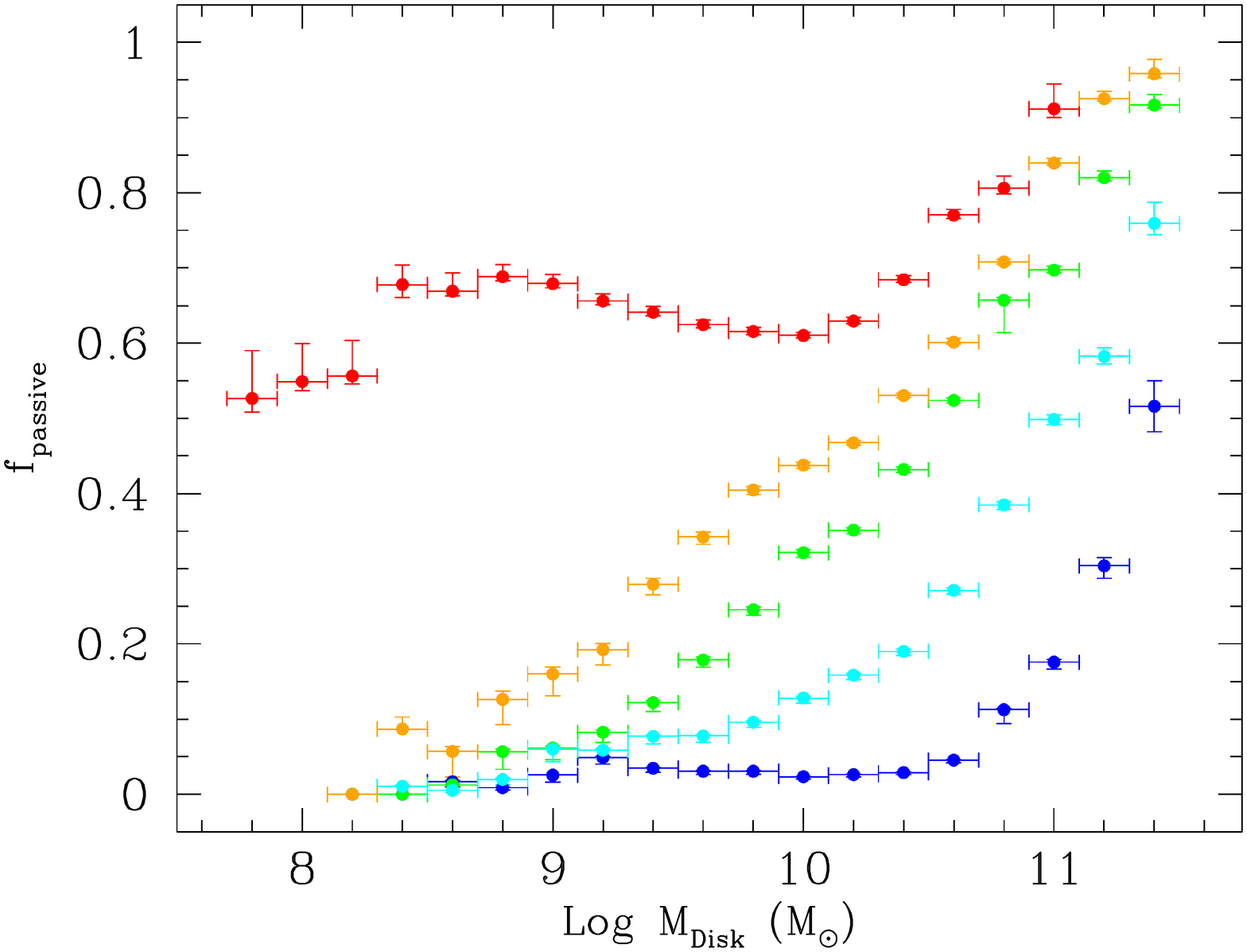}}
\caption{Passive fraction split into bulge-to-total stellar mass ratio (B/T) range as indicated on the top left panel, for (from {\it top left to bottom right}): bulge mass, stellar mass, halo mass, the over-density of galaxies evaluated at the 5th nearest neighbour ($\delta_{5}$), and disk mass. The 1$\sigma$ error on the passive fraction computed via the jack-knife technique, and bin size, are displayed as the error bars. The plots are organised in terms of the amount of variation experienced in the passive fraction across B/T ranges at fixed other galaxy properties, from least ({\it top left}) to most ({\it bottom right}) variation. It is clearly evident that all of the metrics are subdivided by B/T structure, but by far the least variation is seen with bulge mass. Interestingly, both halo mass and total stellar mass show a drastic variation in their passive fraction relationships with varying B/T structure.}
\end{figure*}

In general, if $\delta f_{\rm p}^{\rm max}(\alpha,\beta)$ $>$ $\delta f_{\rm p}^{\rm max}(\beta,\alpha)$, this implies that there is a greater maximum difference in the passive fraction at a fixed $\alpha$ by varying $\beta$ than maximum difference at a fixed $\beta$ by varying $\alpha$. Thus, we would conclude that $\beta$ is more constraining of the passive fraction than $\alpha$. Likewise, if $A(\delta f_{\rm p}(\alpha,\beta))$ $>$ $A(\delta f_{\rm p}(\beta,\alpha))$, this implies that there is greater variation across the range in $\alpha$ by varying $\beta$ than across the range in $\beta$ by varying $\alpha$. Thus, we would also conclude that $\beta$ is a greater single correlator to the passive fraction than $\alpha$. This gives the general prescription for using these statistics to assess the differential effects on the passive fraction of each variable in terms of the remaining variables.

In the first column of Tab. 3 we show the results for $A(\delta f_{\rm p})$ applied to the configurations in Fig. 10. For both $\delta f_{\rm p}^{\rm max}$ and $A(\delta f_{\rm p})$ we arrange the variables into order of how much variation they experience with stellar mass, with equivalent results. From least to most variation these are: halo mass, bulge mass, B/T, disk mass, $\delta_{5}$ (the same ordering as in Fig. 10). This suggests that B/T, disk mass, and $\delta_{5}$ are not strong contenders for the most constraining single variable of $f_{\rm passive}$. However, both halo mass and bulge mass show strong correlations with the passive fraction at a fixed stellar mass, indicating that they are worthy of further study. Since there are strong inter-correlations between stellar mass and halo mass, it is logical to explore how things differ at a fixed halo mass instead of stellar mass. We explore this in the next section.

\subsection{Differential effects on the passive fraction for centrals at fixed halo mass}

Following the prescription of the previous section, but now splitting by halo mass instead of stellar mass, we assess the impact on the passive fraction of each of the remaining variables in turn in Fig. 11. This figure is organised (from top left to bottom right) in terms of how much variation in the passive fraction at a given observable is engendered through varying the halo mass. We find that there is less variation with halo mass at a fixed bulge mass than for any other observable considered, and note that varying the halo mass by up to three orders of magnitude does not significantly affect the passive fraction at a fixed bulge mass. This implies that the passive fraction is tightly coupled to the bulge mass, and perhaps further suggests that bulge mass is more important than halo mass in assigning the passive fraction. 
Interestingly, where there is variation in the passive fraction at a fixed bulge mass through varying the halo mass, we find that higher mass halos harbour {\it less} passive galaxies for their bulge mass, in the range $M_{\rm bulge} \sim 10^{9} - 10^{10} M_{\odot}$. This suggests that whatever mechanism, related to the bulge, which causes quenching is more efficient in lower mass halos; and further rules out halo mass quenching as a primary driver to star formation cessation at these masses. The passive fraction at a fixed stellar mass is slightly more affected by varying the halo mass than at a fixed bulge mass, however, there is still a tight correlation evident here. 

For the other variables (B/T, $\delta_{5}$, disk mass) we find a huge variation in the passive fraction at fixed values of these parameters across the ranges in halo mass explored. This result reflects the result with stellar mass shown in the previous section (see Fig. 10), and strongly suggests that the morphology, disk mass, and local density of central galaxies is sub-dominant to knowledge of the halo, bulge, or stellar mass in determining the passive fraction.

As in the previous section, we compute the maximum difference, $\delta f_{\rm p}^{\rm max}$ (defined in eq. 8), between the extremes of the binnings, but this time for halo mass instead of stellar mass. We also compute the area enclosed between the extremes of the passive fraction, $A(\delta f_{\rm p})$ (defined in eq. 9), also split by halo mass. The results for these statistics are shown in the second column of Tab. 2 \& 3, respectively. We find that bulge mass is statistically slightly more constraining of the passive fraction than stellar mass, when split by halo mass. Furthermore, it is striking how much variation there is in the passive fraction at a fixed B/T ratio with halo mass, further demonstrating that morphology (or structure) is not the primary correlator of the quenching process.

Comparing stellar mass split by halo mass (Fig. 11, top middle panel) to halo mass split by stellar mass (Fig. 10, top left panel), we see that there is more variation in the former than the latter (see Tab. 2 \& 3 for the statistics). This implies that halo mass is more constraining of the passive fraction than stellar mass (as has been argued for in Woo et al. 2013). However, since we have also determined that bulge mass is a better single correlator than stellar mass, we must now investigate which is dominant out of the halo and the bulge.

\subsection{Differential effects on the passive fraction for centrals at fixed B/T structure}

We have already established that B/T structure is not the primary driver of the passive fraction, due to the large variation in the $f_{\rm passive}$ - B/T relation when split by stellar or halo mass (see Figs. 10 \& 11). However, we have not yet explored how varying structure at a fixed stellar mass or halo mass affects the passive fraction. In Fig. 12 we show directly the effect of varying the B/T ratio at fixed other galaxy properties (for bulge mass, stellar mass, halo mass, $\delta_{5}$, disk mass) on the passive fraction. Perhaps surprisingly, we observe a huge variation in the passive fraction across different B/T morphologies at a fixed stellar mass and at a fixed halo mass. This implies that neither stellar mass nor halo mass alone can constrain the passive fraction accurately. In fact all of the metrics we explore show some significant variation with galaxy structure in terms of their passive fractions. However, clearly the least variation seen is for bulge mass, and this motivates considering the 
bulge as the primary indicator of star formation quenching. This notwithstanding, the fact that there is some noticeable variation at a fixed bulge mass with B/T range further suggests that the disk mass may be a significant secondary correlator to the passive fraction (even though as a primarily correlator it always performs poorly, see Figs. 10 - 12 and Tab. 2 \& 3). We investigate this possibility explicitly in the next section.

We compute the maximum difference in the passive fraction, $\delta f_{\rm p}^{\rm max}$, between extremes of B/T range, and further compute the area enclosed between these extremes $A(\delta f_{\rm p})$ (see eq. 8 \& 9 for the definitions) for each of the other galaxy observables. We show these results in the third column of Tab. 2 \& 3, respectively. For both of these metrics, we find that there is significantly less variation in the passive fraction across B/T values at a fixed bulge mass, than for stellar mass or halo mass. Further, halo mass and stellar mass both exhibit statistically less variation than for disk mass and $\delta_{5}$ across differing B/T ranges. Applying a structural decomposition to the passive fraction correlations with the various masses (bulge, stellar, halo) has revealed a profound connection between the bulge and the passive fraction, which transcends that of both stellar and halo mass. This motivates a redefinition of `stellar mass-quenching' (Peng et al. 2010, 2012) as `bulge 
mass-quenching', since bulge mass is more constraining of the passive fraction than stellar mass. Moreover, `halo mass-quenching' (as in Woo et al. 2013) is also found to be sub-dominant to `bulge mass-quenching'. In the next section we compare directly the role of these three masses in driving the passive fraction.

In principle we could go on and calculate the variation in each of the rest of the variables, and thus show completely how much each varies as a function of the remainder in terms of the passive fraction (an exercise we have done but do not show for the sake of brevity). This is ultimately superfluous, however, since we have already shown that $\delta_{5}$ is a very poor correlator with the passive fraction for centrals (see. Fig. 9), and bulge mass and disk mass are represented in B/T and total stellar mass already.

\subsection{Bulge Mass is King}

\begin{figure*}
\rotatebox{270}{\includegraphics[height=0.49\textwidth]{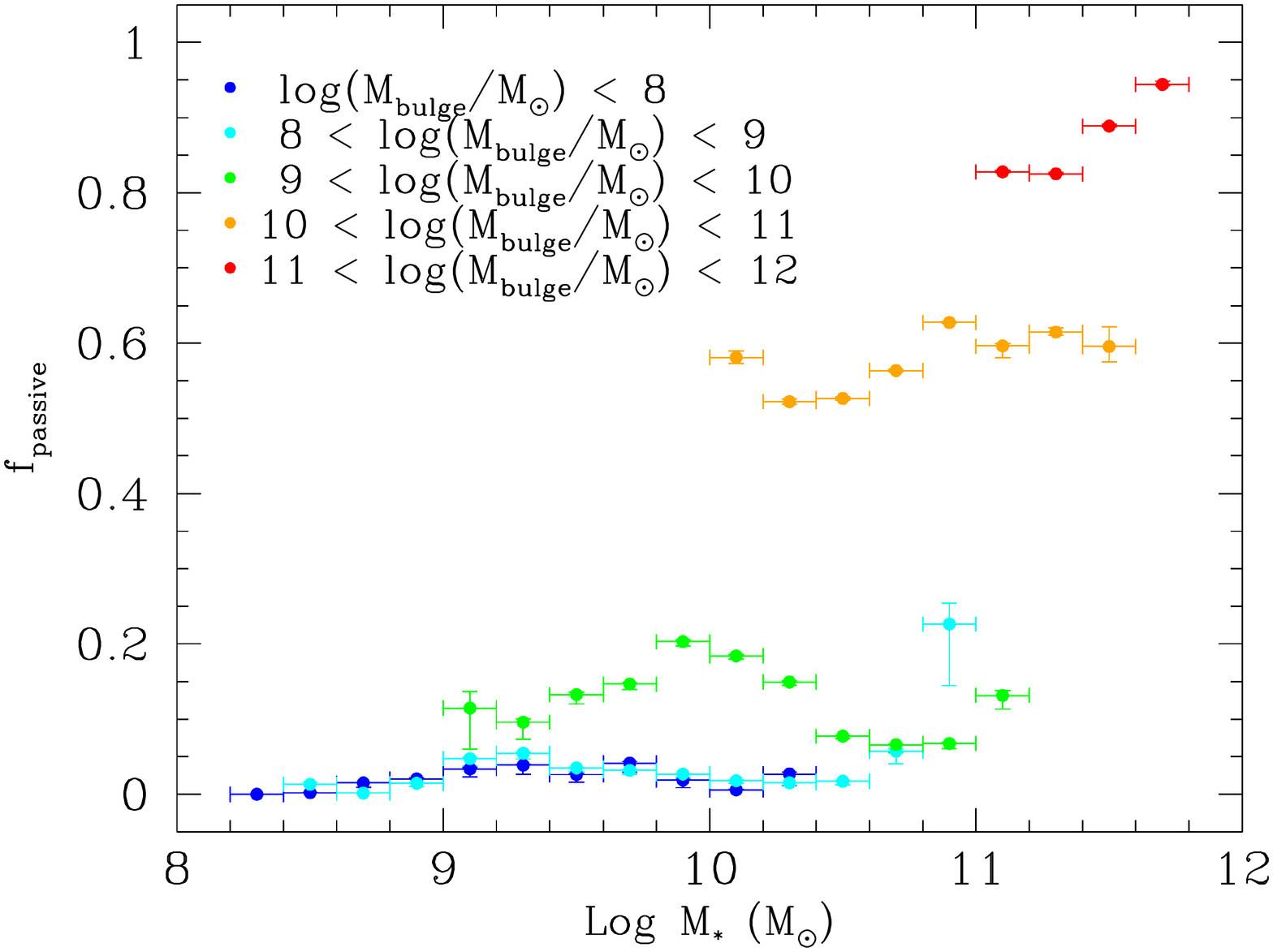}}
\rotatebox{270}{\includegraphics[height=0.49\textwidth]{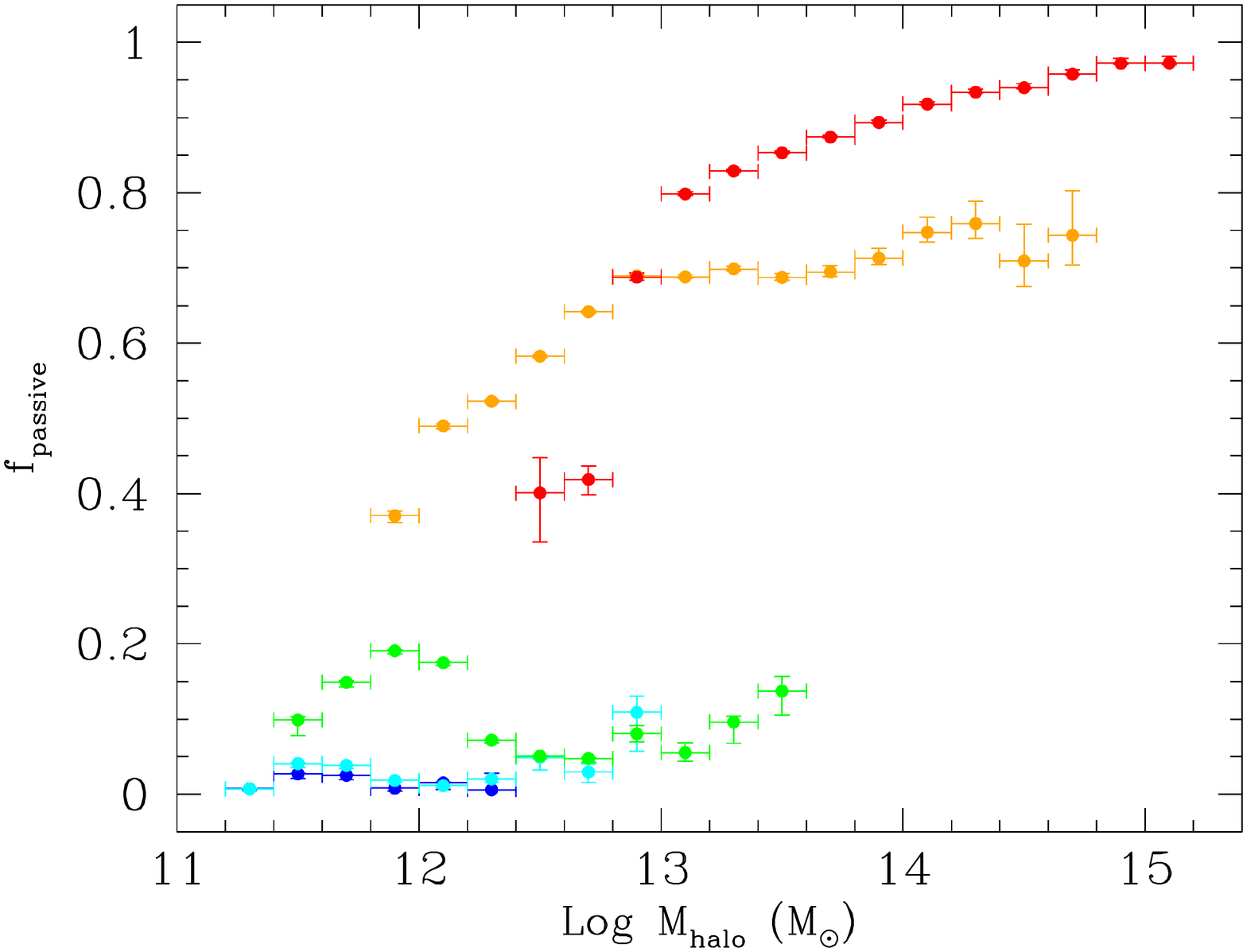}}
\caption{Passive fraction vs. stellar mass ({\it left panel}) and halo mass ({\it right panel}) both divided by bulge mass range as indicated on the stellar mass plot. The 1$\sigma$ error on the passive fraction computed via the jack-knife technique, and bin size, are displayed as the error bars. Note the large variation in the passive fraction across bulge mass ranges at both a fixed stellar mass and halo mass. This can be compared to the variation in the passive fraction at a fixed bulge mass with stellar mass (Fig. 10 {\it top middle panel}) and halo mass (Fig. 11 {\it top left panel}). There is distinctly less variation in the passive fraction at a fixed bulge mass with varying stellar and halo mass ranges than at a fixed halo or stellar mass with varying bulge mass ranges.}
\end{figure*}

\begin{figure*}
\rotatebox{270}{\includegraphics[height=0.49\textwidth]{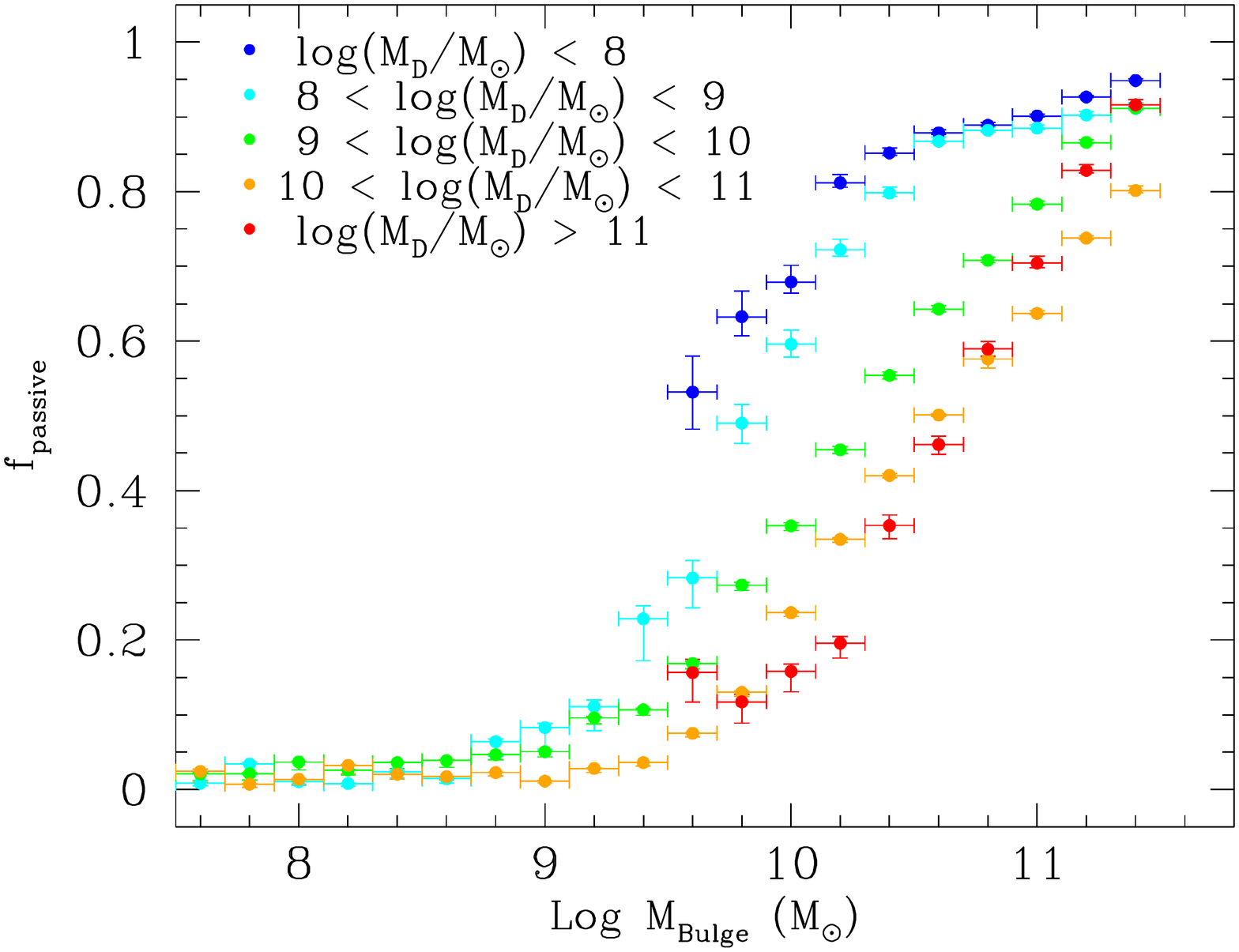}}
\rotatebox{270}{\includegraphics[height=0.49\textwidth]{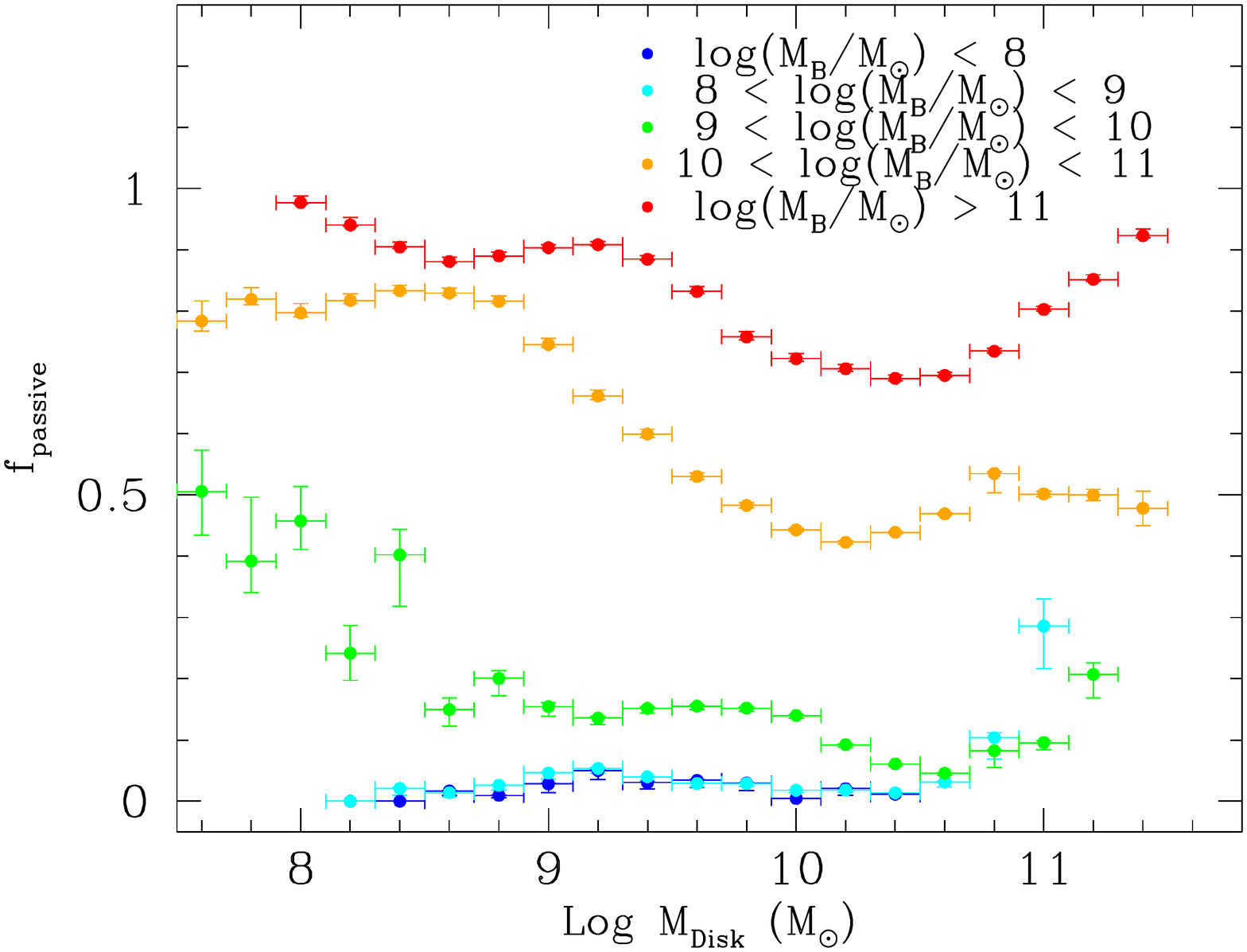}}
\caption{{\it Left panel}: Passive fraction vs. bulge mass divided by disk mass range, as indicated on the plot. {\it Right panel}: Passive fraction vs. disk mass divided by bulge mass range, as indicated on the plot. The 1$\sigma$ error on the passive fraction computed via the jack-knife technique, and bin size, are displayed as the error bars. Note that there is larger variation in the passive fraction at a fixed disk mass with varying bulge mass than at a fixed bulge mass with varying disk mass, which demonstrates that bulge mass is more constraining of the passive fraction than disk mass. However, there is still significant variation in the passive fraction at a fixed bulge mass due to variation in disk mass, which implies that disk mass is a significant secondary correlator to the passive fraction.}
\end{figure*}

In the previous three sub-sections we have explored the differential effects of intrinsic and environmental correlators to the passive fraction for central galaxies at fixed stellar mass, halo mass, and B/T. Generally, bulge mass, halo mass, and stellar mass show tighter correlations with less variation in the other variables, compared to B/T, disk mass, and $\delta_{5}$ (see Figs. 10 - 12, and Tab. 2 \& 3). Out of the top three most constraining variables, the different masses show varying tighter dependencies in terms of the other variables with which they are split. Since there are strong inter-relations between all of the variables in our sample (with some examples shown in Figs. 2 - 4), attempting to combine the results into some form of average is difficult, and ultimately poorly motivated. As an alternative, we posit that it is the maximum variation a given variable experiences in its passive fraction relationship with any of the remaining variables which reveals how fundamental that first variable is.
 
Accordingly, in Tab. 2 \& 3 we rank each of the variables (bulge mass, stellar mass, halo mass, B/T, disk mass, and $\delta_{5}$) according to how much variation they experience in the passive fraction with the most dividing of the remaining variables.

For the top three correlators to the passive fraction (bulge mass, stellar mass, and halo mass) we find that the structural parameter, B/T, provides the greatest variation in the area contained within the extremes passive fraction (see Tab. 3). Ordering by the variation with B/T yields that bulge mass is the single tightest correlator to the passive fraction, with stellar mass, then halo mass as the next in line. A similar result is found for the maximum difference in the passive fraction (see Tab. 2). Note that for total stellar mass and halo mass this gives a slightly different ordering than with the direct comparison (see \S 4.4). This is probably due to the fact that they are similarly constraining of the passive fraction, with only slight differences, and the error on the ranking methods lead to this discrepancy. Bulge mass, however, does significantly better at constraining the passive fraction than both halo and total stellar mass. All three of these masses correlate much stronger than B/T to 
the passive fraction, which itself shows the greatest variation across differing halo and stellar masses. Disk mass and $\delta_{5}$ are both very poorly constraining of the passive fraction, showing large variation with stellar mass, halo mass, and B/T. 

To test explicitly the dominance of bulge mass over stellar mass and halo mass in constraining the passive fraction, we show $f_{\rm passive}$ vs. stellar mass (left panel) and halo mass (right panel) divided into ranges of bulge mass in Fig. 13. There is a large amount of variation in the passive fraction across varying bulge mass ranges at both fixed stellar mass and halo mass. In fact the relationships between passive fraction and stellar and halo mass are quite flat at fixed bulge mass. Comparing this to the variation in the passive fraction at a fixed bulge mass with varying stellar mass (Fig. 10 top middle panel) and halo mass (Fig. 11 top left panel) reveals that there is dramatically more variation in the passive fraction with varying bulge mass at fixed halo and stellar mass than the other way around. The $f_{\rm passive}$ - $M_{\rm bulge}$ relation remains very tight and steep even when viewed in narrow bins of both stellar mass and halo mass. 

Statistically, we find that the total area enclosed by the passive fraction from the extremes of varying bulge mass across the full range in halo mass is $A(\delta f_{\rm p})$ = 0.31$^{+0.02}_{-0.02}$, with a maximum difference of $\delta f_{\rm p}^{\rm max}$ = 0.74$^{+0.02}_{-0.02}$. For stellar mass varied by bulge mass range we find an area enclosed by the passive fraction of $A(\delta f_{\rm p})$ = 0.24$^{+0.02}_{-0.01}$, with a maximum difference of $\delta f_{\rm p}^{\rm max}$ = 0.70$^{+0.02}_{-0.01}$. Comparing these values to the values for bulge mass split by halo mass ($A(\delta f_{\rm p})$ = 0.10$^{+0.01}_{-0.01}$; $\delta f_{\rm p}^{\rm max}$ = 0.42$^{+0.02}_{-0.01}$), and stellar mass ($A(\delta f_{\rm p})$ = 0.10$^{+0.01}_{-0.01}$; $\delta f_{\rm p}^{\rm max}$ = 0.48$^{+0.02}_{-0.03}$), reveals that variation at a fixed bulge mass with halo mass and stellar mass leads to significantly smaller areas contained and lower maximum differences in the passive fraction than with varying bulge mass at 
fixed stellar and halo masses. Thus, we conclude that 
whatever process(es) give rise to the transition from the blue cloud to the red sequence in central galaxies are coupled primarily to bulge mass, over halo mass, stellar mass, and B/T.

Given that we have established that there is least variation in the passive fraction at a fixed bulge mass in terms of the remainder of our observables, we are now in a position to assess what is next most important. At a fixed bulge mass, we have seen that there is some residual widening of the passive fraction relation due to both stellar mass and B/T ratio. This immediately suggests that the disk properties may be the root cause of the variation at a fixed bulge mass. To test this we show the $f_{\rm passive}$ - $M_{\rm bulge}$ relation in Fig. 14 (left panel), split by disk mass. 
There is a larger variation at a fixed bulge mass with varying disk mass than for any other variable (compare to Figs. 10 -- 12, bulge mass plots), implying that disk mass is the next most constraining variable after bulge mass, even though as a contender for the dominant position it performed particularly poorly (see, e.g., bottom right panels of Figs. 11 \& 12). 

Statistically, we find that the area contained within the $f_{\rm passive}$ - $M_{\rm bulge}$ relation sub-divided by disk mass is $A(\delta f_{\rm p})$ = 0.20$^{+0.02}_{-0.02}$, with a maximum difference $\delta f_{\rm p}^{\rm max}$ = 0.62$^{+0.02}_{-0.01}$. These are significantly higher values than for any other variable used to sub-divide the bulge mass - passive fraction relation (see Tab. 2 \& 3, respectively). Thus, we statistically confirm our by-eye evaluation that disk mass engenders the greatest variation in the passive fraction at a fixed bulge mass. Hence, disk mass is the best choice for the second most important indicator of the passive fraction after bulge mass.

On the right hand panel of Fig. 14 we show the $f_{\rm passive}$ - $M_{\rm disk}$ relation split by bulge mass range, and note that there is far more variation with bulge mass at a fixed disk mass than the other way around. Thus, observationally, if one desires to know the likelihood that a given galaxy will be actively forming stars or not, there would be a greater chance of answering correctly if one has the bulge mass as opposed to either the halo mass, total stellar mass, or B/T morphology alone. This is equivalent to attesting that knowledge of the disk, its existence and extent, is sub-dominant to knowledge of the bulge in determining the global star forming properties of the whole system. 
However, once the bulge mass is known, the single most constraining of the remaining variables is actually the disk mass. Interestingly, knowledge of the halo mass is not particularly constraining at all, once the inter-correlation with bulge mass is accounted for; and stellar mass and B/T ratio are only useful insofar as they correlate with the bulge mass, primarily, and disk mass, secondarily.

Therefore, we are led to conjecture that the bulge determines the star forming properties of the whole galaxy, with the specifics of the disk being, to first order, ignorable if one wishes to determine whether or not a galaxy will be forming stars actively or not. This is quite a surprising result. Naively, one might expect that, since most star formation occurs in the disks of regular galaxies, the disk properties (i.e. stellar mass and mass ratio contribution) would play a leading role in determining the global star formation properties of the whole galaxy. However, we find that this is not the case. This notwithstanding, at a fixed bulge mass the disk mass does provide the greatest perturbation upon the dominant relation of $f_{\rm passive}$ - $M_{\rm bulge}$ (Fig 14, left panel). This is not simply a reflection of stellar mass being the second most constraining variable (see Tabs. 2 \& 3), because an increase in total stellar mass is found to cause an increase in the passive fraction, yet an 
increase in total stellar mass {\it at a fixed bulge mass} actually decreases the passive fraction. That is, once the dominant trend of the bulge mass - passive fraction relation is accounted for, further increase in stellar mass increases the SFRs of galaxies due to increasing their disk components. Thus, if bulge mass is known, the next most significant correlator is presence of disk (which can be more constructively thought of as B/T ratio or disk mass than total stellar mass). This does not cause a conflict with the rankings established above since they are an ordering for the best {\it single} correlator to the passive fraction, and the disk mass (or B/T) is only useful in conjunction with the bulge mass.

\begin{table}
 \caption{$\delta f_{\rm p}^{\rm max}$. The Maximum Difference Between Extremes of the Passive Fraction (at fixed $\alpha$-variable from varying the $\beta$-variable).}
 \label{tab2}
 \begin{tabular}{@{}ccccc}
  \hline
\hline
Rank$^{*}$ & $\alpha$ \textbackslash  $\beta$ : & $M_{*}$ & $M_{\rm Halo}$ & B/T \\ 
\hline

1 & $M_{\rm Bulge}$ & 0.48$^{+0.02}_{-0.03}$  & 0.42$^{+0.02}_{-0.01}$ & 0.42$^{+0.01}_{-0.01}$ \\\\

2 & $M_{*}$ & - & 0.55$^{+0.03}_{-0.02}$ &  0.68$^{+0.01}_{-0.01}$ \\\\

3 & $M_{\rm Halo}$ & 0.34$^{+0.01}_{-0.01}$ & - & 0.70$^{+0.01}_{-0.01}$ \\\\

4 & B/T & 0.74$^{+0.01}_{-0.07}$ & 0.82$^{+0.02}_{-0.01}$ & - \\\\

5 & $\delta_{5}$ & 0.80$^{+0.01}_{-0.01}$  & 0.90$^{+0.03}_{-0.01}$ & 0.90$^{+0.04}_{-0.04}$ \\\\

6 & $M_{\rm Disk}$ & 0.80$^{+0.01}_{-0.01}$  & 0.96$^{+0.01}_{-0.01}$ &  0.74$^{+0.03}_{-0.01}$ \\

\hline
\end{tabular}

\vspace{0.5cm}

$^{*}$ The ranking is ordered from lowest to highest maximum difference, evaluated for each $\alpha$-variable from the most dividing of the $\beta$-variables (see \S 4.6).

\end{table}

\begin{table}
 \caption{$A(\delta f_{\rm p})$. The Area Enclosed Between Extremes of the Passive Fraction (at fixed $\alpha$-variable from varying the $\beta$-variable).}
 \label{tab3}
 \begin{tabular}{@{}cccccc}
  \hline
\hline
Rank$^{*}$ & $\alpha$ \textbackslash  $\beta$ : & $M_{*}$ & $M_{\rm Halo}$ & B/T \\ 
\hline

1 & $M_{\rm Bulge}$ & 0.10$^{+0.01}_{-0.01}$  & 0.10$^{+0.01}_{-0.01}$ & 0.16$^{+0.02}_{-0.02}$  \\\\

2 & $M_{*}$ & -- & 0.13$^{+0.01}_{-0.01}$ & 0.32$^{+0.02}_{-0.03}$  \\\\

3 & $M_{\rm Halo}$ & 0.07$^{+0.01}_{-0.01}$ & -- & 0.39$^{+0.02}_{-0.02}$ \\\\

4 & B/T & 0.46$^{+0.02}_{-0.02}$ & 0.62$^{+0.03}_{-0.04}$ & -- \\\\

5 & $\delta_{5}$ & 0.56$^{+0.02}_{-0.02}$  & 0.68$^{+0.03}_{-0.03}$ & 0.59$^{+0.05}_{-0.03}$ \\\\

6 & $M_{\rm Disk}$ & 0.57$^{+0.02}_{-0.02}$  & 0.76$^{+0.02}_{-0.02}$ & 0.60$^{+0.02}_{-0.01}$ \\

\hline
\end{tabular}

\vspace{0.5cm}

$^{*}$ The ranking is ordered from lowest to highest area enclosed, evaluated for each $\alpha$-variable from the most dividing of the $\beta$-variables (see \S 4.6).

\end{table}

\section{Discussion}
\subsection{Comparison to Other Observational Studies}

The existence of two distinct populations of galaxies, when divided by integrated galaxy colour, sSFR, or as in this work $\Delta$SFR, is apparent from the bimodality of the distribution of these metrics (e.g. Figs. 6 \& 7). Much previous work has investigated the origin and nature of this bimodality in galaxy properties (including Strateva et al. 2001, Kauffmann et al. 2003, Baldry et al. 2004, Brinchmann et al. 2004, Ellis et al. 2005, Baldry et al. 2006, Driver et al. 2006, Drory \& Fisher 2007, Bell et al. 2008, Bamford et al. 2009, Cameron et al. 2009, Cameron \& Driver 2009, Peng et al. 2010, Wuyts et al. 2011, Peng et al. 2012, Bell et al. 2012, Cheung et al. 2012, Barro et al. 2013, Fang et al. 2013, Mendel et al. 2013, and Woo et al. 2013, Omand et al. 2014). 
Various internal structural properties of galaxies, such as total stellar mass, S\'{e}rsic index, central light concentration, B/T luminosity ratio, as well as environmental features, such as halo mass, local galaxy surface number density, cluster-centric distance, and central - satellite divisions, have been shown to correlate with the bimodality of galaxies' star forming properties, and/or global colour. Beyond this, several attempts to construct a phenomenological model for the cessation of star formation (or `quenching') have been made, most notably by Baldry et al. (2006) and Peng et al. (2010, 2012). 
Great success has been achieved in predicting the probability of a given galaxy being passive from these models, yet one may wish to go further and understand what physical process(es) are giving rise to this transformation of galaxies from the star forming main sequence to the passive red sequence.  

In this work we have examined the dependence of the passive fraction on total stellar mass, halo mass, environment from local over-density and central - satellite divisions, galaxy structure from B/T stellar mass ratio, and bulge and disk masses separately. Furthermore, due to our large sample of over half a million galaxies we have examined the differential effects of these potential drivers for the cessation of star formation, e.g. we examine the impact of varying B/T structure on the passive fraction at a fixed stellar mass for central galaxies, where additional environmental effects are negligible. We find agreement with Peng et al. (2010, 2012) in noting a positive correlation between stellar mass and the passive fraction, which is separable from the dependence of the passive fraction on environmental metrics, such as central - satellite division (see Fig. 8). 
However, we extend this analysis by looking at the role of B/T structure and bulge mass and disk mass, separately, at a fixed total stellar mass, as well as by considering the differential effect of varying halo mass (as in Woo et al. 2013). 

We find that, for centrals, there is least variation in the passive fraction at a fixed bulge mass, through varying the remainder of the observables studied in this work. This implies that bulge mass is the tightest single correlator to the passive fraction, better than halo mass, total stellar mass, B/T structure, disk mass, or $\delta_{5}$ (see Tab. 1 \& 2). We see explicitly in Fig. 13 that the $f_{\rm passive} - M_{*} / M_{\rm halo}$ relations are quite flat at a fixed bulge mass, whereas the relationship between $f_{\rm passive} - M_{\rm bulge}$ remains tight and steep at fixed stellar mass (Fig. 10, top middle panel) and halo mass (Fig. 11, top left panel). 
Superficially, this disagrees with Woo et al. (2013) who have argued for halo mass being the primary driver for quenching centrals, and also disagrees with Peng et al. (2010, 2012) who have argued for total stellar mass being the primary intrinsic driver of quenching. However, these previous works do not have a structural index parameter 
and so the comparison is incomplete. 

If we restrict our investigation to halo mass, stellar mass and $\delta_{5}$ (as in Woo et al. 2013) for central galaxies, i.e. disregarding our bulge + disk decompositions and hence bulge mass, disk mass, and B/T, we recover an equivalent result. In fact we find that there is less variation at a fixed halo mass than for stellar mass or $\delta_{5}$, through varying the remainder of the variables. If we restrict our investigation further and only consider stellar mass and $\delta_{5}$ for centrals (as in Peng et al. 2012) we find that there is a tight $f_{\rm passive} - M_{*}$ relation and that additional variation with $\delta_{5}$ is negligible, in close accord with the prior study. 
Finally, if we restrict the stellar mass range studied (as in Driver et al. 2006) and allow ourselves to vary B/T (which correlates well with S\'{e}rsic index, which they use) we find a strong correlation between structure and the passive fraction (e.g. Fig. 10, top right panel) and a strong bimodality in colour (or $\Delta$SFR)/ morphology, which is consistent with their result. Therefore, our results are perfectly consistent to much previous work, however, we extend this by uncovering the dominant role of bulge mass in constraining the passive fraction through a full structural analysis.

The importance of bulge mass in constraining the passive fraction agrees well with Cheung et al. (2012) and Fang et al. (2013), who consider the central surface mass density, as a proxy for bulge mass, and note that this is more constraining of the passive fraction than stellar mass alone. Our result also agrees well with Bell et al. (2008), who find that a high S\'{e}rsic index bulge is {\it always} found in truly passive systems, and with Drory \& Fisher (2007) who find that classical pressure supported bulges (as opposed to pseudo-bulges) are implicated in forming passive (red) galaxies. 
Moreover, recent work by Mendel et al. (2013) established that recently quenched galaxies also have higher S\'{e}rsic indices (and are visually more likely early-type) than star forming galaxies, implying that the presence and growth of bulge structure occurs contemporaneously to the cessation in star formation. Earlier work by Kauffmann et al. (2003) and Brinchmann et al. (2004) also note that higher concentration and S\'{e}rsic index systems are preferentially passive at a fixed stellar mass, suggesting the ultimate role of the bulge in determining the passive fraction. 
Perhaps our greatest contribution in this work is demonstrating that it is bulge {\it mass} specifically, rather than B/T structure or S\'{e}rsic index, which is the tightest single correlator to the passive fraction, and in further establishing that bulge mass is a stronger driver to the passive fraction than total stellar mass or even halo mass for centrals. In principle this fact can be quite constraining to the proposed theoretical models for quenching (see \S 5.2 \& 5.3).

The fact that the passive fraction is more constrained by bulge mass than total stellar mass or halo mass for central galaxies, motivates a reformulation of Peng et al. (2010, 2012) `stellar mass-quenching' as `bulge mass-quenching'. Moreover, we also conclude that `bulge mass-quenching' is more significant than `halo mass-quenching' (e.g. Woo et al. 2013) for central galaxies. However, the root physical cause for the tight correlation between bulge mass and the passive fraction is not yet established. We argue in the next section that the tight correlation between bulge mass and central supermassive black hole mass (Haring \& Rix et al. 2004) offers a tempting solution through AGN feedback, particularly through the radio-mode (e.g. Croton et al. 2006). `Bulge mass-quenching' may therefore ultimately be the observational corollary of radio-mode AGN feedback. Thus, feedback from the AGN dominates for centrals over virialised shock heating of infalling gas into the halo (which correlates 
primarily to halo mass, e.g. Dekel et al. 2009), and morphological quenching from the bulge component stabilizing giant molecular cloud collapse in the disk (which correlates primarily to B/T, e.g. Martig et al. 2009).

An interesting feature of the $f_{\rm passive} - M_{*}$ relation (see Fig. 12 top middle panel) is that there appears to be a threshold in stellar mass at which galaxies start to shut off their star formation (as also seen in Geha et al. 2012), but that this threshold value varies as a function of B/T structure. Indeed, if we look at this in terms of bulge mass (see Fig. 12 top left panel) we see that the threshold mass varies much less with structure. This implies that the process which drives the shutting down of star formation in central galaxies is only viable at bulge masses above $\sim$ $10^{9}$ $M_{\odot}$. Perhaps coincidentally, this is exactly where the lowest mass central black holes are detected (e.g. Greene et al. 2010), potentially further implicating AGN feedback mechanisms in our new `bulge-mass quenching' picture. 

Once the dominance of bulge mass is taken into account, the next most constraining variable that we have considered is actually disk mass. That is, the passive fraction experiences more variation at a fixed bulge mass with varying disk mass, than for any other variable, including halo mass and $\delta_{5}$ (see left panel of Fig. 14, and compare to the bulge mass plots in Figs. 10 -- 12). This suggests that whatever process(es) give rise to star formation quenching in centrals, which must correlate closely with bulge mass, must also not destroy the disk components of galaxies, such that varying disk mass can construct a perturbation upon the dominant $f_{\rm passive} - M_{\rm bulge}$ relation. In the next section we construct a plausibility argument for the most likely theoretical explanation to the observational trends seen in this paper.

\subsection{Potential Theoretical Explanations I: Radio-Mode AGN Feedback}

For central galaxies, we find that bulge mass is more constraining of the passive fraction than halo mass, total stellar mass, or B/T structure alone (see Tab. 1 \& 2). This requires a dominant mechanism for the cessation of star formation which is coupled intrinsically to the process which builds up the bulge, or some other phenomenon which is tightly coupled to the bulge. A natural solution is that of the supermassive black hole - bulge mass relation (Haring \& Rix 2004), and the link between black hole mass and the total integrated energy released in forming the black hole. It has been known for some time that the energy released by AGN over their lifetimes can be as great as a hundred times the binding energy of a given host galaxy (Silk \& Rees 1998, Fabian et al. 1999), and that at least half of all massive galaxies will harbour bright Seyfert luminosity AGN at some point since z = 3 (Bluck et al. 2011). 
This energy can provide the feedback mechanism needed to shut off the star formation in such a way that bulge mass is found to be dominant in determining the star forming properties, i.e. passive fraction, of galaxies observationally.

We present a sketch for the conceptual framework in which radio-mode AGN feedback could naturally lead to similar results to those found in this work. First it is important to note that the use of AGN feedback in all cosmological simulations of galaxy formation and evolution (both from a semi-analytical and hydrodynamical approach) is required to explain the steepness of the stellar mass function at high masses, and thus shut off star formation in these systems, preventing super-massive galaxies from forming (Croton et al. 2006, Bower et al. 2006, de Lucia et al. 2006, de Lucia \& Blaizot 2007, Bower et al. 2008, Guo et al. 2011, Vogelberger et al. 2013).  

In most contemporary theoretical frameworks (e.g. Croton et al. 2006, Bower et al. 2008, Guo et al. 2011) it is radio-mode, not quasar-mode, AGN feedback which is expected to have the dominant impact on the star formation history of most masses of galaxies. It is pertinent to note here that it is not the instantaneous accretion rate or current luminosity of an AGN which is expected in these theoretical models to provide the energy for the radio-mode AGN feedback process, but rather the total integrated energy in forming the black hole over cosmic time. AGN are highly variable and stochastic systems and it is reasonable to expect that any strong direct observational signal of this gentle and long lasting feedback to be washed out by this variability. 
Moreover, since AGN and new stars both need gas in order to form and be active, it is logical that bright AGN would preferentially reside in gas rich star forming systems (as found by Diamond-Stanic et al. 2012). Nonetheless, the integrated effect of an AGN inputting energy into the halo of a galaxy at much lower luminosities than that applicable to the quasar-mode regime could still have an effect on the star forming efficiency of the galaxy over long time periods. In such a scenario we will show that it is black hole mass, and thus bulge mass, which will be the key observable for AGN driven cessation of star formation, not black hole accretion rate or AGN luminosity.

The total energy, $E_{BH}$, released in forming a black hole of mass, $M_{BH}$, can be given by the integral over time, from the formation time to the present, of the time-dependent total bolometric luminosity of the black hole accretion disk, $L_{BH}$. The bolometric luminosity is coupled to the accretion rate of the back hole (see e.g. Bluck et al. 2011), and can be calculated from the temporal derivative of the energy-mass equivalence formula ($E = Mc^{2}$). Thus we can write:

\begin{eqnarray}
E_{BH} \approx \int^{t_{0}}_{t_{\rm form}} L_{BH}(t) dt = \int^{t_{0}}_{t_{\rm form}} \eta c^{2} \frac{dM_{BH}(t)}{dt} dt \nonumber \\
= \eta c^2 \big[M_{BH}(t_{0}) - M_{BH}(t_{\rm form})\big] \approx \eta c^2 M_{BH}(t_{0})
\end{eqnarray}

\noindent Where $\eta$ is the fraction of accreted matter converted into energy via the accretion process. $\eta$ is permitted to vary from 0 - 0.37 on theoretical grounds, see Thorne (1974), and is observationally found to reside at a value around 0.1, with no indication of redshift or luminosity dependence (e.g. Elvis et al. 1994, Elvis et al. 2002, Bluck et al. 2011). By assuming time independence of $\eta$, it is trivial to show that the total energy released in forming a black hole is proportional to its mass, hence:

\begin{equation}
E_{BH} \propto M_{BH} 
\end{equation}

\noindent Therefore, the total energy released from forming a black hole must also be proportional to the mass of the bulge or spheroid, due to the observed bulge mass -- black hole mass relationship (Magorrian et al. 1999, Ferrarese \& Merritt 2000, Gebhardt et al. 2000, Haring \& Rix 2004):

\begin{equation}
M_{BH} \propto M^{\alpha}_{\rm bulge} 
\end{equation}

\noindent where $\alpha$ is determined observationally. Thus, the total possible energy available for AGN feedback, $E_{BH}$, is also proportional to the mass of the bulge component, or spheroid, in log-space.

It is reasonable to assume that, within the radio-mode AGN feedback driven quenching model, the probability of a galaxy being passive will be positively correlated with the amount of energy available to feedback on the system, above some activation threshold. The total energy available to feedback on the system will be some fraction of the total energy released by the black hole formation and growth. Thus, we can write: 

\begin{equation}
P_{\rm passive} \sim E_{\rm feedback} - \phi_{\rm Act} = \kappa E_{BG} - \phi_{\rm Act}
\end{equation}

\noindent where $\kappa$ is the fraction of energy released in forming the black hole that is available, via momentum and energy conserving modes, to couple back to the galaxy and cause the AGN feedback on star formation. $\phi_{\rm Act}$ is the activation energy, i.e. the amount of energy required before an effect on the passive fraction is noticeable, which will vary as a function of halo mass for central galaxies. 

In the radio-mode paradigm the feedback energy will couple negatively to the star formation in the galaxy via gently heating the gas in the halo in which the galaxy resides over long periods of time, shutting down the supply routes of the cold gas used as fuel for star formation. Putting all of this together, and assuming that the probability of being passive is roughly equivalent to the volume corrected fraction of passive systems in a given population (i.e. ${f_{\rm passive} \sim P_{\rm passive}}$), we arrive at:

\begin{equation}
f_{\rm passive} \sim M_{\rm bulge} - \phi^{'}_{\rm Act}
\end{equation}

\noindent which is essentially the main observational result of this paper (see Figs. 10 - 12, and \S 4.6). Note that we can constrain the mass activation threshold ($\phi^{'}_{\rm Act}$) to be at $M_{\rm bulge} \sim 10^{9-9.5} M_{\odot}$. We can also constrain the functional form of the $f_{\rm passive} - M_{\rm bulge}$ relationship to be roughly logarithmic, i.e. 

\begin{equation}
f_{\rm passive} \approx \gamma \big( {\rm log}(M_{\rm bulge}) - {\rm log}(\phi^{'}_{\rm Act}) \big)
\end{equation}

\noindent where the constant of proportionality, $\gamma$, is found to be approximately 0.5 $\pm$ 0.2. However, in Fig. 14, which plots the passive fraction against bulge mass, and divides by disk mass range, we do still see some significant variation in the passive fraction at a fixed bulge mass. This variation is less than for the other internal structural parameters studied in this work, but nonetheless it is still desirable to understand where the scatter comes from. 

Yet again, AGN radio-mode feedback offers a plausible explanation. In this paradigm the gas in the galaxy is never pushed out or heated substantially, only the gas in the halo is gently heated over long time scales. This results in the supply routes of cold gas into the galaxy being shut down, but the gas already present will remain, and if conditions are right continue to form stars. 
Since more disk dominated galaxies are thought to have larger cold gas reservoirs than more spheroidal galaxies (e.g. Young et al. 1989), it is natural to expect that the effects of radio-mode quenching will take longer to be manifest in more disky systems. This will result in galaxies having lower passive fractions for their bulge mass at higher disk masses, as is seen explicitly in Fig. 14.

\subsection{Potential Theoretical Explanations II: Other Possibilities}

In addition to AGN radio-mode feedback, at least four other physical processes have been proposed to affect the bulge mass - passive fraction inter-dependency for central galaxies. We assess these in turn here: 

1) The bulge component stabilizing disk instabilities, which lead to the formation of giant molecular clouds (GMCs), which are the sites of star formation in galaxies (as in Martig et al. 2009). In this scenario tidal torques are evoked to stabilize the collapse of star forming regions, and these have a 1/$r^{3}$ fall off, which implies that morphology (or B/T) would be a better indicator than bulge mass itself. This is because at a fixed B/T ratio, increasing the bulge mass will also increase the disk mass and hence radius to the GMC, such that the increase in size dominates over the increase in mass on the tidal force. We observe that higher bulge masses at a fixed B/T are more (not less) passive (see Fig. 12 top left panel). Thus, we can rule out morphological quenching as a primary driver of the passive fraction. 

2) Central galaxies with higher mass bulges reside in higher mass halos for their total stellar masses (e.g. Cole et al. 2000). This results in an increase in the importance of halo-mass quenching (as in Dekel \& Birnboim 2006, Dekel et al. 2009) for these systems. However, in this scenario the single tightest correlator to the passive fraction would be expected to be halo mass, and yet we find that there is greater variation in the passive fraction at a fixed halo mass with varying bulge mass, than at a fixed bulge mass with varying halo mass (compare Fig. 11 top left panel to Fig. 13 right panel). Therefore, halo mass quenching cannot be the primary driver of the passive fraction for central galaxies. 

3) Quasar-mode AGN feedback from catastrophic bursts of AGN activity in major gas-rich galaxy mergers (as in Di Matteo et al. 2005, Hopkins et al. 2006a,b, Hopkins et al. 2008). Even though the merger rate at low redshifts is very low (e.g. De Propris et al. 2007), the potential increase in the merger rate with redshift (Bluck et al. 2009, Lin et al. 2008, Lopez-Sanjuan et al. 2012, Bluck et al. 2012, Man et al. 2012) may indicate that the quasar-mode feedback route may have been more significant in the earlier Universe. That said, whatever process(es) shut of the star formation in galaxies must not destroy the (sometimes substantial) disk components, and this favours a more gentle process, such as radio-mode AGN feedback. Even if the disks reform (as argued for in Hopkins et al. 2013), these reformed disk would be actively forming stars, not passive as seen in many cases in this paper. Thus, quasar-mode feedback cannot be the only cause of the trends witnessed here, with radio-mode feedback needed to keep 
passive systems from accreting gas and reforming star-forming disks.

4) Galaxy mergers and interactions can result in the cold gas reservoirs of galaxies being depleted in periods of enhanced star formation (e.g. Ellison et al. 2008, Scudder et al. 2012, Patton et al. 2013), as well as growing the central black hole (e.g. Sanders et al. 1988, Di Matteo et al. 2005, Ellison et al. 2011), which can lead to more effective radio-mode AGN feedback and larger bulge components as well (e.g. Hopkins et al. 2010). However, this is likely better thought of as a set of physical processes, with distinct and, in principle, separable effects. These include environment (e.g. tidal interactions and harassment), enhanced stellar, supernova, and AGN feedback, halo-mass growth, and bulge growth modes in addition to the depletion of cold gas reservoirs. These may all induce a transition from the blue cloud to the red sequence, and further will lead to correlations with the bulge. 
However, it is clear that major merging is not essential for a galaxy to become passive because large intact passive disks are found in our sample (see, e.g., Fig. 10 top right panel). Furthermore, since minor mergers tend to actually bring gas into the system, it is difficult to see how these could directly lead to an increase in the passive fraction. Of course, indirectly, insofar as they result in an increase in halo mass, bulge mass, and central black hole mass they can couple to 1), 2), and 3), which have already been discussed above and ruled out as the primary drivers of the passive fraction.

We are led to conjecture that by far the most obvious and likely explanation to the tight $f_{\rm passive} - M_{\rm bulge}$ relationship witnessed in this paper is AGN feedback through the radio mode. This is because the other proposed mechanisms for achieving star formation cessation are not fully consistent with our data. In particular, we find that bulge mass is a tighter correlator than halo mass, ruling out halo mass-quenching (e.g. Dekel \& Birnboim 2006) as a dominant route. We also find that morphology is not the primary driver of the passive fraction, which rules out morphological quenching (Martig et al. 2009). The fact that large extended passive disks are found to exist in our sample further rules our major merging and the AGN quasar-mode as likely drivers of the passive fraction. So, we are left with radio-mode AGN feedback (e.g. Croton et al. 2006, Bower et al. 2008) which is found to couple to bulge mass (see eq. 10 and surrounding text), and fully allows for a perturbation on this trend from 
varying disk mass (as seen in Fig. 14).

\section{Conclusions}

In this paper we have studied the star forming properties of over half a million SDSS galaxies at z = 0.02 - 0.2, with stellar masses in the range 8 $< {\rm log(M_{*}/M_{\odot}}) <$ 12. Our sample contains a large variety of morphological types from pure spheroids to disks, via a sequence in bulge-to-total stellar mass ratios, and encompasses a range in halo mass and local galaxy density of over four orders of magnitude. We use the SDSS DR7 (Abazajian et al. 2009) with spectroscopically determined SFRs from Brinchmann et al. (2004). We construct a definition of passive from the distance to the star forming main sequence, $\Delta$SFR, that each galaxy resides at. Specifically, we define a galaxy to be passive if it is forming stars at a rate a factor of ten lower than similar stellar mass and redshift galaxies which are actively star forming. This metric clearly distinguishes star forming from passive galaxies (see Fig. 6).

We construct the passive fraction, which is the fraction of passive galaxies to total galaxies in a given sub-sample, weighted by the inverse of the volume over which galaxies of given magnitudes and colours would be visible in the SDSS, thus removing observer bias. In particular, we use bulge + disk decompositions in ugriz photometry (from Simard et al. 2011) and in stellar mass (from Mendel et al. 2014), and group membership and halo masses from Yang et al. (2007, 2009). We investigate the effects on the passive fraction of varying total stellar mass, halo mass, bulge mass, bulge-to-total stellar mass ratio (B/T), disk mass, local galaxy over-density, and central - satellite divisions. We summarise our key results here.

\noindent {\bf For the whole sample:}

\noindent $\bullet$ We find that the SFR - $M_{*}$ relation is not dependent on B/T structure or local galaxy density for star forming galaxies (Fig. 5), and, thus, that stellar mass and redshift are the only drivers of this relation.

\noindent $\bullet$ The $\Delta$SFR parameter cleanly divides galaxies into two populations, a star forming and passive peak. The star forming peak is dominated by lower stellar and halo masses, lower B/T ratios, and lower local over-densities, whereas the passive peak is dominated by higher stellar and halo masses, higher B/T ratios, and higher local over-densities (see Figs. 6 \& 7).

\noindent $\bullet$ We find positive correlations between the passive fraction and total halo mass, stellar mass, and B/T structure for both central and satellite galaxies (Fig. 8). 

\noindent $\bullet$ We also find that satellite galaxies have higher passive fractions for their stellar masses and B/T structures than central galaxies. This effect is more pronounced at low stellar masses, suggesting that the environmental processes which cause cessation in star formation are more effective in lower mass galaxies. 

\noindent $\bullet$ At a fixed halo mass we find that centrals are more passive than satellites, most probably due to their relatively higher stellar masses. This suggests that intrinsic drivers to the passive fraction dominate over environmental effects.

For the remainder of our analyses in this paper we select only central galaxies where additional environmental effects to halo mass are found to be negligible.

\noindent {\bf For central galaxies:}

\noindent $\bullet$ We find a correlation between B/T structure and the passive fraction, but note that there is large variation at a fixed B/T with both stellar and halo mass (Figs. 10 \& 11).

\noindent $\bullet$ We also find a strong correlation between the passive fraction and total stellar and halo mass, but we discover that this correlation varies substantially with B/T structure (Fig. 12), such that more spheroidal (higher B/T) galaxies are more frequently passive for their stellar or halo mass than more disk-like (lower B/T) galaxies.

\noindent $\bullet$ The passive fraction - bulge mass relation exhibits the least variation in terms of the other variables considered, suggesting that bulge mass is the primary driver of the passive fraction. In fact we find far more variation at a fixed stellar or halo mass with varying bulge mass range than at a fixed bulge mass with varying stellar or halo mass range (see Fig. 13 and compare to Figs. 10 \& 11).

\noindent $\bullet$ We quantify the maximum difference in passive fraction ($\delta f_{p}^{\rm max}$) between the highest and lowest binnings of each variable in turn at fixed other galaxy properties (see Tab. 1). We also compute the area enclosed ($A(\delta f_{p})$) between the highest and lowest binnings of each variable at fixed other galaxy properties (see Tab. 2). 

\noindent $\bullet$ From the $\delta f_{p}^{\rm max}$ and $A(\delta f_{p})$ statistics we find that there is significantly less variation at a fixed bulge mass than for any other variable, including halo mass, stellar mass, and B/T structure. 

\noindent $\bullet$ At a fixed bulge mass, there is greater variation in the passive fraction with disk mass than for any other variable. This implies that the disk mass is the next most important variable to know after bulge mass, ahead of, e.g, halo mass and local galaxy density.

We construct a plausibility argument for the explanation of our observed trends through radio-mode AGN feedback in \S 5.2. In particular, we demonstrate that the total energy available to feedback on galaxies from the AGN is proportional to the black hole mass, and hence also tightly correlated with the bulge mass (e.g. Haring \& Rix 2004). Thus, bulge mass is expected to correlate better with the passive fraction than current instantaneous AGN luminosity or accretion rate. We go on to consider whether our results are consistent with various other proposed quenching mechanisms in \S 5.3. We conclude that the tighter correlation between passive fraction and bulge mass, than with halo mass or B/T structure, rule our halo mass quenching (e.g. Dekel \& Birnboim 2006) and morphological quenching (e.g. Martig et al. 2009) as dominant routes for star formation cessation. 
We also argue that the presence of substantial passive disk components in our sample rules out major merging and AGN quasar-mode feedback as the dominant quenching mechanism, preferring the gentler radio-mode process.   

In summary, we explore the origin of galaxy bimodality through a detailed analysis of the star forming properties of over half a million local SDSS galaxies, for which we have bulge + disk decompositions. We evaluate how the passive fraction varies as a function of both intrinsic galaxy structural properties, and of environment. Our most important result is that the passive fraction for central galaxies is more tightly coupled to bulge mass, than any other intrinsic property, including total stellar mass, halo mass, and B/T structure. This requires that the physical mechanism(s) for inducing the cessation of star formation must be strongly coupled to the bulge. Finally, we argue that radio-mode AGN feedback offers the most plausible theoretical framework for an explanation of this observational result.

\section*{Acknowledgments}

This paper has benefited from many lively and interesting discussions over the past year. In particular we thank Chris Conselice, Richard Ellis, Carlos Frenk, Phil Hopkins, John Kormendy, Jon Loveday, Jillian Scudder, Peter Thomas, and Paul Torrey. We thank the anonymous referee for a very positive and helpful review. We also gratefully acknowledge funding from the National Science and Engineering Research Council (NSERC) of Canada, particularly for Discovery Grants awarded to SLE and DRP. ES gratefully acknowledges the Canadian Institute for Advanced Research (CIfAR) Global Scholar Academy, and ES and JM gratefully acknowledge the Canadian Institute for Theoretical Astrophysics for partial funding.

Funding for the SDSS and SDSS-II has been provided by
the Alfred P. Sloan Foundation, the Participating Institutions, the National Science Foundation, the U.S. Department of Energy, the National Aeronautics and Space Administration, the
Japanese Monbukagakusho, the Max Planck Society, and the
Higher Education Funding Council for England. The SDSS
Web Site is http://www.sdss.org/

The SDSS is managed by the Astrophysical Research Con-
sortium for the Participating Institutions. The Participating
Institutions are the American Museum of Natural History, Astrophysical Institute Potsdam, University of Basel, University of Cambridge, Case Western Reserve University, University of Chicago, Drexel University, Fermilab, the Institute for Advanced Study, the Japan Participation Group, Johns
Hopkins University, the Joint Institute for Nuclear Astrophysics, the Kavli Institute for Particle Astrophysics and
Cosmology, the Korean Scientist Group, the Chinese Academy
of Sciences (LAMOST), Los Alamos National Laboratory,
the Max-Planck-Institute for Astronomy (MPIA), the Max-Planck-Institute for Astrophysics (MPA), New Mexico State
University, Ohio State University, University of Pittsburgh, University of Portsmouth, Princeton University, the United States Naval Observatory, and the University of Washington.

\appendix 

\section{Testing B/T Recovery with Model Galaxies}

\begin{figure*}
\rotatebox{270}{\includegraphics[height=0.33\textwidth]{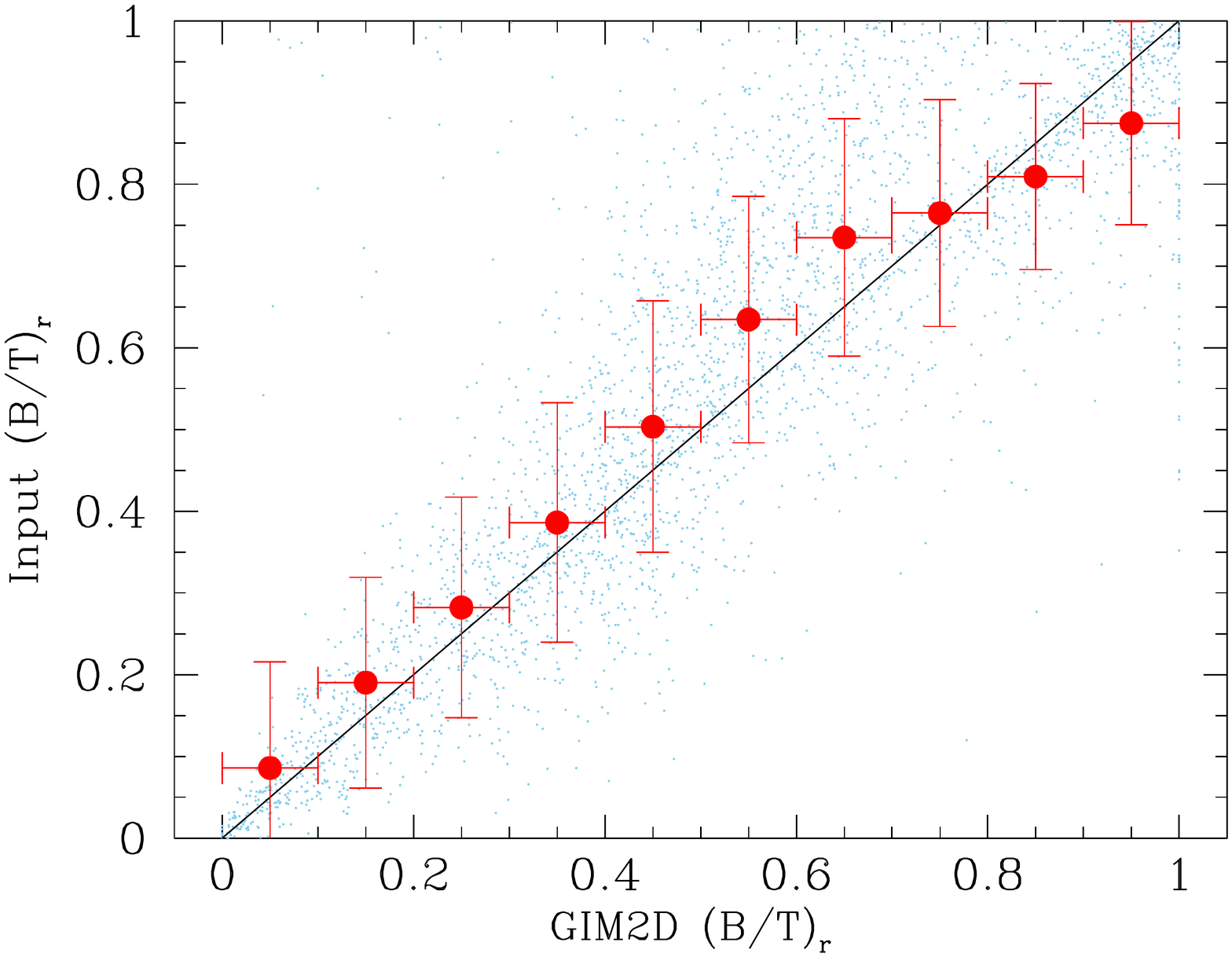}}
\rotatebox{270}{\includegraphics[height=0.33\textwidth]{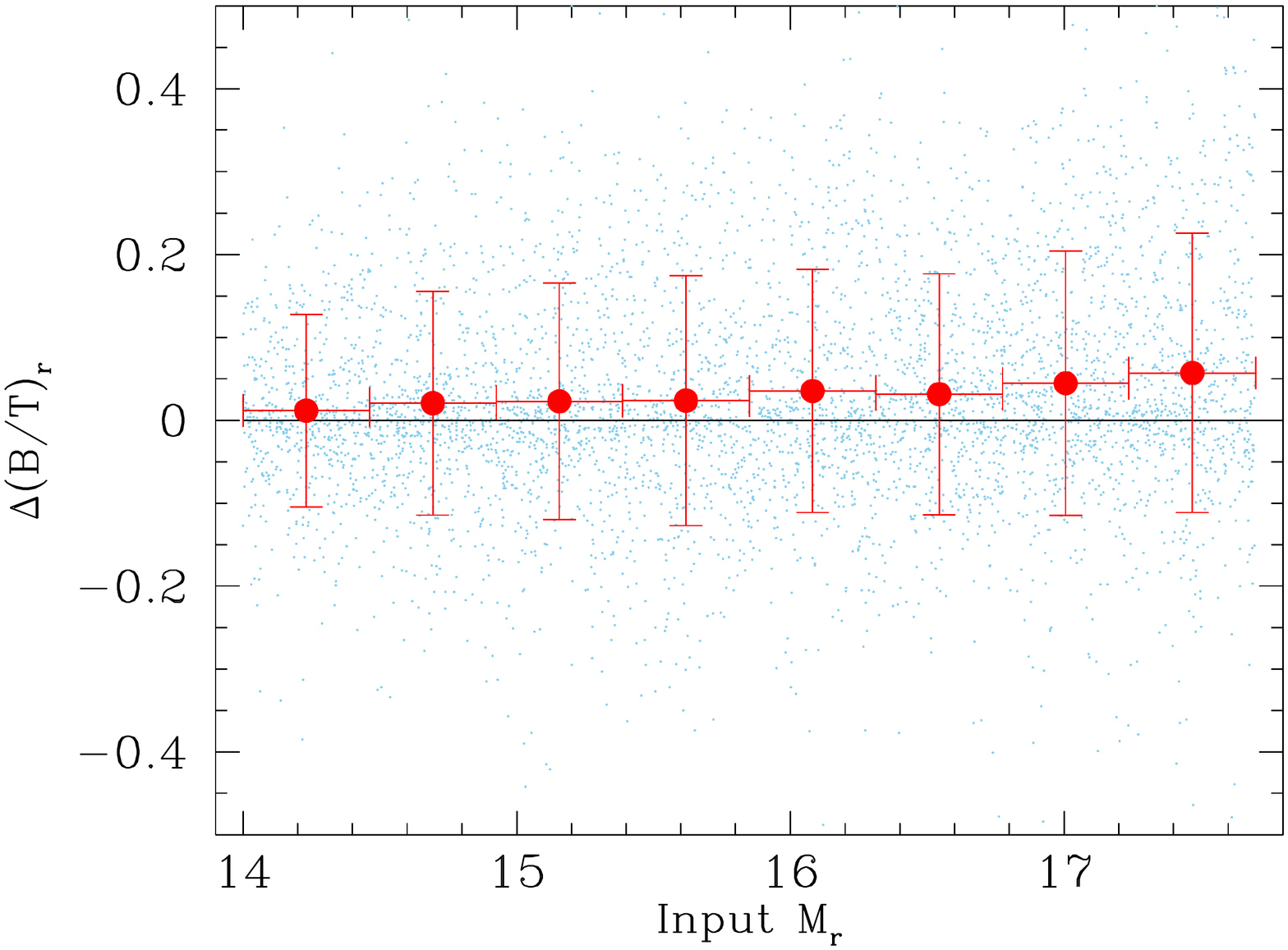}}
\rotatebox{270}{\includegraphics[height=0.33\textwidth]{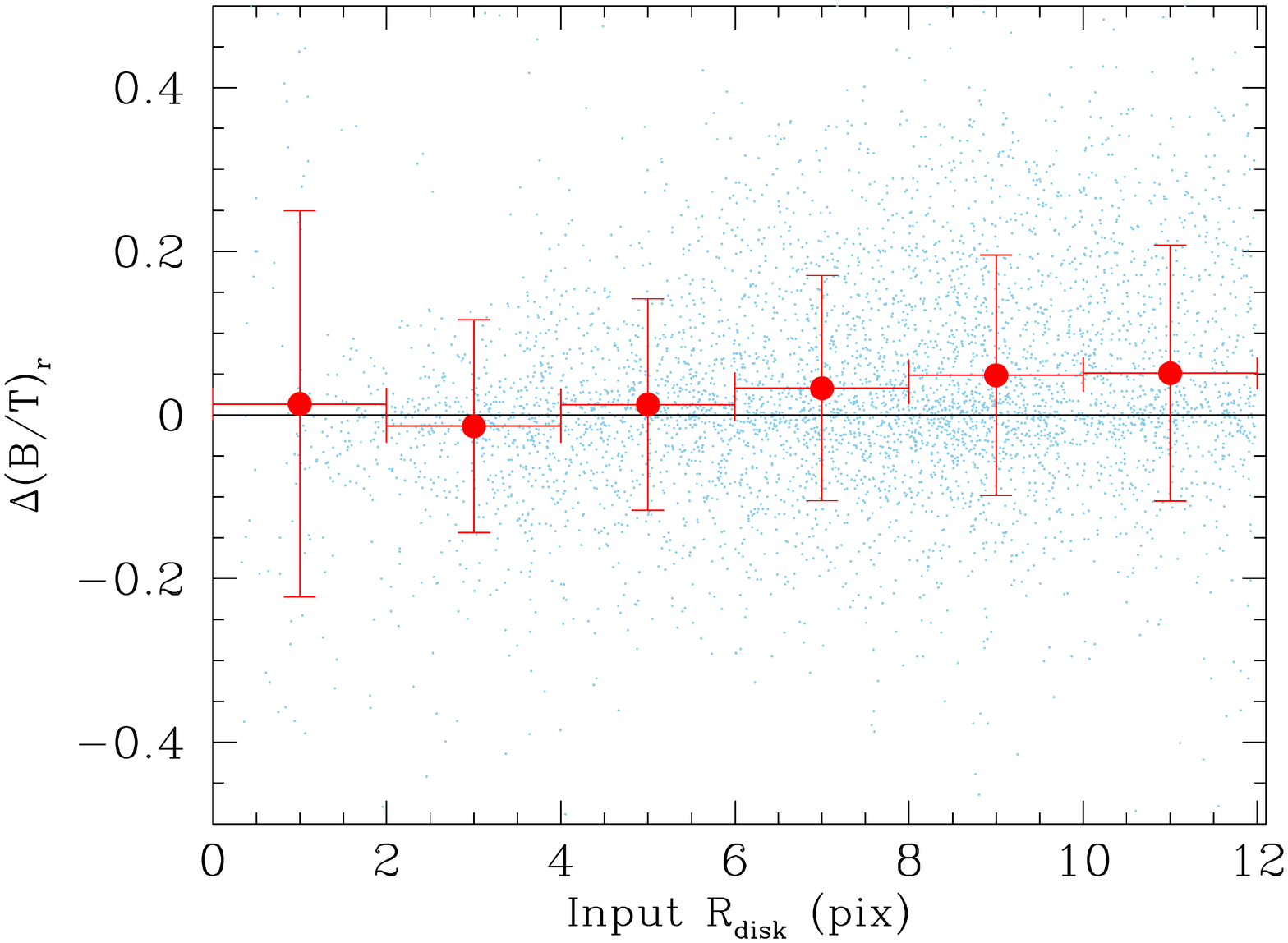}}
\caption{{\it Left-hand panel}: Simulated input B/T ratio vs. GIM2D recovered B/T ratio, both computed in the r-band. The simulated (model) galaxies are constructed from a $n_{s}$ = 4 bulge + $n_{s}$ = 1 disk profile, linearly combined to give a range in B/T values, and a range in intrinsic galaxy properties, similar to that observed in the SDSS. {\it Middle panel}: The difference between input and output B/T ratio ($\Delta$(B/T)$_{r}$ = (B/T)$_{\rm r,sims}$ - (B/T)$_{\rm r,GIM2D}$) vs. input apparent r-band magnitude, M$_{r}$. {\it Right-hand panel}: $\Delta$(B/T)$_{r}$ vs. input size of the disk component, R$_{\rm disk}$, in pixels. In all plots the red dots indicate the mean of the bin, with the error bars showing the full 1 $\sigma$ dispersion around the mean. Only one in every 20 points are shown at random in light blue for clarity. There is generally very good retrieval of simulated B/T ratios by the GIM2D code across a range in magnitudes, sizes and B/T values similar to the SDSS galaxies studied 
in this paper.}
\end{figure*}

In this appendix we construct a variety of tests to probe the reliability of our bulge + disk decompositions, and assess their suitability for performing scientific studies such as the one presented in the main body of this paper.

We use a suite of simulated mock galaxies to see how the GIM2D code performs in recovering known B/T ratios for galaxies embedded in the sky background of the SDSS, and with similar apparent magnitudes and sizes to the galaxies studied in this paper. Two distinct types of model galaxies are used. First, we construct composite systems from a $n_{s}$ = 4 bulge + $n_{s}$ = 1 disk and add these together in various combinations, spanning the full B/T range. In total there are $\sim$ 25,000 simulated galaxies with (B/T)$_{r}$ = 0 - 1, apparent r-band magnitude M$_{r}$ = 14 - 17.7, and with half-light disk radii = 0.1 - 12 pixels. We restrict the sample to those mock galaxies with half light radii $r_{\rm half}$ $<$ 15 pixels, without embedded disks, i.e. we further require that $r_{\rm bulge} < 1.67 \times r_{\rm disk}$. An initial analysis of this data set is provided in Mendel et al. (2014) with relation to the use of the photometric B/T determinations from Simard et al. (2011), and the further implementation 
of this photometric input in the derivation of stellar masses for bulge and disk components. Here we restrict our analysis to (the representative) r-band decompositions, which scale well with 
stellar mass for most systems.

In Fig. A1 we present comparisons of model input parameters to output r-band bulge-to-total light ratios, (B/T)$_{r}$. On the left-hand panel of Fig. A1 we plot the input simulated (B/T)$_{r}$ against the output, recovered, (B/T)$_{r}$ from GIM2D fitting. To guide the eye we add the one-to-one line. We also show the position of the mean and the standard deviation about the mean for a range of binnings. It is clear that there is generally very good retrieval of the simulated input (B/T)$_{r}$ for the full range of values. The average standard deviation about the mean is 0.17, and we can use this as an approximate 1 $\sigma$ error estimate on a single B/T value. This level of accuracy is more than sufficient to proceed with the investigations attempted in this work. Since we always bin B/T values together, and further do not look at differences in B/T less than 0.1, we are benefited greatly by the large number of galaxies in our sample, and the 1/$\sqrt{N}$ reduction in error on the mean of the binning. 
Even without this statistical advantage, we could still proceed with the bulk of analyses presented in the paper, because sensitivities in B/T of $<$ 0.25 would still convincingly separate out morphologies and variation in bulge and disk mass properties.

The standard deviation around the 1:1 line is not the only indicator of the success of the fits. One must also look at the mean (or median) recovered value for each binning. Here again we find a very good agreement across our full simulated sample, with typical deviations from the 1:1 line of $\pm$ 0.05. This is a success rate far above the minimum required for robustness in the results of this paper. We further investigate whether there are any dependencies on the the deviation from perfect (B/T)$_{r}$ recovery which scale with apparent magnitude (middle panel in Fig. A1) and with disk size (right panel in Fig. A1). Here we find typically good agreement at all magnitudes and sizes, with $\Delta$(B/T)$_{r}$ (= (B/T)$_{\rm r,sim}$ - (B/T)$_{\rm r,GIM2D}$) giving absolute values $<$ 0.08 in all of the binnings considered. We do notice a slight increase in  $\Delta$(B/T)$_{r}$ with both increasing apparent r-band magnitude and increasing disk size. This is as expected, since with less bright 
observed galaxies and with galaxies with their light spread over larger surface areas (higher disk radii) the fits have to contend with lower surface brightness and hence less well resolved features. This notwithstanding, the general agreement is excellent and more than sufficient to utilise the bulge + disk photometric decompositions for composite systems, throughout the magnitude and size range targeted in this study.

However, not all galaxies are genuinely two-component systems, with a well behaved bulge (with $n_{s} \sim$ 4) and a well behaved disk (with $n_{s} \sim$ 1). In order to test the overall accuracy of our decomposition methods we must also test the case where we fit a galaxy with GIM2D for two-components, but it has only one component. To investigate this we construct another sample of $\sim$ 65,000 simulated galaxies, this time with a pure single S\'{e}rsic profile with $n_{s}$ = 0.5 - 8. Again these span a similar range in apparent magnitude and recovered size (in pixels) as the SDSS galaxies in our sample. For their analysis we embed the systems into the SDSS sky background and perform the GIM2D photometric fitting exactly as is done with the observations. 

The results of the fitting recovery can be seen in Fig. A2. The colours indicate the input S\'{e}rsic index (with redder colours indicating higher S\'{e}rsic indices), and the distribution shows the recovery of (B/T)$_{r}$ by GIM2D. Ostensibly, since all of the galaxies considered in this plot are modeled from a pure single S\'{e}rsic profile all of their B/T values should be either zero or one. 
Of course, in the case where $n_{s} \sim$ 3 by far the most common physical cases from real data are for bulge + disk systems, and thus the sense in which a galaxy can be a pure (single component) system with an intermediate $n_{s}$ is somewhat misleading. For the remainder of this discussion we will only consider the high and low S\'{e}rsic index systems, which we may assume ought to have recovered GIM2D (B/T)$_{r}$ values of either one (for high $n_{s}$) or zero (for low $n_{s}$).

For the disks (with input $n_{s} <$ 2), we find that the recovered (B/T)$_{r}$ has a distribution which is peaked at zero, and has a power law tail to higher values. Almost 80 \% of input disks are recovered with (B/T)$_{r} < 0.2$, and over 95 \% of disks are recovered with (B/T)$_{r} < 0.4$. This is a fairly good result from the point of view of deliberately fitting a pure disk erroneously with a two component fit. In general we learn that we can recover pure disks fairly well, and at worst some galaxies would be scattered to the next bin over, but that happens in less than $\sim$ 20 \% of cases. The results for input spheroids (with $n_{s}$ $>$ 4, orange line, or $>$ 6, red line in Fig. A2) are a little more complicated.

We find that the output GIM2D (B/T)$_{r}$ distribution for input pure high S\'{e}rsic index model galaxies is bimodal, with one peak at (B/T)$_{r}$ = 1 which is very narrow, and another (potentially worrying) peak around (B/T)$_{r}$ = 0.75, which is considerably broader. This suggests that some spheroids might be decomposed into bulge + disk systems erroneously by our fitting, and thus, that there may be `false disks' in our final sample. The identification and removal of these cases is the subject of the next section of this appendix. 
For now it is pertinent to note that this effect does not represent a catastrophic failure of the fitting of Sloan galaxies, because the output B/T values are still quite high, i.e. input pure spheroids are either identified as pure spheroids or (at worst) identified as composites with high B/T. Potentially we could proceed with our analyses with this knowledge and source of bias, however we will attempt to do better and actually remove these `false disks'. It should be noted here that this is not a new or unique problem, and the existence of false disks in B/T decompositions of modest depth and resolution data is well known (see, e.g., Bamford et al. 2009, Cameron et al. 2009, Cameron \& Driver 2009). 

\begin{figure}
\rotatebox{270}{\includegraphics[height=0.49\textwidth]{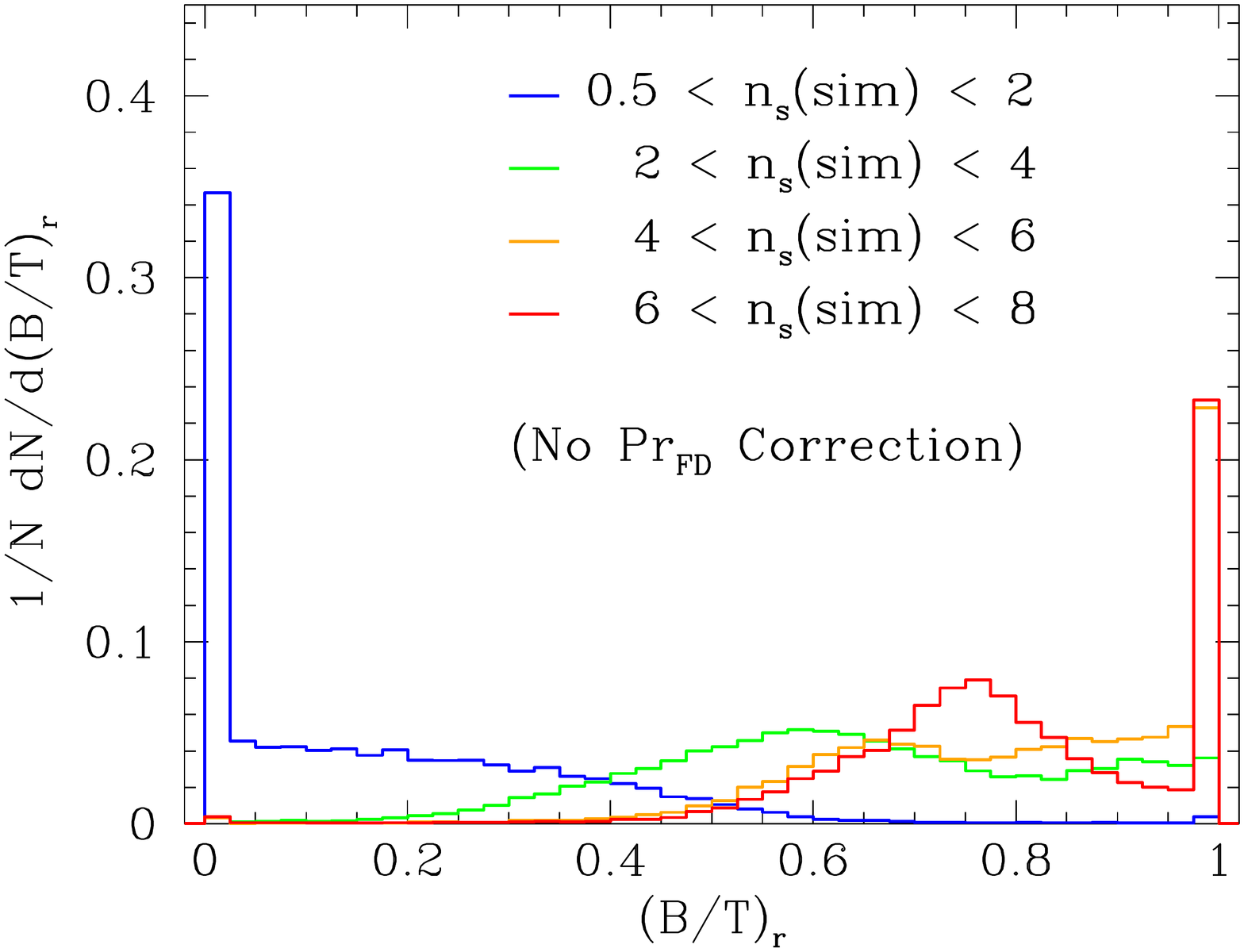}}
\caption{Normalised recovered (GIM2D) B/T ratio distributions in the r-band for pure S\'{e}rsic model galaxies. The colours indicate the input (single) S\'{e}rsic index ($n_{s}$) range, as shown on the plot. This figure shows the uncorrected raw distributions (and may be compared to Fig. B3 for corrected versions). In general high $n_{s}$ model galaxies have high recovered B/T ratios, and low $n_{s}$ model galaxies have low recovered B/T ratios, with middling values in $n_{s}$ leading to middling values in B/T. This is as expected. However, note the dual peak at high $n_{s}$ values (in red), which is the source of `false disks' in the simulated sample.}
\end{figure}

\section{Removing False Disks}

\begin{figure}
\rotatebox{270}{\includegraphics[height=0.49\textwidth]{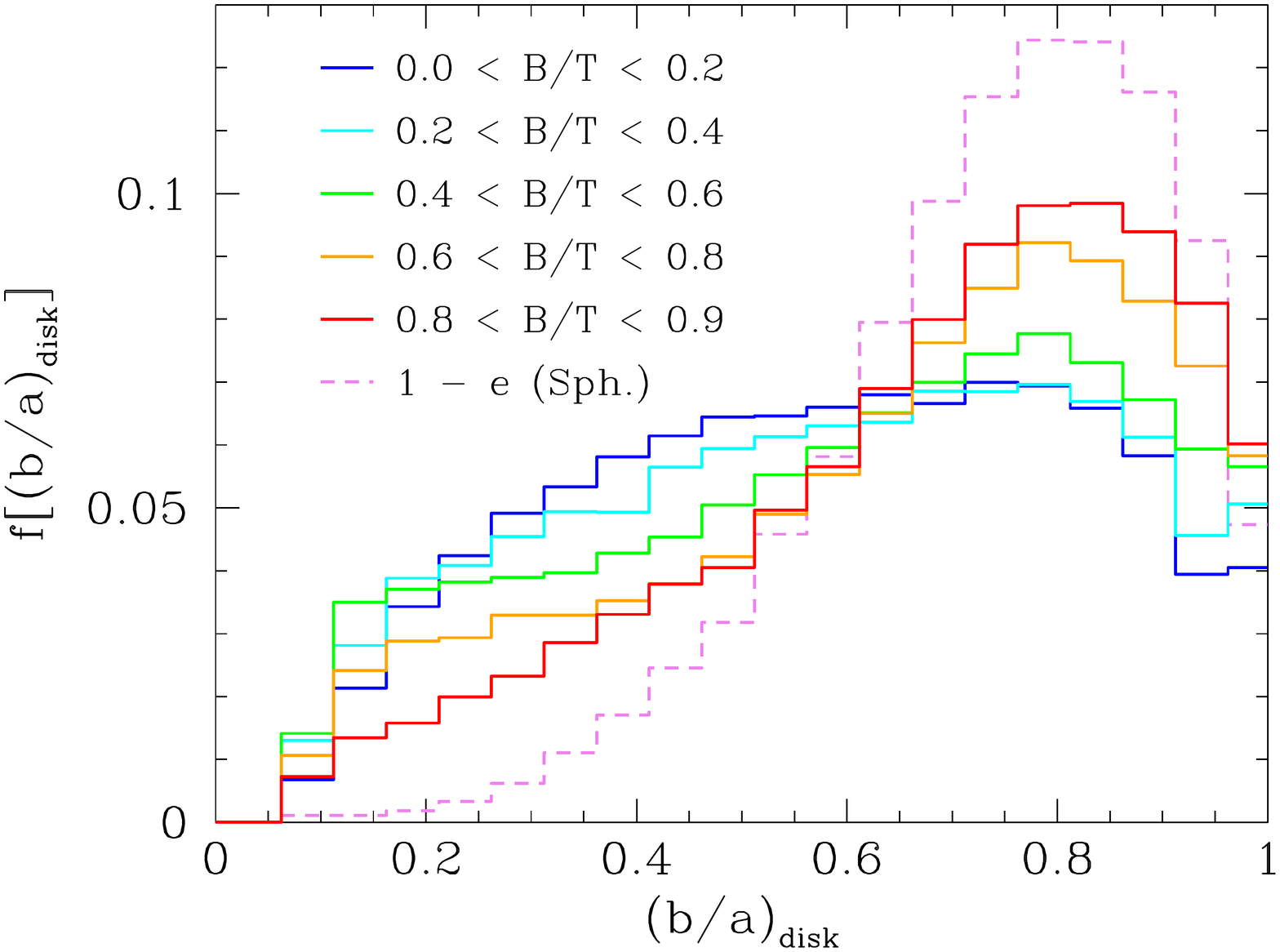}}
\caption{Normalised distribution of semi-minor to semi-major disk axis ratio (b/a) for observed SDSS galaxies in our sample. The distribution is split by measured B/T ratio (by stellar mass). For comparison, the effective b/a ratio for elliptical galaxies and bulges is shown in purple, dotted line. Note how at low B/T ratios the distribution in b/a is quite flat, which is expected since the variation is purely a function of viewing angle. Whereas at higher B/T values the distribution in b/a becomes more peaked at higher b/a values, like with the spheroids. This is a classic test for false disk components in an observed sample.}
\end{figure}

The problem of false disks in the decompositions can be seen directly in the observational data, without recourse to simulations. The classic test (see e.g. Simard et al. 2011) is to plot the distribution of the semi-minor to semi-major axis ratio (b/a) of the disk component, and to look at this in bins of B/T (by light and/or stellar mass). For real disks the distribution should be broadly flat, because to first order disks are circular and the variation in this axis ratio is purely a function of the inclination of the viewing angle. Since galaxies are not ordered in any preferential direction (a consequence of isotropy), no peak in the b/a distribution for disks should be present. 
For spheroids, however, this is clearly not the case, and one would expect the distribution to be quite different with a peak at high b/a. If bulges (or spheroids) are having some of their light erroneously decomposed into disks, then their `disk' b/a distribution would become skewed as a result, in effect becoming more peaked at high b/a and suppressed at lower b/a. From the literature and our simulation results we also expect this to be more common in higher B/T systems (e.g. Simard et al. 2011).

In Fig. B1 we plot the b/a distribution for disk components in bins of B/T (by stellar mass), and for a comparison also plot the effective b/a ratio for spheroids (= 1 - e, where e is the ellipticity). It is clear that at low B/T values the distribution is fairly flat, as expected for real disks, but as we go to higher bins of B/T the distribution of b/a becomes more peaked at high b/a values, and in that sense becomes more similar to the spheroid distribution than the disk one. This suggests that at moderately high B/T values we sometimes include pure spheroids in our composite sample, just as was seen in the previous section with the simulated galaxy suite. The main focus of this section is to develop a new technique to identify and ultimately correct these false disks. 

\begin{figure*}
\rotatebox{270}{\includegraphics[height=0.49\textwidth]{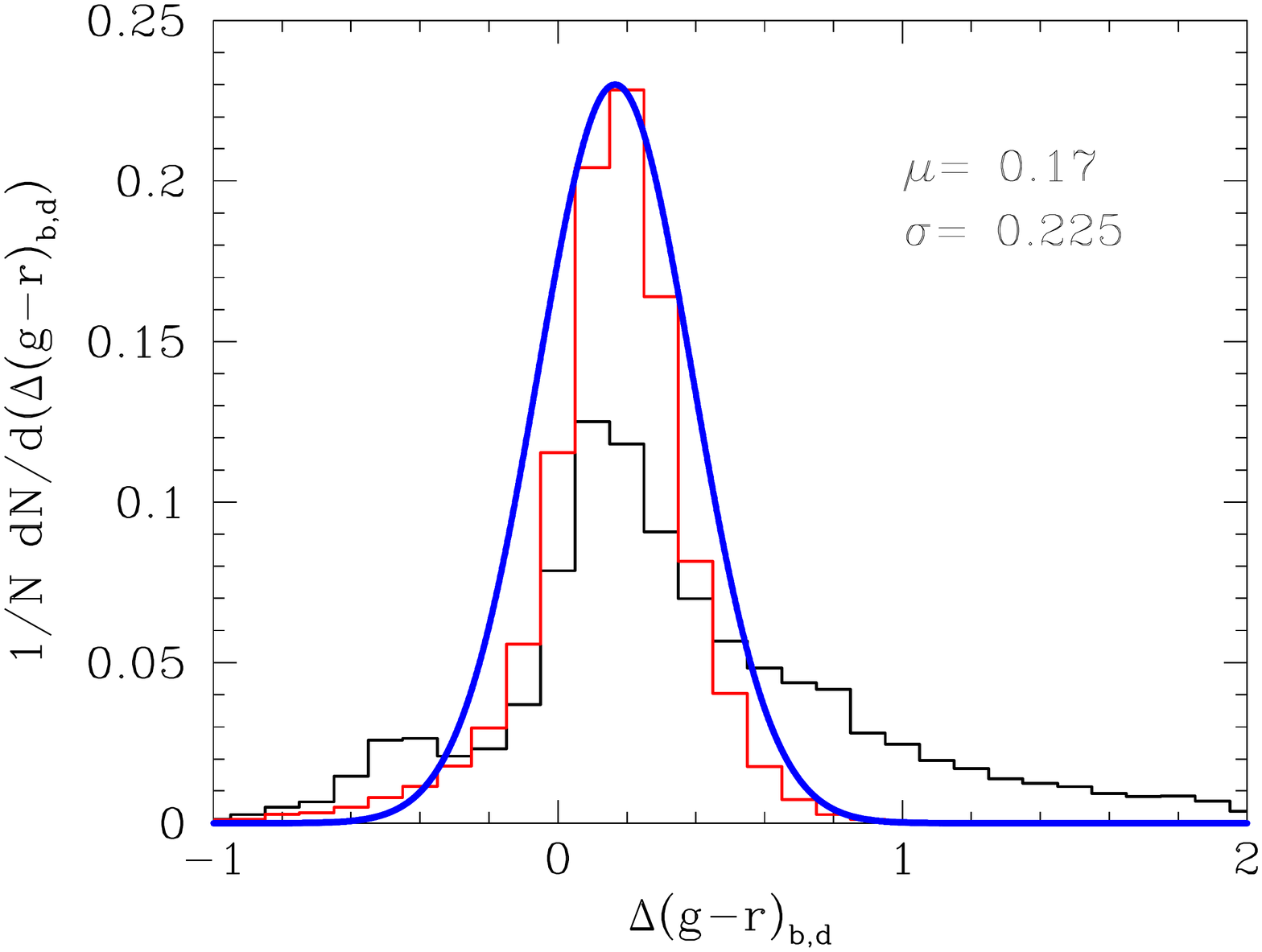}}
\rotatebox{270}{\includegraphics[height=0.49\textwidth]{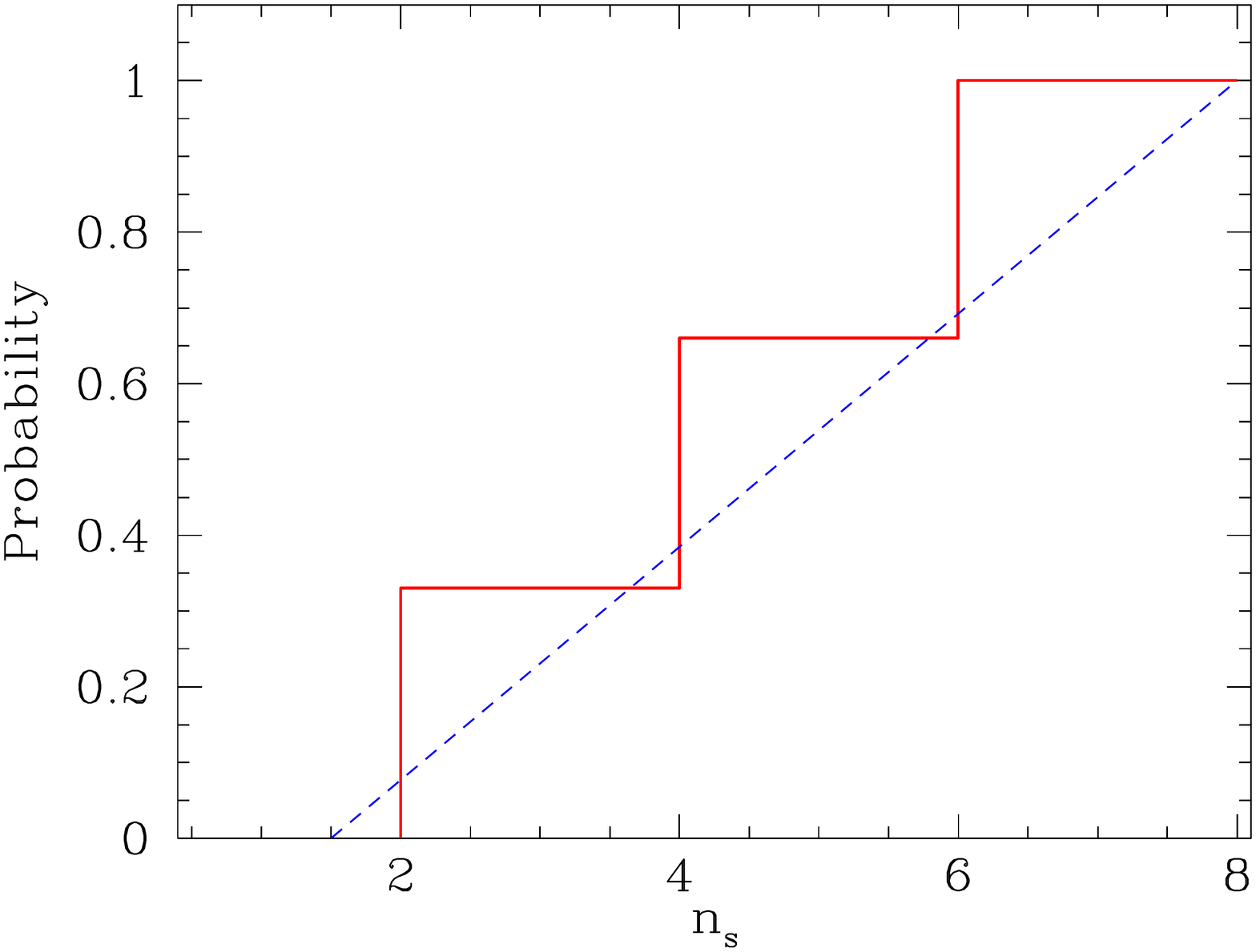}}
\caption{{\it Left-hand panel}: Normalised distribution of the colour gradient, defined in eq. B1. The black histogram is for the full SDSS galaxy sample, whereas the red histogram represents only those galaxies which are highly likely to have false disks, due to their S\'{e}rsic indices and B/T ratios. The solid blue line shows a Gaussian function normalised to the red histogram, and slightly widened. This is used as a tool to construct a probability of a given decomposed disk component being false, see \S App. B. {\it Right-hand panel}: Examples of two functions used to model the S\'{e}rsic index, $n_{s}$, dependence on the probability of a galaxy harbouring a false disk component.}
\end{figure*}

\begin{figure*}
\rotatebox{270}{\includegraphics[height=0.49\textwidth]{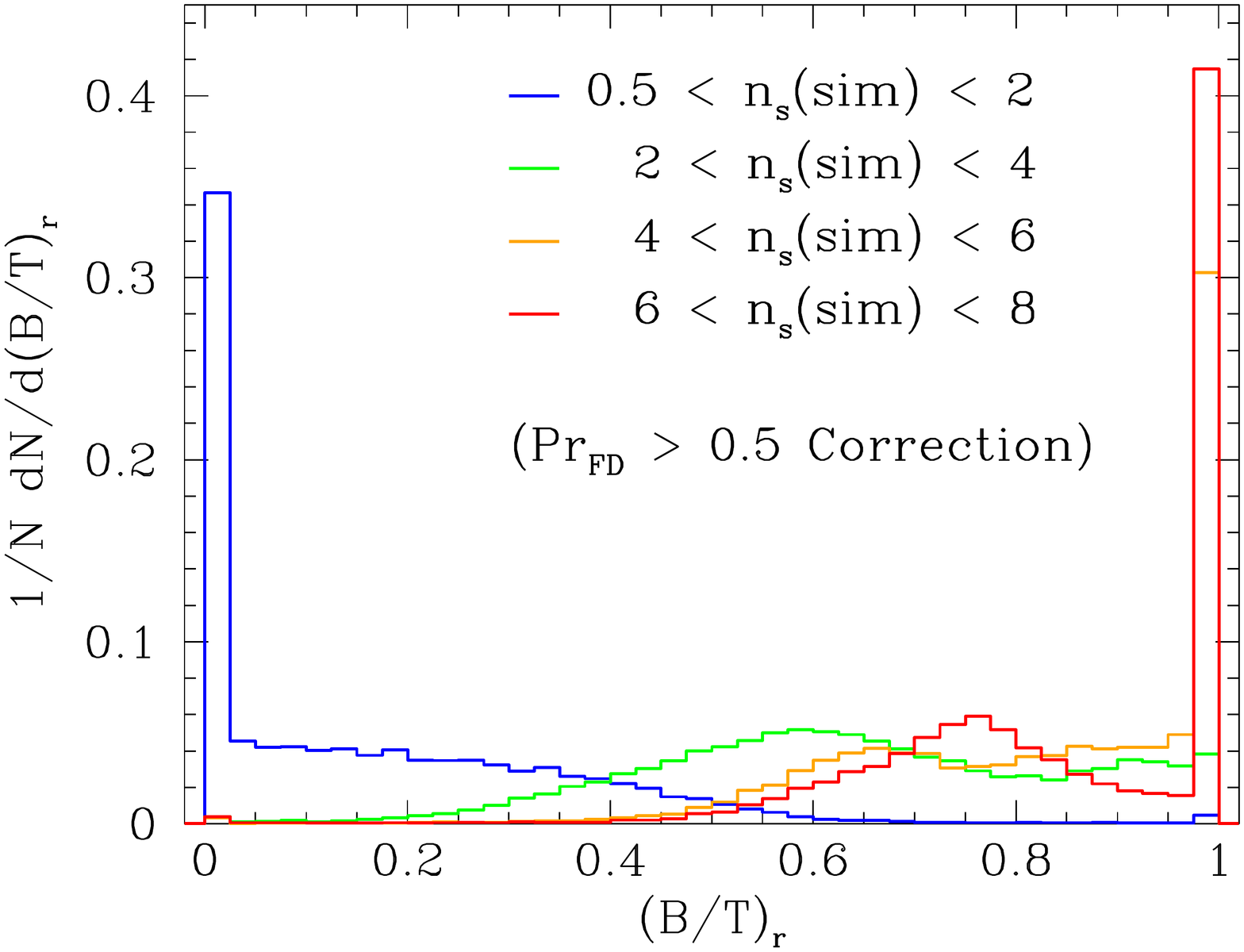}}
\rotatebox{270}{\includegraphics[height=0.49\textwidth]{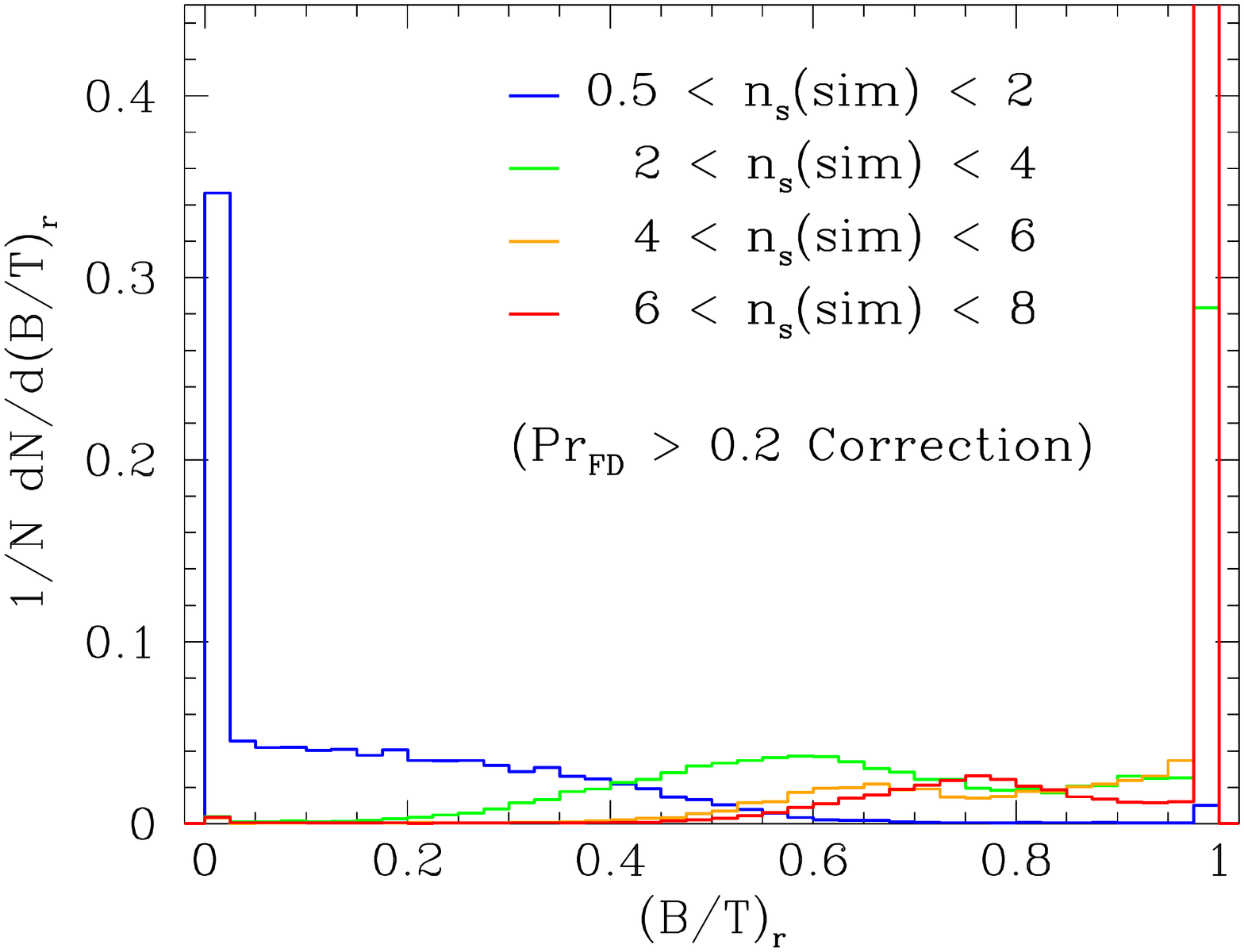}}
\caption{Normalised recovered (GIM2D) B/T ratio distributions in the r-band for pure S\'{e}rsic model galaxies. The colours indicate the input (single) S\'{e}rsic index ($n_{s}$) range, as shown on each plot. These plots show the same distribution as in Fig. A2, but for two levels of correction for `false disks' (as indicated on the plots), see \S App. B for details. Note how the lower (and wider) B/T value peak for high $n_{s}$ model galaxies decreases in prominence with increasing strength of `false disk' cut made, from left to right.}
\end{figure*}

At this juncture it is pertinent to note that in Simard et al. (2011) the b/a distributions for even high B/T galaxies were found to be fairly flat when a $P_{pS}$ cut was made, see \S 2.2. Thus, if we select only those galaxies which are statistically better fit with two-components as opposed to one, then the fraction of false disk interlopers is negligible. The trouble with such a method for false disk removal, however, is that the $P_{pS}$ statistic varies as a strong function of apparent magnitude and effective size of the galaxies being fit. Therefore, it can lead to a biased sample in terms of stellar mass, and B/T morphology, which can be extremely difficult to correct for. So, it is desirable to look for means to correct for false disks without using this statistic. Ultimately, there are trade-offs between completeness and purity in the recovered sample which must be carefully thought about when constructing a passive fraction, as in this paper.

Perhaps an obvious place to start with identifying false disks is to look at their colours, and note the population which is particularly red compared to the main population. Although this route might well be very successful in other studies, we choose not to use colour as a direct index of probability of a given disk being false because our current concern in this paper is investigating the star forming properties of galaxies. Thus, it is very important not to introduce any biases into the star formation rates of the parent sample, such as through assuming that disks would not be redder than some value. Instead we consider the colour gradient of galaxies identified as composites in our sample as a possible tool to identify false disks. We define this colour gradient as:

\begin{equation}
\Delta(g-r)_{\rm b,d} = (g-r)_{\rm bulge} - (g-r)_{\rm disk}
\end{equation}

\noindent which measures the difference in g-r colour between bulge and disk components in the same galaxy. We show the distribution of this colour gradient on the left-hand panel of Fig. B2, for our full sample (in black) and a selected sub-sample of composites with `suspicious' disks (in red). For the `suspicious' disks sub-sample, we consider those galaxies with B/T = 0.6 - 0.9 (where the `false disks' are found in the simulations of \S App. A) and with very high S\'{e}rsic indices ($n_{s} >$ 5) for the whole galaxy, which many would assume to be actually spheroids. 
Immediately it can be seen in Fig. B2 that the full sample extends out to much higher differences in colour than the restricted (possible false disk) sample, which have colours of bulges and disks which are closer to each other. Note, however, that they are not peaked around a $\Delta(g-r)_{\rm b,d}$ of zero. This is because, even for pure spheroids, there are observed colour gradients within galaxies, whereby the outer regions are bluer than the inner ones (e.g. Bernardi et al. 2003). 

\begin{figure}
\includegraphics[width=0.49\textwidth]{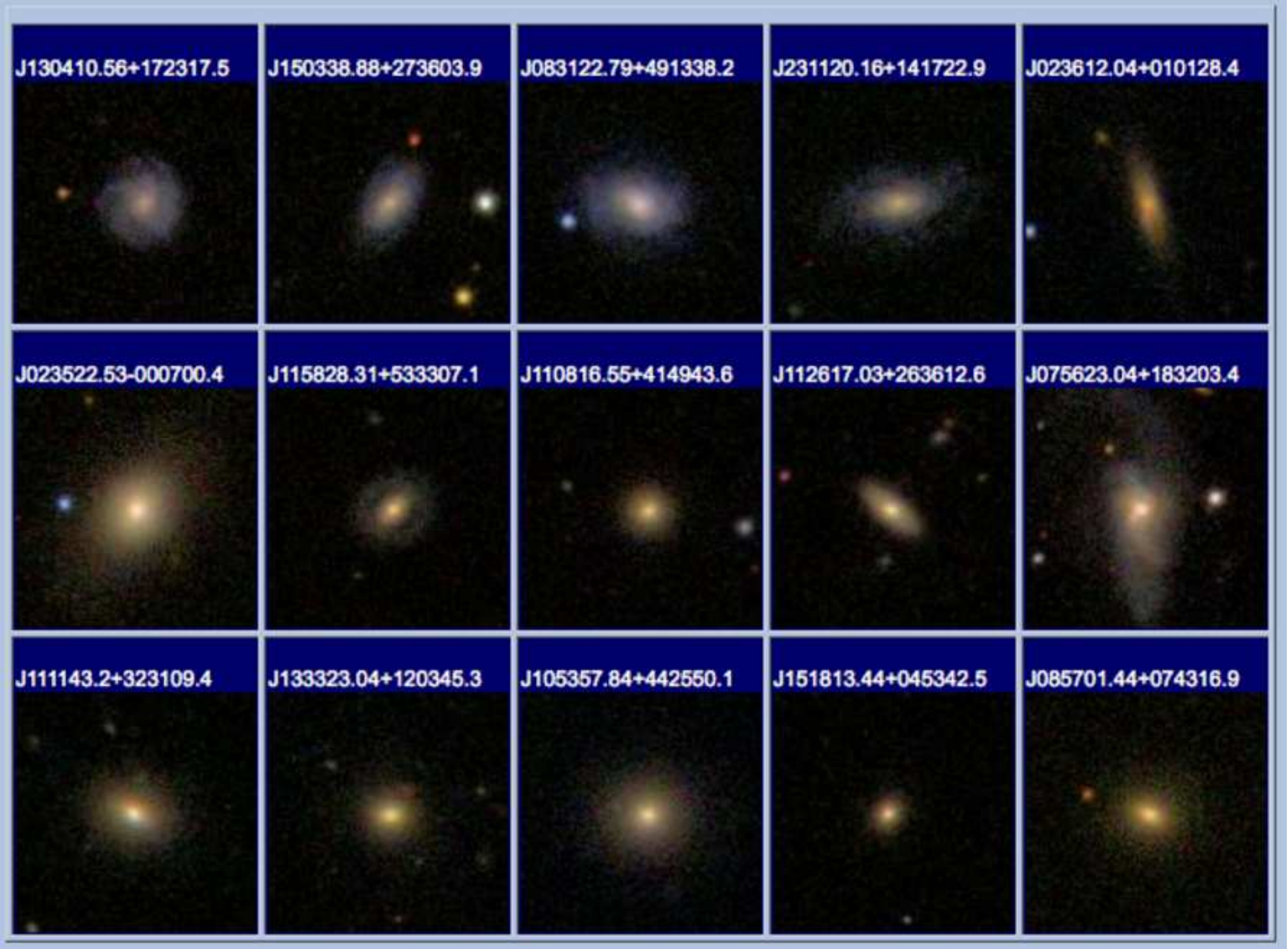}
\caption{A collage of SDSS galaxy images from our sample with z $<$ 0.1 and log($M_{*}/M_{\odot}$) $\sim$ 10.5. Crucially, each galaxy has a measured B/T stellar mass ratio of $\sim$ 0.7, indicating that it ought to be bulge dominated, but have a definite disk component. Note that this is the range of interest for defining `false disks', see Fig. A2. {\it Top row} is for $\rm Pr_{FD}$ $<$ 0.2 galaxies. In this group almost all galaxies looked at contain an obviously real disk component. {\it Middle row} is for 0.2 $<$ $\rm Pr_{FD}$ $<$ 0.5 galaxies. In this range there is a mix of real and false disks. {\it Bottom row} is for $\rm Pr_{FD}$ $>$ 0.5 galaxies. Here there are almost no real disks observed.}
\end{figure}

\begin{figure*}
\rotatebox{270}{\includegraphics[height=0.33\textwidth]{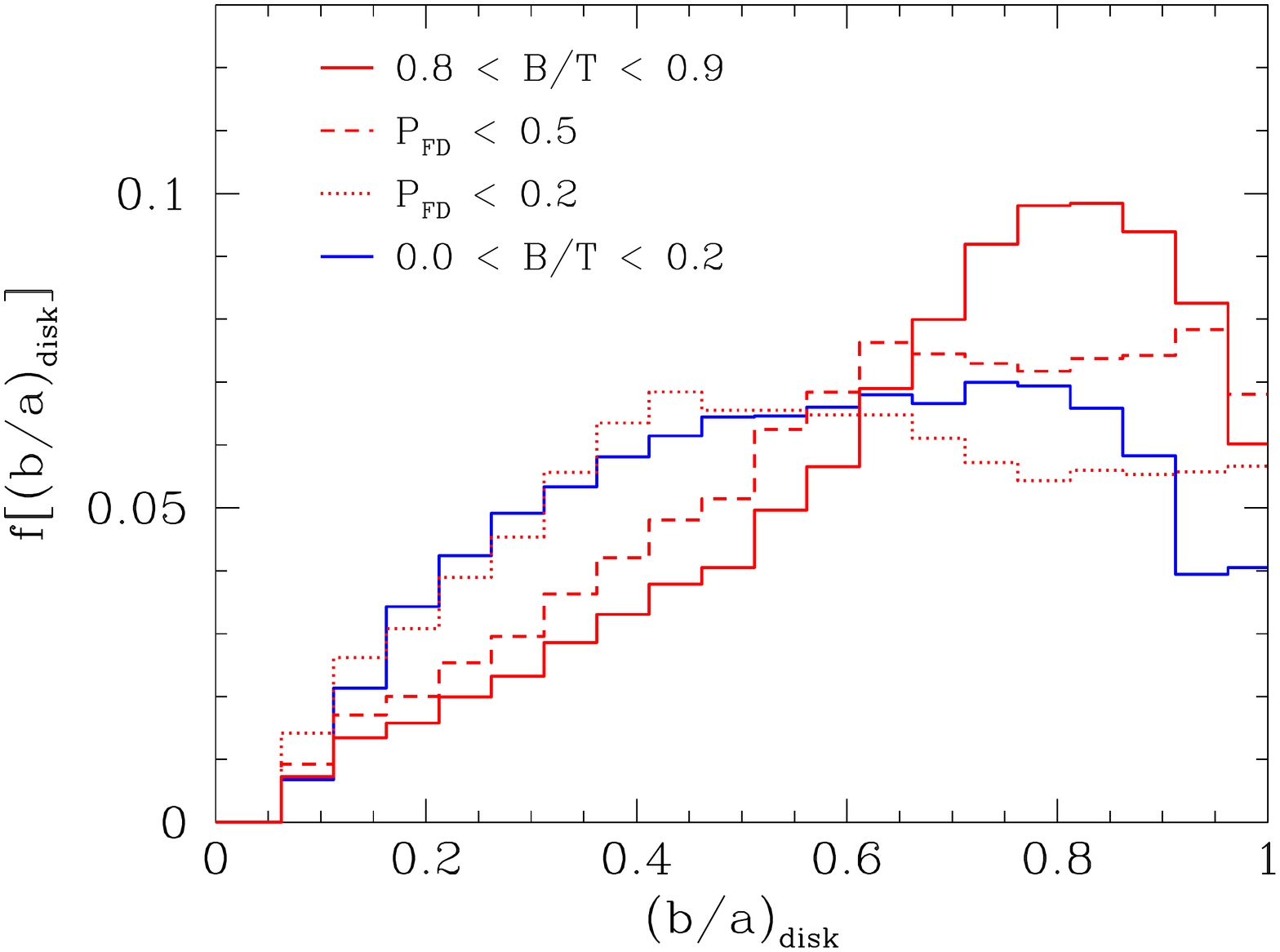}}
\rotatebox{270}{\includegraphics[height=0.33\textwidth]{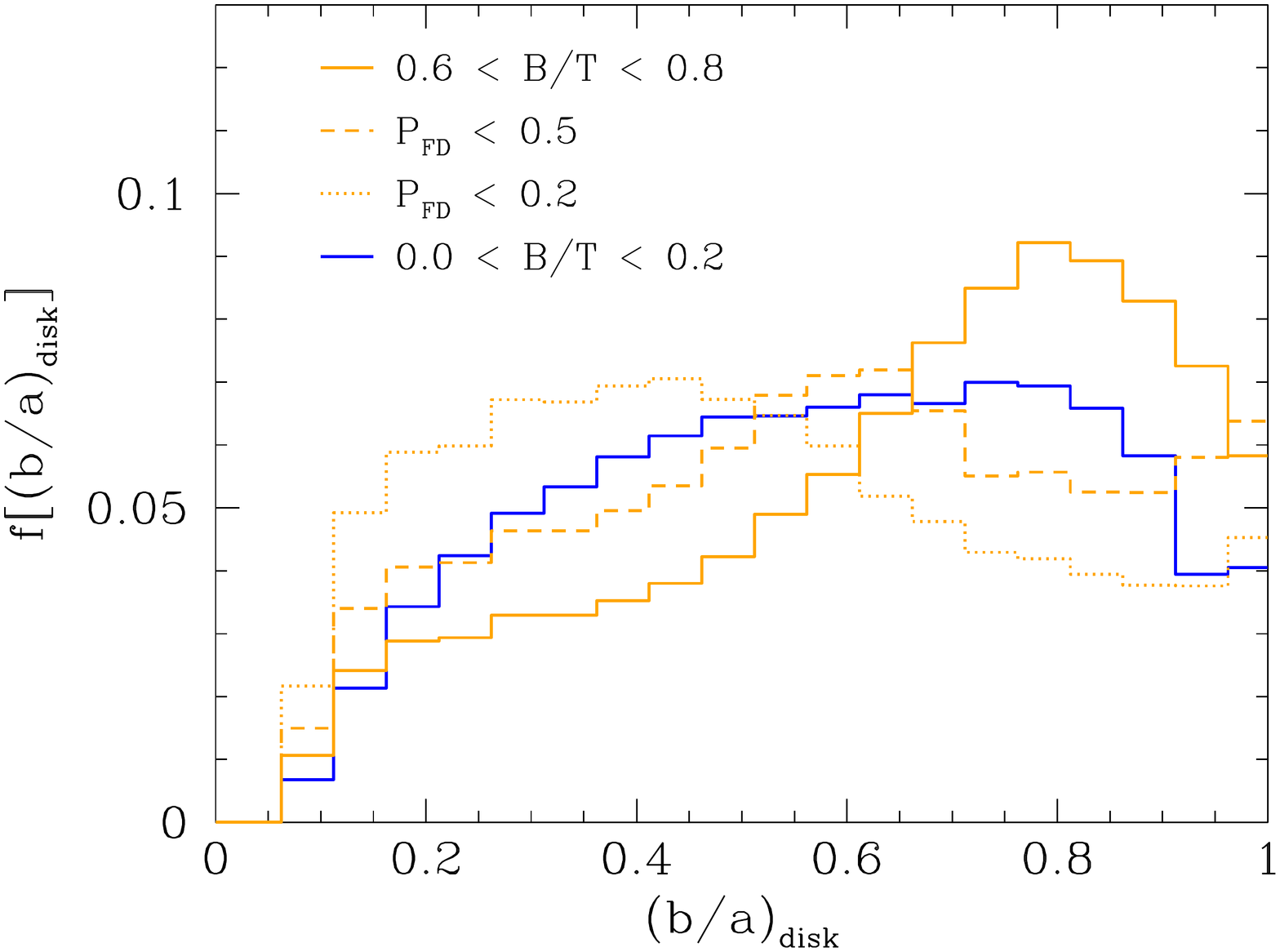}}
\rotatebox{270}{\includegraphics[height=0.33\textwidth]{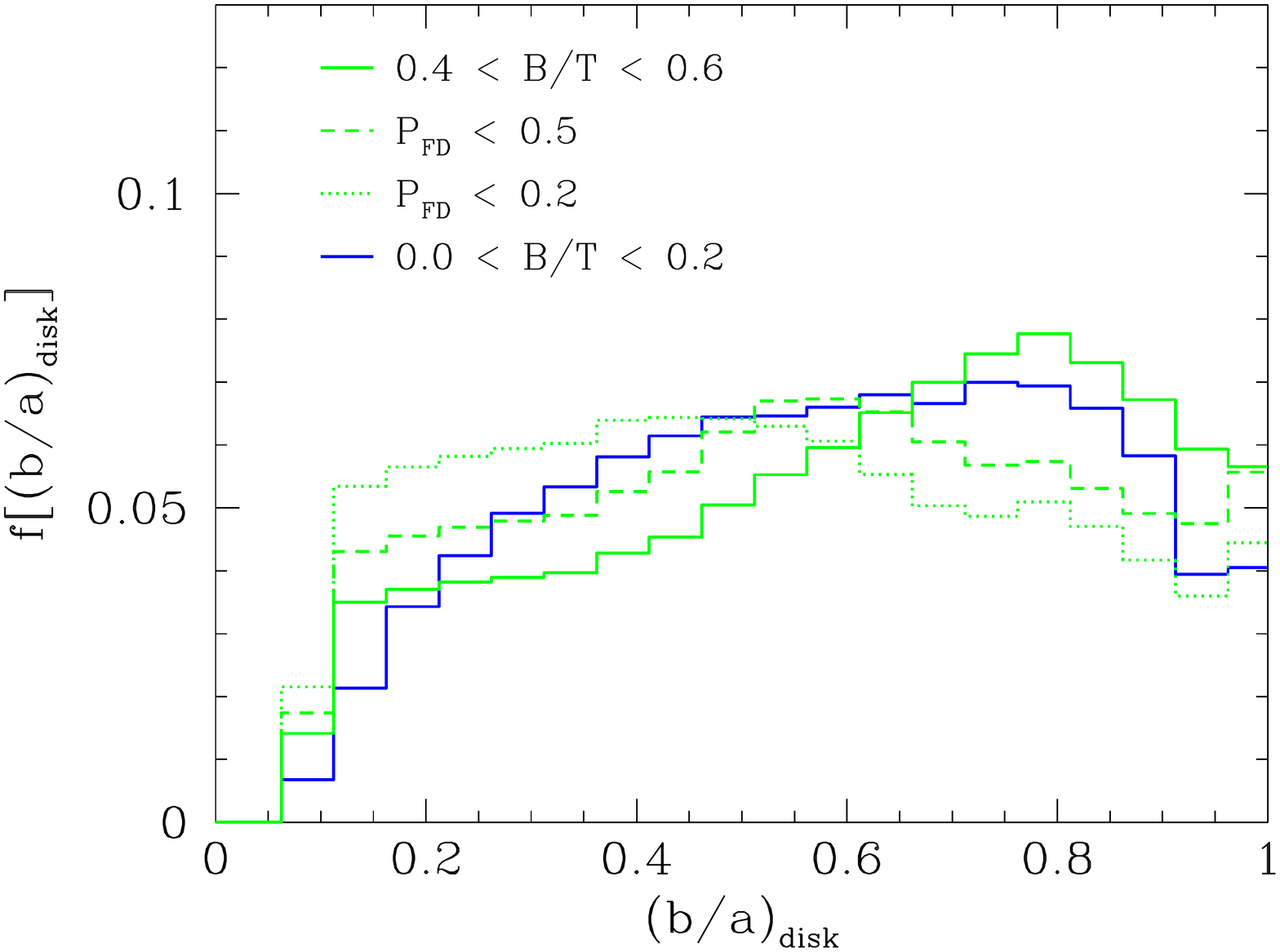}}
\rotatebox{270}{\includegraphics[height=0.33\textwidth]{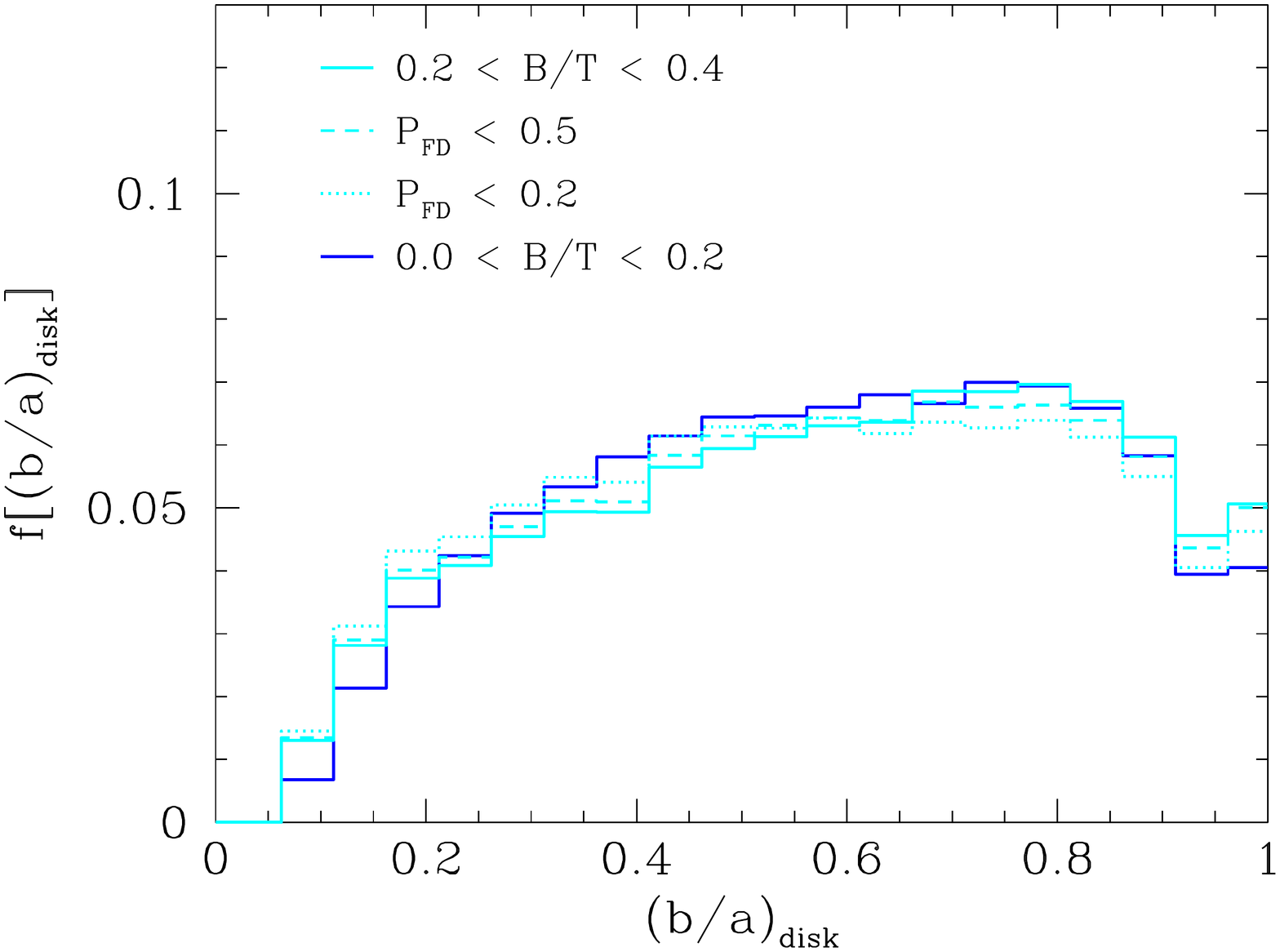}}
\rotatebox{270}{\includegraphics[height=0.33\textwidth]{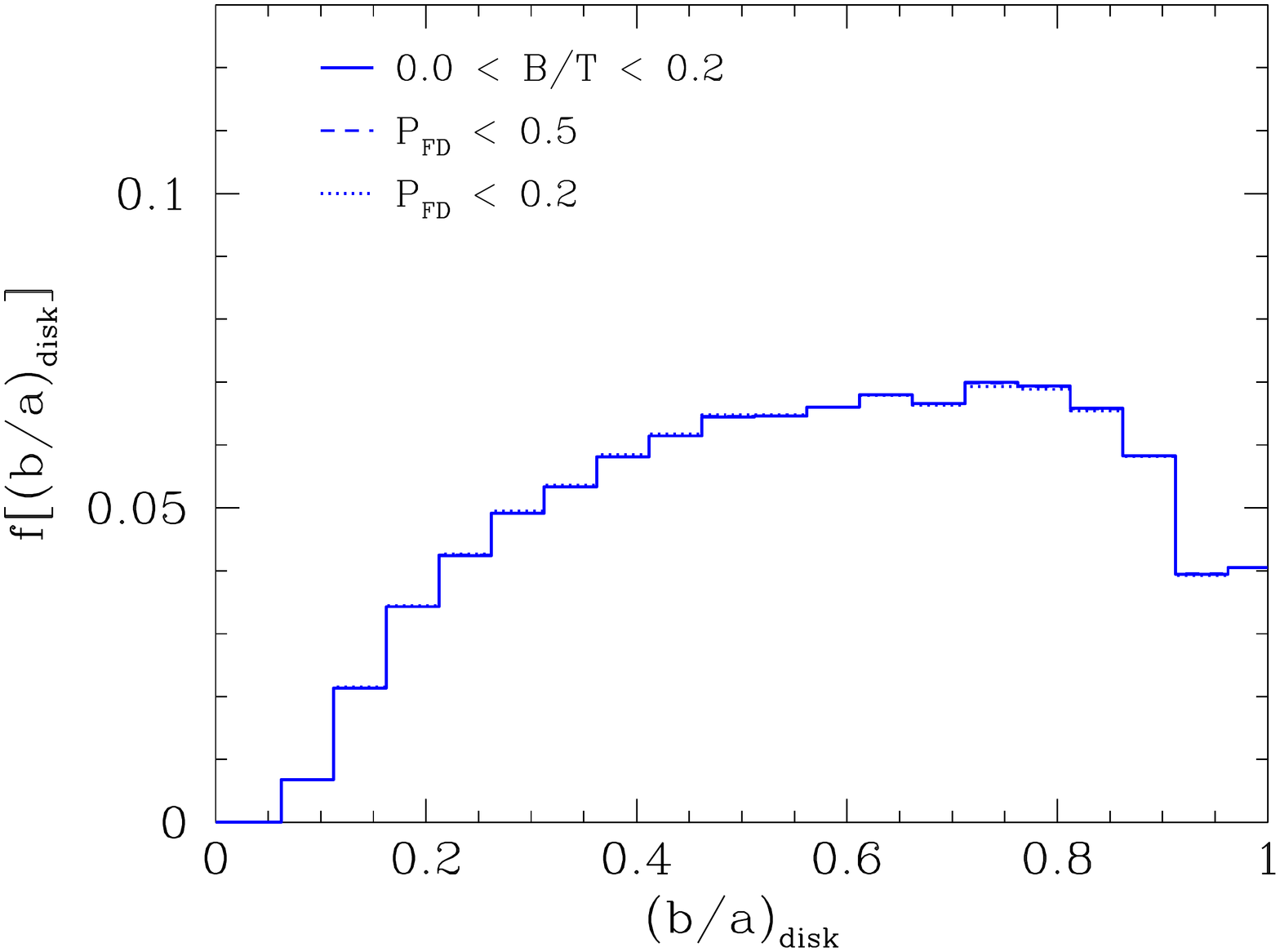}}
\caption{The normalised distributions of the semi-minor to semi-major disk axis ratio (b/a) for a range in B/T (by stellar mass) ratios, as in Fig. B 1. In this figure we additionally show two cuts for `false disk' correction (with probability of false disk ${\rm Pr_{FD}}$ $<$ 0.5 \& 0.2, defined in eq. B3), and show the revised histograms (dotted and dashed as indicated) on each B/T range b/a plot along with the original histogram. For comparison the, ostensibly flat, b/a distribution for disk dominated systems (with B/T $<$ 0.2) is shown. Note how at low B/T values the ${\rm Pr_{FD}}$ corrections made have no effect, which is good, whilst at higher B/T values the corrections serve to flatten the b/a distributions, as intended. For intermediate B/T ranges, the higher of the two cuts to false disks may over-correct, which can be seen by it causing a slight peak at lower b/a values. }
\end{figure*}

We fit a Gaussian to the colour gradient distribution of the suspicious disks, and then allow this to be broadened slightly (the only effect of this is to increase the absolute value of our final statistic). We will use this as one of our metrics for determining likely false disks. We assume that as one moves (in either direction) away from the peak value of the `suspicious' disk set the probability of being a false disk declines. Explicitly, we define the probability of being a false disk from this colour gradient metric alone as:

\begin{equation}
\rm {Pr_{FD}(\Delta(g-r)_{\rm b,d}) = exp \big(-\frac{1}{2} \big(\frac{\Delta(g-r)_{\rm b,d} - \mu}{\sigma}\big)^{2} \big) }
\end{equation}

\noindent where we take the mean $\mu$ = 0.17 and the standard deviation $\sigma$ = 0.225. This expression represents the Gaussian shown in blue in Fig. B2, but peaked now to unity for use as a likelihood function. It can be used to estimate how likely a given disk is to be false, based solely on the difference in colour of its apparent disk to its bulge. Obviously, such a cut would be incomplete as many real disks would be removed by any cut of this function as well, which can clearly be seen by the fact that the whole distribution of galaxies also peaks near the mean value of this function (in Fig. B2). Thus, additional metrics will be required.

We also consider the distribution of b/a for spheroids (from their ellipticities) in Fig. B1, and conclude that galaxies with false disks would have b/a disk distributions likely peaked where the spheroidal distribution peaks. Therefore, we take the distribution of b/a in spheroids as a probability function for disks of a given b/a being false, $\rm {Pr_{FD}(b/a)}$, by normalising the maximum value to unity. Finally, we add in the expected correlation with increased S\'{e}rsic index from a single fit, constructing yet another probability distribution function, $\rm{Pr_{FD}(n_{s})}$, see right hand panel of Fig. B2. 
We model this dependence as a step function and as a simple power law, but note here that the only real difference this makes is in the absolute value of the final probability at which we will place our final cut(s) at. Bringing this together, we construct a final probability of a given decomposition having a false disk element, $\rm {Pr_{FD}}$, by convolving the three probability indicators outlined 
above. Specifically, we calculate:

\begin{equation}
\rm { Pr_{FD} = Pr_{FD}(b/a) \times Pr_{FD}(\Delta(g-r)_{\rm b,d}) \times Pr_{FD}(n_{s}) }
\end{equation}

\noindent where each of these components are defined above. Although these metrics are not strictly independent, and hence the use of multiplicative convolution would technically require a subtractive co-variance term, we are only attempting to get an approximate indicator of when a given galaxy is likely harbouring a false disk in its decomposition. Moreover, we use this probability as a phenomenological tool only, tuned to the data, as outlined below. Thus, in the absolute sense, whether or not $\rm{Pr_{FD}}$ is in fact interpretable as a true probability is moot, yet ultimately unimportant for our purposes.

We show the effect of selecting out galaxies above a range in $\rm{Pr_{FD}}$ values on the pure single S\'{e}rsic model data in Fig. B3. In this test we remove galaxies with $\rm{Pr_{FD}} > $ 0.2 and 0.5 and replace them back into the sample as pure spheroids with B/T = 1. We see that it is possible to correct for virtually all false disks at a cut at 0.2 (right panel), but that one can still do a good job of reducing the fraction of false disk interlopers with a more modest cut at 0.5 (left panel). In fact the bimodality of the recovered B/T distribution for high S\'{e}rsic index galaxies is almost completely removed by a judicious cut in our probability function, $\rm{Pr_{FD}}$. However, one must also consider whether this (perhaps over-zealous) cut could result in throwing out potentially real disks as well.

To test this, we investigate the fraction of genuine bulge + disk composite systems in the simulations which are erroneously corrected for having a false disk. We find that whilst for a cut in $\rm{Pr_{FD}}$ at 0.5 less than 10 \% of real disks are mistakenly reassigned to a B/T of one, at a cut of 0.2 as much as 50 \% of the corrected disks may have actually been real. How to proceed comes down to a choice in philosophy: is it more important to be complete or pure in terms of sample selection. Clearly, both are desirable but this is an extremely difficult thing to achieve in practice. 
For a visual indication of this effect, see Fig. B4, which shows how galaxies actually look at a stellar mass $M_{*} \sim 10^{10.5} M_{\odot}$ and with B/T $\sim$ 0.7 in three ranges of  $\rm{Pr_{FD}}$. Clearly, at $\rm{Pr_{FD}} < $ 0.2 all of the galaxies do contain real disks, and this fact in bourne out by looking at hundreds of these cases in detail by eye. Alternatively, at $\rm{Pr_{FD}} > $ 0.5 there are virtually no real disks 
evident. Perhaps unsurprisingly, from our definition of the false disk probability metric, at 0.2 $< \rm{Pr_{FD}} < $ 0.5 there are some galaxies which do harbour false disks and some which do not.

To see what the effect of various cuts is on the real data, we plot again the b/a distributions in Fig. B5, and separate this time by both B/T range and a $\rm{Pr_{FD}}$ cut as indicated. Using a cut in $\rm{Pr_{FD}}$ value to correct for false disks in general flattens the distributions, as we would hope. Specifically, we see very little effect at low B/T from our correction which is expected. At higher B/T the various $\rm{Pr_{FD}}$ cuts improve the peaked b/a distributions. A cut at 0.5 is sufficient to flatten the distributions in b/a up to a B/T binninng of 0.8 or so, but to flatten the highest B/T `disks' one would have to go to 0.2 or so. This more exclusive cut appears to over-correct at lower B/T values, and results in a (slight) peak at lower b/a for corrected disks in this regime.  

Throughout the results section of this paper we have employed a cut in the probability of false disk, $\rm{Pr_{FD}}$, at 0.5 whereby galaxies with higher probabilities are replaced to a B/T = 1 and a $M_{\rm bulge} = M_{*}$. This cut provides a good compromise between purity and completion which is appropriate for a study like this, where inclusivity is important due to the need to construct relative fractional abundances of galaxies with star forming properties as a function of structural and environmental indices. 
The recovery of pure spheroids as pure spheroids (in the highest B/T bin) in the simulations is as good as it is for disks when we apply this cut. Moreover, only around 10 \% of corrected galaxies are actually not false by our estimates at this threshold. In the next section we show the general invariance of the main results of this paper to the exact cut on the probability of false disk which we make, and in particular demonstrate that for both a `complete' (but not pure) and a `pure' (but not complete) sub-set we recover the same differential results between the potential drivers of star formation cessation, which are the primary results of this work.

\section{Stability of the Main Results}

\begin{figure*}
\rotatebox{270}{\includegraphics[height=0.33\textwidth]{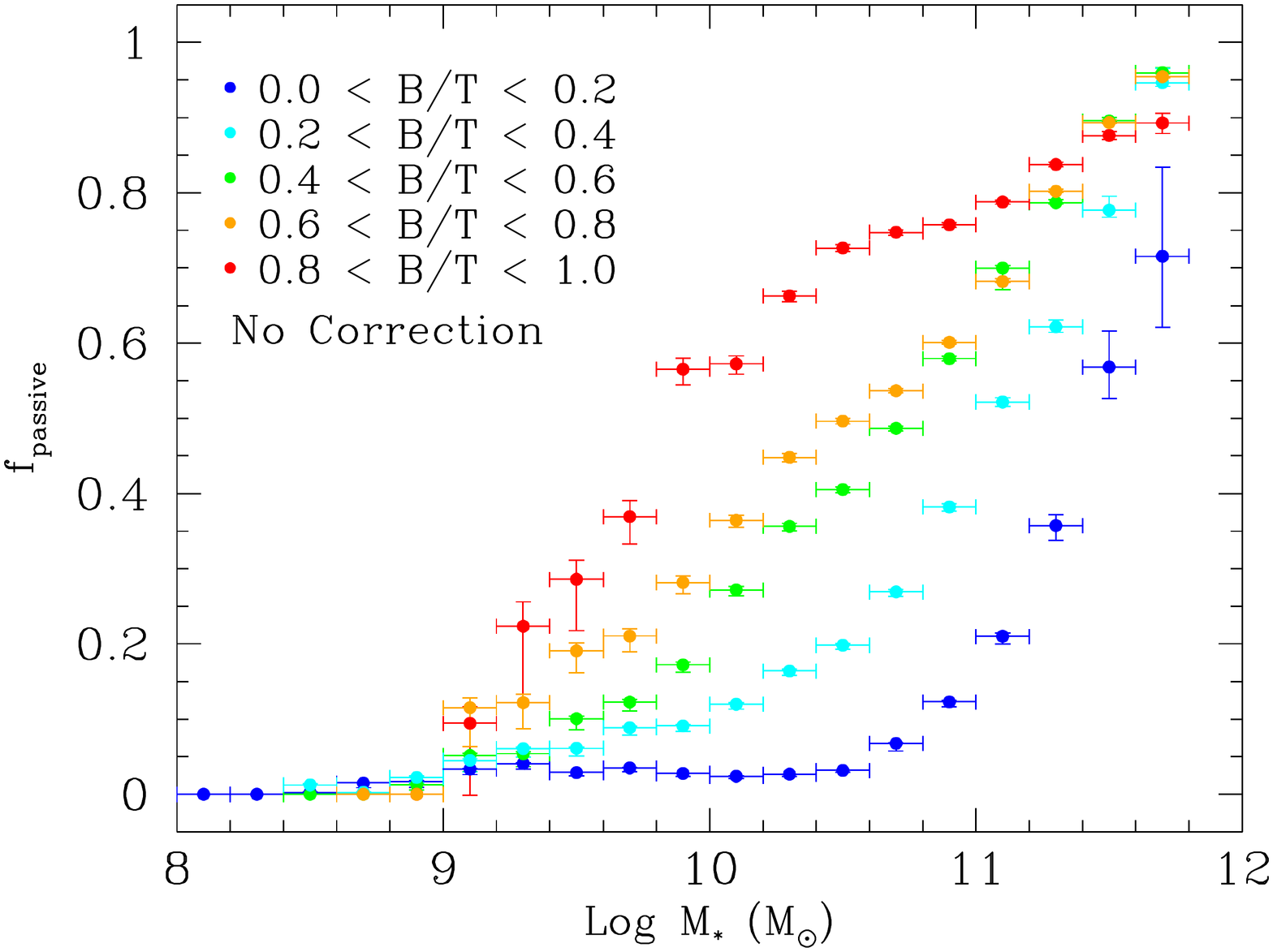}}
\rotatebox{270}{\includegraphics[height=0.33\textwidth]{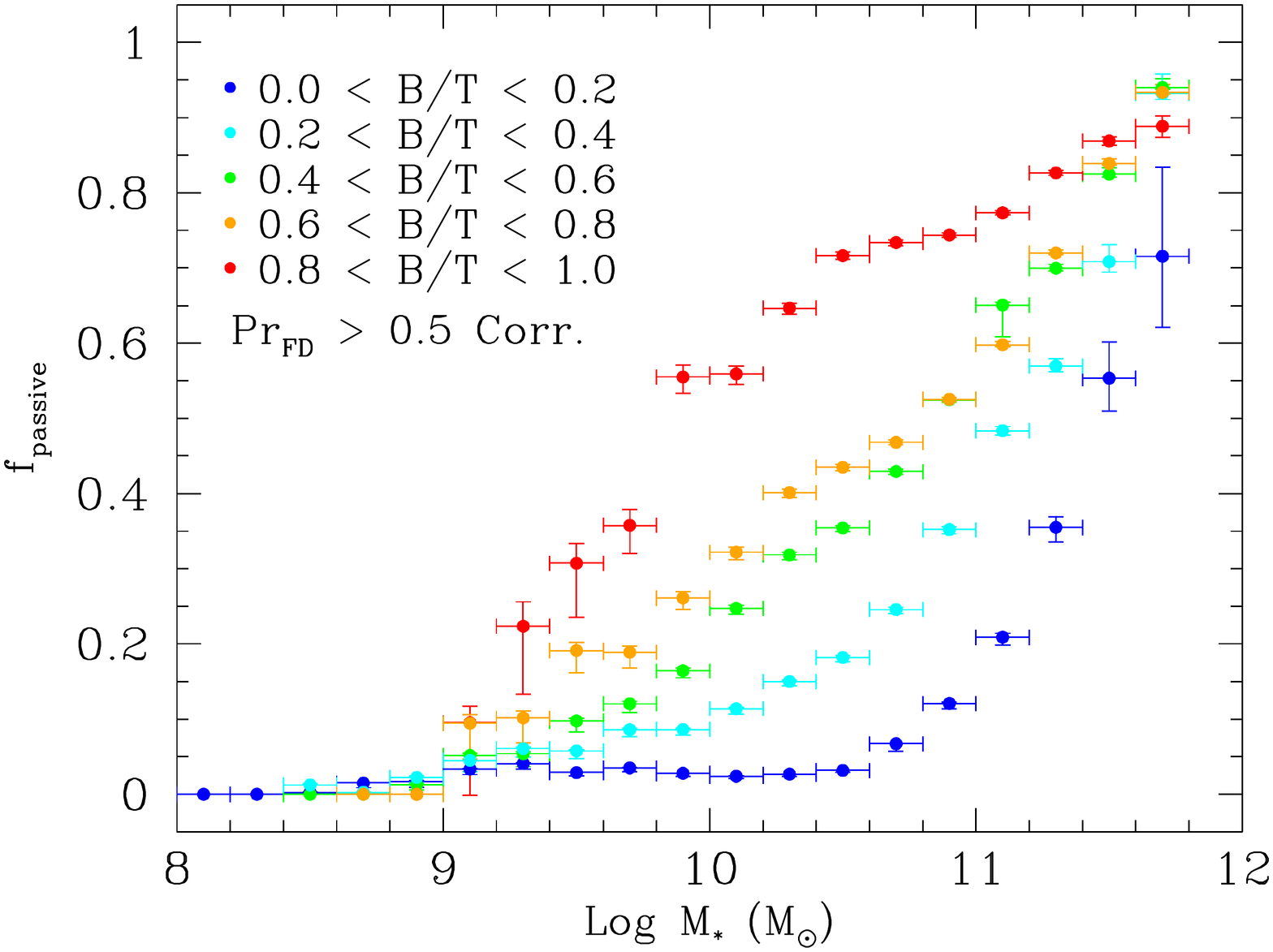}}
\rotatebox{270}{\includegraphics[height=0.33\textwidth]{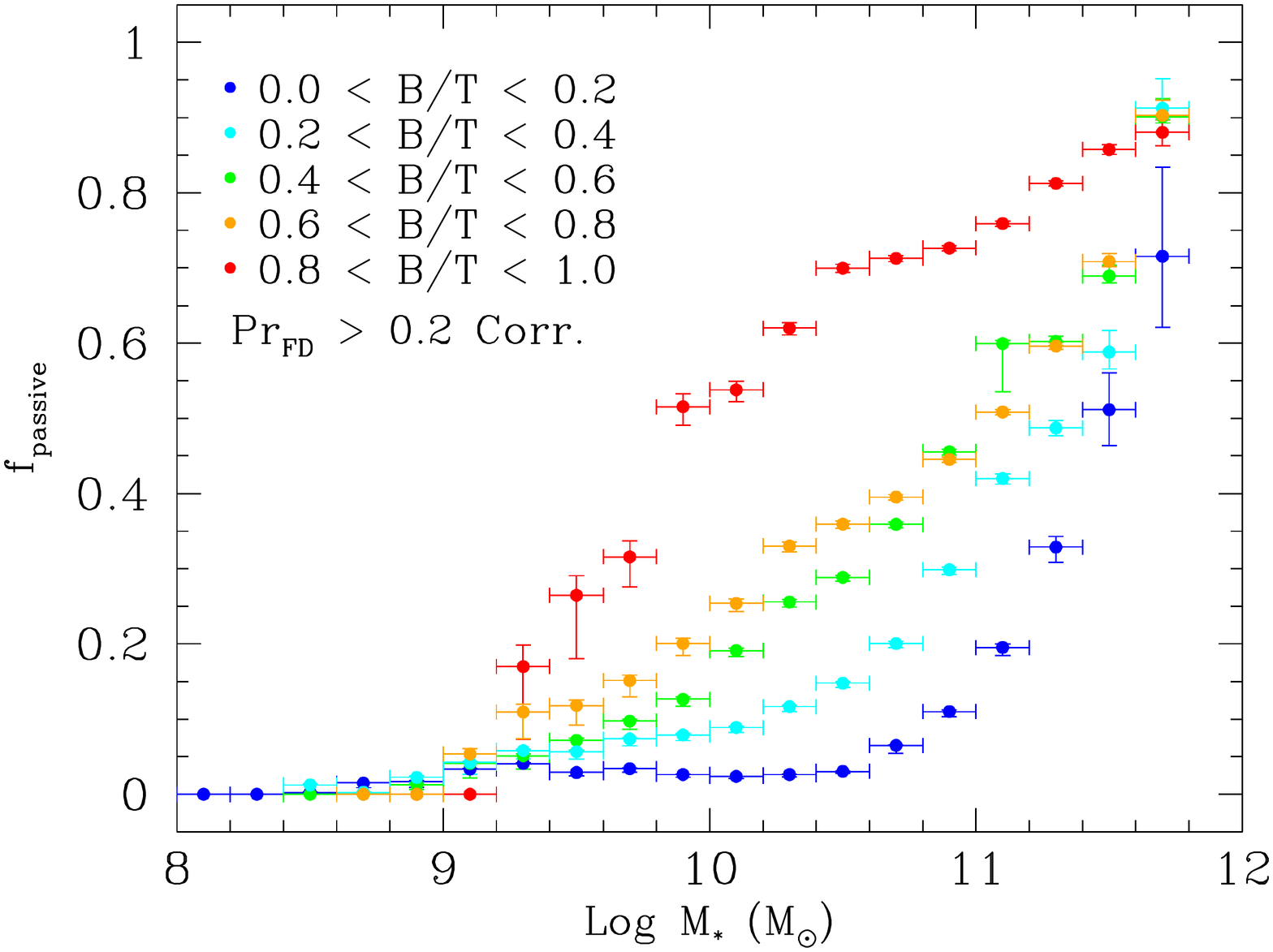}}
\rotatebox{270}{\includegraphics[height=0.33\textwidth]{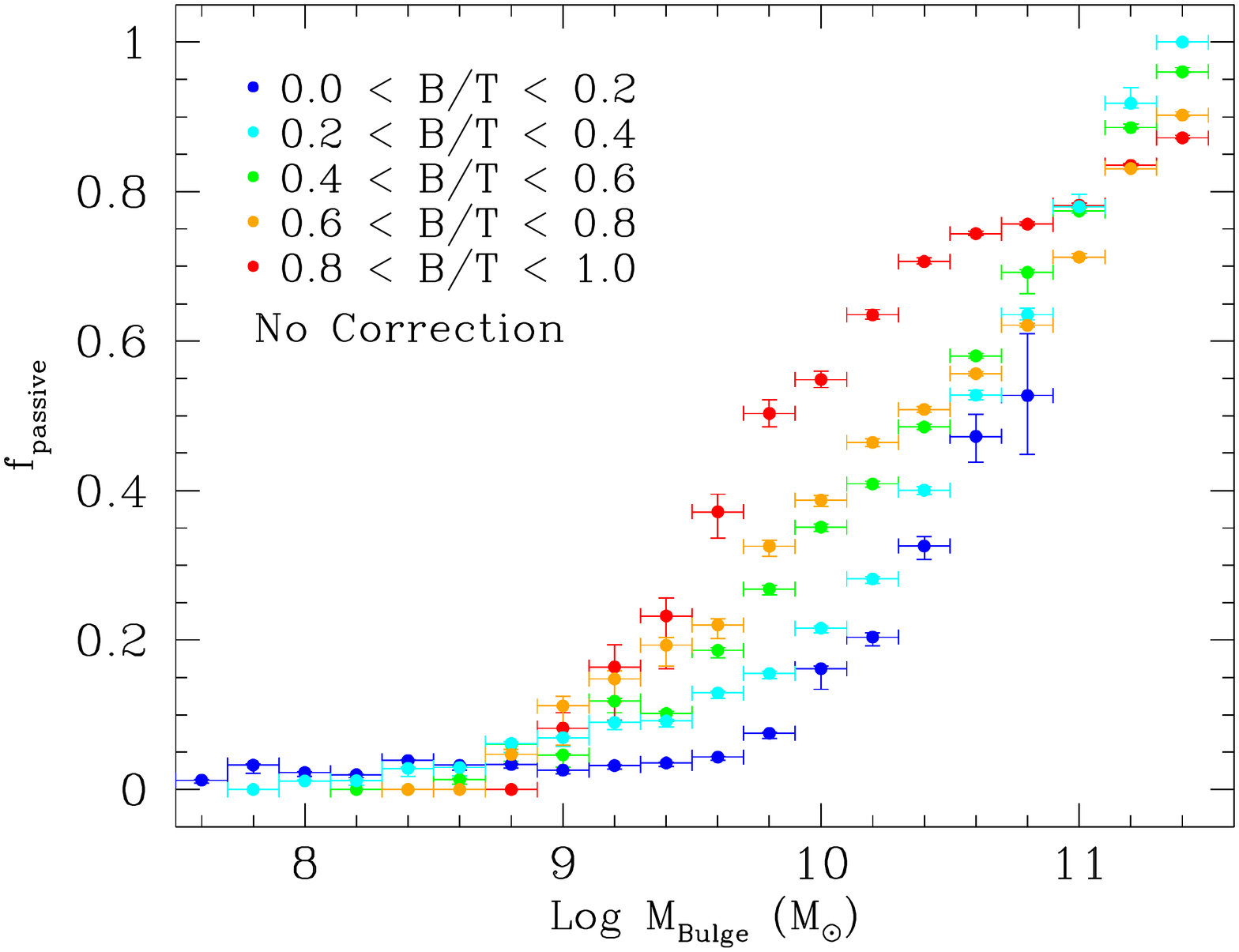}}
\rotatebox{270}{\includegraphics[height=0.33\textwidth]{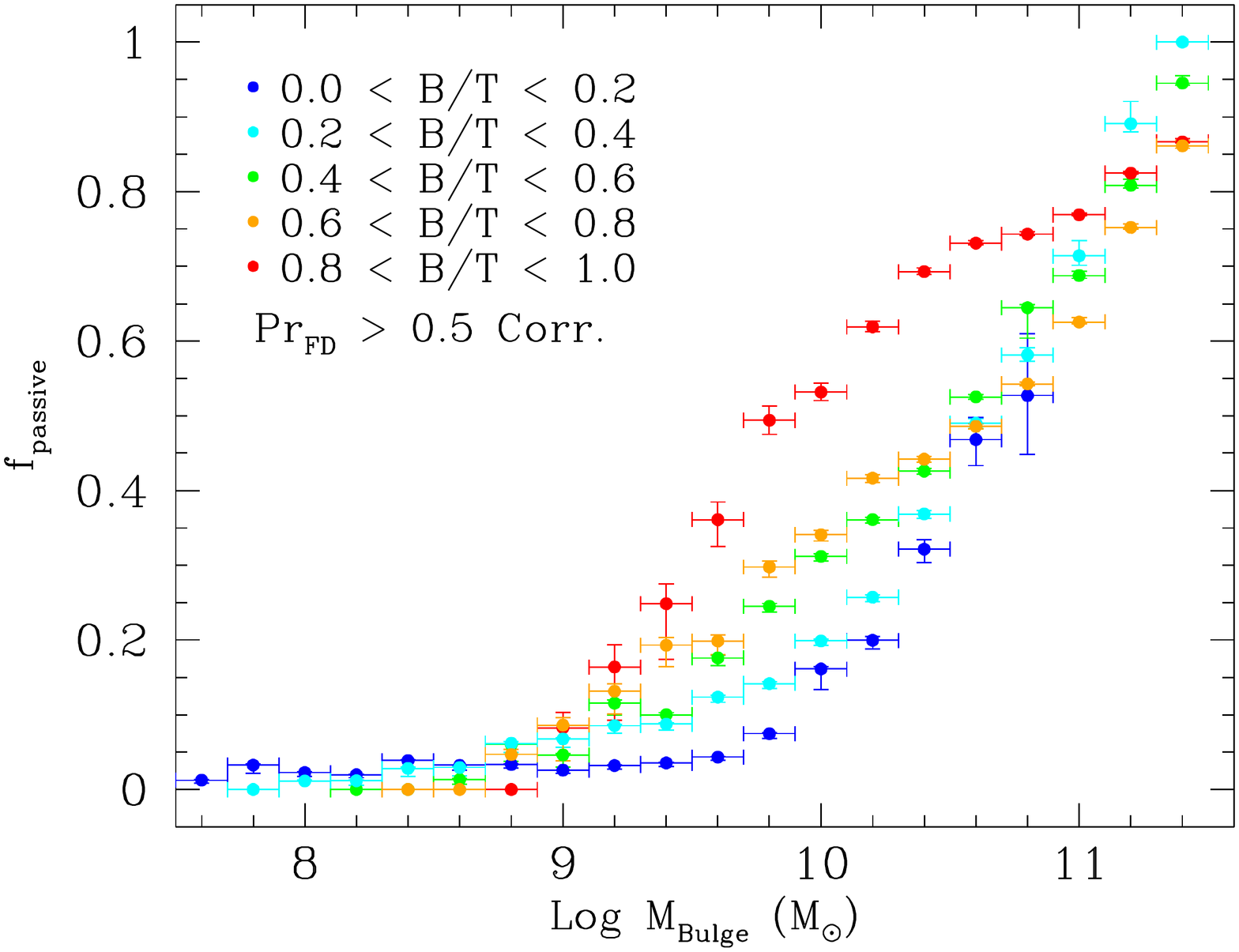}}
\rotatebox{270}{\includegraphics[height=0.33\textwidth]{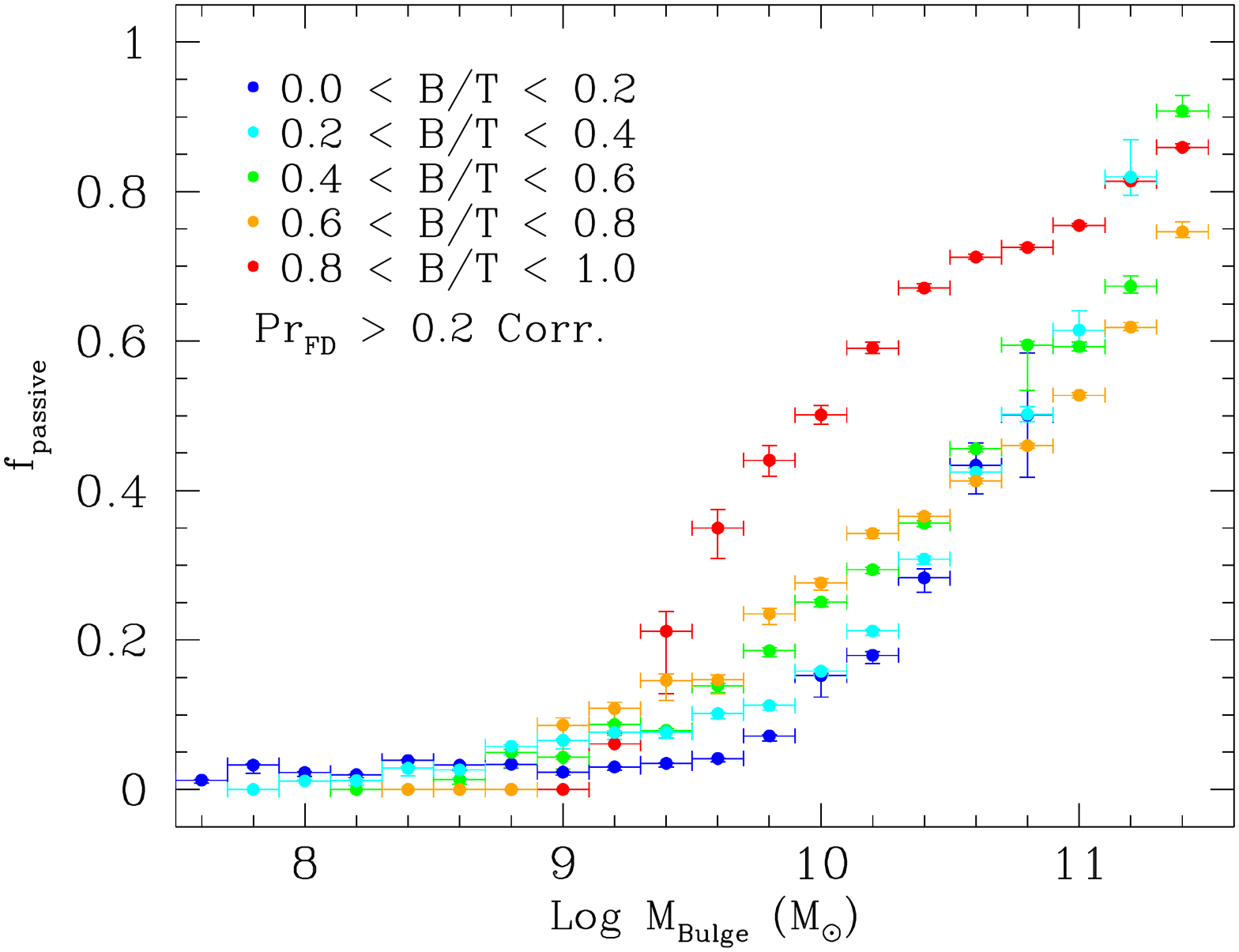}}
\rotatebox{270}{\includegraphics[height=0.33\textwidth]{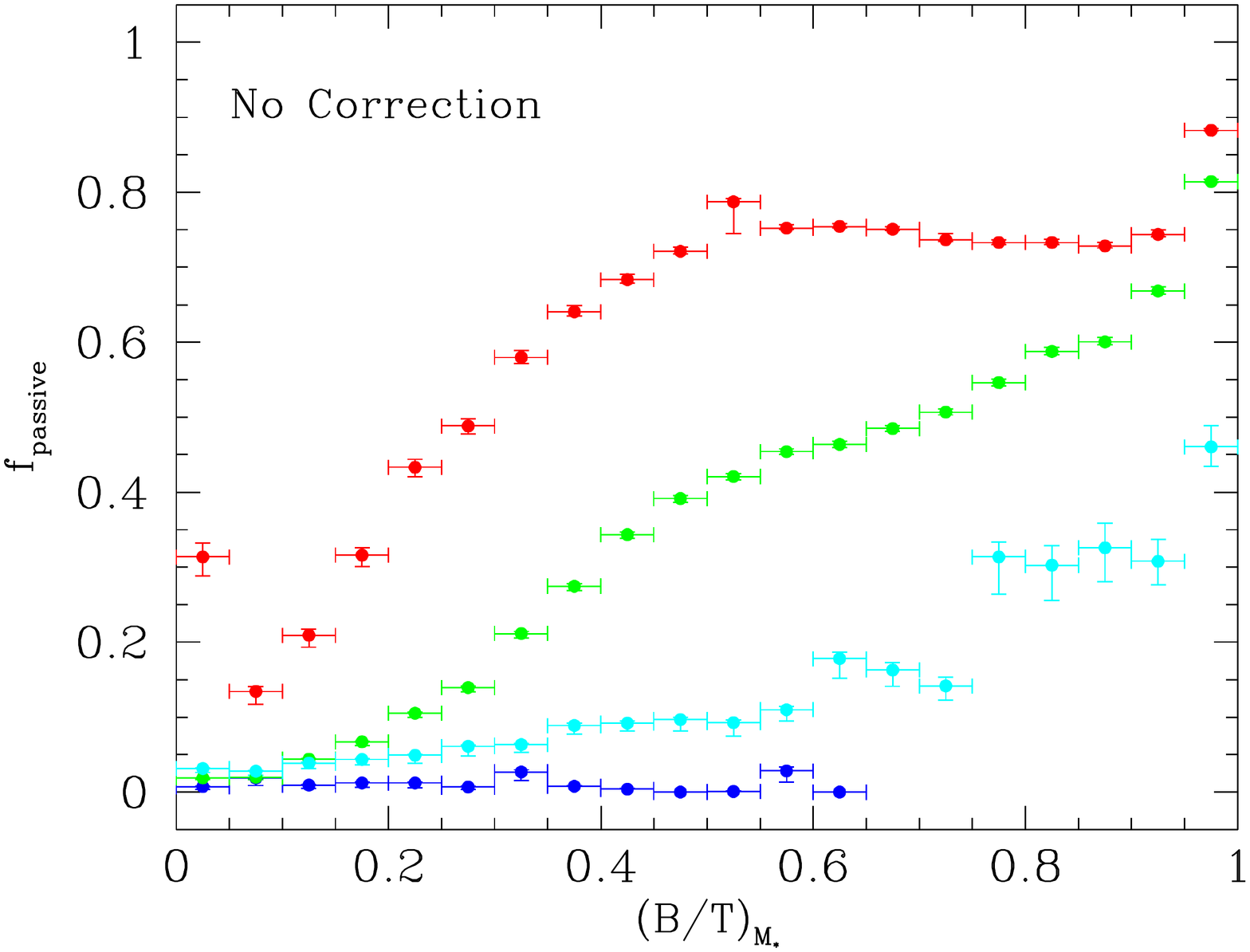}}
\rotatebox{270}{\includegraphics[height=0.33\textwidth]{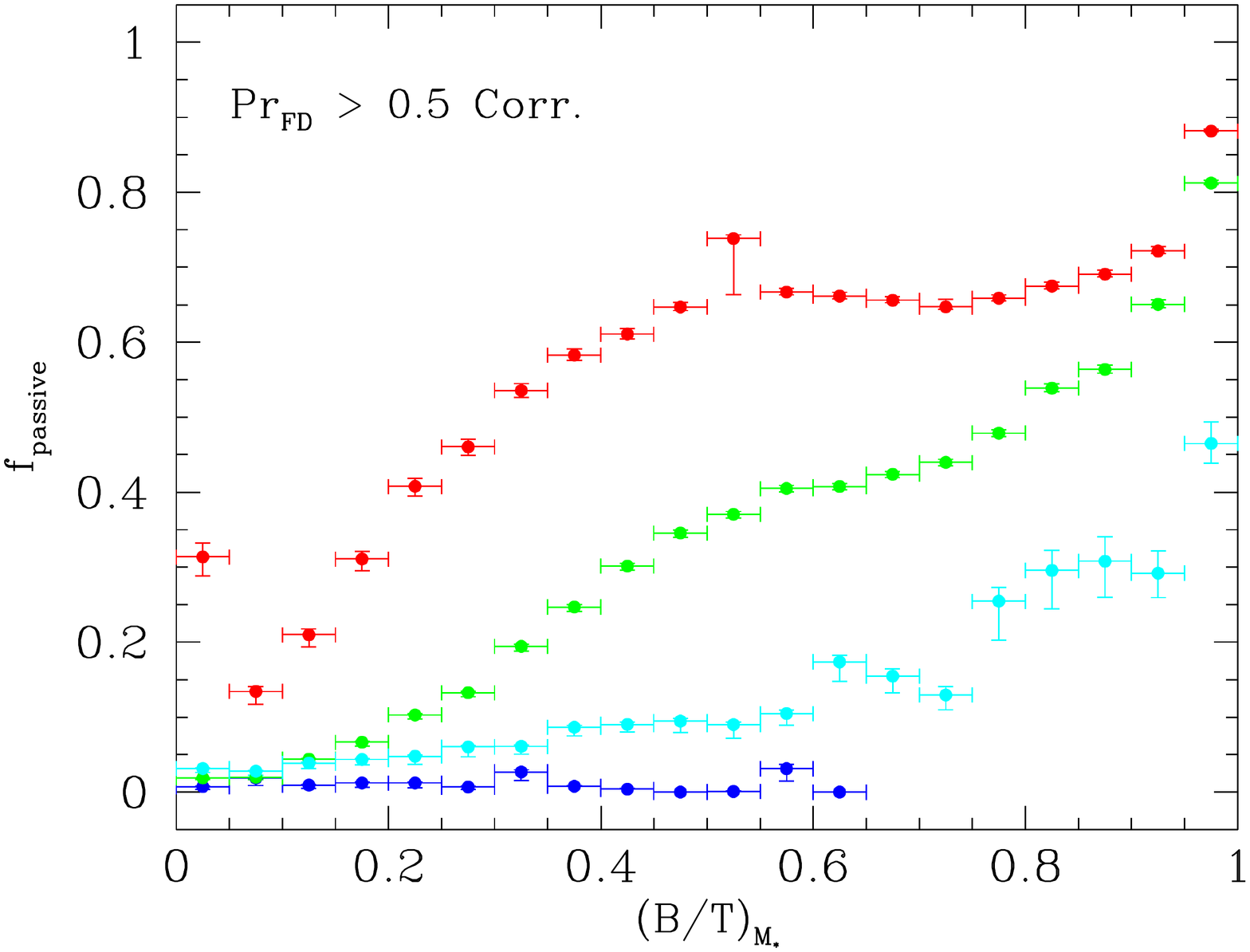}}
\rotatebox{270}{\includegraphics[height=0.33\textwidth]{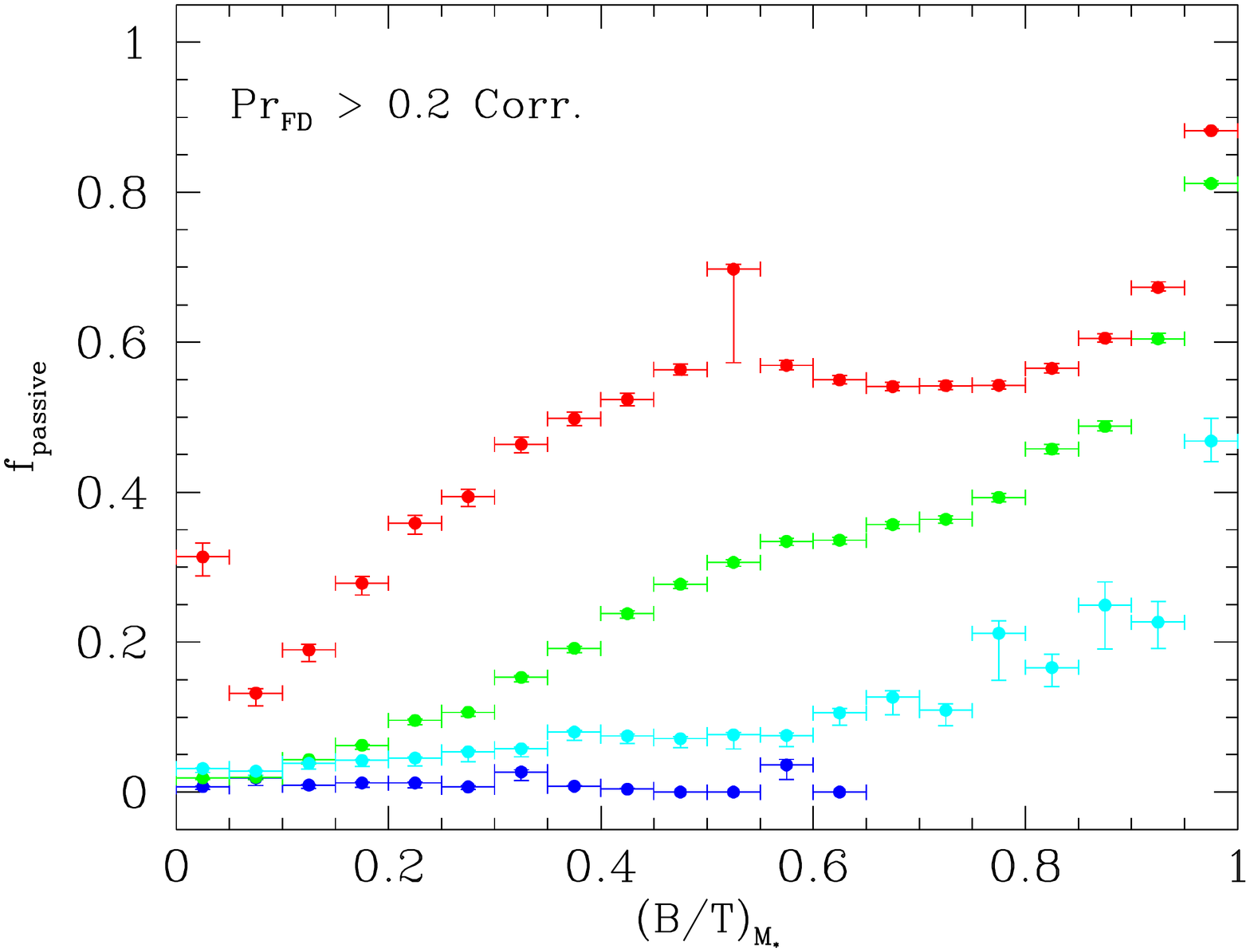}}
\rotatebox{270}{\includegraphics[height=0.33\textwidth]{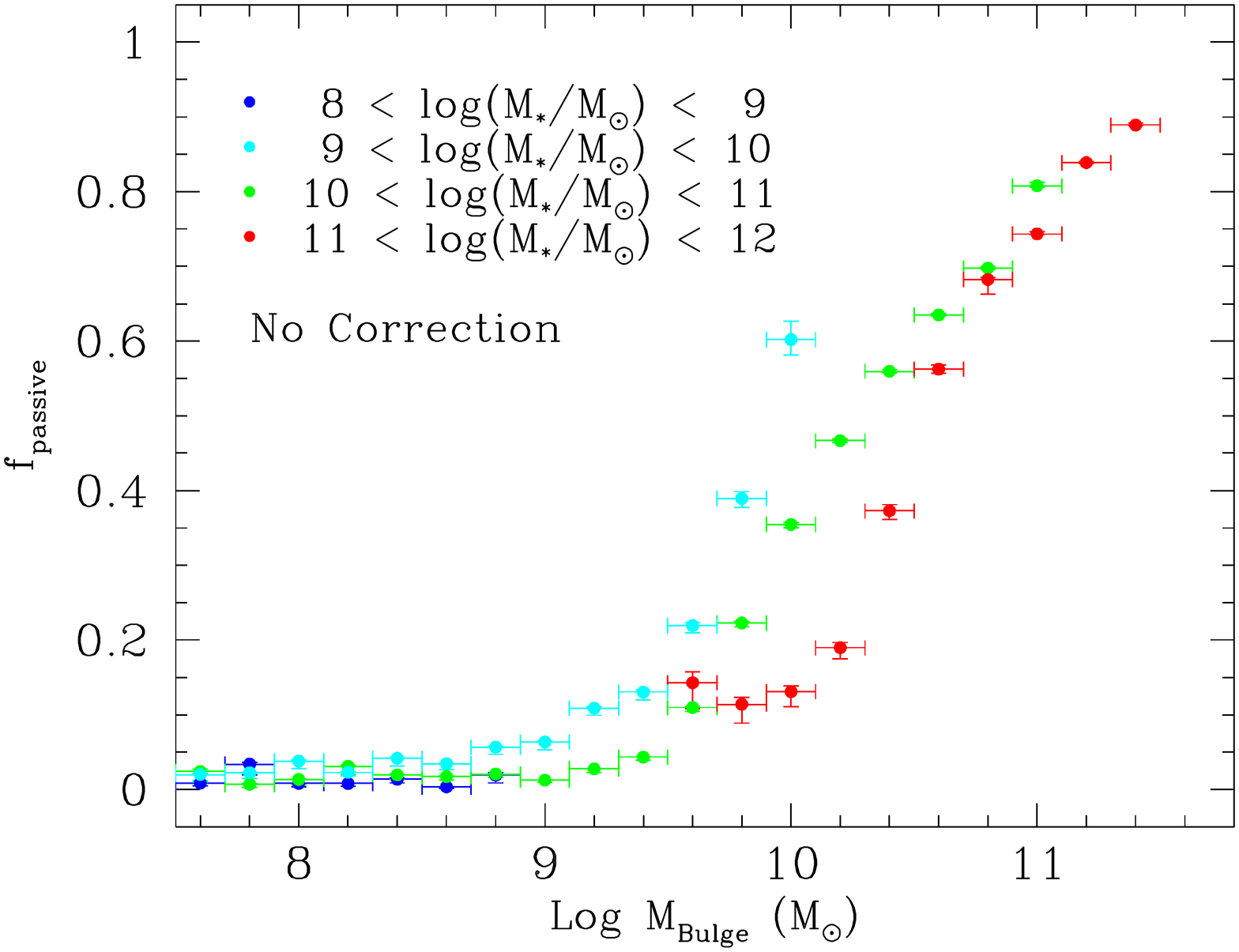}}
\rotatebox{270}{\includegraphics[height=0.33\textwidth]{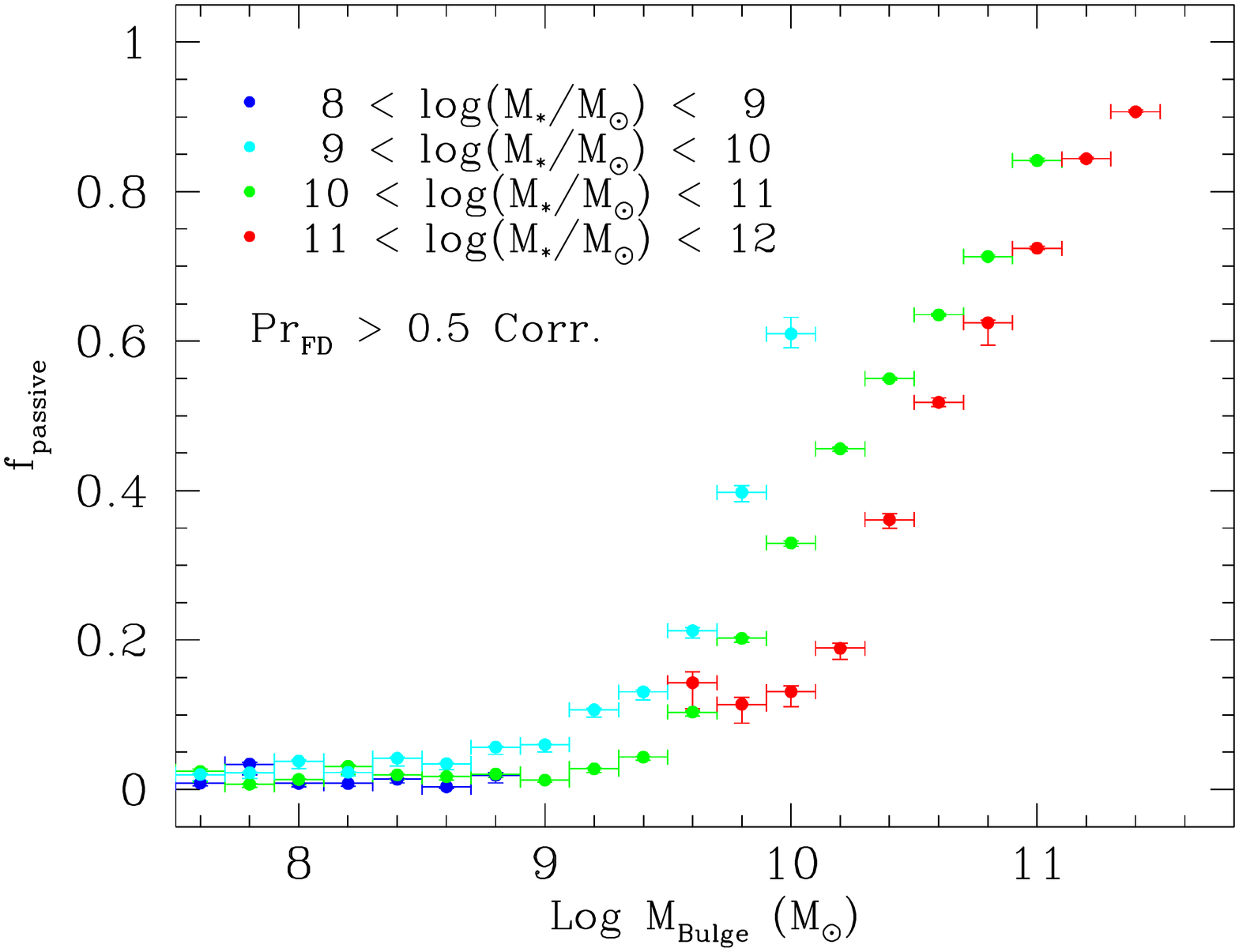}}
\rotatebox{270}{\includegraphics[height=0.33\textwidth]{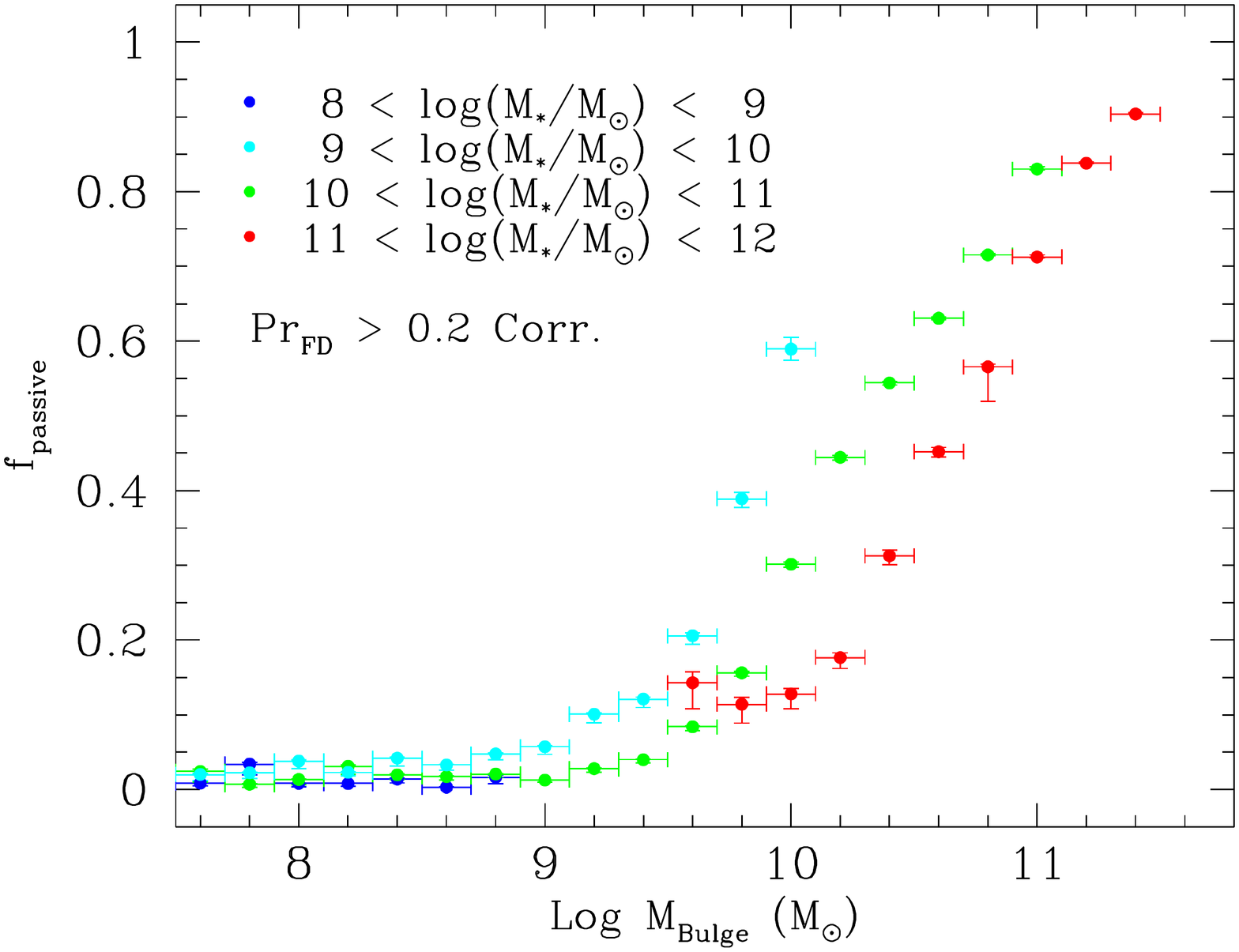}}
\caption{Each column of plots represents a rendering of several of the main science plots for this paper, first seen in Figs. 11 \& 12, with a different cut made on the probability of the disk component being false, ${\rm Pr_{FD}}$. The {\it left hand column} shows plots with no correction for false disks, thus the sample is complete but not pure. The {\it middle column} shows plots with a compromise cut with ${\rm Pr_{FD}}$ $>$ 0.5 galaxies removed from the sample and replaced back in with B/T = 1 and $M_{\rm bulge}$ = $M_{*}$. The {\it right hand column} shows plots with a verbose cut at ${\rm Pr_{FD}}$ $>$ 0.2, which forms a pure, but not complete, sub-sample. The {\it top row} shows the passive fraction - total stellar mass relation, with the {\it second row} down showing the passive fraction - bulge mass relation, both of which are sub-divided by B/T morphology as indicated. 
The {\it third row} down shows the passive fraction - B/T morphology relation, with the {\it bottom row} showing the passive fraction -bulge mass relation, both of which are subdivided by total stellar mass range as indicated on the lower plots. Note that the structure of these relationships remains invariant under the various cuts for false disks shown here, e.g. bulge mass remains a tighter correlator to the passive fraction than either total stellar mass or morphology, independent of the ${\rm Pr_{FD}}$ cut made.}
\end{figure*}

In this section we establish the invariance of our primary conclusions to various cuts and selections made on the data in the main analysis of this paper. In particular, we consider three possible cuts on our probability of false disk parameter, ${\rm Pr_{FD}}$, which represent an over zealous cut at 0.2, whereby almost no false disks will remain, but many actual disks will be erroneously corrected; a compromise cut at 0.5 (used throughout the main body of this paper) which we argue does a good job of removing many, but not all, false disks while preserving the vast majority of real disks; and finally, we investigate the use of not applying any cut at all, i.e. we allow all galaxies to enter our sample as measured by the GIM2D fit, which will provide a complete, but not pure sample.

In Fig. C1 we show the effects of these three ${\rm Pr_{FD}}$ cuts on a selection of the primary science plots which are affected by B/T structure, first shown in Figs. 10 - 12. The left column is for no correction, the middle column is for the compromise correction at 0.5, and the right column shows the verbose cut at 0.2. Although the exact depiction of the data for each cut varies, the global trends, and importantly their relative structure remains invariant. For example, for all ${\rm Pr_{FD}}$ cuts we find a positive correlation of the passive fraction with total stellar mass, bulge mass, and B/T morphology. We also find very similar differential effects, whereby, e.g., galaxies with higher B/T morphologies at a fixed stellar mass have higher passive fractions, and galaxies with higher stellar masses at a fixed B/T morphology also have higher passive fractions. This can be seen clearly by looking at the top row and the third row of Fig. C1.

One of the most important results to come from our analyses in this paper is the observation that the passive fraction varies less with B/T morphology at a fixed bulge mass than at a fixed total stellar mass. This fact is preserved in all renderings of our ${\rm Pr_{FD}}$ cuts (see the top row of Fig. C1, and the one below it). Complementary to this, we find a huge diversity in the passive fraction across different total stellar masses at a fixed B/T morphology, but relatively very little diversity in the passive fraction across different stellar masses at a fixed bulge mass. Again this is true for all ${\rm Pr_{FD}}$ cuts (see bottom two rows in Fig. C1). 
This indicates that our primary results are not a product of poor bulge + disk decompositions in an isolated sub-sample of galaxies, but instead are generally present whether we include no false disks (but remove many real ones) or we remove no real disks (but include several false ones), or even if we adopt a compromise cut, as throughout the main body of this paper. 

The decision to include AGN, with spectroscopically deduced SFRs from the D$_{n}$4000 - sSFR relation (see \S 2.1 and Brinchmann et al. 2004) could potentially introduce a systematic bias into our analysis. To test this, we show in Fig. C2 the passive fraction relation with bulge mass, total stellar mass, and halo mass, each subdivided into B/T structural range. These plots are first shown in Fig. 12, and represent the most constraining to the passive fraction. Here, however, we remove all AGN via the Stasinska et al. (2006) line cut on the BPT emission line diagram above a S/N threshold of 1. This ensures that there is negligible contribution to the emission lines from AGN and hence that the SFRs may be computed accurately. Although the specific details of the plots change slightly, the main features of these plots are unchanged, which implies that the inclusion of AGN has not biased our sample in such a way as to mimic the trends witnessed throughout this paper. 

Another area of possible systematic bias comes from the fibre correction implemented in transforming spectroscopic-derived SFRs into global (total) SFRs. This process is well tested and calibrated in Brinchmann et al. (2004), however, since our analysis finally yields that it is the bulge (and hence central region) of galaxies which is most important in determining the passive fraction of whole galaxies, it is natural to question whether a problem with the fibre correction could mimic this science result. We test this by constructing an analogy to the passive fraction presented in the main body of this paper, using only the photometry of the galaxies, which thus has no fibre correction implemented. Specifically, we look at the g-r colour (computed from global galaxy photometry) as a function of total stellar mass in Fig. C3. Here we divide galaxies into red and blue subsets by calculating whether they lie above or below (respectively) the green valley line (presented in Mendel et al. 2014). We define a 
galaxy to be red if its global g-r colour, (g-r)$_{\rm Galaxy}$:

\begin{equation}
(g-r)_{\rm Galaxy} > 0.06 \times {\rm log(M_{*}/M_{\odot})} - 0.01 \hspace{0.2cm} .
\end{equation}

\noindent These galaxies are shown in red in Fig. C3, and the rest are shown in blue. If we compare this plot to Fig. 6 (top-left panel), we can see there is a clear analogy in methodology. Although this approach throws out much of the information contained in the detailed spectra of these galaxies, it does contain an essential advantage for our current purposes: It allows us to look at the global colours of galaxies (and by inference their star forming properties) without cause for fibre corrections.

We define the red fraction in exact analogy to the passive fraction (in eq. 7). Explicitly, we calculate:

\begin{equation}
f_{\rm red} = \frac{{\sum\limits_{i=1}^{N_{\rm red}}} 1/V_{\rm max,i}({\rm red)}}{{\sum\limits_{j=1}^{N_{\rm all}}} 1/V_{\rm max,j}({\rm all})} 
\end{equation}

\noindent where $V_{max}$ is the spacial volume over which a galaxy of given ugriz magnitudes and colours would be detected in the SDSS, and thus make it into our analysis (see Mendel et al. 2014 for full details). This is used to weight the fractional contribution of each galaxy, effectively removing observer bias. We are now in a position to replicate our passive fraction analysis for the red fraction, avoiding the fibre correction but ignoring much of the valuable information contained within the spectra on the star formation rates of our galaxies. 

In Fig. C4 we present the red fraction dependence on bulge mass, total stellar mass, and halo mass subdivided by B/T morphology. This figure can be compared with the top row in Fig. 12 for the spectroscopic passive fraction. Comparing these figures, we see a close resemblance in form with the red fraction to the equivalent plots made with the passive fraction. In particular, we see a monotonic rise in the red fraction with total stellar mass, halo mass, and bulge mass, after some initial threshold at $\sim$ $10^{9}$ $M_{\odot}$ in bulge mass, which is not strongly morphology dependent, just as with the passive fraction. 
Furthermore, we witness a fanning out of the red fraction - stellar mass and halo mass sequences based on B/T structure, whereby more spheroid dominated systems are redder for their stellar and halo masses than more disk dominated systems. We also note that the variation in the red fraction with B/T morphology is less pronounced with bulge mass than with total stellar and halo mass. Thus, our primary conclusions are not affected by any potential bias or systematic effect introduced through the aperture correction of Brinchmann et al. (2004).

In summary, in this appendix we have investigated the impact of the various sample selections, conversions, and bulge + disk decomposition systematics on the stability of our main science results. We have found that for all corrections to the false disk problem, including a verbose cut leaving virtually no false disks in the sample, the primary science results remain unchanged. Further, we have demonstrated that potential biases from the fibre correction procedure or the inclusion of AGN in our parent sample do not lead to effects which could mimic our results. 
Moreover, we establish that for a photometric only approach to the passive fraction (the red fraction) we recover qualitatively very similar results to what we found with the spectroscopically derived passive fraction, with no fibre corrections. Taken together these checks on the data and scientific method used represent a crucial confidence test, which our primary result, the dominance of the bulge mass in determining the passive fraction, passes successfully.

\begin{figure*}
\rotatebox{270}{\includegraphics[height=0.33\textwidth]{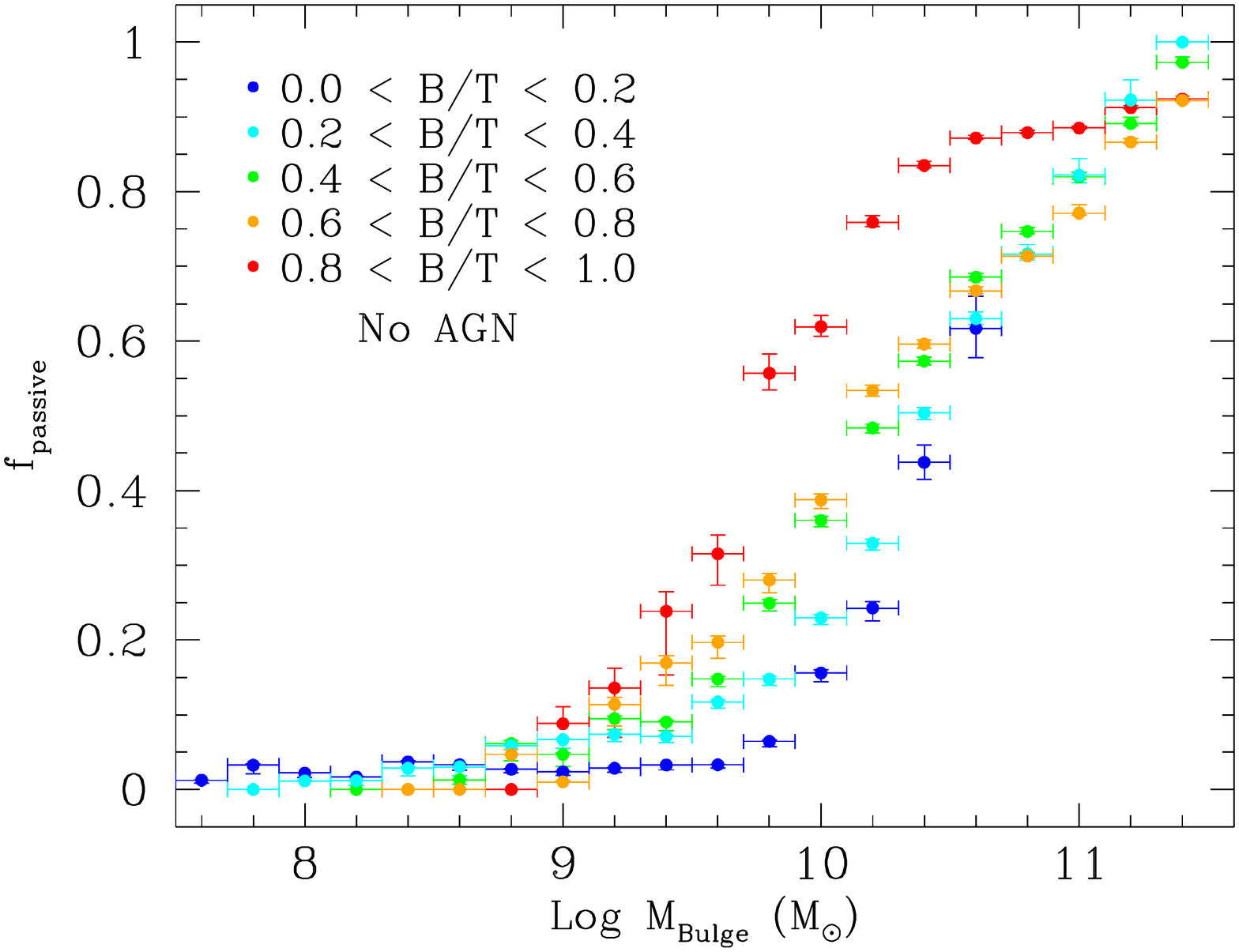}}
\rotatebox{270}{\includegraphics[height=0.33\textwidth]{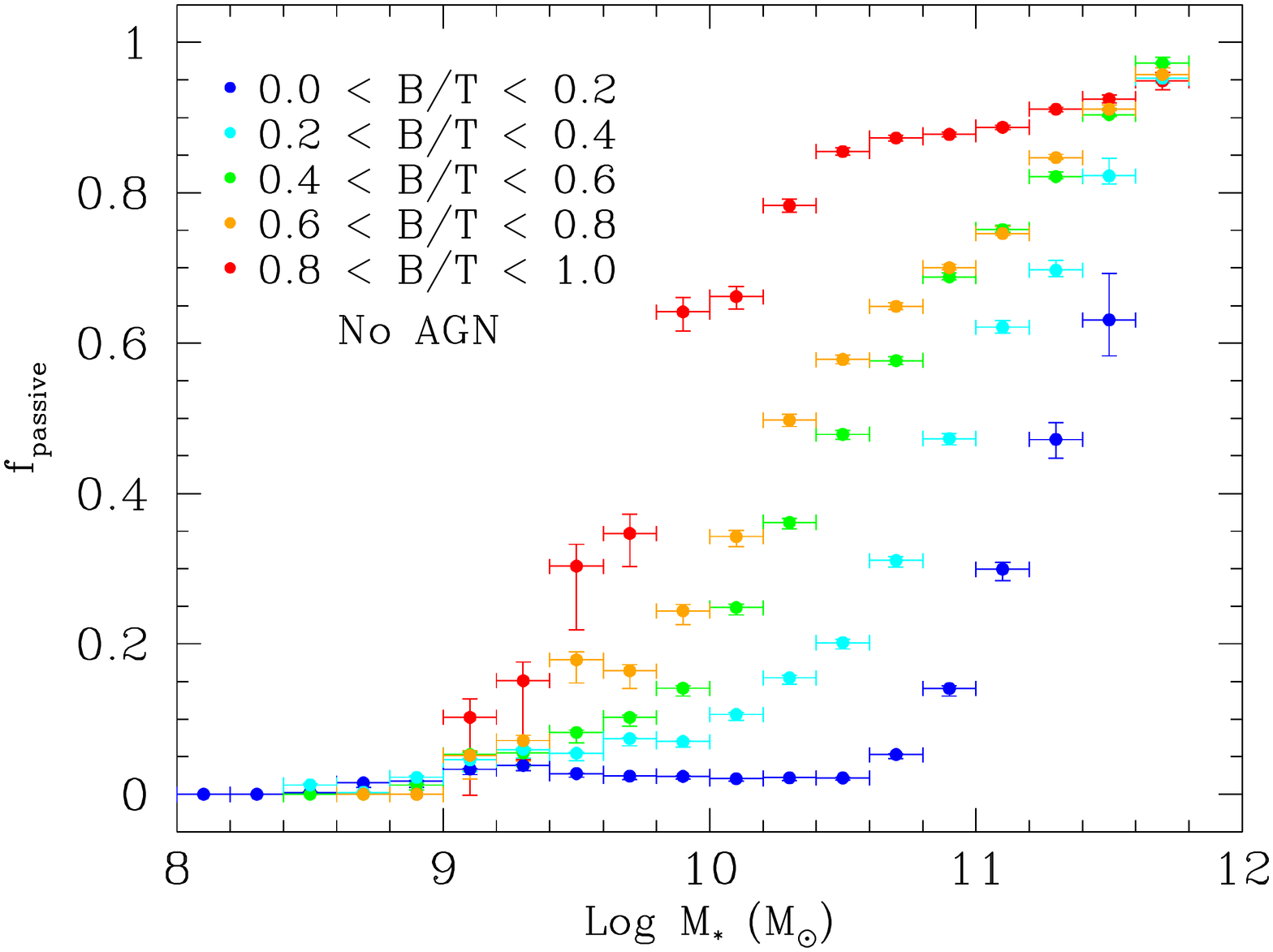}}
\rotatebox{270}{\includegraphics[height=0.33\textwidth]{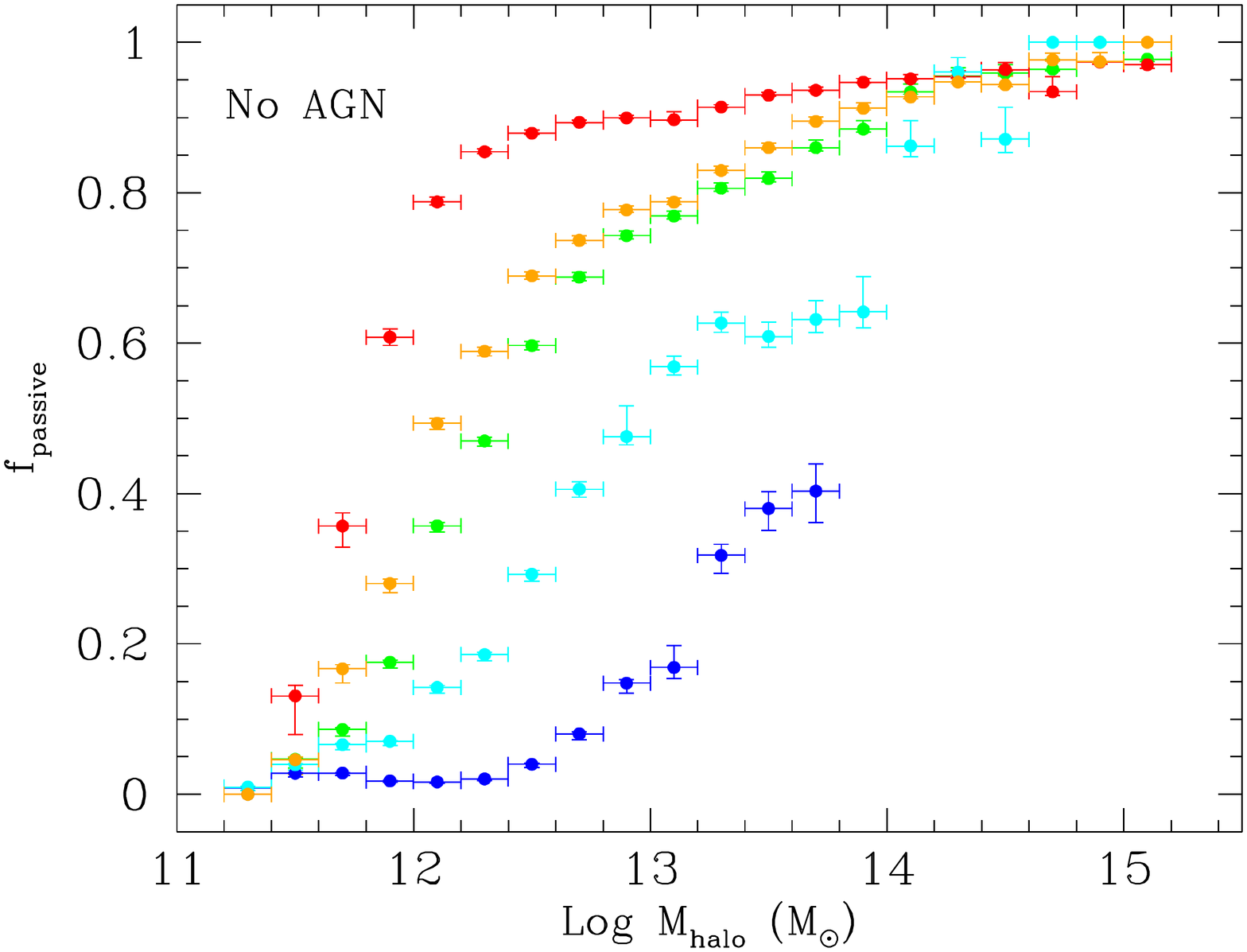}}
\caption{The passive fraction for galaxies without AGN vs. bulge mass ({\it left panel}); total stellar mass ({\it middle panel}); and halo mass ({\it right panel}). Each plot is split by B/T structural range as indicated on the left and middle plots. This figure may be compared to Fig. 12 ({\it top row}) which shows the same relationships with AGN included. Note that there is no significant difference in the main features of these plots, which implies that including AGN, and thus using indirect measures of SFR for these galaxies, does not adversely affect our conclusions. Specifically, note the significantly greater variation with B/T range at a fixed stellar and halo mass than at a fixed bulge mass.}
\end{figure*}

\begin{figure}
\rotatebox{270}{\includegraphics[height=0.49\textwidth]{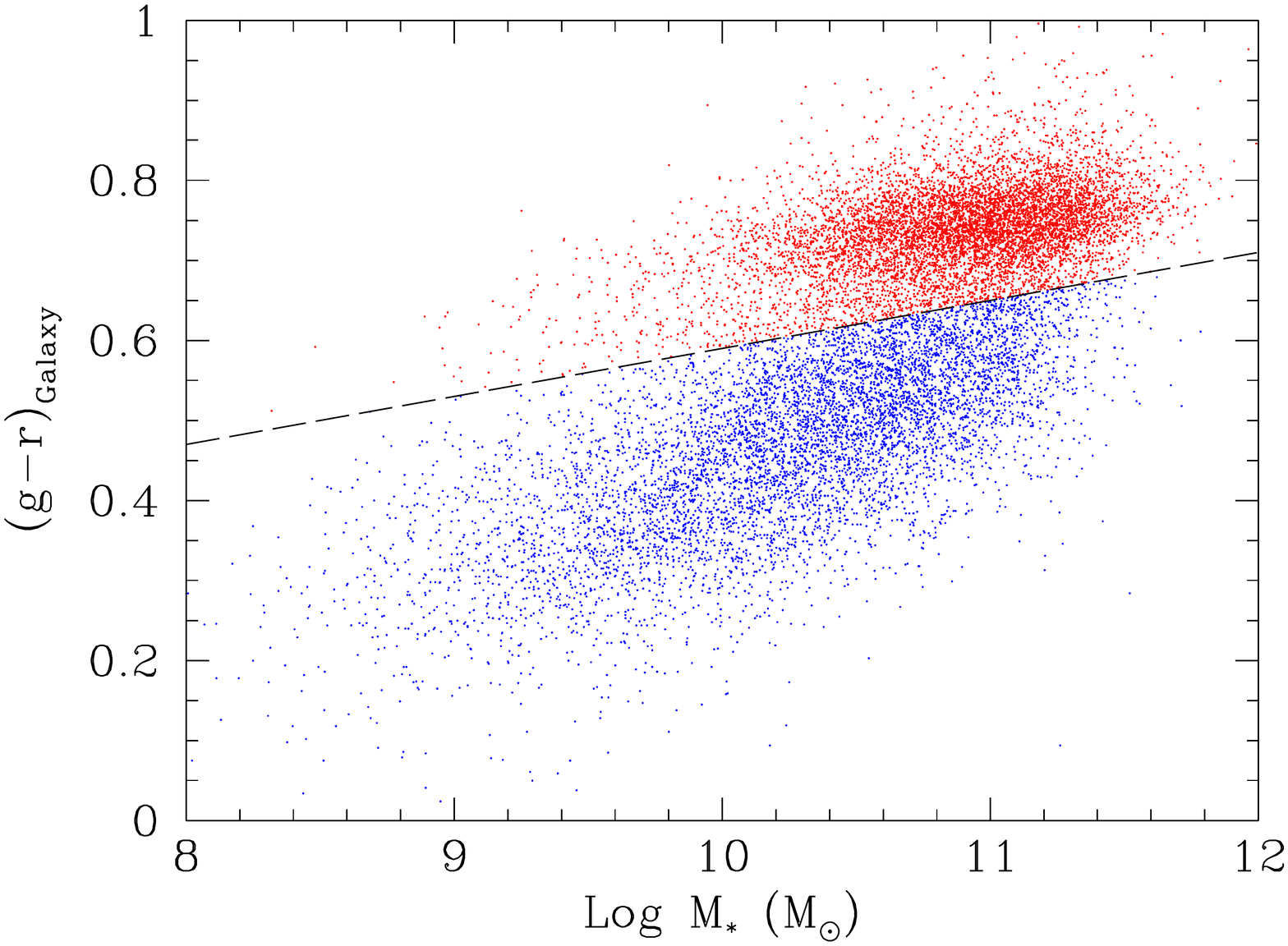}}
\caption{g-r colour for the whole galaxy vs. total stellar mass. The red sequence is divided from the blue cloud by a simple line cut defined by (g-r)$_{\rm Galaxy}$ = 0.06 $\times$ log($M_{*}/M_{\odot}$) - 0.01, as in Mendel et al. (2014). This provides an alternative way to define the passive fraction to the $\Delta$SFR method used throughout the bulk of this paper. Crucially, the red fraction derives no input from the spectra of the galaxies, so no aperture correction is needed.}
\end{figure}

\begin{figure*}
\rotatebox{270}{\includegraphics[height=0.33\textwidth]{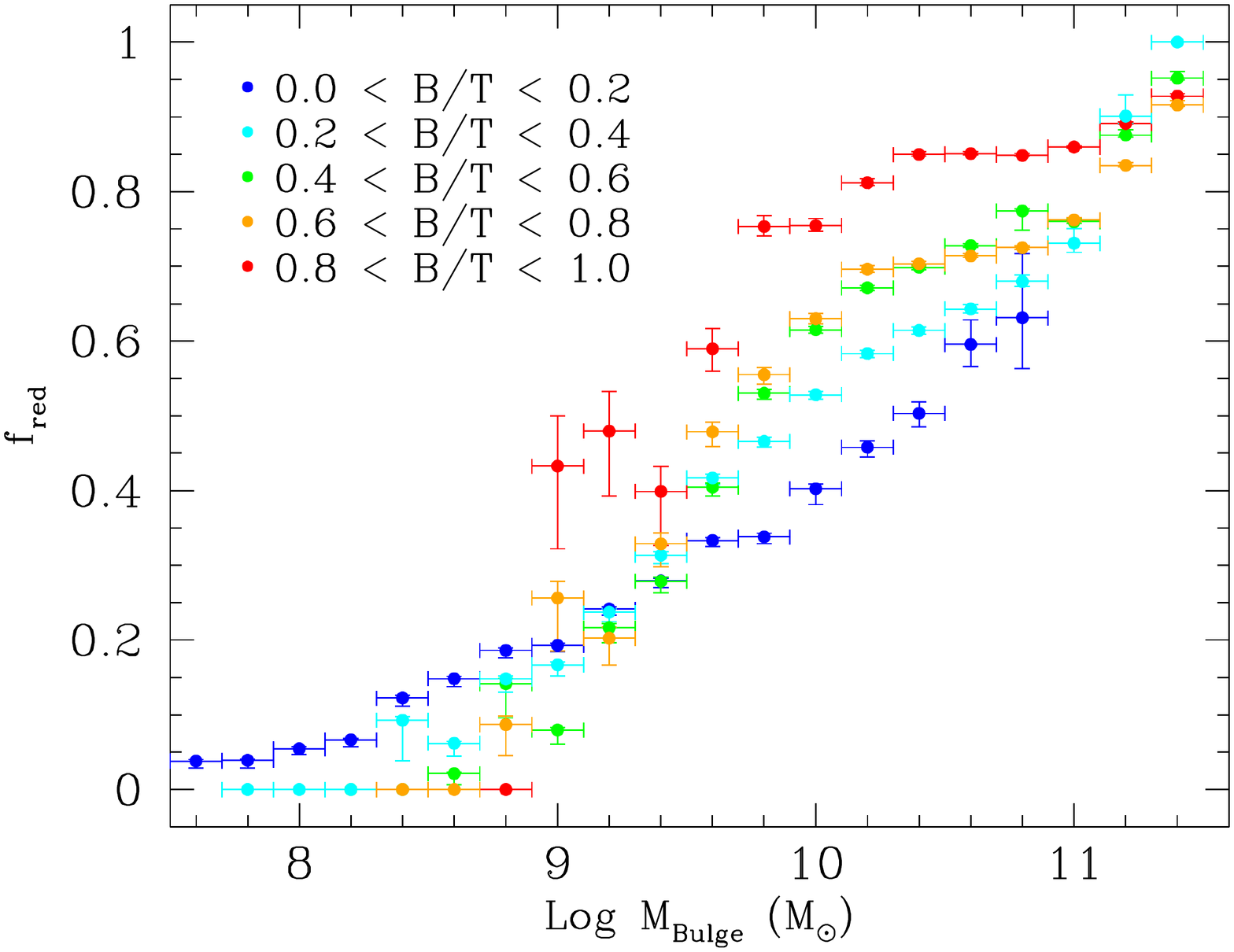}}
\rotatebox{270}{\includegraphics[height=0.33\textwidth]{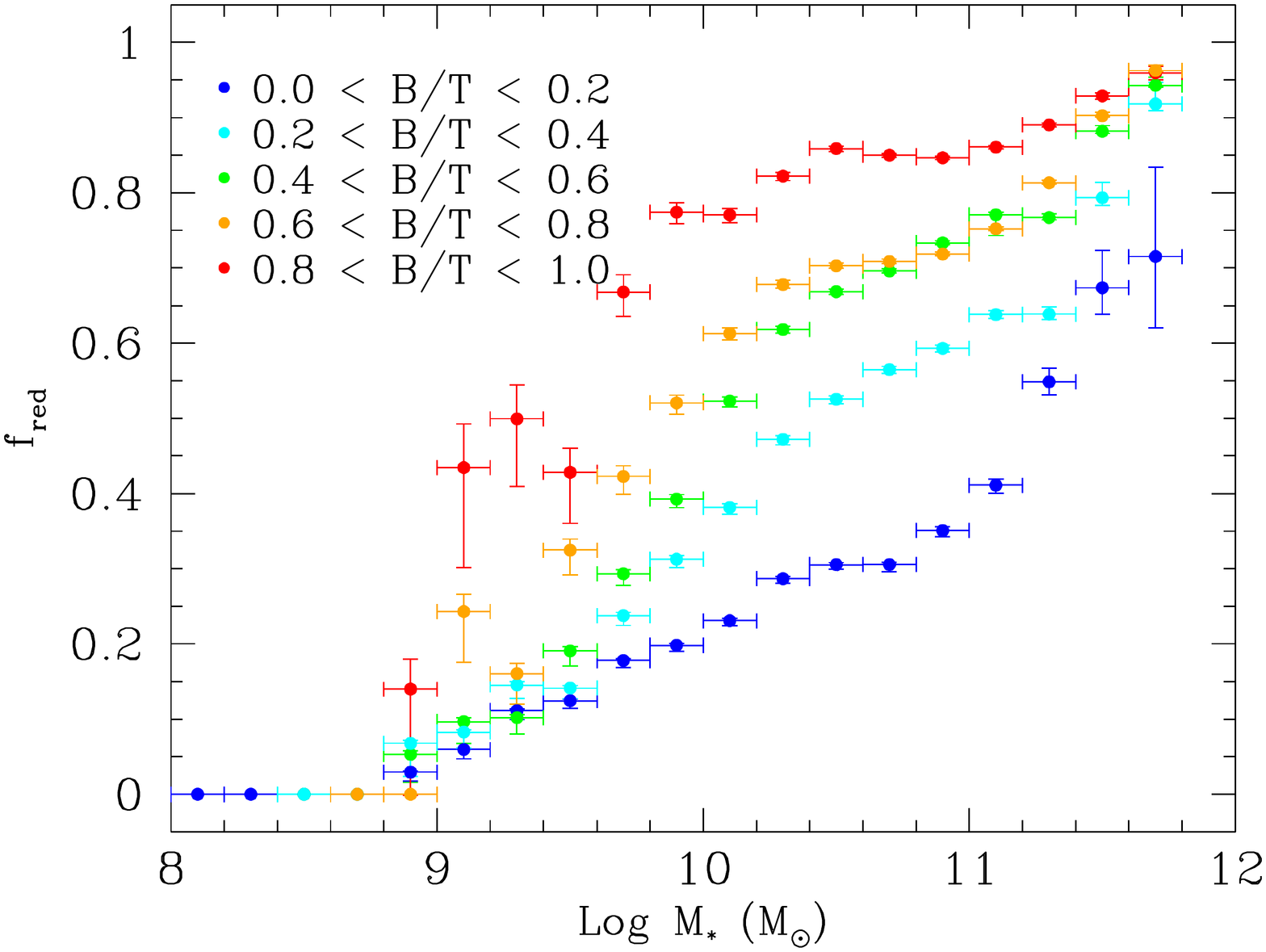}}
\rotatebox{270}{\includegraphics[height=0.33\textwidth]{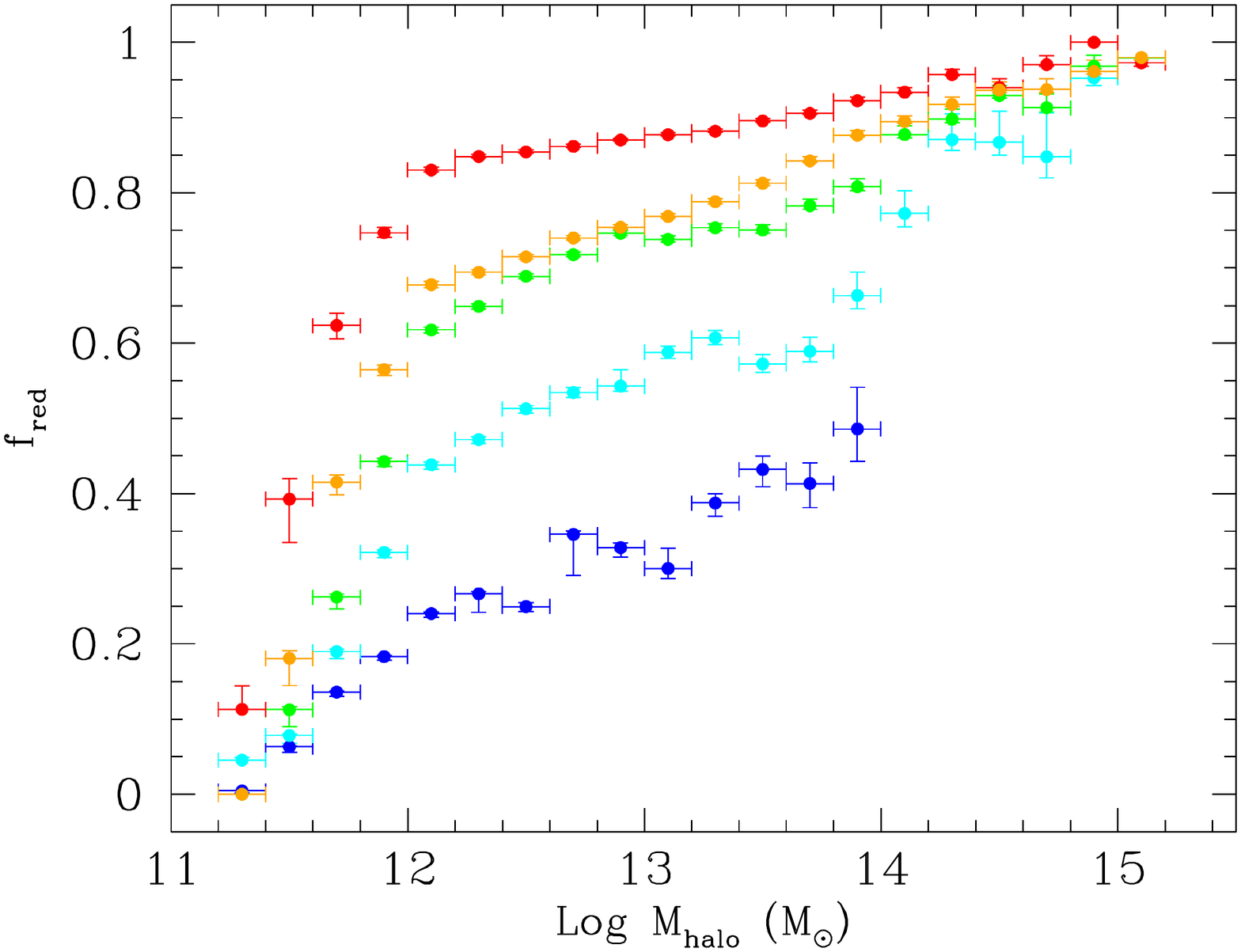}}
\caption{The red fraction vs. bulge mass ({\it left panel}); total stellar mass ({\it middle panel}); and halo mass ({\it right panel}). Each plot is split by B/T structural range as indicated on the left and middle plots. As with the passive fraction, we see least variation across B/T range with bulge mass, with significantly more variation at a fixed stellar and halo mass. This provides a check that the conversion from fibre to global SFRs used in the main body of this paper does not adversely affect our conclusions.}
\end{figure*}

\section{Classical vs. Free S\'{e}rsic Bulges}

Throughout the main body of this paper we utilise bulge + disk decompositions from $n_{s}$ = 4 bulge + $n_{s}$ = 1 disk profile GIM2D fitting of the ugriz SDSS data, with stellar masses of the components deduced via SED fitting of the light components separately (see \S 2.2, Simard et al. 2011, Mendel et al. 2013). In this final appendix we consider whether the choice of fixing the bulge component to a classical De Vaucouleurs (1948) profile may introduce any biases into our analyses. Simard et al. (2011) construct bulge + disk decompositions in the g- and r-bands for SDSS galaxies for free S\'{e}rsic index bulges as well as for the $n_{s}$ = 4 bulges already considered in this work. In this appendix we compare the B/T ratios from these two methods to ascertain whether any systematic differences may affect our conclusions. For the sake of brevity we restrict our investigation to the representative r-band case, which scales well with bulge-to-total stellar mass ratios (as outlined in \S App. A).

In Fig. D1 we plot the relationship between r-band bulge-to-total light ratio constructed from a classical ($n_{s}$ = 4) and a free S\'{e}rsic ($n_{s}$ = 0.5 - 8) bulge profile. In both cases the disk component is treated as a pure exponential (with $n_{s}$ = 1). On each of the panels the small data points are colour coded by the free S\'{e}rsic index value of the bulge, with redder colours indicating higher values of $n_{\rm s,free}$. On the left-hand panel of Fig. D1, where we include the full SDSS sample, we find that there is generally quite good agreement between the two profile cases except at low (B/T)$_{\rm r}$[$n_{s}$=4] values. The discrepancy at low B/T, however, is clearly due to very low Sersic index bulges (blue points). If we remove the cases with $n_{\rm s,free}$ $<$ 2 (which we may assume ought to have been fit as disks), we recover a much tighter dependence (see middle panel of Fig. D2). 
Here we note a slight tendency for the highest free S\'{e}rsic bulges (red points) to lie above the 1:1 line. This is expected since bulges with very steep S\'{e}rsic profiles fit as $n_{s}$ = 4 bulges tend to leave some light unaccounted for which can lead to erroneous `false' disks being fit (see \S App. A \& B for full details on this issue). However, if we correct for the `false' disks using our new methods (outlined in \S App. B) we find that the accord between free S\'{e}rsic and $n_{s}$ = 4 bulges is excellent (see right hand panel of Fig. D1). 

The average difference between (B/T)$_{\rm r}$[$n_{s}$=4] and (B/T)$_{\rm r}$[free $n_{s}$] is $\sim$ 0.02, which is significantly smaller than the bin size, and far smaller than any binnings we make in this paper. The average standard deviation around the mean is $\sim$ 0.1. We may take this as a reasonable estimate for the random error on any given galaxy's measured B/T ratio. In this work we are interested in the mean properties of large samples of galaxies, thus reducing the random error on our results further by $\sigma/\sqrt{N}$. Thus, we are left only with the {\it very small} systematic offset. This strongly implies that there would be no significant differences to any of our results based on the choice of $n_{s}$ = 4 instead of free S\'{e}rsic bulges, provided we make a correction for false disks and require that all bulges have $n_{s}$ $>$ 2. 
Even without these reasonable steps we would still recover a good correlation between the methods and acceptable errors, with the only appreciable offsets found at very low B/T values.

Since the S\'{e}rsic index of bulges is found to systematically increase with stellar mass (e.g. Graham 2001, Graham \& Worley 2008) it is important to check that there are no systematic differences in our measured B/T values across stellar mass range. Further, since the resolution of similar mass galaxies declines with redshift it is also desirable to check for any systematic deviations between the free S\'{e}rsic and $n_{s}$ = 4 models across redshift range. We plot the difference between $n_{s}$ = 4 and free S\'{e}rsic bulge-to-total light ratios (${\rm \Delta (B/T)_{r} [n_{\rm 4} - n_{\rm free}]}$) explicitly as a function of stellar mass and redshift in Fig. D2. We find that these distributions are very flat and extremely close to zero in value, indicating that there are no differential systematic effects engendered in our bulge + disk decompositions across redshift or stellar mass ranges. 
We also find that restricting the redshift range of our galaxy sample to z$_{\rm spec}$ $<$ 0.1 (instead of 0.2 throughout the main body of this work) does not adversely affect any of our conclusions, and leaves all of the science plots in this paper virtually identical. Therefore, we conclude that our bulge + disk decompositions are reliable at the redshift and stellar mass ranges studied in this paper, and moreover that no significant biases or systematics are introduced through the use of $n_{s}$ = 4 bulges.

\begin{figure*}
\rotatebox{270}{\includegraphics[height=0.33\textwidth]{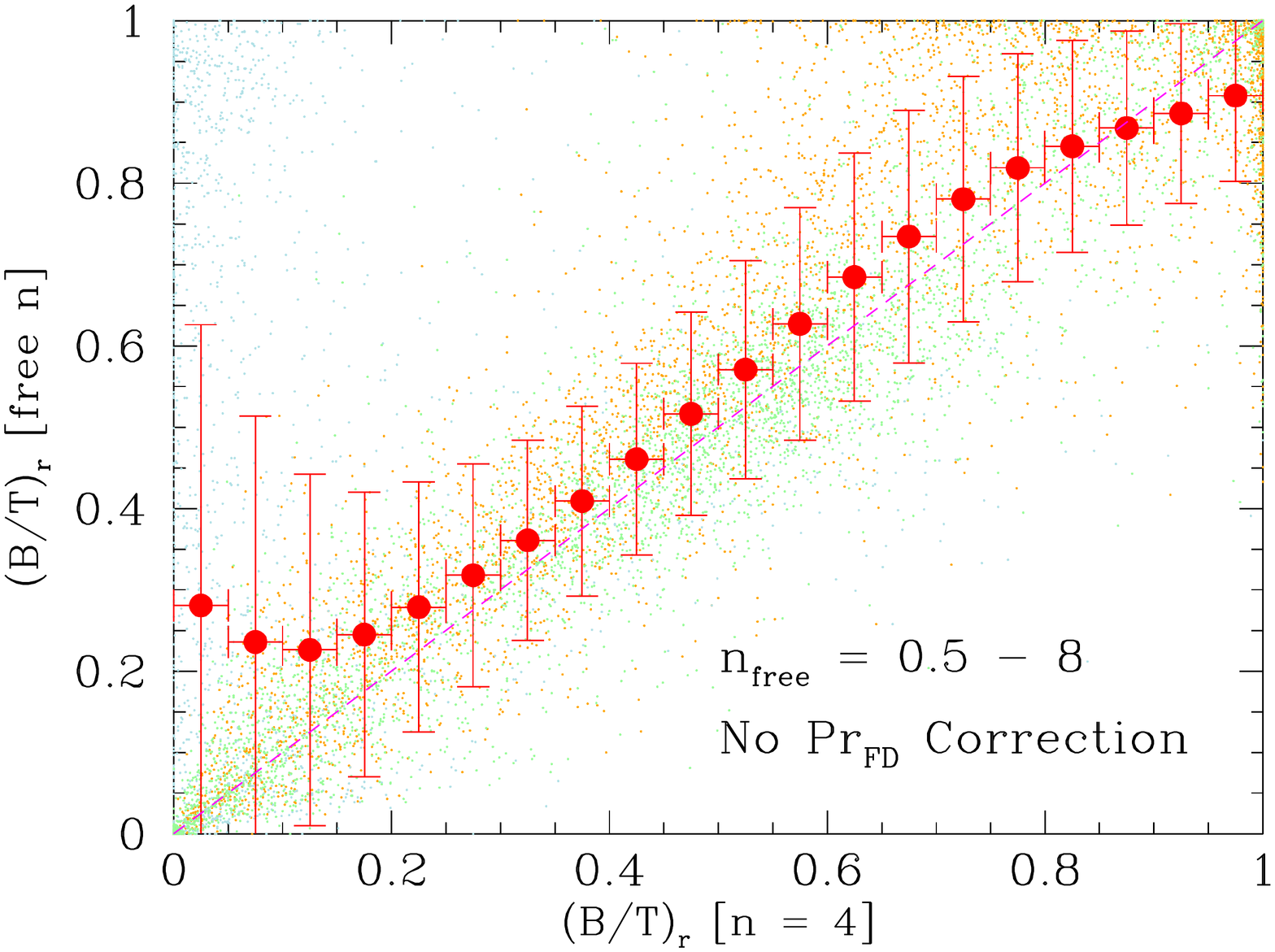}}
\rotatebox{270}{\includegraphics[height=0.33\textwidth]{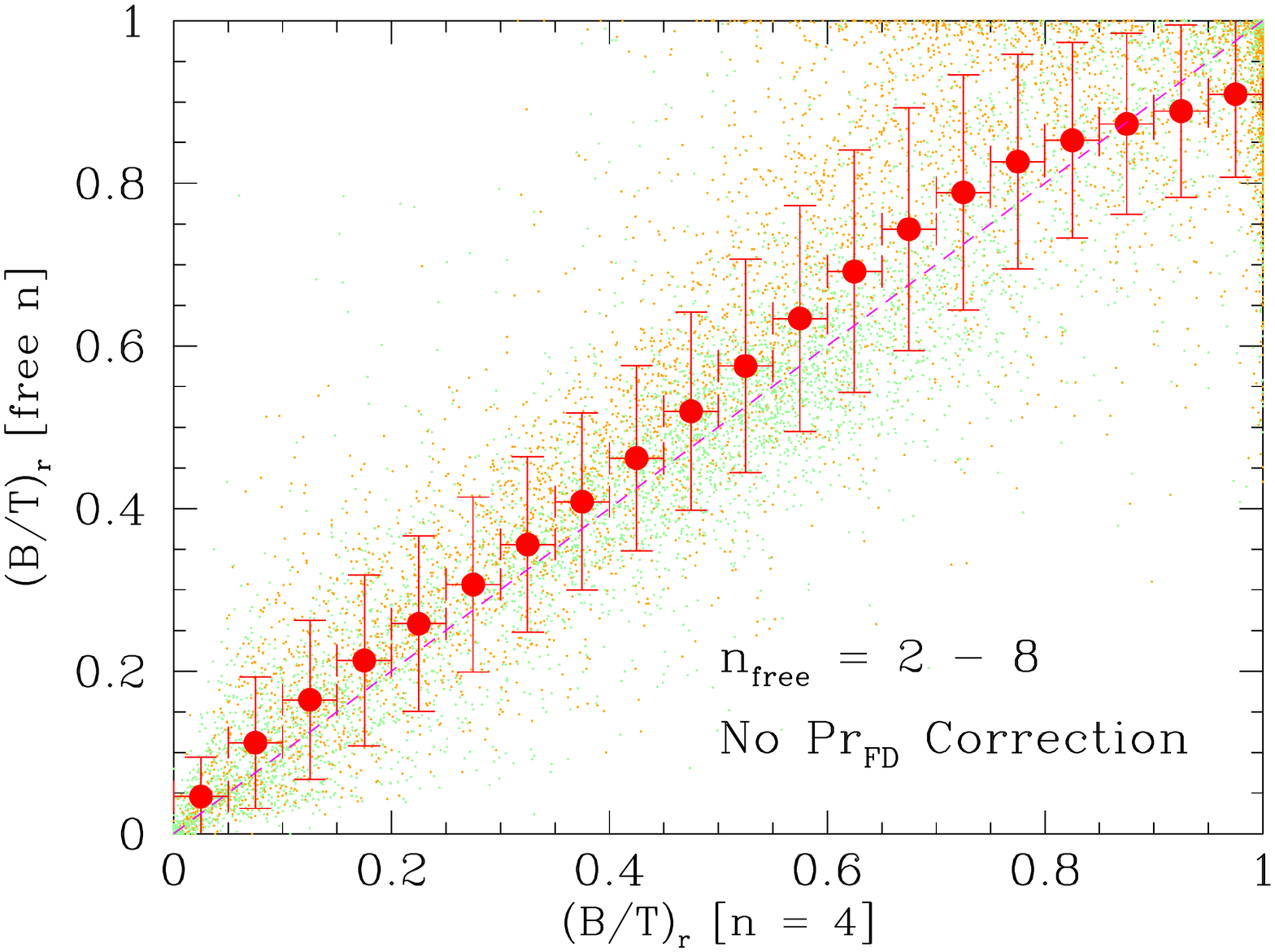}}
\rotatebox{270}{\includegraphics[height=0.33\textwidth]{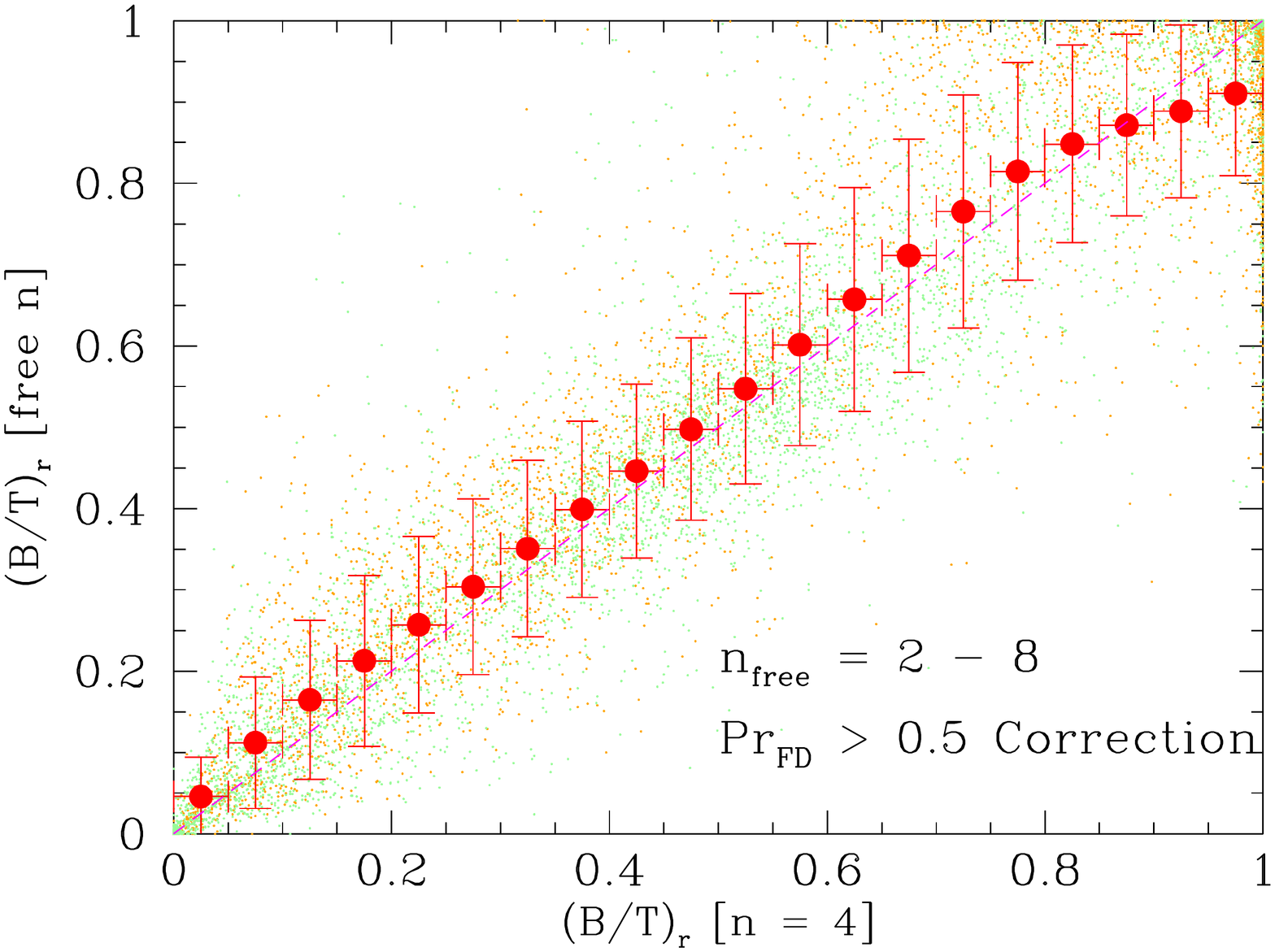}}
\caption{Bulge-to-total light ratio comparison between $n_{s}$ = 4 classical bulge and free S\'{e}rsic bulge GIM2D fitting of our SDSS galaxy sample, both computed in the r-band. For both methods disks are fitted with exponentially declining ($n_{s}$=1) S\'{e}rsic profiles. {\it Left panel}: shows the relationship for the full free S\'{e}rsic range from $n_{s}$ = 0.5 - 8, and with no correction made for `false' disks (see \S App. B). {\it Middle panel}: Shows the relationship with very low S\'{e}rsic index bulges ($n_{s}$ $<$ 2) removed, i.e. those cases which are better fit as a disk; also with no correction made for `false' disks. {\it Right panel}: shows the same as the middle panel but with `false' disks corrected (as in \S App. B). 
Note that the agreement between (B/T)$_{\rm r}$[free n] and (B/T)$_{\rm r}$[n=4] is very good when we require that bulges do not have disk-like profiles, and becomes excellent if we additionally correct for `false' disks. The filled red circles are mean values, with their error bars given as the full 1$\sigma$ range. 
The 1:1 line is shown as a dashed magenta line on each panel for comparison. The small dots are individual data points coloured by the free S\'{e}rsic index of the bulge, with redder colours indicating higher S\'{e}rsic indices. Note that the discrepancy between the two methods at low (B/T)$_{\rm r}$[n=4] (in the left panel) is driven entirely by disks being fitted as bulges (shown in blue). High S\'{e}rsic index bulges (shown in red) lie preferentially above the 1:1 line (in the middle panel), but these minor outliers are almost entirely corrected for by the `false' disk removal (see right panel).}
\end{figure*}

\begin{figure*}
\rotatebox{270}{\includegraphics[height=0.49\textwidth]{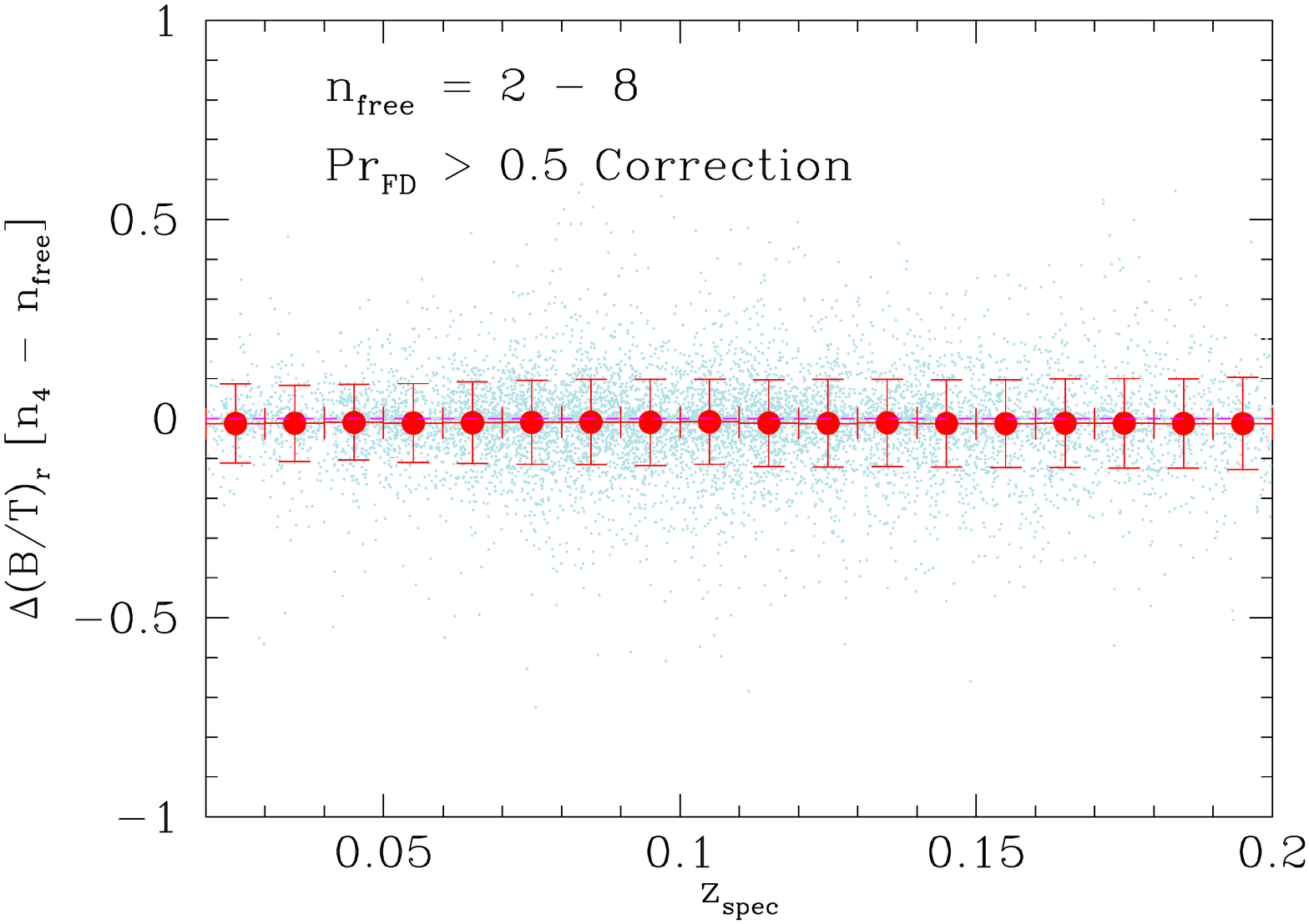}}
\rotatebox{270}{\includegraphics[height=0.49\textwidth]{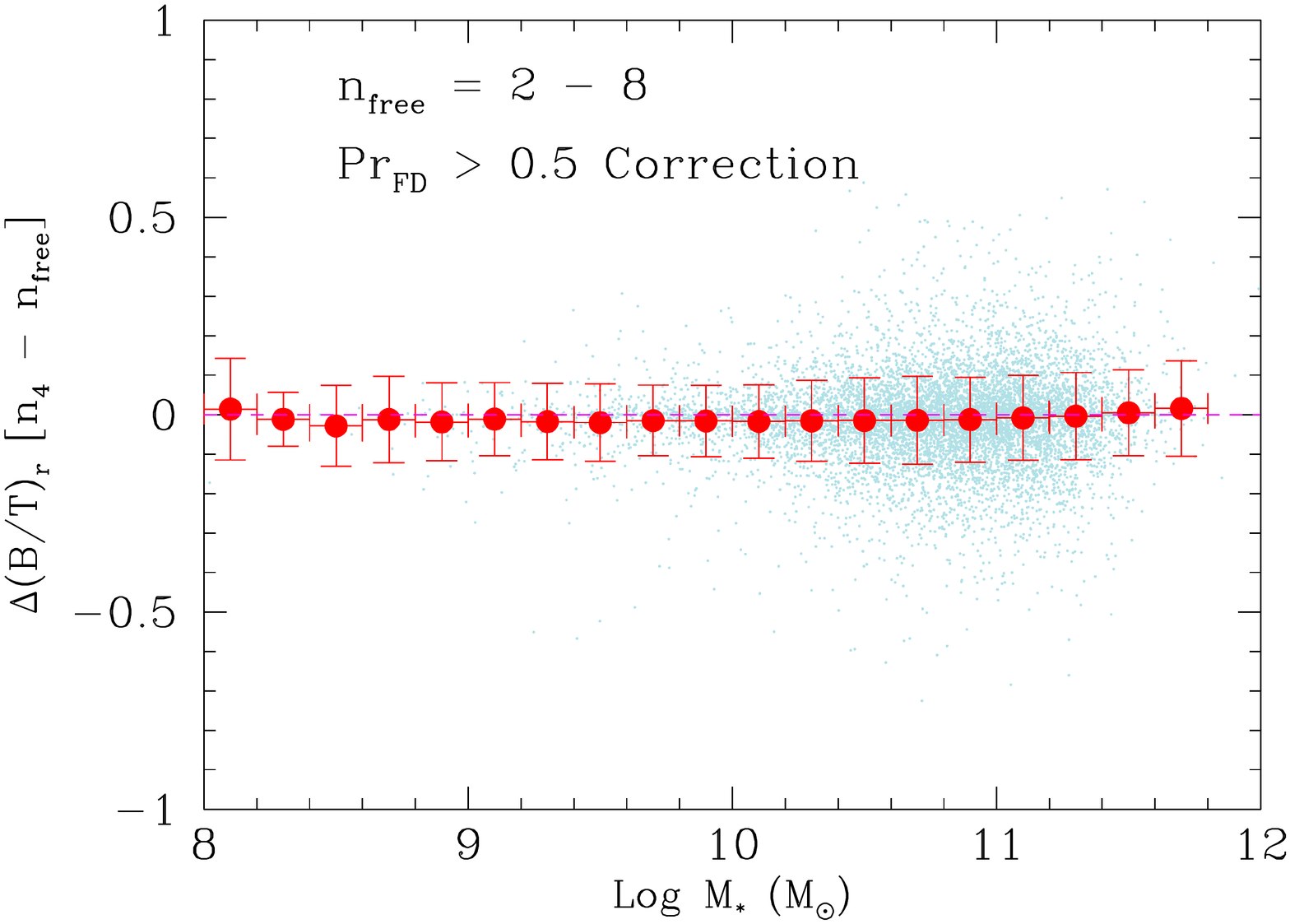}}
\caption{The difference between bulge-to-total r-band light ratio computed from $n_{s}$=4 classical bulge and free S\'{e}rsic bulge GIM2D fitting (${\rm \Delta (B/T)_{r} [n_{4} - n_{free}]}$) vs. redshift ({\it left panel}) and total stellar mass ({\it right panel}). The filled red circles are mean values, with their error bars given as the full 1 $\sigma$ range. The small blue dots indicate the position of the actual data, with only one in every twenty points shown for clarity. The ${\rm \Delta (B/T)_{r} [n_{4} - n_{free}]}$=0 line is shown as a dashed magenta line for comparison. In both plots the free S\'{e}rsic bulge is allowed to vary from $n_{s}$ = 2 - 8, preventing disks being fit as bulges. A correction for `false' disks is made exactly as described in detail in \S App. B. 
The agreement between the two methods is excellent, and there is clearly no differential redshift or stellar mass effects (i.e. the distributions are both flat and very close to zero). Thus, we conclude that free  S\'{e}rsic bulge fitting would not change any of the results in this paper in any significant way.}
\end{figure*}

\end{document}